\pdfoutput=1

\documentclass[a4paper,11pt]{article} 

\usepackage{atlasphysics}
\usepackage[displaymath]{lineno}
\usepackage{jheppub} 

\usepackage{graphicx} 
\usepackage{epstopdf}
\usepackage{subfigure}
\usepackage{amsmath}
\usepackage{amssymb}
\usepackage{multirow}
\usepackage{lscape}
\usepackage{lineno}
\usepackage{rotating}
\usepackage{afterpage}
\usepackage{slashed}
\usepackage{bm}
\usepackage{url}
\usepackage{xspace}
\usepackage{booktabs}
\usepackage[section]{placeins}
\usepackage{flafter}

\usepackage{hyperref}
\hypersetup{
  colorlinks=true,
  linkcolor=red,
  citecolor=red,
  urlcolor=blue
}

\newcommand{\seventev}    {\ensuremath{7~\mathrm{TeV}}\xspace}

\newcommand{\sqsseven}    {\ensuremath{\sqrt{s} = \seventev}\xspace}

\newcommand{\Herwig}    {\textsc{HERWIG}\xspace}
\newcommand{\herwigpp}  {\textsc{HERWIG}++\xspace}
\newcommand{\Herwigpp}  {\herwigpp}
\newcommand{\geant}     {\textsc{Geant4}\xspace}
\newcommand{\jimmy}     {\textsc{JIMMY}\xspace}
\newcommand{\Alpgen}    {\textsc{ALPGEN}\xspace}
\newcommand{\Mcatnlo}   {\textsc{MC@NLO}\xspace}
\newcommand{\Acermc}    {\textsc{AcerMC}\xspace}

\newcommand{\Pythia}    {\textsc{PYTHIA}\xspace}
\newcommand{\Powheg}    {\textsc{POWHEG}\xspace}
\newcommand{\PowhegBox} {\Powheg-BOX\xspace}
\newcommand{\PowPythia} {\textsc{POWHEG}+\Pythia}
\newcommand{\HJimmy}    {\textsc{HERWIG}+\jimmy}

\newcommand{\LO}        {\ensuremath{\mathrm{LO}}\xspace}
\newcommand{\NLO}       {\ensuremath{\mathrm{NLO}}\xspace}

\newcommand{\W}         {\Wboson\xspace}
\newcommand{\Z}         {\Zboson\xspace}

\newcommand{\munu}      {\ensuremath{\mu\nu}\xspace}
\newcommand{\Wmunu}     {\ensuremath{W\ra\munu}\xspace}
\newcommand{\Wjets}     {\Wboson+jets\xspace}
\newcommand{\Zjets}     {\Zboson+jets\xspace}

\newcommand{\antikt}{anti-$k_{t}$}
\newcommand{\Antikt}{Anti-$k_{t}$}

\newcommand{\Fastjet} {{\sc FastJet}\xspace}

\newcommand{\invpb}     {~\ensuremath{{\rm pb}^{-1}}\xspace}

\newcommand{\invfb}     {~\ensuremath{{\rm fb}^{-1}}\xspace}

\newcommand{\pileup}    {pile-up\xspace}
\newcommand{\Pileup}    {Pile-up\xspace}

\newcommand{\JES}   {{\rm JES}}
\newcommand{\JMS}   {{\rm JMS}}
\newcommand{\EMJES} {{\rm EM+JES}}

\newcommand{\LCWJES}{{\rm LCW+JES}}
\newcommand{\EM}    {{\rm EM}}

\newcommand{\LCW}   {{\rm LCW}}

\newcommand{\Etmiss}   {\ensuremath{{E}_{\mathrm{T}}^{\mathrm{miss}}}}

\newcommand{\rtrksubjet}       {\ensuremath{r^{\rm subjet}_{\rm trk}}\xspace}
\newcommand{\Rtrk}             {\ensuremath{R_{\rm trk}}\xspace}
\newcommand{\Rtrksubjet}       {\ensuremath{\Rtrk^{\rm subjet}}\xspace}

\newcommand{\DeltaR}     {\ensuremath{\Delta R}}
\newcommand{\Dy}         {\ensuremath{\Delta y}\xspace}

\newcommand{\Dphi}       {\ensuremath{\Delta\phi}\xspace}

\newcommand{\DyDphi}     {\ensuremath{\sqrt{(\Dy)^{2} + (\Dphi)^{2}}}\xspace}

\newcommand{\DeltaRydef} {\ensuremath{\DeltaR = \DyDphi}\xspace}

\newcommand{\JVF}       {\ensuremath{JVF}\xspace}

\newcommand{\tWb}     {\ensuremath{t \ra Wb}\xspace}

\newcommand{\insitu}   {{\textit{in situ}}\xspace}
\newcommand{\Insitu}   {{\textit{In situ}}\xspace}
\newcommand{\pp}       {\ensuremath{pp}\xspace}

\newcommand{\ptjet}    {\ensuremath{\pt^\mathrm{jet}}\xspace}

\newcommand{\Wpt}      {\ensuremath{\pt^{\Wboson}}\xspace}
\newcommand{\toppt}    {\ensuremath{\pt^{\rm t}}\xspace}

\newcommand{\pttrk}   {\ensuremath{\pt^{\mathrm{track}}}} 
 
\newcommand{\pTtrkjet}{\ensuremath{p^{\mathrm{track \; jet}}_{\rm T}}}

\newcommand{\sumPtTrkSq}  {\ensuremath{\sum(\pt^{\rm \ track})^{2}}\xspace}

\newcommand{\etajet}{\ensuremath{\eta}}
\newcommand{\phijet}{\ensuremath{\phi}}

\newcommand{\topos}{topo-clust\-ers}

\newcommand{\Npv}     {\ensuremath{N_{\rm PV}}\xspace}

\newcommand{\avgmu}   {\ensuremath{\langle \mu \rangle}\xspace}

\newcommand{\ATLAS}   {ATLAS}

\newcommand{\LHC}    {LHC}

\newcommand{\parentjet }{{\rm parent\, jet}\xspace}

\input{boostedsymbols.tex}

\newcommand{\comment}[1]{}

\newcommand{\T}               {\rule{0pt}{2.6ex}}
\newcommand{\B}               {\rule[-1.2ex]{0pt}{0pt}}

\newcommand{\figref}[1]       {figure~\ref{fig:#1}}
\newcommand{\Figref}[1]       {Figure~\ref{fig:#1}}
\newcommand{\figsref}[2]      {figures~\ref{fig:#1} and \ref{fig:#2}}
\newcommand{\Figsref}[2]      {Figures~\ref{fig:#1} and \ref{fig:#2}}
\newcommand{\figrange}[2]     {figures~\ref{fig:#1}--\ref{fig:#2}}

\newcommand{\equref}[1]       {eq.~(\ref{eq:#1})}

\newcommand{\tabref}[1]       {table~\ref{tab:#1}}

\newcommand{\secref}[1]       {section~\ref{sec:#1}}
\newcommand{\Secref}[1]       {Section~\ref{sec:#1}}

\newcommand{\intlumi}{$4.7 \pm 0.1$\invfb}

\usepackage{preprintcover}  
\PreprintCoverPaperTitle{Performance of jet substructure techniques for large-$\boldsymbol{R}$ jets in proton--proton collisions at $\boldsymbol{\sqrt{s}=7}$~TeV using the ATLAS detector}
\PreprintIdNumber{CERN-PH-EP-2013-069}
\PreprintCoverAbstract{This paper presents the application of a variety of techniques to study
  jet substructure. The performance of various modified jet algorithms, or jet 
  \emph{grooming} techniques, for several jet types and event topologies is investigated
  for jets with transverse momentum larger than 300~GeV. Properties of jets 
  subjected to the mass-drop filtering, trimming, and pruning algorithms are 
  found to have a reduced sensitivity to multiple proton--proton interactions, are
  more stable at high luminosity and improve the physics potential of searches for 
  heavy boosted objects. Studies of the expected discrimination power of jet mass 
  and jet substructure observables in searches for new physics are also presented. 
  Event samples enriched in boosted \W and \Z bosons and top-quark pairs are used to 
  study both the individual jet invariant mass scales and the efficacy of algorithms 
  to tag boosted hadronic objects. The analyses presented use the full 2011 ATLAS 
  dataset, corresponding to an integrated luminosity of \intlumi\ from proton--proton 
  collisions produced by the Large Hadron Collider at a centre-of-mass energy of 
  \sqsseven.
}
\PreprintJournalName{JHEP}

\title{Performance of jet substructure techniques for large-$\boldsymbol{R}$ jets in proton--proton collisions at $\boldsymbol{\sqrt{s}=7}$~TeV using the ATLAS detector}

\author{The ATLAS Collaboration}
                             
\abstract{This paper presents the application of a variety of techniques to study
  jet substructure. The performance of various modified jet algorithms, or jet 
  \emph{grooming} techniques, for several jet types and event topologies is investigated
  for jets with transverse momentum larger than 300~GeV. Properties of jets 
  subjected to the mass-drop filtering, trimming, and pruning algorithms are 
  found to have a reduced sensitivity to multiple proton--proton interactions, are
  more stable at high luminosity and improve the physics potential of searches for 
  heavy boosted objects. Studies of the expected discrimination power of jet mass 
  and jet substructure observables in searches for new physics are also presented. 
  Event samples enriched in boosted \W and \Z bosons and top-quark pairs are used to 
  study both the individual jet invariant mass scales and the efficacy of algorithms 
  to tag boosted hadronic objects. The analyses presented use the full 2011 ATLAS 
  dataset, corresponding to an integrated luminosity of \intlumi\ from proton--proton 
  collisions produced by the Large Hadron Collider at a centre-of-mass energy of 
  \sqsseven.
}

\begin{document}

\maketitle
\flushbottom

\clearpage

\section{Introduction}
\label{sec:introduction}

The dominant feature of high-energy proton--proton (\pp) collisions at the Large Hadron Collider (LHC) is the production of highly collimated sprays of energetic hadrons, called jets, that originate from the quarks and gluons in the primary collisions. The large centre-of-mass energy at the LHC enables the production of Lorentz-boosted heavy particles, whose decay products can be reconstructed as one large-area jet. The study of the internal structure of jets goes beyond the four-momentum description of a single parton and yields new approaches for testing Quantum Chromodynamics (QCD) and for searching for new physics in hadronic final states. However, many of the new tools developed for the study of jet substructure at the LHC have only recently been validated with data in a hadron--hadron collider environment. For example, the effect of multiple \pp\ interactions on large-area jet measurements has not been extensively studied experimentally.

This paper presents a comprehensive set of studies designed to establish the efficacy, accuracy, and precision of several of the tools available for determining and analysing the internal structure of jets at the LHC. New jet algorithms and strategies, referred to as jet grooming, that refine the definition of a jet in a high-luminosity environment, are studied using data taken at a centre-of-mass energy of \sqsseven during 2011. A variety of techniques and tagging algorithms intended to improve the mass resolution in the reconstruction of boosted objects that decay hadronically are studied in the data both in inclusive jet samples and in samples enriched in events containing boosted \W/\Z bosons and top quarks. Evaluations of the systematic uncertainties for jet mass measurements are presented for a variety of jet algorithms. Comparisons of the discrimination between signal and background provided by various observables are also evaluated for a selection of models of new physics containing boosted hadronic particle decays.

The organization of the paper is as follows. In this section, motivation for the use of new jet reconstruction techniques for Lorentz-boosted particles is given and the jet algorithms and jet substructure variables that are used in the analyses presented here are defined.  \Secref{Detector} provides descriptions of the ATLAS detector and the Monte Carlo simulations, and \secref{recocalib} defines the jet reconstruction and calibration procedures that are used throughout.  The latter section includes a discussion of the jet mass scale and the subjet energy scale, which are important ingredients in the jet grooming algorithms.  \Secref{pileup} describes studies of the effect of jet grooming on jet properties in the presence of \pileup, which represents a major experimental challenge at the present and future LHC machine.

Studies of the performance of the various jet algorithms are conducted with three classes of event samples in both data and Monte Carlo simulation in \secref{boostedwtop}:  inclusive jet events, which are dominated by light-quark or gluon jets whose properties are defined primarily by soft gluon emission; boosted hadronically decaying \W and \Z bosons, which form jets that are dominated by two high-\pt components; and top-quark decays, where the \W boson decays hadronically, which form jets that have three prominent components (due to the $b$-quark in addition to the \W). In \secref{boostedwtop:mc}, the effect on jet resolution of the various jet grooming algorithms is compared in simulated events separately for signal (\W, \Z, and top jets) and background from light-quark and gluon jets.  The discrimination between background and signal is then studied using a number of grooming configurations by comparing jet properties for the different types of events before and after grooming. This is followed in \secref{boostedwtop:inclusivedatamc} by a direct comparison of multiple Monte Carlo predictions and inclusive jet data.  Lastly, \secref{boostedwtop:ttbar} presents jet grooming studies on boosted top-quark events.  Finally, conclusions are drawn in \secref{Conclusions}.

\subsection{Motivation}
\label{sec:intro:motivation}

The centre-of-mass energy of the LHC has opened new kinematic regimes to experimental study. 
The new phase space available for the production of Standard Model (SM) particles with significant Lorentz boosts, or even new massive particles that decay to highly boosted SM particles, necessitates new techniques to conduct measurements in novel final states.
For example, when sufficiently boosted, the decay products of \W bosons~\cite{Cui2010}, top quarks~\cite{Thaler2008, TopJetsLHC}, and Higgs bosons~\cite{Butterworth:2008iy} can become collimated to the point that standard reconstruction techniques begin to fail.
When the separation of the quarks in these boosted topologies becomes smaller than the radius parameter of the jets, they often fail to be individually resolved by standard jet algorithms and configurations.
Moreover, the high-luminosity conditions at the LHC can further degrade even the most complex procedures for reconstructing decays of boosted hadronic objects. 
Multiple \pp interactions per bunch crossing (\pileup) produce soft particles unrelated to the hard scattering that can contaminate jets in the detector considerably more than at previous hadron--hadron colliders.
In events where boosted particle decays are fully contained within individual large-radius jets, a diminished mass resolution due to \pileup may dramatically weaken sensitivity to new physics processes.
It is crucial that the above issues be addressed together, as the efficacy of a given technique for boosted object reconstruction may depend critically on its vulnerability to experimental conditions.

\begin{figure}[t]
  \begin{center}
  \subfigure[\tWb]{
    \includegraphics[width=0.47\textwidth]{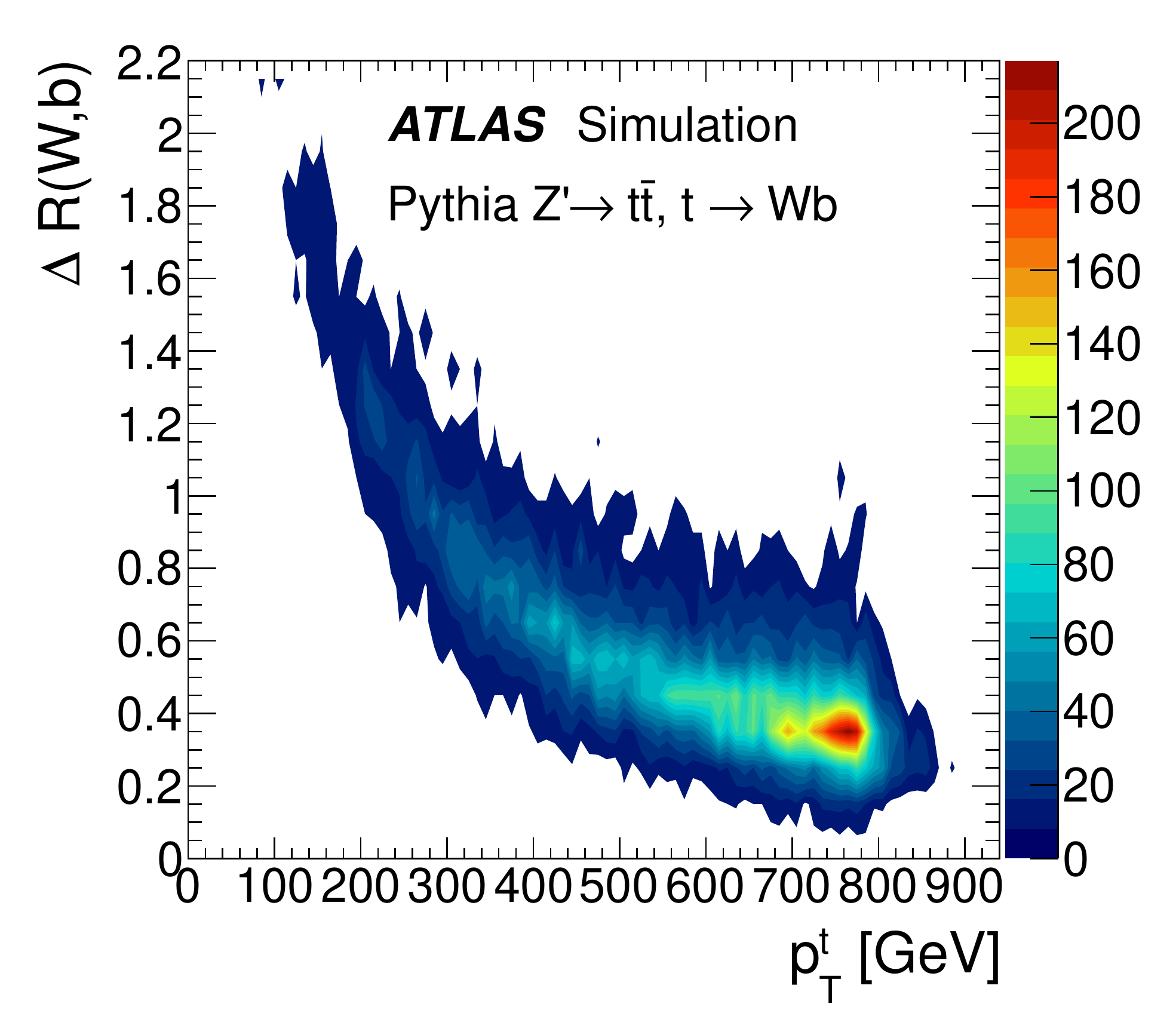}
    \label{fig:intro:TopptdrWb}
    }
  \subfigure[\Wqq]{
    \includegraphics[width=0.47\textwidth]{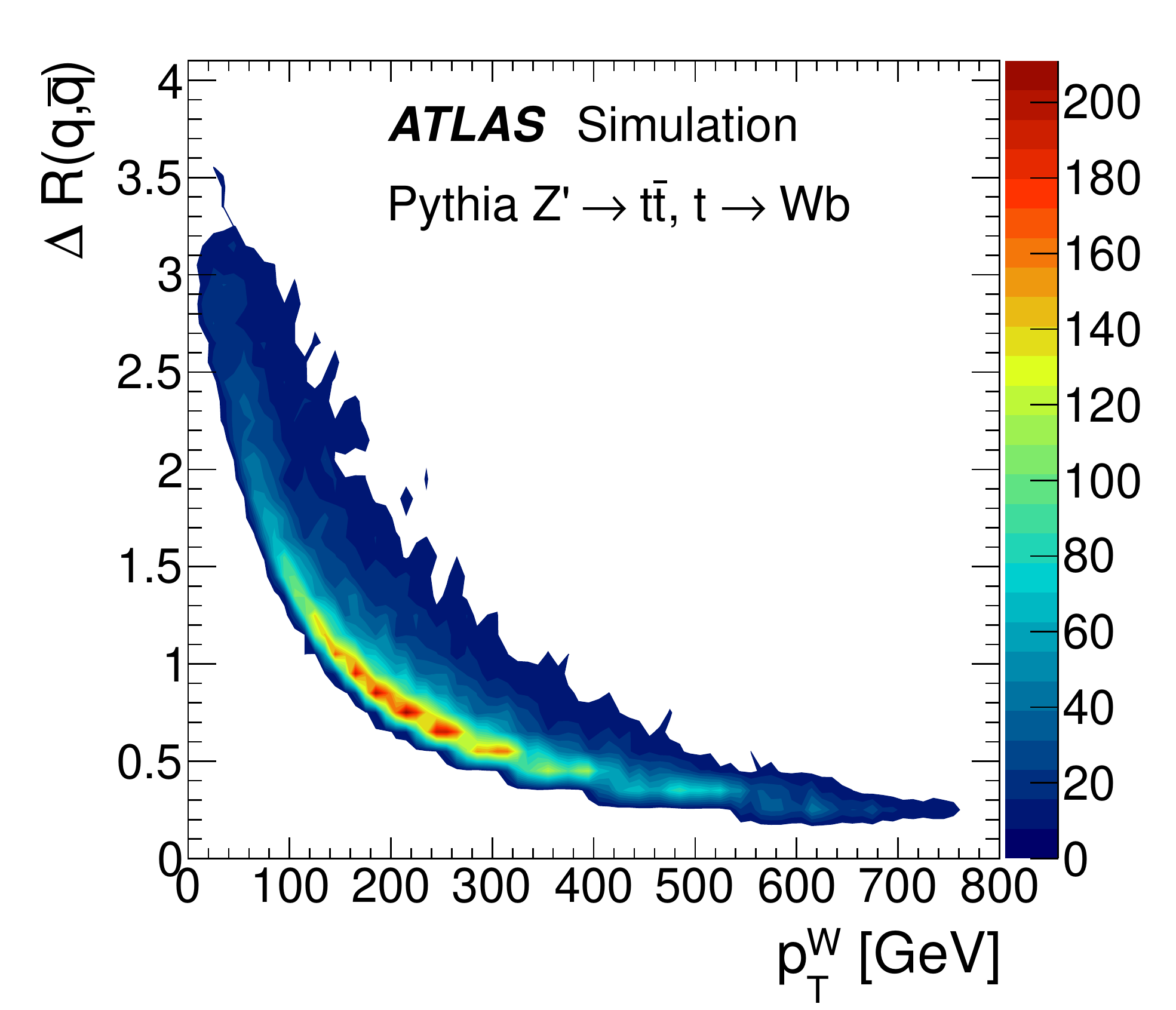}
    \label{fig:intro:Wptdrqq}
    }
  \end{center}
  \caption{\subref{fig:intro:TopptdrWb} The angular separation between the \W boson and 
           $b$-quark in top decays, \tWb, as a function of the top-quark transverse momentum
           (\toppt) in simulated \Pythia~\cite{pythia} \Zprimett ($\massZprime=1.6$~TeV) events.
           \subref{fig:intro:Wptdrqq} The angular distance between the light quark and 
           anti-quark from \tWb decays as a function of the \pt of the \W boson (\Wpt). 
           Both distributions are at the generator level and do not include effects
           due to initial and final-state radiation, or the underlying event.}
  \label{fig:intro:ptdr}
\end{figure}

One example of a new physics process that may produce heavy objects with a significant Lorentz boost is the decay of a new heavy gauge boson, the \Zprime, to top-quark pairs.
\Figref{intro:ptdr} shows the angular separation between the \Wboson\ and $b$ decay products of a top quark in simulated \Zprimett ($\massZprime=1.6$~TeV) events, as well as the separation between the light quarks of 
the subsequent hadronic decay of the $W$ boson.
In each case, the angular separation of the decay products is approximately

\begin{equation}
  \DeltaR \approx \frac{2 m}{\pt},
  \label{eq:ruleofthumb}
\end{equation}

\noindent where \DeltaRydef, and \pt\ and $m$ are the transverse momentum and the mass, respectively, of the decaying particle.\footnote{The \ATLAS{} coordinate system is a right-handed system with the $x$-axis pointing to the centre of the \LHC{} ring and the $y$-axis pointing upwards. 
The polar angle $\theta$ is measured with respect to the \LHC{} beam-line. The azimuthal angle $\phi$ is measured with respect to the $x$-axis. 
The rapidity is defined as $y = 0.5 \times {\rm ln} [ ( E + p_z )/( E - p_z ) ]$,  where $E$ denotes the energy and $p_z$ is the component of the momentum along the beam direction. 
The pseudorapidity $\eta$ is an approximation for rapidity $y$ in the high-energy limit, and it is related to the polar angle $\theta$ by $\eta = -\ln \tan(\theta/2) $. 
Transverse momentum and energy are defined as $\pt = p \times \sin \theta$ and $ E_{\rm T} = E \times \sin \theta$, respectively.}
For $\Wpt>200$~GeV, the ability to resolve the individual hadronic decay products 
using standard narrow-radius jet algorithms begins to degrade, and when \toppt is greater than $300$~GeV, 
the decay products of the top quark tend to have a separation $\DeltaR<1.0$. 
Techniques designed to recover sensitivity in such cases focus on \largeR jets in order to maximize efficiency. In this paper, \textit{large-$R$} refers to jets with a radius parameter $R\geq1.0$.
At \sqsseven, nearly one thousand SM \ttbar\ events per \invfb are expected with \toppt greater than $300$~GeV.
New physics may appear in this region of phase space, the study of which was limited by integrated luminosity and available energy at previous colliders.

A single jet that contains all of the decay products of a massive particle has significantly different properties than a jet of the same \pt originating from a light quark. 
The characteristic two-body or three-body decays of a high \pT\ vector boson or top quark result in a hard substructure that is absent from typical high \pT\ jets formed from gluons and light quarks. These subtle differences in substructure can be resolved more clearly by removing soft QCD radiation from jets. 
Such adaptive modification of the jet algorithm or selective removal of soft radiation during the process of iterative recombination in jet reconstruction is generally referred to as \textit{jet grooming}~\cite{Butterworth:2008iy, pruning2009, Krohn2010}.

\begin{figure}[!ht]  
  \begin{center}
        \includegraphics[width=0.45\textwidth]{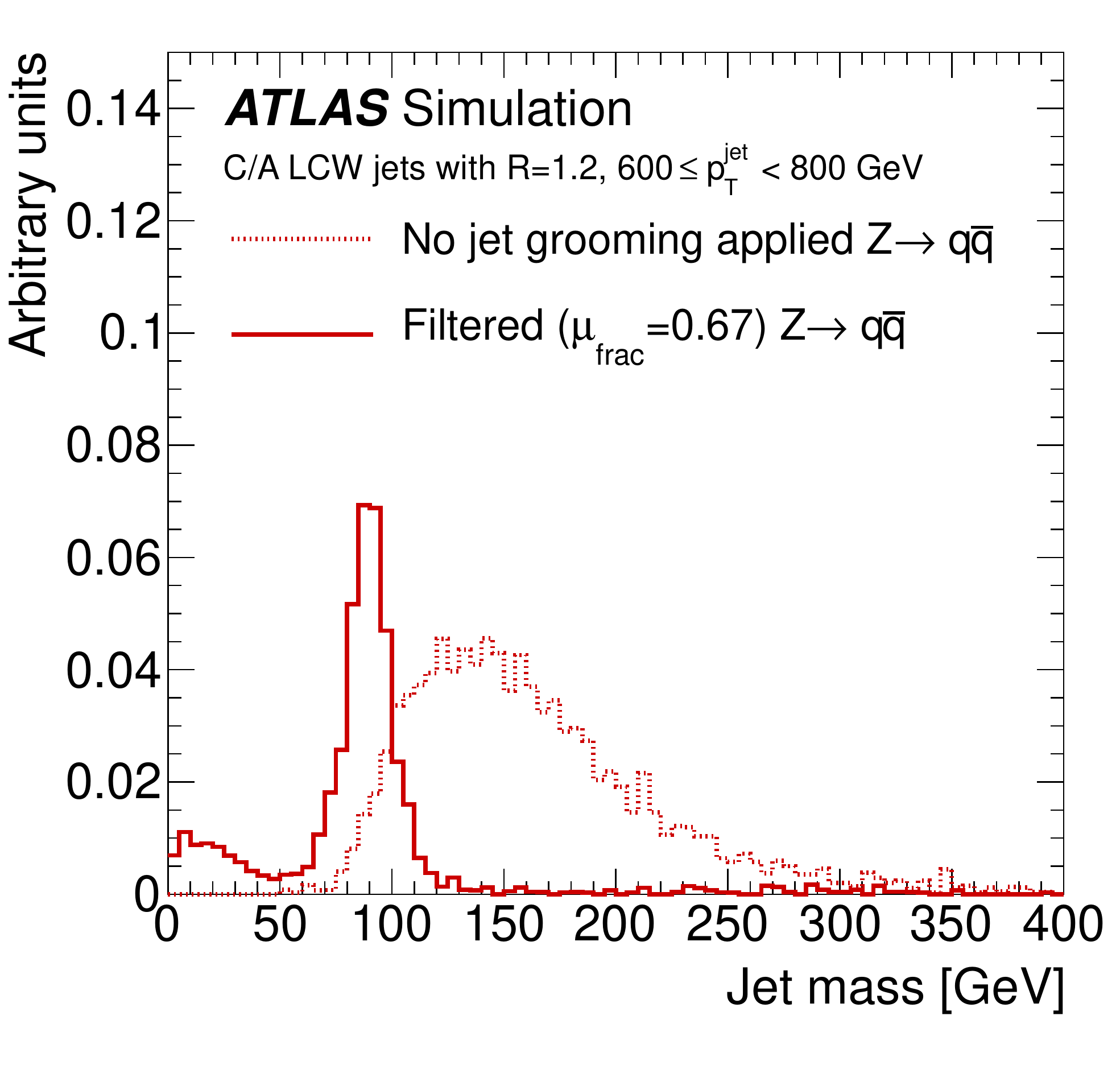}
  \end{center}
  \caption{
    Single-jet invariant mass distribution for \ca (\CamKt) $R=1.2$ jets in 
    simulated events containing highly boosted hadronically decaying \Z bosons before 
    and after the application of a grooming procedure referred to as 
    \textit{mass-drop filtering}. The technical details of this figure are explained 
    in \secref{intro:definitions}. The normalization of the groomed distribution includes 
    the efficiency of mass-drop filtering with respect to the ungroomed \largeR jets 
    for comparison. The local cluster weighting (LCW) calibration scheme is 
    described in \secref{recocalib:mccalib}.
  \label{fig:moneyplot}} 
\end{figure}

Recently many jet grooming algorithms have been designed to remove contributions to a given jet that are irrelevant or detrimental to resolving the hard decay products from a boosted object (for recent reviews and comparisons of these techniques, see for example refs.~\cite{BOOST2010, BOOST2011}).
The structural differences between jets formed from gluons or light quarks and individual jets originating from the decay of a boosted hadronic particle form the basis for these tools. 
The former are characterized primarily by a single dense core of energy surrounded by soft radiation from the parton shower, hadronization, and underlying event (UE) remnants~\cite{QGopal, Dasgupta:2007wa, SchwartzQG}. 
Jets containing the decay products of single massive particles, on the other hand, can be distinguished by hard, wide-angle components representative of the individual decay products that result in a large reconstructed jet mass mass, as well as typical kinematic relationships among the hard components of the jet~\cite{Thaler2008, Thaler:2010tr, Butterworth:2008iy, YSplitter, Cui2010, Hook2011, Jankowiak2011, Soper2011, Chen2012}.
Grooming algorithms are designed to retain the characteristic substructure within such a jet while reducing the impact of the fluctuations of the parton shower and the UE, thereby improving the mass resolution and mitigating the influence of \pileup. These features have only recently begun to be studied experimentally~\cite{Abbott:1999gj, JetMassAndSubstructure, CMSHadronicTTResonance2011, CMSHadronicWZResonance2011, RPVGluino7TeV, ATLASHTTandTTT2012, Aad:2012meb} and have been exploited heavily in recent studies of the phenomenological implications of such tools in searches for new physics~\cite{Butterworth:2008iy, Plehn:2010st, Butterworth2007, Butterworth2009, Kim2010, Kribs2010a, Soper2010, Bai2011a, Berger2011a, Fan2011, Yang2011, Hook2012}.
A groomed jet can also be a powerful tool to discriminate between the often dominant multi-jet background and the heavy-particle decay, which increases signal sensitivity. \Figref{moneyplot} demonstrates this by comparing the invariant mass distribution of single jets in events containing highly boosted hadronically decaying \Z bosons before and after the application of a grooming procedure referred to as \textit{mass-drop filtering}. In this simulated $\Zboson\rightarrow q\bar{q}$~sample described in \secref{data-mc}, \pileup events are also included. Prior to the application of this procedure, no distinct features are present in the jet mass distribution, whereas afterwards, a clear mass peak that corresponds to the \Z boson is evident.

\subsection{Definitions}
\label{sec:intro:definitions}

\subsubsection{Jet algorithms}
\label{sec:intro:jetalgs}

In this paper, three jet algorithms are studied: the \akt algorithm~\cite{Cacciari:2008gp}, the \ca (\CamKt) algorithm~\cite{Dokshitzer:1997in, Wobisch:1998wt}, and the \kt algorithm~\cite{Catani:1991hj, Ellis1993, Catani1993}. 
These are implemented within the framework of the \Fastjet software~\cite{Cacciari200657, Fastjet}. 
They represent the most widely used infrared and collinear-safe jet algorithms available for hadron--hadron collider physics today.
Furthermore, in the case of the \kt and \CamKt algorithms, the clustering history of the algorithm -- that is, the ordering and structure of the pair-wise subjet recombinations made during jet reconstruction -- provides spatial and kinematic information about the substructure of that jet.
The \akt algorithm provides jets that are defined primarily by the highest-\pt constituent, yielding stable, circular jets. 
The compromise is that the structure of the jet as defined by the successive recombinations carried out by the \akt algorithm carries little or no information about the \pt\ ordering of the shower or wide angular-scale structure. 
It is, however, possible to exploit the stability of the \akt algorithm \textit{and} recover meaningful information about the jet substructure: \akt jets are selected for analysis based on their kinematics (\eta\ and \pt), and then the jet constituents are reclustered with the \kt algorithm to enable use of the \kt-ordered splitting scales described in \secref{intro:observables}. 
The four-momentum recombination scheme is used in all cases and the jet finding is performed in rapidity--azimuthal angle ($y$--$\phi$) coordinates. Jet selections and corrections are made in pseudorapidity--azimuthal angle ($\etajet$--$\phijet$) coordinates.

\subsubsection{Jet properties and substructure observables}
\label{sec:intro:observables}

Three observables are used throughout these studies to characterize jet substructure and distinguish massive boosted objects from gluons or light quarks: mass, \kt splitting scales, and $N$-subjettiness.

\paragraph{Jet mass:}

The jet mass is defined as the mass deduced from the four-momentum sum of all jet constituents. Depending on the input to the jet algorithm (see \secref{recocalib:inputs}), the constituents may be considered as either massive or massless four-momenta.

\paragraph{$\bm{k_{t}}$ splitting scales:}
   
The \kt splitting scales~\cite{ButterworthSplit2002} are defined by reclustering the constituents of a jet with the \kt recombination algorithm, which tends to combine the harder constituents last. 
At the final step of the jet recombination procedure, the \kt distance measure, \dij, for the two remaining proto-jets (intermediate jet-like objects at each stage of clustering), referred to as subjets in this case, can be used to define a \textit{splitting scale} variable as:

\begin{equation}
  \Dij = \text{min}(p_{\mathrm{T}i}, p_{\mathrm{T}j})\times \Delta R_{ij},
  \label{eq:grooming_ktsplitting}
\end{equation}
\noindent
where $\Delta R_{ij}$ is the distance between the two subjets in $\eta - \phi$ space. 
With this definition, the subjets identified at the last step of the reclustering in the \kt algorithm provide the \DOneTwo observable.
Similarly, \DTwoThr characterizes the splitting scale in the second-to-last step of the reclustering. 
The parameters \DOneTwo and \DTwoThr can be used to distinguish heavy-particle decays, 
which tend to be reasonably symmetric when the decay is to like-mass particles, from the largely asymmetric splittings that originate from QCD radiation in light-quark or gluon jets.
The expected value for a two-body heavy-particle decay is approximately $\DOneTwo\approx m_{\mathrm{particle}}/2$, whereas 
jets from the parton shower of gluons and light quarks tend to have smaller values of the splitting scales and to exhibit 
a steeply falling spectrum for both \DOneTwo and \DTwoThr (see for example \figref{groomed_jets_split_compare_twoprong}).

\paragraph{$\bm{N}$-subjettiness:}

The $N$-subjettiness variables \tauN~\cite{Thaler:2010tr, Thaler:2011gf} are observables related to the subjet multiplicity. 
The \tauN variable is calculated by clustering the constituents of the jet with the \kt algorithm and requiring that exactly $N$ subjets be found. 
This is done using the exclusive version of the \kt algorithm~\cite{Catani1993} and is based on reconstructing clusters of particles in the jet using all of the jet constituents.
These $N$ final subjets define axes within the jet. 
The variables \tauN are then defined in \equref{grooming_nsubj} as the sum over all constituents $k$ of the jet, such that

\begin{equation}
  \tauN = \frac{1}{d_{0}} \sum_k p_{\mathrm{T}k} \times \text{min}(\delta R_{1k}, \delta R_{2k},...,\delta R_{Nk})~\text{,~~with} 
  ~~~d_{0}\equiv\sum_{k} p_{\mathrm{T}k}\times R
  \label{eq:grooming_nsubj}
\end{equation}
%
where $R$ is the jet radius parameter in the jet algorithm, $p_{\mathrm{T}k}$ is the \pT\ of constituent $k$ and $\delta R_{ik}$ is the distance from subjet $i$ to constituent $k$.
Using this definition, \tauN describes how well jets can be described as containing $N$ or fewer \kt\ subjets by assessing the degree to which constituents are aligned with the axes of these subjets for a given hypothesis $N$. 
The ratios $\tau_2/\tau_1$ and $\tau_3/\tau_2$ can be used to provide discrimination between jets formed from the parton shower of initial-state gluons or light-quarks and jets formed from two hadronic decay products (from $Z$-bosons, for example) or three hadronic decay products from boosted top quarks. 
These ratios are herein referred to as \tauTwoOne and \tauThrTwo respectively. For example, $\tauTwoOne \simeq 1$ corresponds to a jet that is very well described by a single subjet whereas a lower value implies a jet that is much better described by two subjets than one.


\subsubsection{Jet grooming algorithms}
\label{sec:intro:grooming}

Three jet grooming procedures are studied in this paper. 
Mass-drop filtering, trimming, and pruning are described, and performance measures related to each are defined.
The different configurations of the grooming algorithms described in this section are summarized in \tabref{grooming_configs}. Additionally, a technique to tag boosted top quarks using the mass-drop filtering method is introduced.
Unless otherwise specified, the jet \pT\ reported for a groomed jet is that which is calculated after the grooming algorithm is applied to the original jet. 

\paragraph{Mass-drop Filtering:}

\begin{figure}[!ht]
 \label{fig:cartoons}
  \begin{center}
    \subfigure[The mass-drop and symmetric splitting criteria.]{
      \includegraphics[scale=0.4]{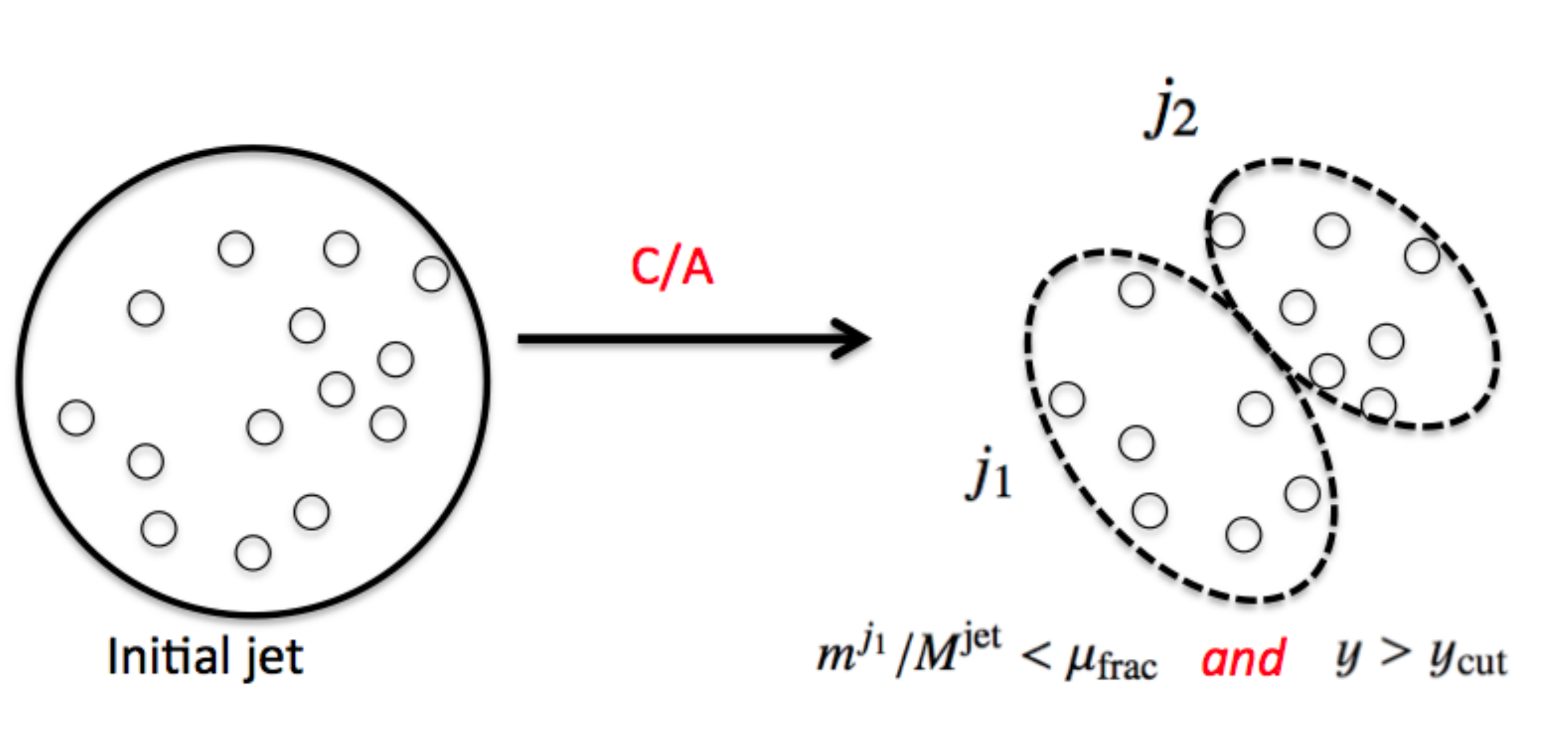}
      \label{fig:cartoon_f1} 
    }
    \subfigure[Filtering.]{
      \includegraphics[scale=0.4]{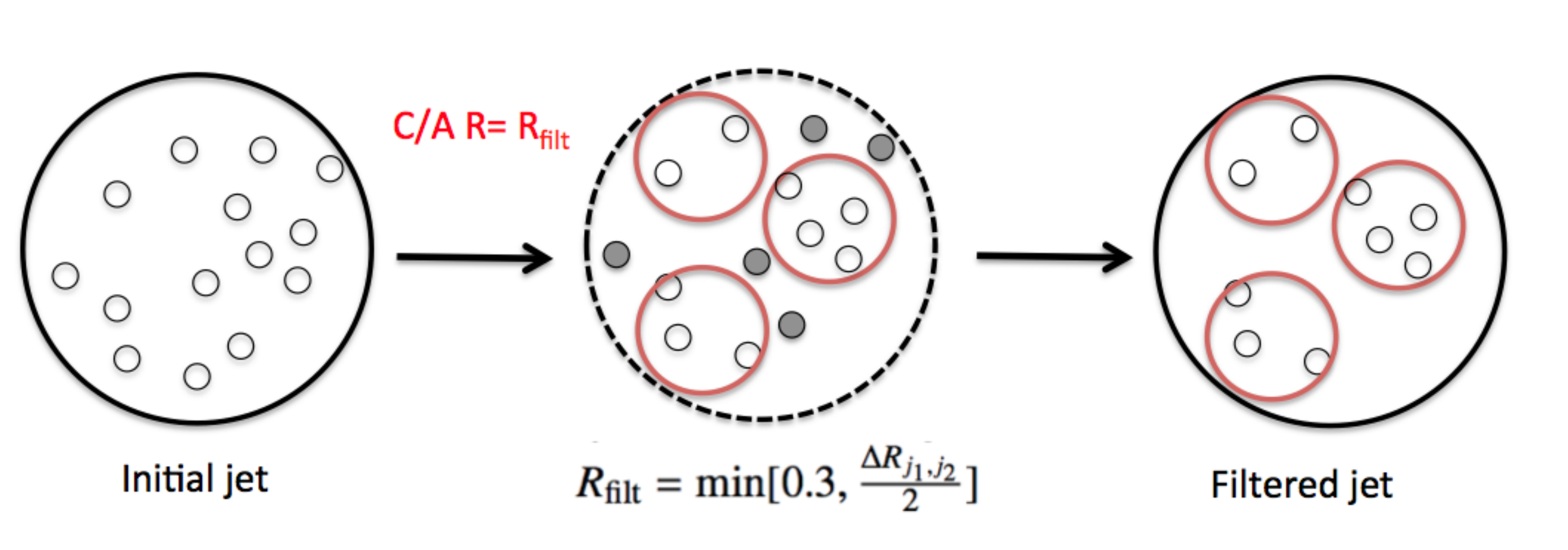}
      \label{fig:cartoon_f2}  
    }
  \end{center}
  \caption{Diagram depicting the two stages of the mass-drop filtering procedure.}
 \end{figure}

%
The mass-drop filtering procedure\footnote{In this paper, mass-drop filtering is often shortened to only \emph{filtering} in figures and captions.} seeks to isolate concentrations of energy within a jet by identifying relatively symmetric subjets, each with a significantly smaller mass than that of the original jet.
This technique was developed and optimized using \CamKt jets in the search for a Higgs boson decaying to two $b$-quarks: \Hbb~\cite{Butterworth:2008iy}. 
The \CamKt algorithm is used because it provides an angular-ordered shower history that begins with the widest combinations when reversing the cluster sequence.
%
%
This provides useful information regarding the presence of potentially large splittings within a jet (see \secref{pileup} and \secref{boostedwtop}).
Although the mass-drop criterion and subsequent filtering procedure are not based specifically on soft-\pt\ or wide-angle selection criteria, the algorithm does retain the hard components of the jet through the requirements placed on its internal structure.
The first measurements of the jet mass of these filtered jets was performed using 35\invpb of data collected in 2010 by the ATLAS experiment~\cite{JetMassAndSubstructure}. 
The mass-drop filtering procedure has two stages:

\begin{itemize}

  \item \textbf{Mass-drop and symmetry} The last stage of the \CamKt clustering is undone. The jet ``splits'' into two subjets, \sjone and \sjtwo, ordered such that the mass of \sjone is larger: $\msubjone > \msubjtwo$. 
The mass-drop criterion requires that there be a significant difference between the original jet mass (\Mjet) and \msubjone after the splitting:
%
\begin{equation}
   \msubjone/\Mjet < \mufrac,
   \label{eq:grooming_filtering:massdrop}
\end{equation}
%
where \mufrac is a parameter of the algorithm.   
The splitting is also required to be relatively symmetric, which is approximated by the requirement that
%
\begin{equation}
  \frac{\text{min}[(\ptsubjone)^2,(\ptsubjtwo)^2]}{(\Mjet)^2}\times \DRsubjets^{2} > \ycut,
  \label{eq:grooming_filtering:symmetry}
\end{equation}
%
where \DRsubjets is a measure of the opening angle between \sjtwo and \sjone, and $\ycut$ defines the energy sharing between the two subjets in the original jet. 
For the analyses presented here, \ycut\ is set to 0.09, the optimal value for identifying two-body decays, obtained in previous studies~\cite{Butterworth:2008iy}. 
To give a sense of the kinematic requirements that this places on a given decay, consider a hadronically decaying \W boson with $\pT^{W}\approx200$~GeV. 
According to the approximation given by \equref{ruleofthumb}, the average angular separation of the two daughter quarks is $\DRsubjets\sim0.8$. 
The symmetry requirement determined by \ycut in \equref{grooming_filtering:symmetry}  thereby implies that the transverse momentum of the softer (in \pT) of the two subjets is greater than approximately 30~GeV. 
Generally, this requirement entails a minimum \pT\ of the softer subjet of $\subjetpt/\ptjet>0.15$, thus forcing both subjets to carry some significant fraction of the momentum of the original jet.
This procedure is illustrated in \figref{cartoon_f1}. If the mass-drop and symmetry criteria are not satisfied, the jet is discarded.

\item \textbf{Filtering} The constituents of \sjone and \sjtwo are reclustered using the \CamKt algorithm 
     with radius parameter $\Rfilt = \text{min} [0.3,\DRsubjets/2]$, where $\Rfilt < \DRsubjets$. 
   The jet is then filtered; all constituents outside the three hardest subjets are discarded. The choice of three allows one additional radiation from a two-body decay to be captured.
   In isolating \sjone and \sjtwo with the \CamKt algorithm, the angular scale of any potential massive particle decay is known.
   By dynamically reclustering the jet at an appropriate angular scale able to resolve that structure, the sensitivity to highly collimated decays is maximized.
   This is illustrated in \figref{cartoon_f2}.
\end{itemize}

In this paper, three values of the mass-drop parameter \mufrac\ are studied, as summarized in \tabref{grooming_configs}. 
The values chosen for \mufrac are based on a previous study~\cite{Butterworth:2008iy} which has shown that $\mufrac = 0.67$ is optimal in discriminating \Hbb\ from background. 
A subsequent study regarding the factorization properties of several groomed jet algorithms~\cite{WalshZubieri} found that smaller values of \mufrac (0.20 and 0.33) are similarly effective at reducing backgrounds, and yet they remain factorizable within the soft collinear effective theory studied in that analysis. 
%

\paragraph{Trimming:}

The trimming algorithm~\cite{Krohn2010} takes advantage of the fact that contamination from \pileup, multiple parton interactions (MPI) and initial-state radiation (ISR) in the reconstructed jet is often much softer than the outgoing partons associated with the hard-scatter and their final-state radiation (FSR). 
The ratio of the \pT\ of the constituents to that of the jet is used as a selection criterion. 
Although there is some spatial overlap, removing the softer components from the final jet  preferentially removes radiation from \pileup, MPI, and ISR while discarding only a small part of the hard-scatter decay products and FSR.
Since the primary effect of \pileup in the detector is additional low-energy deposits in clusters of calorimeter cells, as opposed to additional energy being added to already existing clusters produced by particles originating from the hard scattering process, this allows a relatively simple jet energy offset correction for smaller radius jets ($R=0.4,~0.6$) as a function of the number of primary reconstructed vertices~\cite{jespaper2010}. 
%

\begin{figure}[!ht]
  \begin{center}
    \includegraphics[scale=0.4]{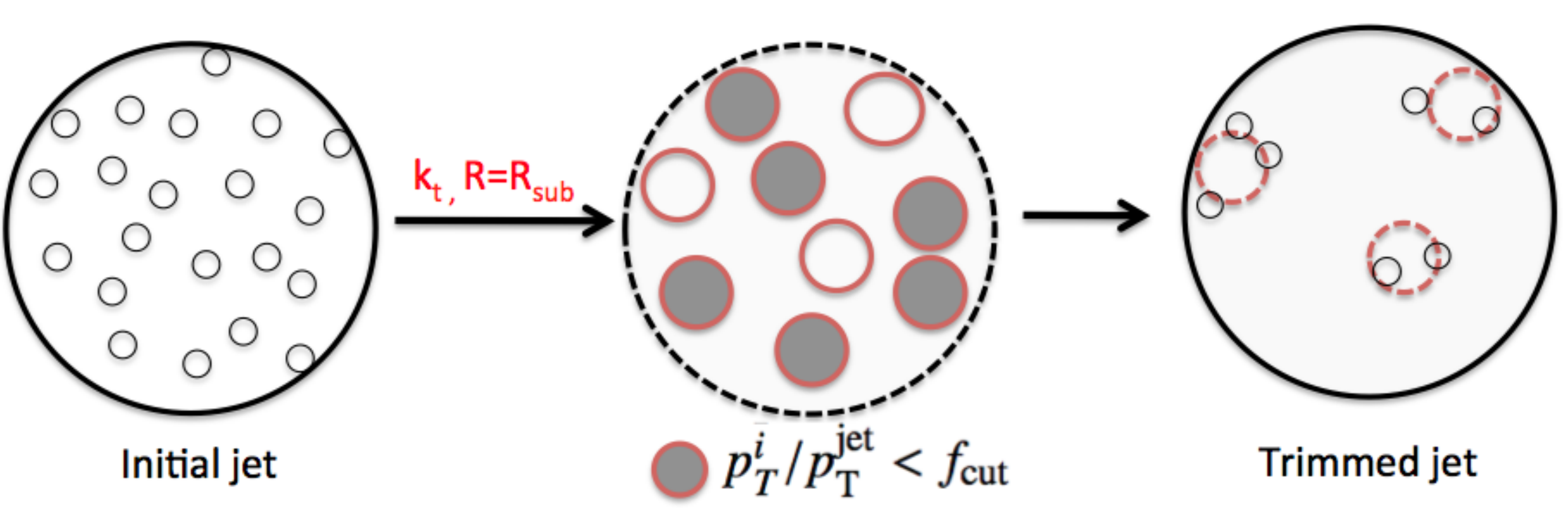}
  \end{center}
  \caption{Diagram depicting the jet trimming procedure.}
  \label{fig:cartoon_t}
\end{figure}

The trimming procedure uses a \kt algorithm to create subjets of size \drsub from the constituents of a jet.  
Any subjets with $\pti/\ptjet < \fcut$ are removed, where \pti\ is the transverse momentum of the $i^{th}$ subjet, and \fcut\ is a parameter of the method, which is typically a few percent. The remaining constituents form the trimmed jet. This procedure is illustrated in \figref{cartoon_t}. Low-mass jets ($\mjet<100$ GeV) from a light-quark or gluon lose typically 30--50\% of their mass in the trimming procedure, while jets containing the decay products of a boosted object lose less of their mass, with most of the reduction due to the removal of \pileup or UE (see, for example, \figsref{groomed_jets_mass_compare_twoprong}{groomed_jets_mass_compare}). The fraction removed increases with the number of \pp interactions in the event.

Six configurations of trimmed jets are studied here, arising from combinations of $\fcut$ and $\drsub$, given in \tabref{grooming_configs}. 
They are based on the optimized parameters in ref.~\cite{Krohn2010} ($\fcut=0.03, \drsub=0.2$) and variations suggested by the authors of the algorithm. 
This set represents a wide range of phase space for trimming and is somewhat broader than considered in ref.~\cite{Krohn2010}.

\paragraph{Pruning:}

The pruning algorithm~\cite{pruning2009, Ellis2009a} is similar to trimming in that it removes constituents with a small relative \pT, but it additionally applies a veto on wide-angle radiation.
The pruning procedure is invoked at each successive recombination step of the jet algorithm (either \CamKt or \kt). It is based on a decision at each step of the jet reconstruction whether or not to add the constituent being considered. As such, it does not require the reconstruction of subjets. For all studies performed for this paper, the \kt algorithm is used in the pruning procedure.
This results in definitions of the terms \textit{wide-angle} or \textit{soft} that are not directly related to the original jet but rather to the proto-jets formed in the process of rebuilding the pruned jet. 
%

\begin{figure}[!ht]
  \begin{center}
    \includegraphics[scale=0.4]{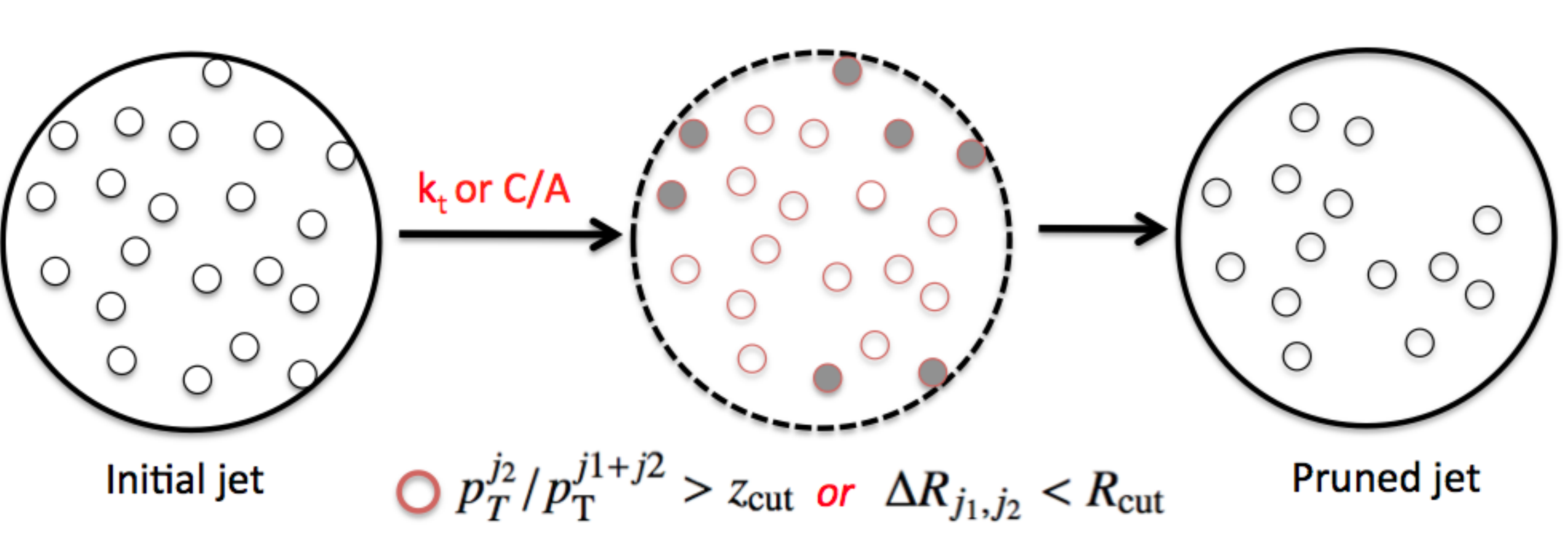}
  \end{center}
  \caption{Diagram illustrating the pruning procedure.}
  \label{fig:cartoon_p}
\end{figure}

The procedure is as follows:

\begin{itemize}
  \item The \CamKt or \kt recombination jet algorithm is run on the constituents, which were found by any jet finding algorithm. 
  
  \item At each recombination step of constituents \sjone and \sjtwo (where $\ptsubjone>\ptsubjtwo$), 
    either $\ptsubjtwo/p_{\mathrm{T}}^{j1+j2} > \zcut$ or $\DRsubjets < \rcut\times(2\Mjet/\ptjet)$ must be satisfied.  Here, $\zcut$ and
   $\rcut$ are parameters of the algorithm which are studied in this paper.

  \item \sjtwo with \sjone are merged if one or both of the above criteria are met, otherwise, \sjtwo is discarded and the algorithm continues.
  
\end{itemize}

The pruning procedure is illustrated in \figref{cartoon_p}. 
Six configurations, given in \tabref{grooming_configs}, based on combinations of \zcut and \rcut are studied here. 
%
%
This set of parameters also represents a relatively wide range of possible configurations.

\begin{table}[!ht]
  \begin{center}
  \begin{tabular}{c|c|l}
    \hline \hline
    Jet finding algorithms used & Grooming algorithm & Configurations considered \\ \hline

     \CamKt & Mass-Drop Filtering & \mufrac = 0.20, 0.33, \textbf{0.67} \\[1.5ex]
     \multirow{2}{*}{\Akt and \CamKt } & \multirow{2}{*}{ Trimming} & \fcut = 0.01, \textbf{0.03}, 0.05\\  
     & & \drsub = \textbf{0.2}, 0.3 \\[1.5ex]
     \multirow{2}{*}{\Akt and \CamKt \B \T} & \multirow{2}{*}{ Pruning \B \T} & \rcut = 0.1, 0.2, 0.3\\
     & & \zcut = 0.05, 0.1 \\[1.5ex]
     \CamKt & \htt & (see \tabref{htt631}) \\[1.5ex]
    \hline \hline
  \end{tabular}
  \end{center}
  \caption{Summary of the grooming configurations considered in this study. 
          Values in boldface are optimized configurations 
          reported in ref.~\cite{Butterworth:2008iy} and ref.~\cite{Krohn2010} 
          for filtering and trimming, respectively.}
  \label{tab:grooming_configs}
\end{table}

\subsubsection{\htt}
\label{sec:intro:htt}

\begin{figure}[ht]
  \begin{center}
    \subfigure[Every object encountered in the declustering process is considered a `substructure object' if it is of sufficiently low mass or has no clustering history.]{
      \includegraphics[width=0.46\textwidth]{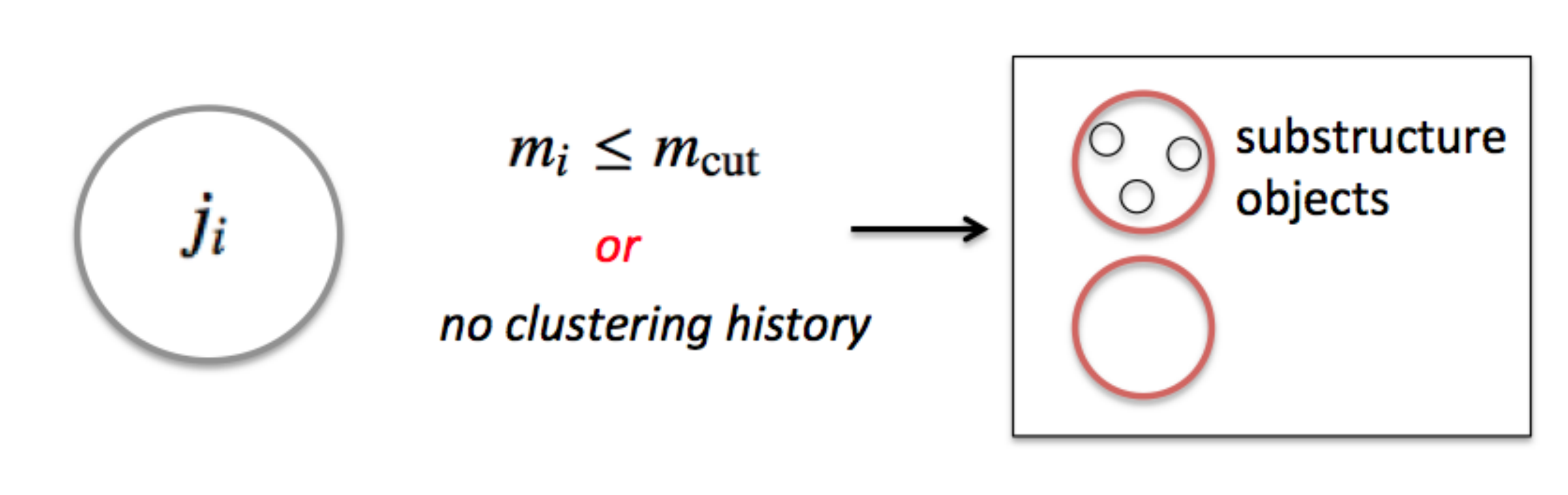}
    } 
    \hfill
    \subfigure[The mass-drop criterion is applied iteratively, following the highest subjet-mass line through the clustering history, resulting in $N_{i}$ substructure objects.]{
      \includegraphics[width=0.46\textwidth]{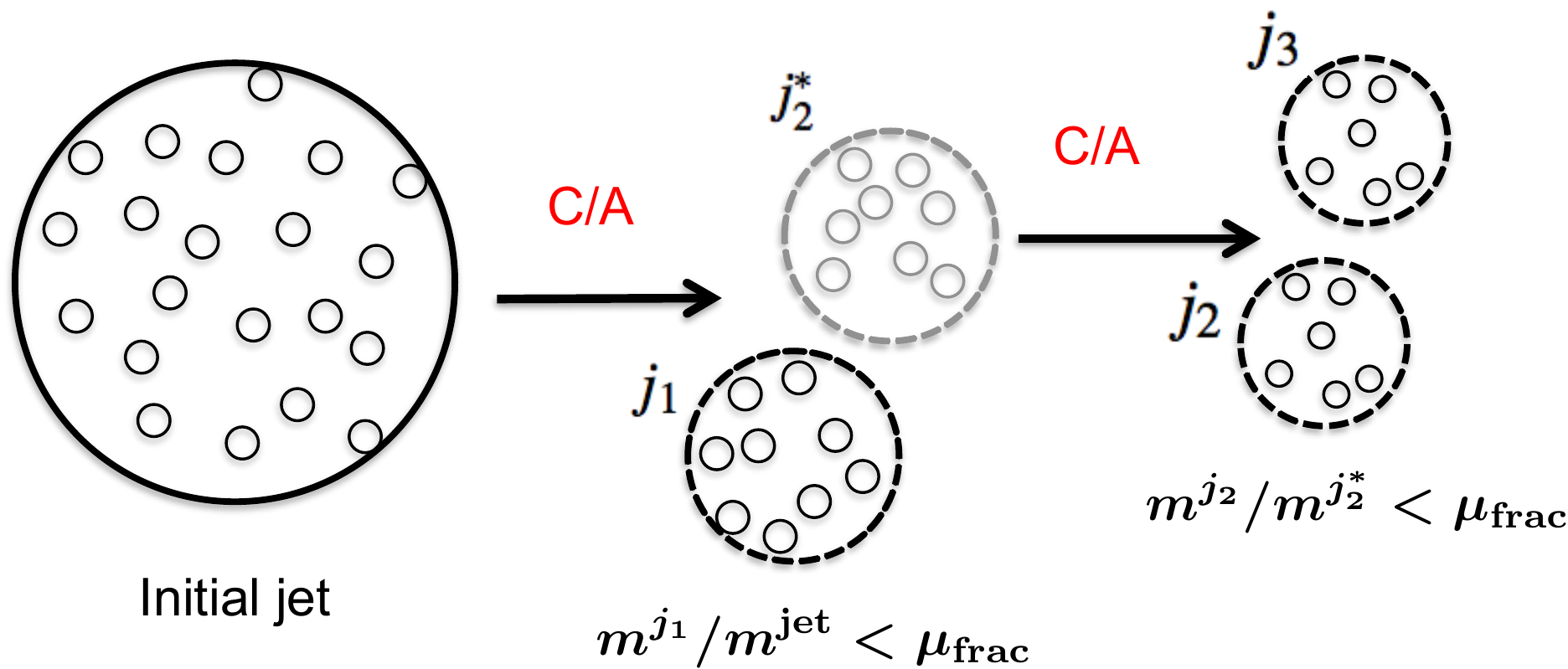}
    }
    \subfigure[For every triplet-wise combination of the substructure objects found in (b), recluster the constituents into subjets and select the \Nsubjet leading-\pt\ subjets, with $3\leq\Nsubjet\leq N_i$ (here, $\Nsubjet=5$).]{
      \includegraphics[width=0.46\textwidth]{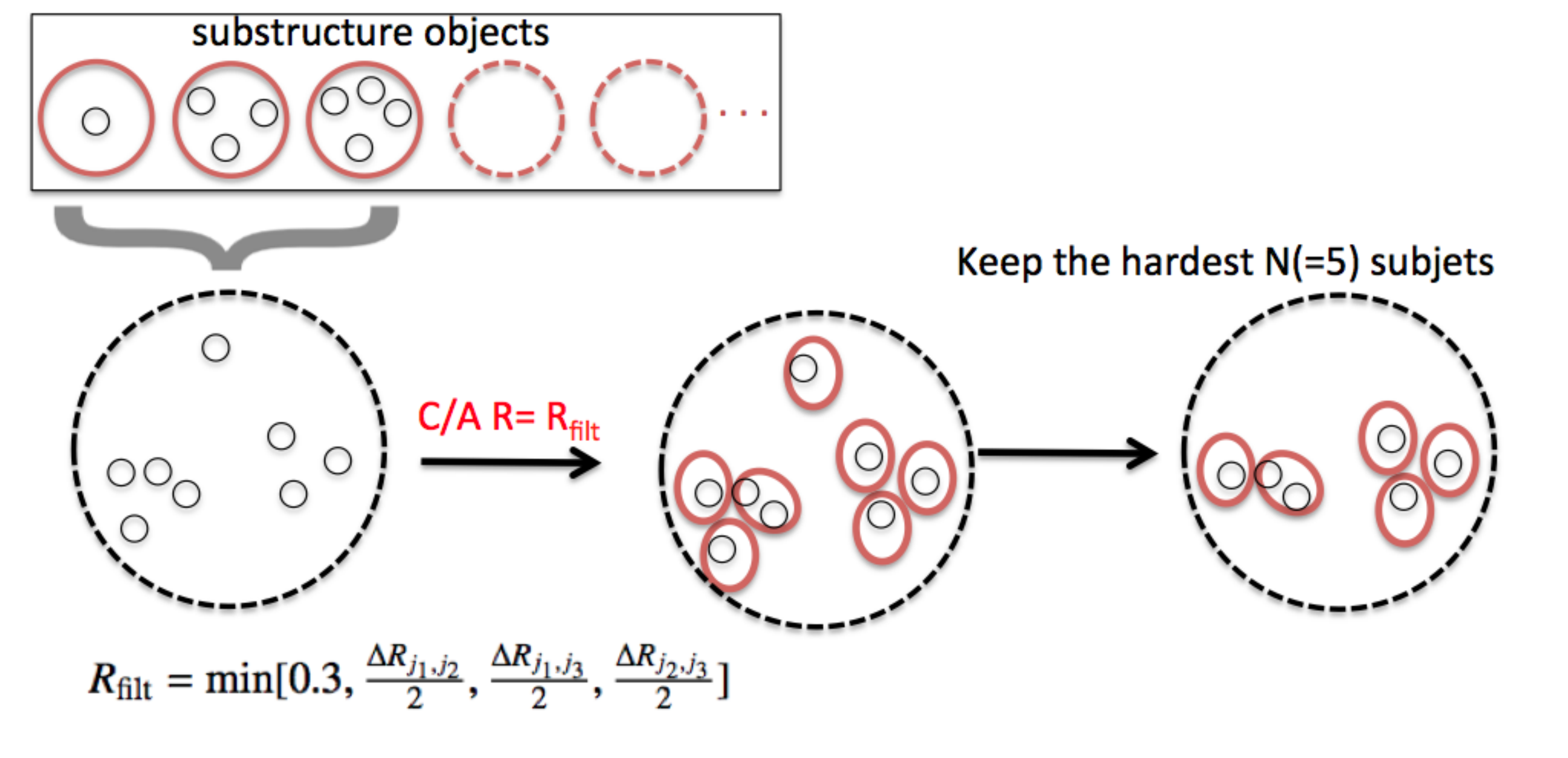}
    } 
    \hfill
    \subfigure[Recluster the constituents of the \Nsubjet subjets into exactly three subjets to make the top candidate for this triplet-wise combination of substructure objects.]{
      \includegraphics[width=0.46\textwidth]{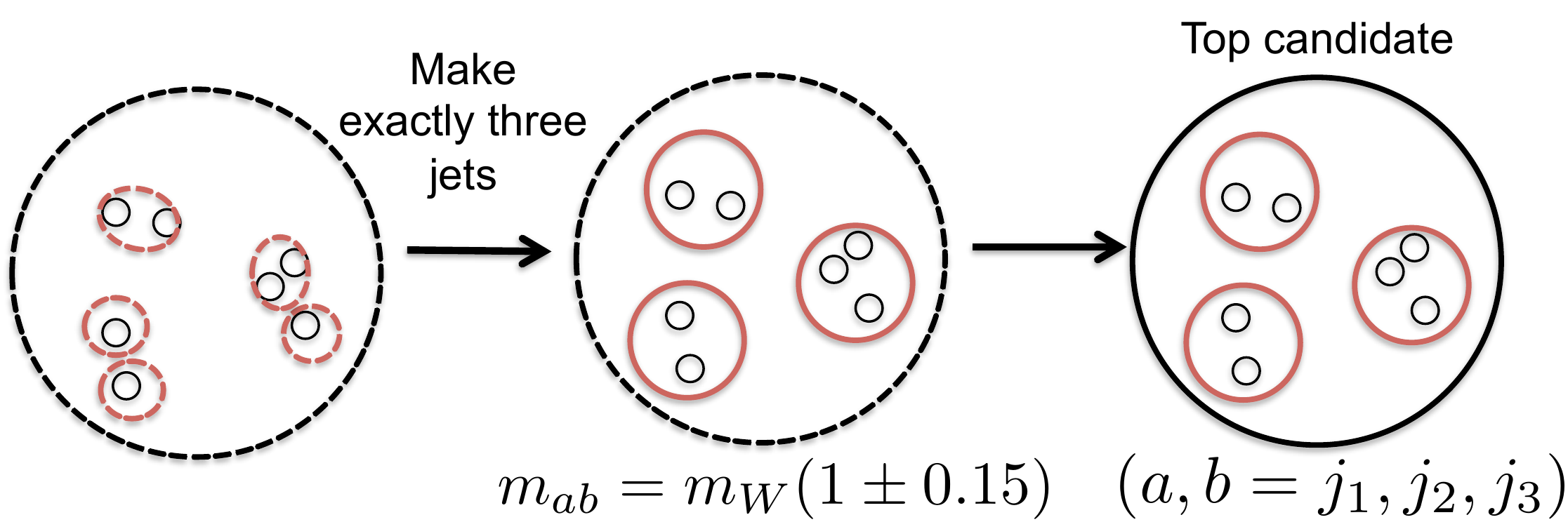}
    }
\end{center}
\caption{The \htt procedure.}
\label{fig:cartoon_htt}
\end{figure}

The \htt algorithm~\cite{Plehn:2010st} is designed to identify a top quark with a hadronically decaying \W boson daughter over a 
large multi-jet background. The method uses the \CamKt jet algorithm and a variant of the mass-drop filtering technique described 
in \secref{intro:grooming} in order to exploit information about the recombination history of the jet. 
This information is used to search for evidence within the jet of the presence of \W decay products as well as an additional 
energy deposition -- the $b$-quark -- that are consistent with the \W and top masses and the expected angular distribution 
of the final-state quarks. The \htt algorithm is optimized for top-quark transverse momentum as low as 200~GeV and therefore 
uses a correspondingly large jet radius parameter. The algorithm proceeds as follows and is illustrated in \figref{cartoon_htt}.

\begin{table}[t]
  \begin{center}
  \begin{tabular}{lccc}
    \hline
    \hline
                            & Default     & Tight & Loose \\ \hline 
      $R$                   & 1.5 or 1.8  &  1.5  &  1.5  \\        
      $m_{\text{cut}}$[GeV] &  30         & 30    & 70    \\        
      $R_{\text{jet}}$      &   0.3       &  0.2  &  0.5  \\        
      \Nsubjet              &   5         &  4    &  7    \\        
      $f_W$[\%]             &  15         & 10    & 20    \\        
    \hline
    \hline
    \end{tabular}
    \end{center}
  \caption{The settings used for studying the performance of the \htt.}
  \label{tab:htt631}
\end{table}

\paragraph{Decomposition into substructure objects:} The mass-drop criterion defined in \equref{grooming_filtering:massdrop} 
is applied to a \largeR C/A jet, where \sjone and \sjtwo are the two subjets from the last stage of clustering, with $\msubjone > \msubjtwo$.
If the criterion is satisfied, the same prescription is followed to split both \sjone and \sjtwo further. The iterative 
splitting continues until the subjets either have masses $m_{i}$ less than a tunable parameter $\mcut$, or represent individual constituents, 
such as calorimeter energy deposits, tracks, or generator-level particles (i.e. no clustering history); see \secref{recocalib:inputs} 
for definitions. This procedure results in $N_i$ subjets. If at any stage $\msubjone>(\Mjet\times\mufrac)$, the mass-drop 
criterion and subsequent iterative declustering is not applied to \sjtwo. The values of \mcut and $R$ studied  are 
summarized in \tabref{htt631}. $R$ values of 1.5 and 1.8, somewhat larger than used generally in mass-drop filtering, 
are chosen based on previous studies~\cite{Plehn:2010st}. When the iterative process of declustering the jet is complete, 
there must exist at least three subjets ($N_i \geq 3$), otherwise the jet is discarded.

\paragraph{Filtering:} All possible combinations of three subjets are formed, and each triplet is filtered one at a time.  
The constituents of the subjets in a given triplet are reclustered into $N_j$ new subjets using the \CamKt algorithm 
with a size parameter $\Rfilt = \text{min} [0.3,\DRsubjets/2]$, where \DRsubjets is the minimum separation between all 
possible pairs in the current triplet. It is therefore possible that $N_j>3$ after the reclustering step. All energy deposits not in the $N_j$ subjets are discarded. 
     
\paragraph{Top mass window requirement:} If the invariant mass of the four-momentum determined by summing the constituents 
of the $N_j$ subjets is not in the range $140\GeV\leq\Mjet<200$~GeV, the triplet combination is ignored. If more than 
one triplet satisfies the criterion, only the one with mass closest to the top-quark mass, \topmass, is used. This triplet (which consists of $N_j\geq 3$ subjets) is thus identified as the top-candidate triplet.

\paragraph{Reclustering of subjets:} From the $N_j$ subjets of the top-candidate triplet, 
\Nsubjet leading-\pt\ subjets are chosen, where \Nsubjet is a parameter satisfying $3 \leq \Nsubjet \leq N_j$. From this set of subjets, 
exactly three jets are built by re-applying the C/A algorithm to the constituents of the 
\Nsubjet subjets, which are exclusively clustered using a distance parameter $R_{\text{jet}}$ listed in \tabref{htt631}. 
This latter step reflects the hypothesis that this is likely to be a top-jet candidate. 
These subjets are calibrated as described in \secref{recocalib:calibca}.

\paragraph{$\bm{W}$ boson mass requirements:} Relations listed in Eqs.~(A.1) of Ref.~\cite{Plehn:2010st} are 
defined using the total invariant mass of the three subjets ($m_{123}$) and the invariant mass $m_{ij}$ formed 
from combinations of two of the three subjets ordered in \pt. These conditions include:

\begin{equation}
 R_- < \frac{m_{23}}{m_{123}} < R_+
 \label{eq:htt_masscut}
\end{equation}
\begin{equation}
 0.2 < \text{arctan}~\frac{m_{13}}{m_{12}} < 1.3.
 \label{eq:htt_masssym}
\end{equation}

\noindent Here, $R_\pm = (1 \pm f_W )(\Wmass/\topmass)$, $f_W$ is a resolution variable (given in \tabref{htt631}), 
and the quantities \Wmass and \topmass denote the \Wboson\ boson and top-quark masses, respectively. If at least one of the criteria in Eqs.~(A.1) of~\cite{Plehn:2010st} is met, the four-momentum addition of the three subjets is considered a candidate top quark.    

\section{The ATLAS detector and data samples}
\label{sec:Detector}
\subsection{The ATLAS detector}
\label{sec:Detector:det}

The ATLAS detector~\cite{detPaper,PerfWithData2010} provides nearly full solid angle 
coverage around the collision point with an inner tracking system covering $|\eta|<2.5$,
electromagnetic and hadronic calorimeters covering $|\eta|<4.9$, and a muon spectrometer 
covering $|\eta|<2.7$. Of the multiple ATLAS subsystems, the most relevant to this analysis are the 
barrel and endcap calorimeters~\cite{LArReadiness, TileReadiness} and the trigger~\cite{TriggerPerf2010}. 



The calorimeter comprises multiple sub-detectors of various designs, spanning the pseudorapidity range up to $|\eta|=4.9$. 
The measurements presented here are performed using data predominantly from the central calorimeters, comprising the 
liquid argon (LAr) barrel electromagnetic calorimeter ($|\eta|<1.475$) and the tile hadronic calorimeter ($|\eta|<1.7$). 
Three additional calorimeter subsystems are located in the higher-$\eta$ regions of the detector: the LAr electromagnetic endcap calorimeter, the LAr hadronic endcap calorimeter, and the forward calorimeter with separate components for electromagnetic and hadronic showers.

Dedicated trigger and data acquisition systems are responsible for the online event selection, which is performed in three stages: Level 1, Level 2, and the Event Filter. The measurements presented in this paper rely primarily on the single-jet and multi-jet triggers implemented at the Event Filter level, which has access to the full detector granularity, and finds multi-jet events with high efficiency. The intermediate trigger levels provide coarser jet finding and sufficient rate reduction to satisfy the trigger and offline selection requirements.

\subsection{Data and Monte Carlo samples}
\label{sec:data-mc}

Data from the entire 2011 ATLAS data-taking period are used, corresponding to \intlumi\ of integrated luminosity~\cite{Aad:2013ucp}. 
All data are required to have met baseline quality criteria and were taken during periods in which almost all of the detector was fully functional. 
Data quality criteria reject events with significant contamination from detector noise or with  issues in the read-out, and are based on assessments for each subdetector individually. 
Multiple proton--proton collisions, or \pileup, result in several reconstructed primary vertices per event. 
The inclusive jet sample that is used for many studies in this paper is selected from the data using a single high-\pt\ jet trigger that requires the leading jet in the event to have $\ptjet>350$~GeV. This trigger threshold was used for the entire 2011 data-taking period and thus represents the full integrated luminosity with negligible inefficiency.

These data are compared to inclusive jet events, which are dominated by light-quark or gluon jets whose properties are defined primarily by soft gluon emission, that are generated by three Monte Carlo (MC) simulation programs: \Pythia 6.425~\cite{pythia}, \Herwigpp~\cite{Herwigpp}, and \PowhegBox 1.0~\cite{Nason:2004rx,Frixione:2007vw,Alioli:2010xa} (patch 4) interfaced to \Pythia 6.425 for the parton shower, hadronization, and UE models. 
Both \Pythia and \Herwigpp use the modified-LO parton distribution function (PDF) set MRST LO*~\cite{PDF-MRST}. 
\PowPythia uses the CT10 NLO PDF~\cite{PDF-CT10} in the matrix element and CTEQ6L1 PDF set~\cite{PDF-CTEQ} for the \Pythia\ parton shower. For both cases, \Pythia is used with the corresponding ATLAS AUET2B tune~\cite{MC11, MC11c} and \Herwigpp uses the so-called \textsc{UE7-2} tune~\cite{Gieseke:2012ft}, which is tuned to UE data from the LHC experiments.
\Pythia or \Herwigpp with \PowPythia provide an important comparison, at least at the matrix-element level, between leading-order (\LO) (\Pythia and \Herwigpp) and next-to-leading-order (\NLO) (\Powheg) calculations. Furthermore, \Pythia and \Herwigpp offer distinct approaches to the modelling of the parton shower, hadronization, and the UE.

Samples of \ttbar\ events are generated with \Mcatnlo v4.01~\cite{mcatnlo} using the CT10 NLO PDF, interfaced to \Herwig v6.520~\cite{herwig} and \jimmy v4.31~\cite{jimmy}.
Alternative samples for the study of systematic uncertainties are generated with \Powheg, with showering  provided by either \Herwig\ or \Pythia.
Samples generated with the \Acermc v3.8~\cite{acer} package, using CTEQ6L1 PDFs, with showering provided by \Pythia\ are also used.
In these samples \Pythia\ parameters have been tuned to increase or decrease the amount of initial- and final-state radiation.
Single-top-quark events in the s-channel and $Wt$ processes are also generated with \Mcatnlo using the 
CT10 NLO PDF set, with only leptonically decaying \W bosons allowed in the final state.  Single-top-quark events 
in the t-channel, where all \W boson decay channels are produced, are generated with \Acermc using the CTEQ6L1 PDF set, 
and are showered using \Pythia\ with the AUET2B tune.

Samples of \Wjets and \Zjets events are produced with the \Alpgen v2.13~\cite{alpgen} generator, using CTEQ6L1 PDFs, interfaced to 
\Herwig\ for parton showering and hadronization.
Samples of diboson production processes ($WW$, $WZ$ and $ZZ$) are produced with the \Herwig\ generator.

After simulation of the parton shower and hadronization, as well as of the UE, events are passed through the full \geant~\cite{Geant4} detector simulation~\cite{simulation}. Following this, the same trigger, event, data quality, jet, and track selection criteria are applied to the Monte Carlo simulation events as are applied to the data.

Boosted particles decaying to hadrons are used for direct comparisons of the performance of the various reconstruction and jet substructure techniques.
For two-pronged decays, a sample of hadronically decaying \Z bosons is generated using the \Herwig\ v6.510~\cite{herwig} event generator interfaced 
with \jimmy\ v4.31~\cite{jimmy} for the UE. A sample of hadronically decaying \W bosons produced using the same configuration as for the \Z boson sample is also used for comparisons of the \htt performance in \secref{ttbar:htt}.
In order to test the performance of techniques designed for three-pronged decays, \ttbar\ events from a non-Standard-Model heavy gauge boson (\Zprime\ with $\massZprime=1.6$~TeV) are generated using the same \Pythia 6.425 tune as above. 
This model provides a relatively narrow \ttbar\ resonance and top quarks with \pT\ $\simeq 800$~\GeV.

\Pileup is simulated by overlaying additional soft \pp collisions, or minimum bias events, which are generated 
with \Pythia 6.425 using the \ATLAS{} {\sc AUET2B} tune \cite{MC11c} and the CTEQ6L1 PDF set. 
The minimum bias events are overlaid onto the hard scattering events according to the measured distribution 
of the average number $\avg{\mu}$ of \pp interactions. 
The proton bunches were organized in trains of 36 bunches with a 50~ns spacing between the bunches. 
Therefore, the simulation also contains effects from out-of-time \pileup, i.e.\ contributions from the collision of bunches neighbouring those where the events of interest occurred. 
Simulated events are reweighted such that the MC distribution of $\avg{\mu}$ agrees with the data, as measured by the luminosity detectors in \ATLAS{}~\cite{Aad:2013ucp}.

\section{Jet reconstruction and calibration}
\label{sec:recocalib}
\subsection{Inputs to jet reconstruction}
\label{sec:recocalib:inputs}

The inputs to jet reconstruction are either stable particles with a lifetime of at least 10~ps (excluding muons and neutrinos) in the case of MC \emph{generator-level jets} (also referred to as \emph{particle jets}), charged particle tracks in the case of so-called \emph{track-jets}~\cite{jespaper2010}, or three-dimensional topological clusters (\topos) in the case of fully reconstructed calorimeter-jets.
Stable particles, such as pions or protons in the simulation, retain their respective masses when input to the jet reconstruction algorithm.
Tracks are assigned the pion mass when used as input to the jet algorithm. 
Quality selections are applied in order to ensure that good quality tracks that originate from the reconstructed hard scattering vertex are used to build track-jets. 
The hard scattering vertex is selected as the primary vertex that has the largest \sumPtTrkSq in the event and that contains at least two tracks. 
The selection criteria are:
%
\begin{itemize}
  \item transverse momentum: $\pttrk>0.5$~GeV;
  \item transverse impact parameter: $|d_{0}|<1.0$~mm;
  \item longitudinal impact parameter: $|z_{0}|\times\sin(\theta)<1.0$~mm;
  \item silicon detector hits on tracks: hits in pixel detector $\geq1$ and in the silicon strip detector $\geq6$;
\end{itemize}
%
where the impact parameters are computed with respect to the hard scattering vertex, and $\theta$ is the angle between the track and the beam. 
In the reconstruction of calorimeter jets, calorimeter cells are clustered together using a three-dimensional topological clustering algorithm that includes noise suppression~\cite{TopoClusters}.
The resulting \topos\ are considered as massless four-momenta, such that $E = |\vec{p}\,|$. They are classified as either electromagnetic or hadronic based on their shape, depth and energy density.
In the calibration procedure, corrections are applied to the energy in order to calibrate the clusters to the hadronic scale.

\subsection{Jet quality criteria and selection}
\label{sec:recocalib:cleaning}

All jets in the event reconstructed with the \akt algorithm with $R=0.4$ and a measured $\ptjet>20$~GeV are required to satisfy the \emph{looser} requirements discussed in detail in ref.~\cite{JetCleaning2011}. 
These selections are designed to retain good quality jets while rejecting as large a fraction as possible of those from non-collision beam background and calorimeter noise.
Jets are required to deposit at least 5\% of their measured total energy in the electromagnetic (EM) calorimeter as well as not more than 99\% of their energy in a single calorimeter layer.

To prevent contamination from detector noise, these jet quality criteria are extended by several requirements applied in a specific detector region.
%
%
%
Any event with an \akt\ $R=0.4$ jet with $\ptjet>20$~GeV that fails the above non-collision beam background or noise rejection requirements is removed from the analysis.

\subsection{Jet calibration and systematic uncertainties}
\label{sec:recocalib:calib}

The precision and accuracy of energy measurements made by the calorimeter system are integral to every physics analysis, and the procedures to calibrate jets are described in ref.~\cite{jespaper2010}. The baseline energy scale of the calorimeters derives from the calibration of the electronic signal arising from the energy deposited by electromagnetic showers measured in beam tests, known as the electromagnetic scale. The hadronic calorimeter has been calibrated with electrons, pions, and muons in beam tests and the energy scale has been validated using muons produced by cosmic rays with the detector \insitu in the experimental hall~\cite{TileReadiness}. The invariant mass of the \Z boson in \Zee\ events measured \insitu in the same data sample studied here is used to adjust the calibration for the EM calorimeters.

\subsubsection{Monte Carlo based calibration}
\label{sec:recocalib:mccalib}

The MC hadronic calibration scheme starts from the measured calorimeter energy at the electromagnetic (\EM) energy scale~\cite{ctb2004electronseoverp, ctb2004electrons, LArTB02uniformity, LArTB02linearity, Tile2002,Pinfold:2008zzb, EndcapTBelectronPion2002, LArTB02muons, Atlaselectronpaper}, which correctly measures the energy deposited by electromagnetic showers. 
A local cluster weighting (\LCW) calibration method classifies topological clusters along a continuous scale as being electromagnetic or hadronic, using shower shapes and energy densities.
Energy corrections are applied to hadronic clusters based on this classification scheme, which is derived from single pion MC simulations and tested \insitu using beam tests. These corrections account for the effects of non-compensation, signal losses due to noise suppression and out-of-cluster effects in building \topos, and energy lost in non-instrumented regions of the calorimeters. 
The results shown here use \LCW{} clusters as input to the jet algorithm. 

The final jet energy scale (\JES{}) calibration is derived as a correction relating the calorimeter's response to the true jet energy. 
It can be applied to \EM{} scale jets, with the resulting calibrated jets referred to as \EMJES, or to \LCW{} calibrated jets, with the resulting jets referred to as \LCWJES{} jets. 
More details regarding the evaluation and validation of this approach for standard \antikt\ $R=0.4, 0.6$ jets can be found in ref.~\cite{jespaper2010}.

The \JES{} correction used here for \largeR jets is derived from a \Pythia\ MC sample including \pileup events. There is no explicit offset correction for \pileup contributions, as in the standard \JES{} procedure~\cite{jespaper2010}.
For standard jet algorithms, the dependence of the jet response on the number of primary vertices (\Npv) and the average number of interactions ($\langle\mu\rangle$) is removed by applying a \pileup offset correction to the \EM{} or \LCW{} scale before applying the \JES{} correction. 
However, no explicit \pileup correction is applied to \largeR jets or to jets with the various grooming algorithms applied.

Since one of the primary goals of the use of \largeR and groomed jet algorithms is to 
reconstruct the masses of jets accurately and precisely, a last step is added to the calibration 
procedure of \largeR jets wherein the mass of the jet is calibrated based on the MC simulation of dijet events. An explicit jet mass calibration is important when using the individual invariant jet mass in physics analyses since it is particularly susceptible to soft, wide-angle contributions that do not otherwise significantly impact the jet energy scale. 
The procedure measures the jet mass response for jets built from \LCW{} clusters after the standard \JES{} calibration.  
The mass response is determined from the mean of a Gaussian fit to the core of the distribution of the reconstructed jet mass divided by the corresponding generator-level jet mass.

\begin{figure}[!ht]
  \centering
  \subfigure[\AKTFat, before calibration]{
    \includegraphics[width=0.45\columnwidth]{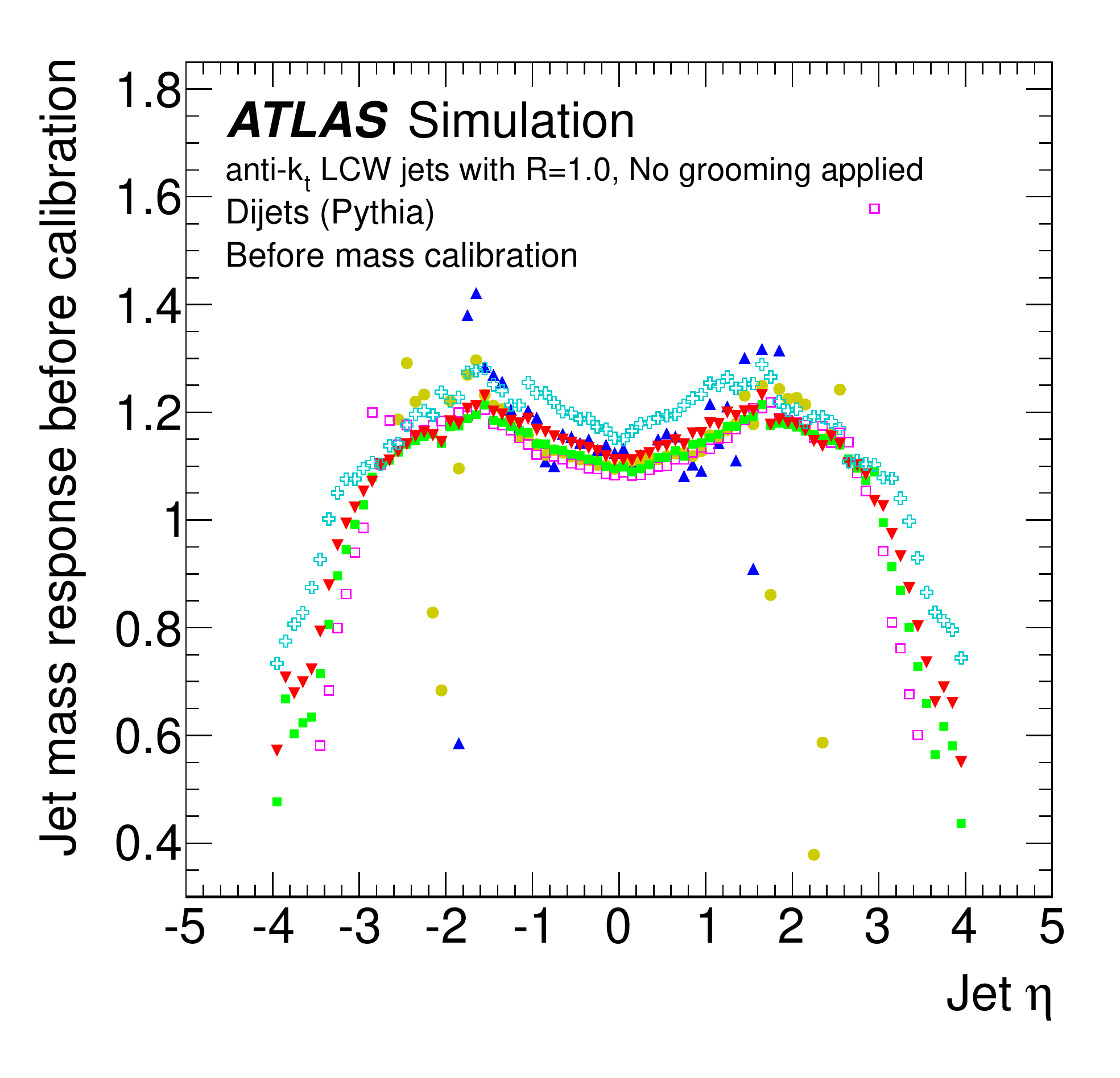}
    \label{fig:jesjms:massscale:massresponse:AKTbefore}}
  \subfigure[\AKTFat, after calibration]{
    \includegraphics[width=0.45\columnwidth]{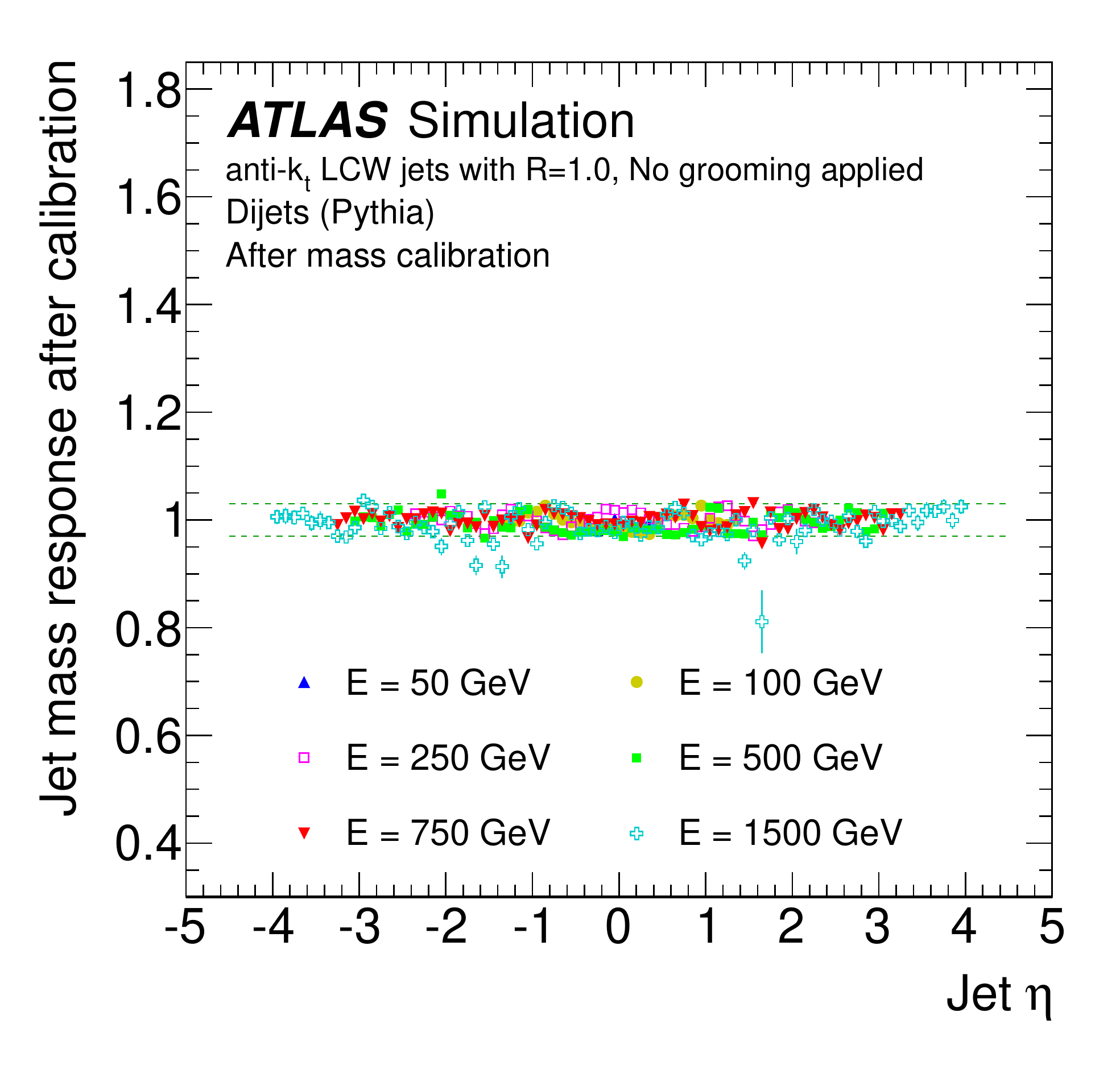}
    \label{fig:jesjms:massscale:massresponse:AKTafter}}
  \caption{Mass response ($m_{\rm reco}/m_{\rm true}$)
           \subref{fig:jesjms:massscale:massresponse:AKTbefore} 
           before and \subref{fig:jesjms:massscale:massresponse:AKTafter} 
           after mass calibration for ungroomed \AKTFat jets. The dotted lines 
           shown in \subref{fig:jesjms:massscale:massresponse:AKTafter}
           represent a $\pm3\%$ envelope on the precision of the final jet 
           mass scale calibration. In each case, the jet energy itself has 
           been calibrated by applying the \JES{} correction.
           }
  \label{fig:jesjms:massscale:massresponse}
\end{figure}

\Figref{jesjms:massscale:massresponse} shows the jet mass response ($m_{\rm reco}/m_{\rm true}$) for several values of jet energy as a function of $\eta$ for \AKTFat jets, before and after calibration to the true jet mass and without jet grooming. In each case, the jet energy itself has been calibrated by applying the \JES{} correction. One can see from this figure that even very high-energy jets near the central part of the detector can have a mean mass scale (or \JMS) differing by up to 20\% from the particle level true jet mass.  
In particular, the reconstructed mass is, on average, greater than that of the particle-level jet due in part to noise and \pileup in the detectors. 
Furthermore, the finite resolution of the detector has a differential impact on the mass response as a function of \eta. 
Following the jet mass calibration, performed also as a function of \eta, a uniform mass response can be restored within 3\% across the full energy and \eta\ range.

\subsubsection{Jet mass scale validation in inclusive jet events using track-jets}
\label{sec:recocalib:massscale}

In order to validate the jet mass measurement made by the calorimeter, calorimeter-jets are compared to track-jets. Track-jets have a different set of systematic uncertainties and allow a reliable determination of the relative systematic uncertainties associated with the calorimeter-based measurement. Performance studies~\cite{dijetdphi} have shown that there is excellent agreement between the measured positions of clusters and tracks in data, indicating no systematic misalignment between the calorimeter and the inner detector.

The use of track-jets reduces or eliminates the impact of additional \pp collisions by requiring the jet inputs (tracks) to come from the hard-scattering vertex. 
The inner detector and the calorimeter have largely uncorrelated instrumental systematic effects, and so a comparison of variables such as jet mass and energy between the two systems allows a separation of physics (correlated) and detector (uncorrelated) effects.  
It is therefore possible to validate the \JES{} and \JMS, and also to estimate directly the \pileup energy contribution to jets. 
This approach was used extensively in the measurement of the jet mass and substructure properties of jets in the 2010 data~\cite{JetMassAndSubstructure} where \pileup was significantly less important and the statistical reach of the measurement was smaller than with the full integrated luminosity of $4.7$\invfb for the 2011 dataset.

The relative uncertainty is determined using the ratio of the transverse momentum of the calorimeter jet, \ptjet, to that of the track-jet, \pTtrkjet. The same procedure is repeated for the jet mass, \Mjet, by using the track-jet mass, \masstrkjet. 
The ratios are defined as

\begin{equation}
  \rtrkjetPt=\frac{\ptjet}{\pTtrkjet}, 
  \qquad 
  \rtrkjetM=\frac{\Mjet}{\masstrkjet},
\end{equation}
\noindent
where the matching between calorimeter and track-jets is performed using a matching criterion of $\DeltaR<0.3$. 
The mean values of these ratios are expected to be well described by the detector simulation if detector effects are well modelled. 
That is to say, even if some underlying physics process is unaccounted for in the simulation, 
as long as this process affects both the track-jet and calorimeter-jet \pt or masses in a similar way, then the 
ratio of data to simulation should be relatively unaffected when averaged over many events.

Double ratios of \rtrkjetM and \rtrkjetPt are constructed in order to evaluate this agreement. 
These double ratios, \RtrkjetPt and \RtrkjetM, are defined as:

\begin{equation}\label{doubleratios}
  \RtrkjetPt = \frac{\rtrkjetPtData}{\rtrkjetPtMC},
  \qquad
  \RtrkjetM = \frac{\rtrkjetMData}{\rtrkjetMMC}.
\end{equation}

The dependence of \RtrkjetPt and \RtrkjetM on \ptjet\ and \Mjet provides a test of the deviation of simulation from data, thus allowing an estimate of the uncertainty associated with the Monte Carlo derived calibration.

\begin{figure}[!ht]
  \centering
  \subfigure[\CAFat]{
    \includegraphics[width=0.44\columnwidth]{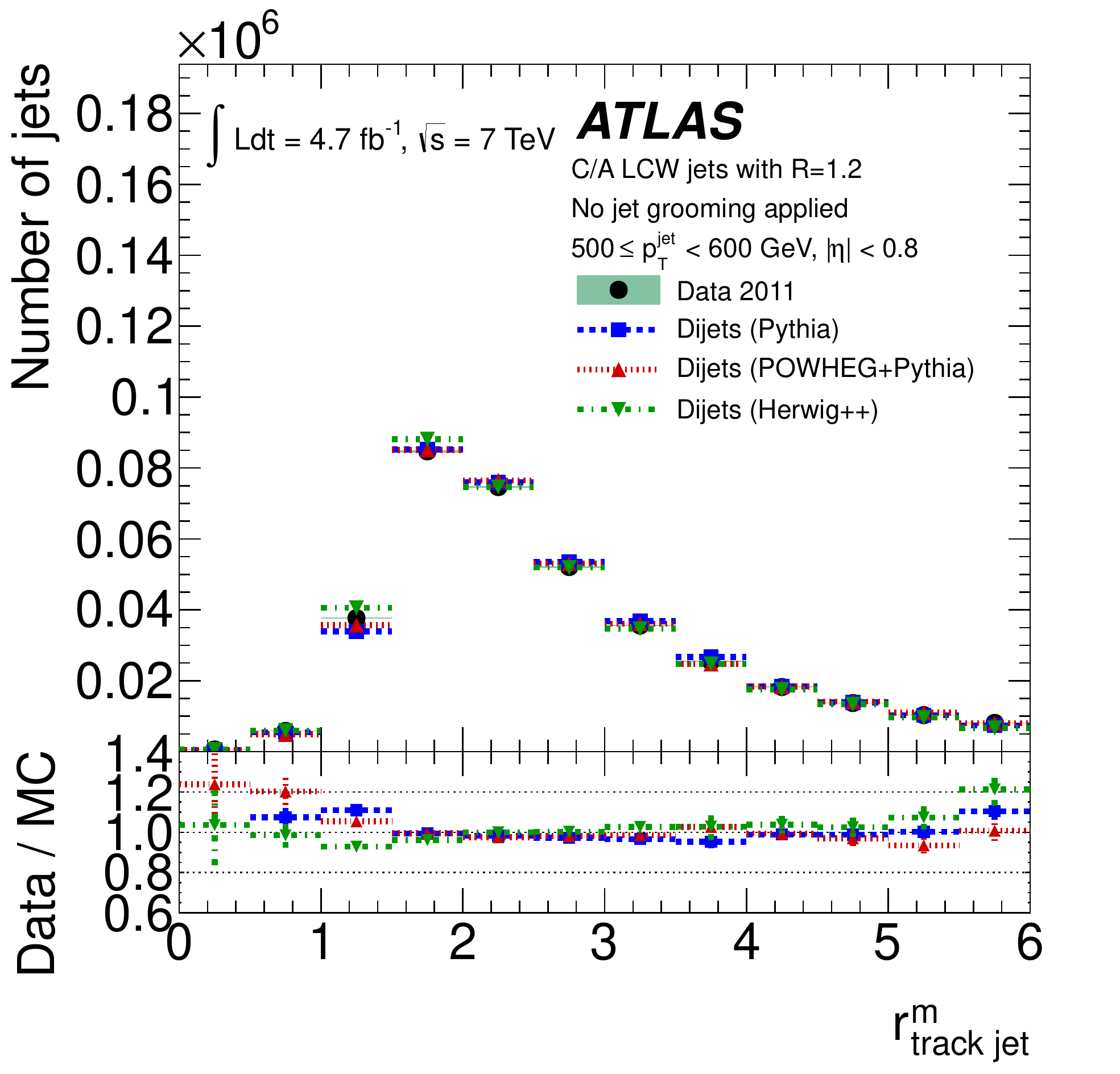}
    \label{fig:jesjms:massscale:ratioProj:CAFat500eta0}}
  \subfigure[\AKTFat (trimmed)]{
    \includegraphics[width=0.44\columnwidth]{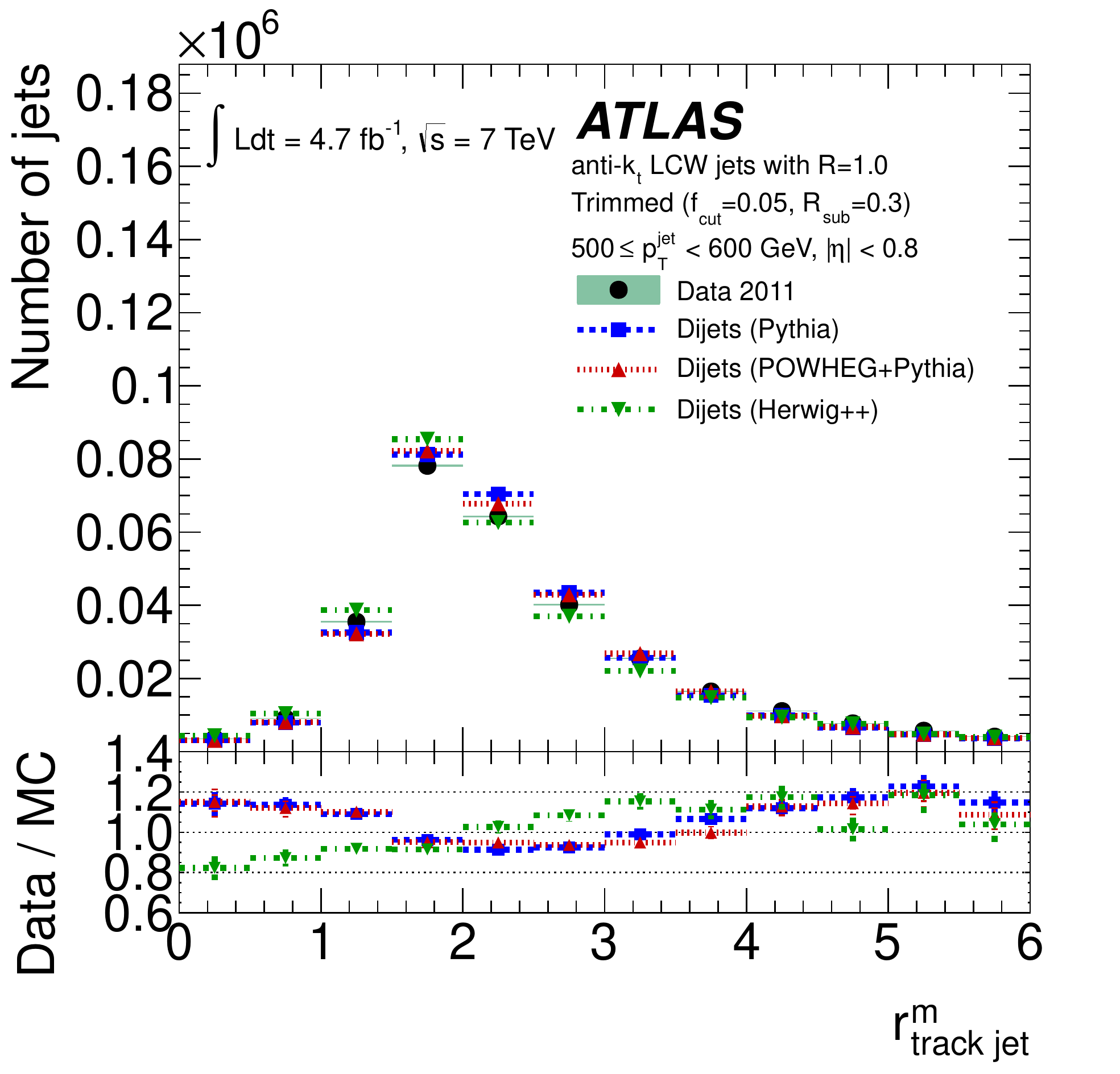}
    \label{fig:jesjms:massscale:ratioProj:AKTTrim500eta0}} 
  \caption{\rtrkjetM distributions for
    \subref{fig:jesjms:massscale:ratioProj:CAFat500eta0} \CAFat jets,
    \subref{fig:jesjms:massscale:ratioProj:AKTTrim500eta0} \AKTFat
    trimmed jets ($\fcut=0.05, \Rsub=0.3$) in the range
    $500\GeV\leq\ptjet<600$~GeV and in the central calorimeter,
    $|\eta|<0.8$. The ratios between data and MC distributions are shown
    in the lower section of each figure. The error bars and bands represent
    the statistical uncertainty only.
  }
  
  \label{fig:jesjms:massscale:ratioProj}
\end{figure}

\Figref{jesjms:massscale:ratioProj} shows the distribution of \rtrkjetM for two jet algorithms and for jets in the range $500\GeV\leq\ptjet<600$~GeV in the central calorimeter region, $|\eta|<0.8$. 
Comparisons between MC simulation and the data are made using \Pythia, \Herwigpp, and \PowPythia, where the distributions are normalized to the number of events observed in the data.
This \ptjet\ range is chosen for illustrative purposes and because of its relevance to searches for boosted vector bosons and top quarks, as the decay products of both are expected to be fully merged into a \largeR jet in this transverse momentum range.
The peak near $\rtrkjetM\approx2$ and the shape of the distribution are both generally well described by the Monte Carlo simulations.
Both the ungroomed and the trimmed \AKTFat distributions show some discrepancies at very low \rtrkjetM, where the description of very soft radiation and hadronization is important, and at high values of \rtrkjetM, above $\rtrkjetM\gtrsim4$. 
The differences are approximately 20\%. 
However, these spectra are used primarily to test the overall scale, so that the important comparison is of the mean values of the distributions, which are quite well described.

The relative systematic uncertainty is first estimated for each MC generator sample as the weighted average absolute deviation of the double ratio, \RtrkjetM, from unity. Measurements of \RtrkjetM are performed in exclusive \ptjet\ and \eta\ ranges. The statistical uncertainty is used as the weight in this case.  The final relative uncertainty is then determined by the maximum of the weighted average deviation among the MC samples considered. Comparisons are made using \Pythia, \Herwigpp, and \PowPythia.

\begin{figure}[!ht]
  \centering
  \subfigure[\CAFat]{
    \includegraphics[width=0.44\columnwidth]{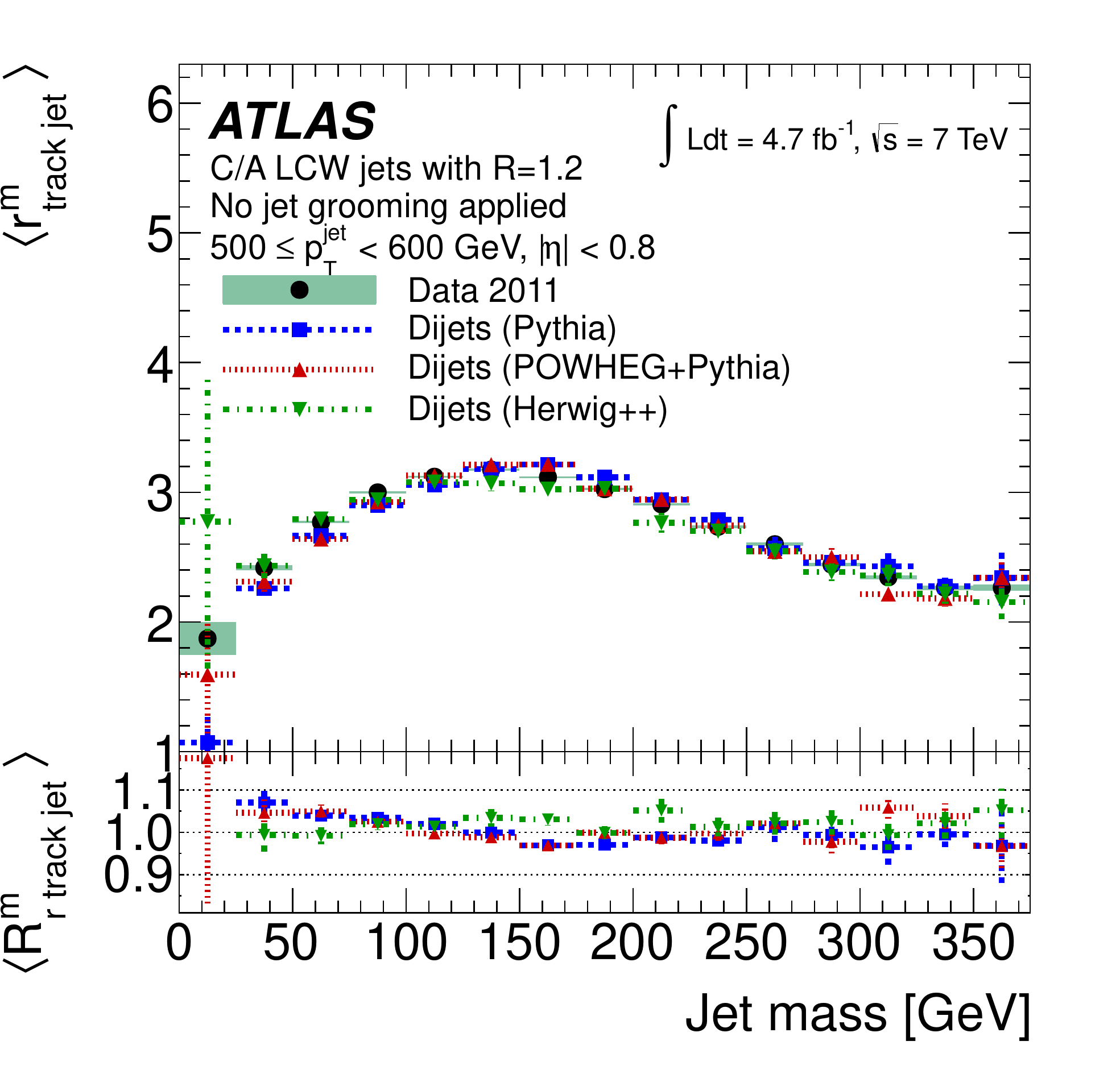}
    \label{fig:jesjms:massscale:ratioVsMass:CAFat500eta0}}
  \subfigure[\AKTFat (trimmed)]{
    \includegraphics[width=0.44\columnwidth]{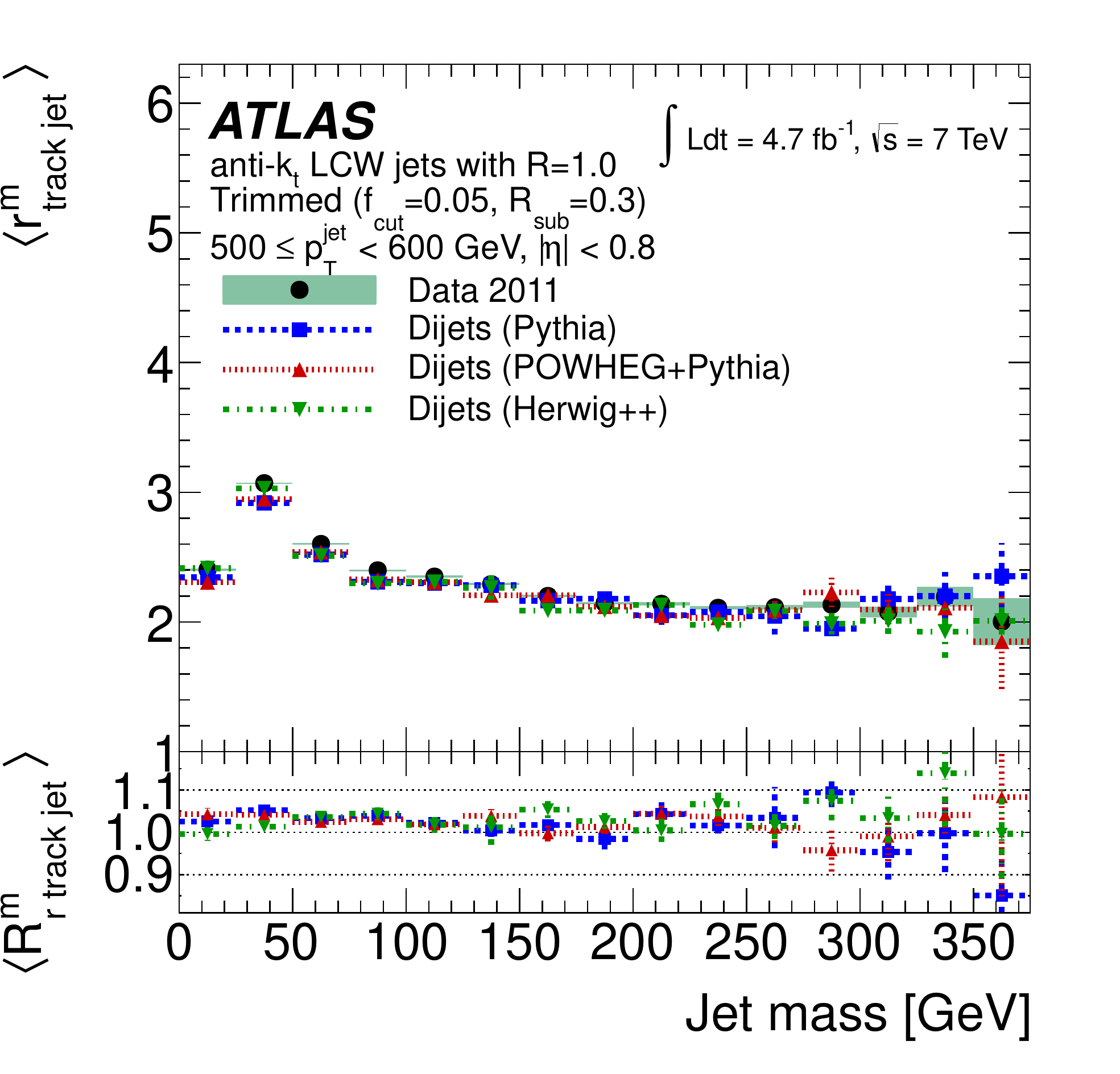}
    \label{fig:jesjms:massscale:ratioVsMass:AKTTrim500eta0}} 

  \caption{Mean values of \rtrkjetM as a function of jet mass for 
      \subref{fig:jesjms:massscale:ratioVsMass:CAFat500eta0} \CAFat jets and
      \subref{fig:jesjms:massscale:ratioVsMass:AKTTrim500eta0} \AKTFat trimmed 
      jets ($\fcut=0.05, \Rsub=0.3$)
      in the range $500\GeV\leq\ptjet<600$~GeV and in the central calorimeter,
      $|\eta|<0.8$. The mean ratios between the data and MC distributions 
      (the double ratios \RtrkjetM) are shown in the lower section of each figure.
      The error bars and bands represent the statistical uncertainty only.
  }
  
  \label{fig:jesjms:massscale:ratioVsMass}
\end{figure}

\Figref{jesjms:massscale:ratioVsMass} presents the distributions of both $\langle \rtrkjetM \rangle$ and the double ratio with respect to MC simulation, $\langle \RtrkjetM \rangle$, for the same algorithms and grooming configurations as shown in \figref{jesjms:massscale:ratioProj}. In the peak of the jet mass distribution, logarithmic soft terms dominate~\cite{TopJetsLHC} and lower-\pT\ particles constitute a large fraction of the calorimeter-jet mass. These particles are bent more by the magnetic field than higher-\pT\ particles, or are not reconstructed as charged-particle tracks, and thus contribute more to the calorimeter-jet mass than the track-jet mass. At much lower calorimeter-jet masses, charged particles can be completely bent out of the jet acceptance, thus reducing the calorimeter-jet mass for a fixed track-jet mass. These effects result in the shape observed in the \rtrkjetM distribution in this region. Higher-mass jets tend to be composed of multiple higher-\pT\ particles that are less affected by the magnetic field and therefore contribute more similarly to the calorimeter-based and track-based mass reconstruction. This results in a flatter and fairly stable \rtrkjetM ratio. This flat \rtrkjetM distribution is present across the mass range for trimmed and mass-drop filtered (not shown) jet masses, as both these algorithms are designed to remove softer particles. Although there is a difference in the phase space of emissions probed at low and high mass, the calorimeter response relative to the tracker response is well modelled by each of the three MC simulations.

The weighted average deviation of \RtrkjetM from unity ranges from approximately 2\% to 4\% for the set of jet algorithms and grooming configurations tested for jets in the range $500\GeV\leq\ptjet<600$~GeV and in the central calorimeter, $|\eta|<0.8$. The results are fairly stable for the slightly less central \eta\ range $0.8\leq|\eta|<1.2$.

\begin{figure}
  \centering
  \subfigure[\AKTFat (no jet grooming)]{
    \includegraphics[width=0.44\columnwidth]{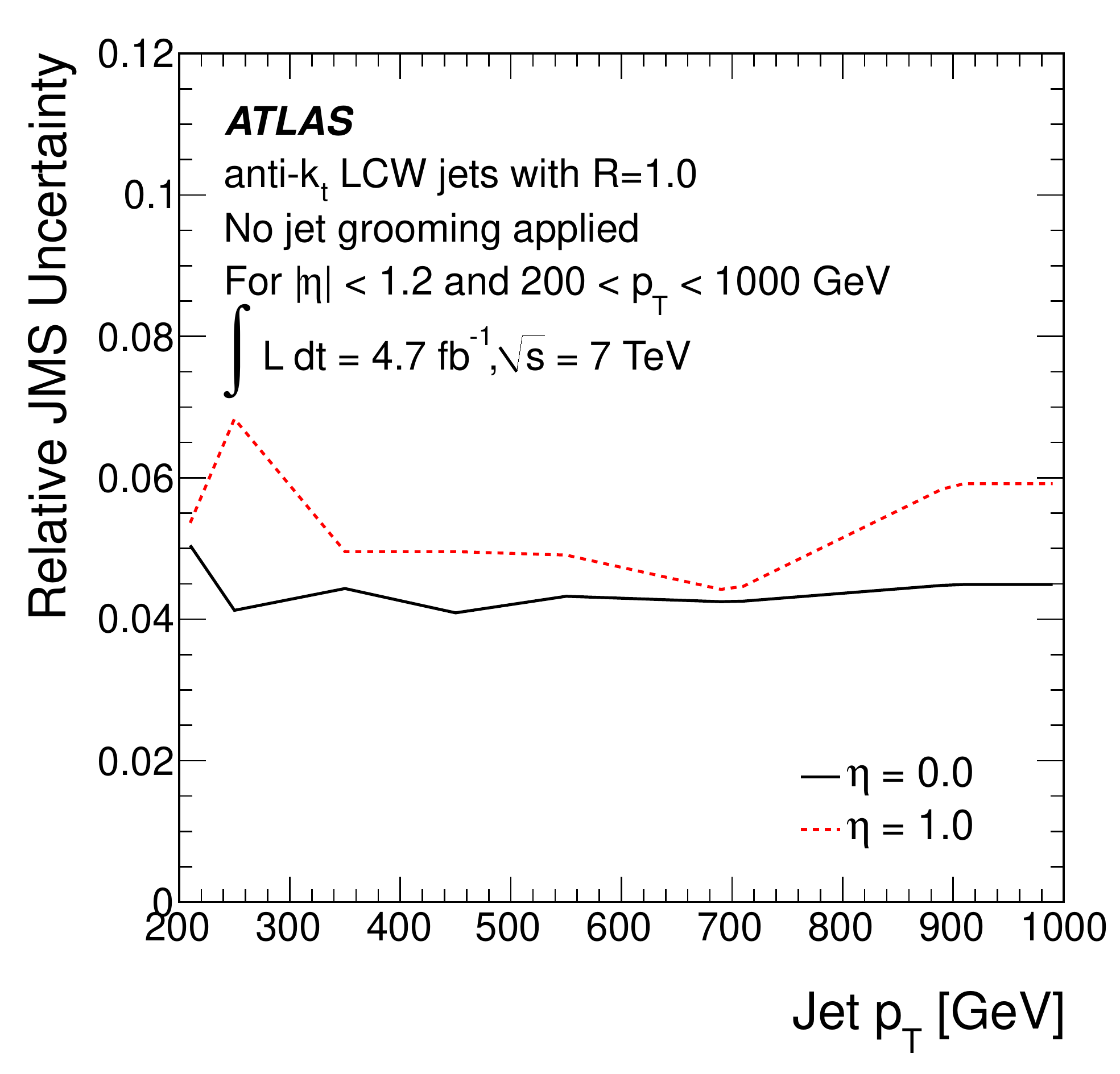}
    \label{fig:jesjms:massscale:AKT}}
  \subfigure[\AKTFat (trimming)]{
    \includegraphics[width=0.44\columnwidth]{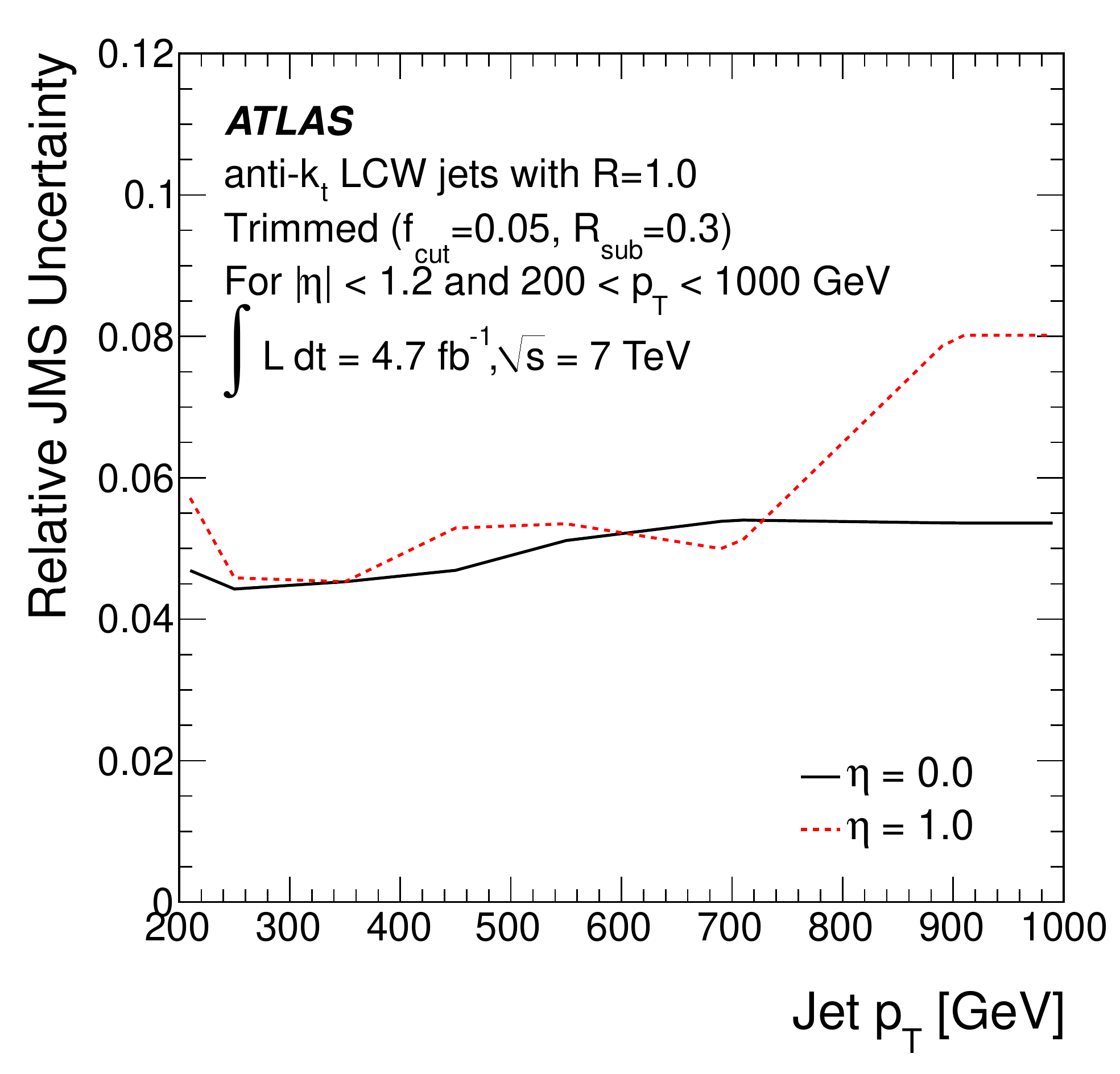}
    \label{fig:jesjms:massscale:AKTtrim}} 
  
  \subfigure[\CAFat (no jet grooming)]{
    \includegraphics[width=0.44\columnwidth]{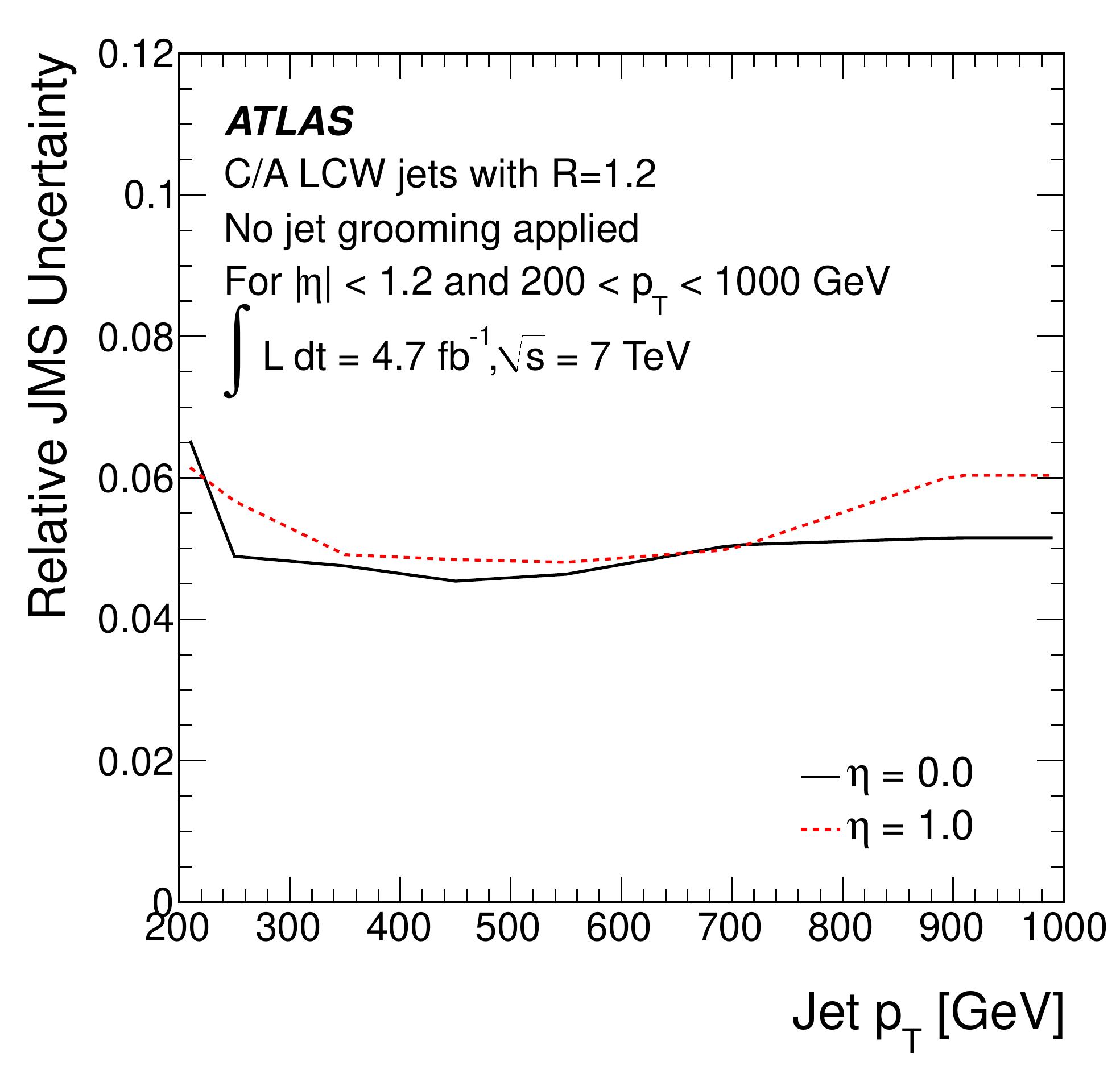}
    \label{fig:jesjms:massscale:CamKt}}
  \subfigure[\CAFat (filtering)]{
    \includegraphics[width=0.44\columnwidth]{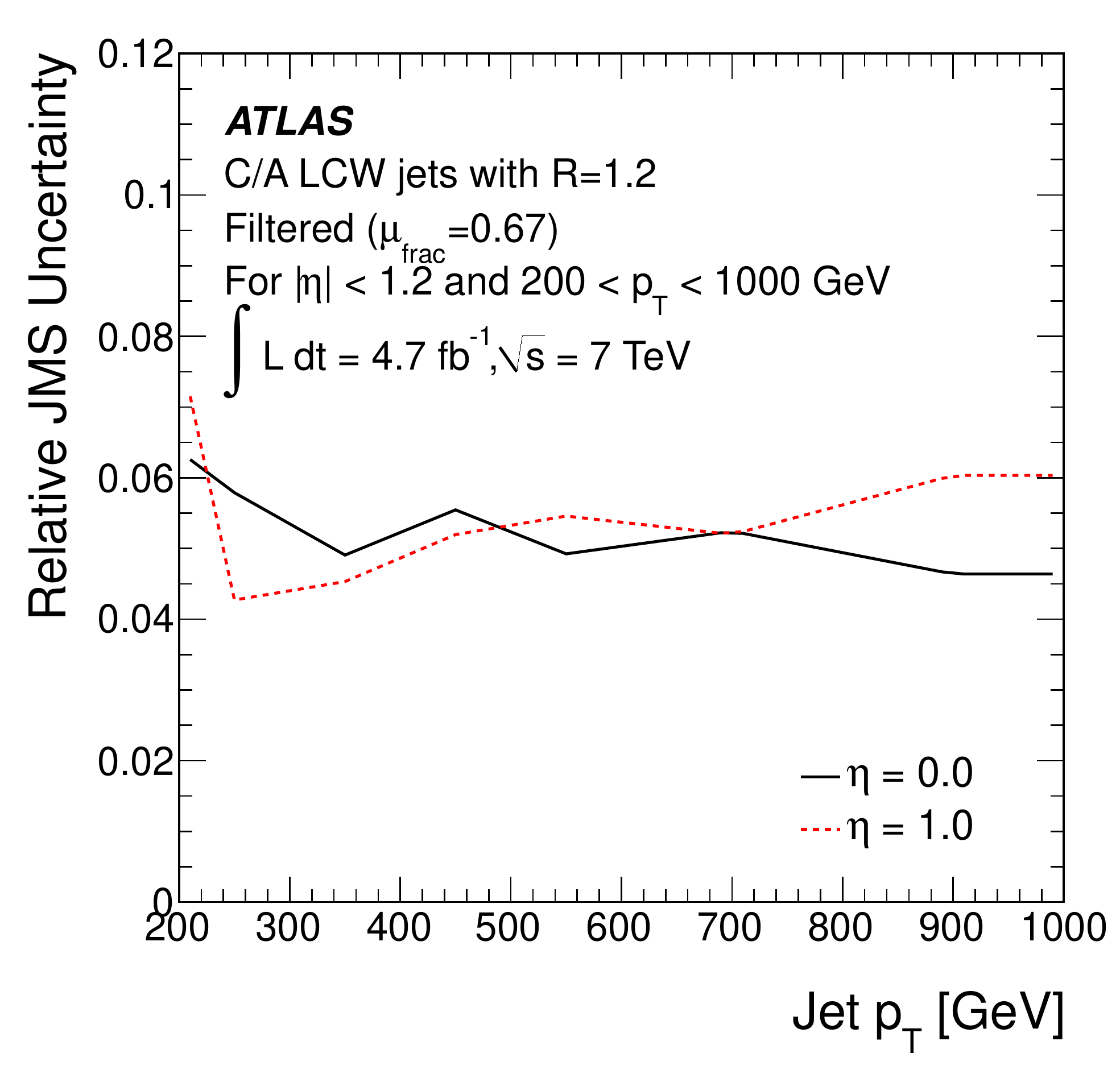}
    \label{fig:jesjms:massscale:CamKtfilt}}
  
  \caption{Summary of the jet mass scale (JMS) relative systematic uncertainties as a function of \ptjet. These uncertainties are determined from track-jet double ratios. For each jet algorithm \subref{fig:jesjms:massscale:AKT} \AKTFat without trimming, \subref{fig:jesjms:massscale:AKTtrim}  \AKTFat with trimming ($\fcut=0.05, \Rsub=0.3$), \subref{fig:jesjms:massscale:CamKt} \CAFat without filtering, and \subref{fig:jesjms:massscale:CamKtfilt} \CAFat with filtering, the two lines shown represent the uncertainty evaluated at $|\eta|=0.0$ (solid) and $|\eta|=1.0$ (dashed). These estimates include a 3\% relative non-closure uncertainty on the MC-based mass scale calibration factors, as well as systematic uncertainties on the impact of the tracking efficiency on the track measurements. 
  }
  
  \label{fig:jesjms:massscale:summary}
\end{figure}

\Figref{jesjms:massscale:summary} presents the full set of jet mass scale systematic uncertainties for various jet algorithms estimated using the calorimeter-to-track-jet double ratios. The total relative uncertainty includes the 3\% uncertainty on the precision of the jet mass scale calibration (see \figref{jesjms:massscale:massresponse:AKTafter}) as well as the uncertainty on the track measurements themselves. The latter uncertainty takes into account the knowledge of tracking inefficiencies and their impact on the \ptjet\ and \massjet measurements using track-jets. Each of these two additional components is assumed to be uncorrelated and added in quadrature with the uncertainty determined solely from the calorimeter-to-track-jet double ratios. The uncertainties are smoothly interpolated between the multiple discrete $\eta$ ranges in which they are estimated. \Figref{jesjms:massscale:summary} shows the uncertainty evaluated at two points $|\eta|=0.0$ (solid) and $|\eta|=1.0$ (dashed) as a function of \ptjet.

The impact of the tracking efficiency systematic uncertainty on \rtrkjetM is evaluated by randomly rejecting tracks used to construct track-jets according to the efficiency uncertainty. This is evaluated as a function of $\eta$ and \massjet for various \ptjet\ ranges. Typically, this results in a 2--3\% shift in the measured track-jet kinematics (both \pT\ and mass) and thus a roughly 1\% contribution to the resulting total uncertainty, since the tracking uncertainty is taken to be uncorrelated to that determined from the double ratios directly. 

The total systematic uncertainty on the jet mass scale is fairly stable near 4--5\% for all jet algorithms up to $\ptjet\approx800$~GeV. At low $\ptjet$, in the range $200\GeV\leq\ptjet<300$~GeV, the average uncertainty for some jet algorithms rises to approximately 5--7\%.
The estimated uncertainty is similar for both the ungroomed and the trimmed or mass-drop filtered jets, except for trimmed \antikt\ jets (see \figref{jesjms:massscale:AKTtrim}) for which the uncertainty in the range $900\GeV\leq\ptjet<1000$~GeV and $|\eta|=1.0$ is approximately 8\%.
%

\subsubsection{Jet mass scale validation using hadronic $W$ decays in \ttbar\ events}
\label{sec:recocalib:wjetsttbar}

An alternative approach to validating the jet mass scale is to study a hadronically decaying particle with a known mass. The most accessible source of hadronically decaying massive particles is events containing top-quark pairs (\ttbar). The \ttbar\ process at the LHC has a relatively large cross-section and the final state contains two \W bosons. About $15\%$ of the time, one of the \W bosons decays to a muon and neutrino, while the other decays to hadrons. This leads to an abundant source of events with a distinctive leptonic signature and a hadronically decaying known heavy particle.

Signal muons are defined as having $\pt > 25$~\GeV\ and $|\eta| < 2.5$, as well as passing a number of quality criteria. Events are required to contain a signal muon and no additional muons or electrons. The missing transverse momentum (\met) is calculated as the negative of the vector sum of the transverse momenta of all physics objects, at the appropriate energy scale, and the transverse momentum of any remaining \topos\ not associated with physics objects in the event. Events are required to have $\met > 25$~\GeV.

In events passing this leptonic selection, a \W candidate is constructed from the signal muon and \met.  In order to reject multi-jet background, this candidate is required to have transverse mass (\mTransv) greater than 40~\GeV.\footnote{Transverse mass \mTransv is defined as $\sqrt{\ET^2 - \pt^2}$ of the vector sum of the four-momentum of the signal muon and \met, assumed to be due to the neutrino.}

Boosted hadronically decaying \W boson candidates are defined as single \largeR jets with $\pt > 200$~\GeV. The hadronic and leptonic \W candidates are required to be separated by at least 1.2 radians in $\phi$ to minimize potential overlap between the decay products of the two \W bosons. In order to further enhance the fraction of top-quark production processes, an \antikt\ jet with $R = 0.4$, $\pt > 20$~\GeV\ and with $\DeltaR > 1.0$ to the hadronic \W candidate is required. This additional jet is also required to be tagged as a $b$-jet by the MV1 algorithm at the 70\% efficient working point~\cite{btagging}.

\begin{figure}[!h]
  \centering
  \subfigure[\CAFat (mass-drop filtered)]{
    \includegraphics[width=0.47\columnwidth]{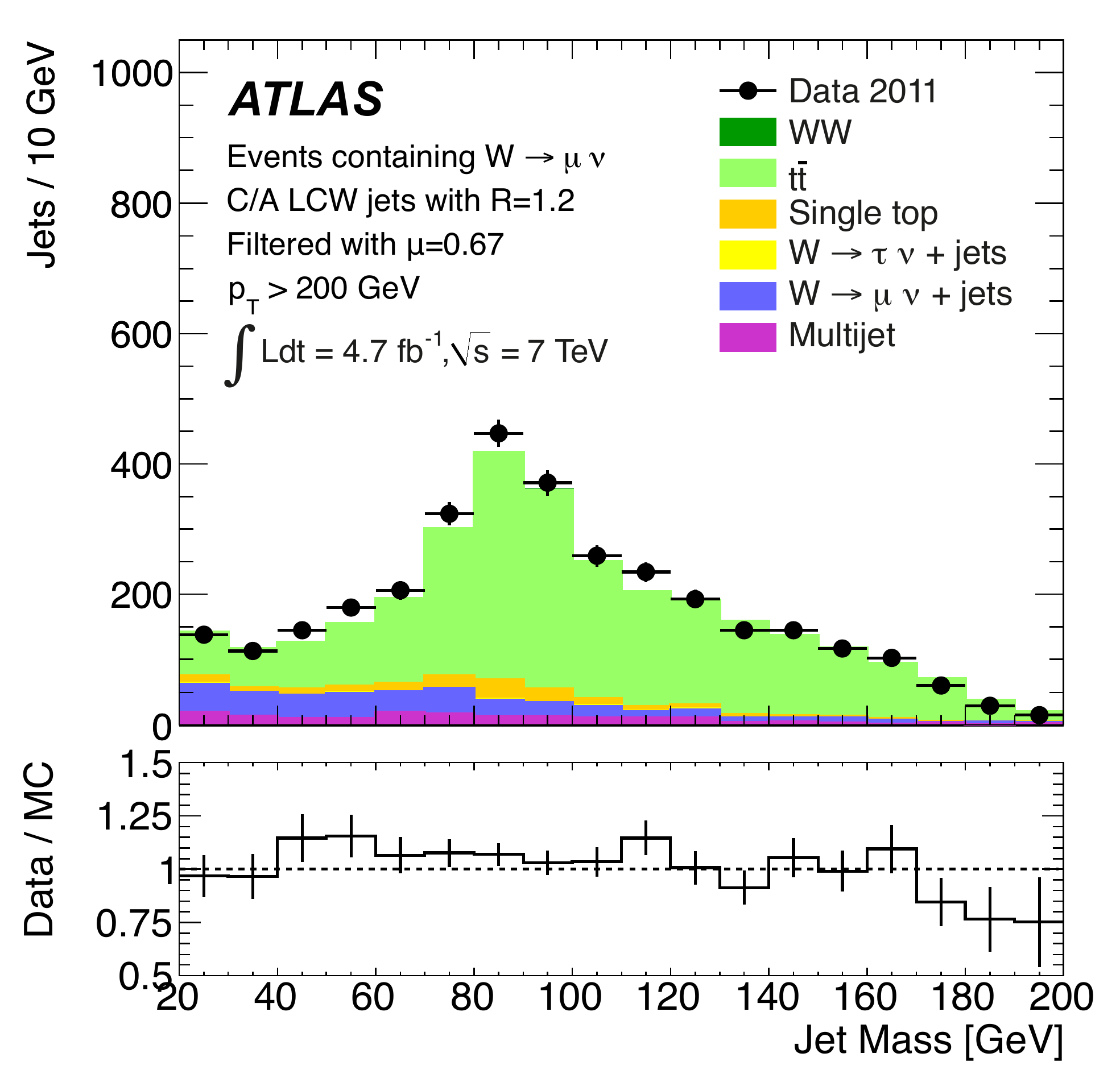}
    \label{fig:wpeakdists:ca}
  }
  \subfigure[\AKTFat (trimmed)]{
    \includegraphics[width=0.47\columnwidth]{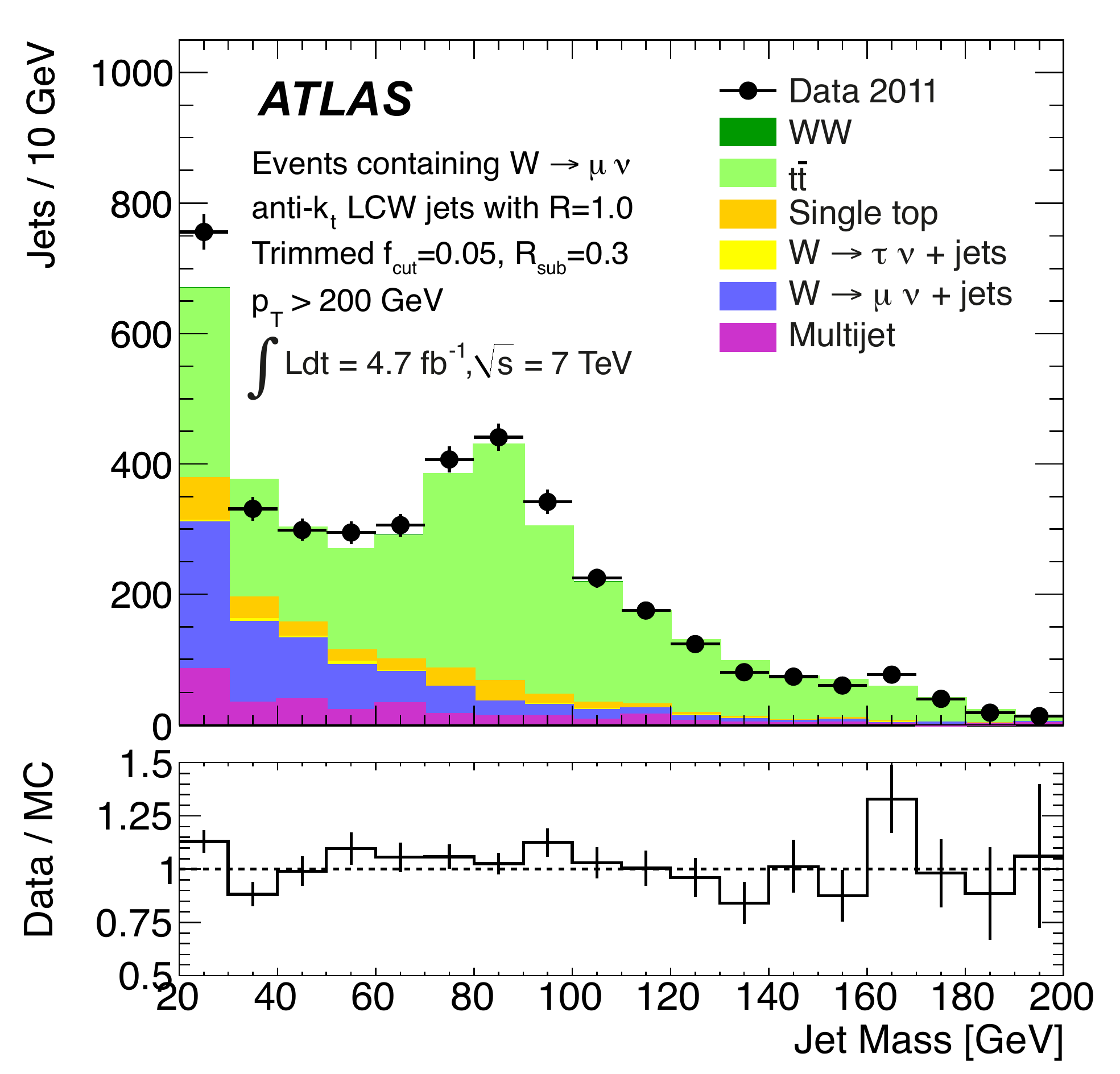}
    \label{fig:wpeakdists:akt}
  }
  \caption{Jet mass distributions for jets with $\pt > 200$~\GeV\ in events 
           containing a $W \rightarrow \mu \nu$ candidate and a $b$-tagged 
           \antikt\ jet. The jets shown are 
           \subref{fig:wpeakdists:ca} \ca\ $R=1.2$ after mass-drop filtering, and 
           \subref{fig:wpeakdists:akt} \antikt\ $R=1.0$ after
           trimming.}   
           
  \label{fig:wpeakdists}
\end{figure}

The resulting mass distributions of hadronic \W candidates for two jet algorithms can be seen in \figref{wpeakdists}. A peak near the \W mass is clearly 
observed for both the mass-drop filtering and trimming algorithms in data and simulated Monte Carlo events.

\begin{figure}[!ht]
  \centering
  \subfigure[Monte Carlo prediction]{
    \includegraphics[width=0.47\columnwidth]{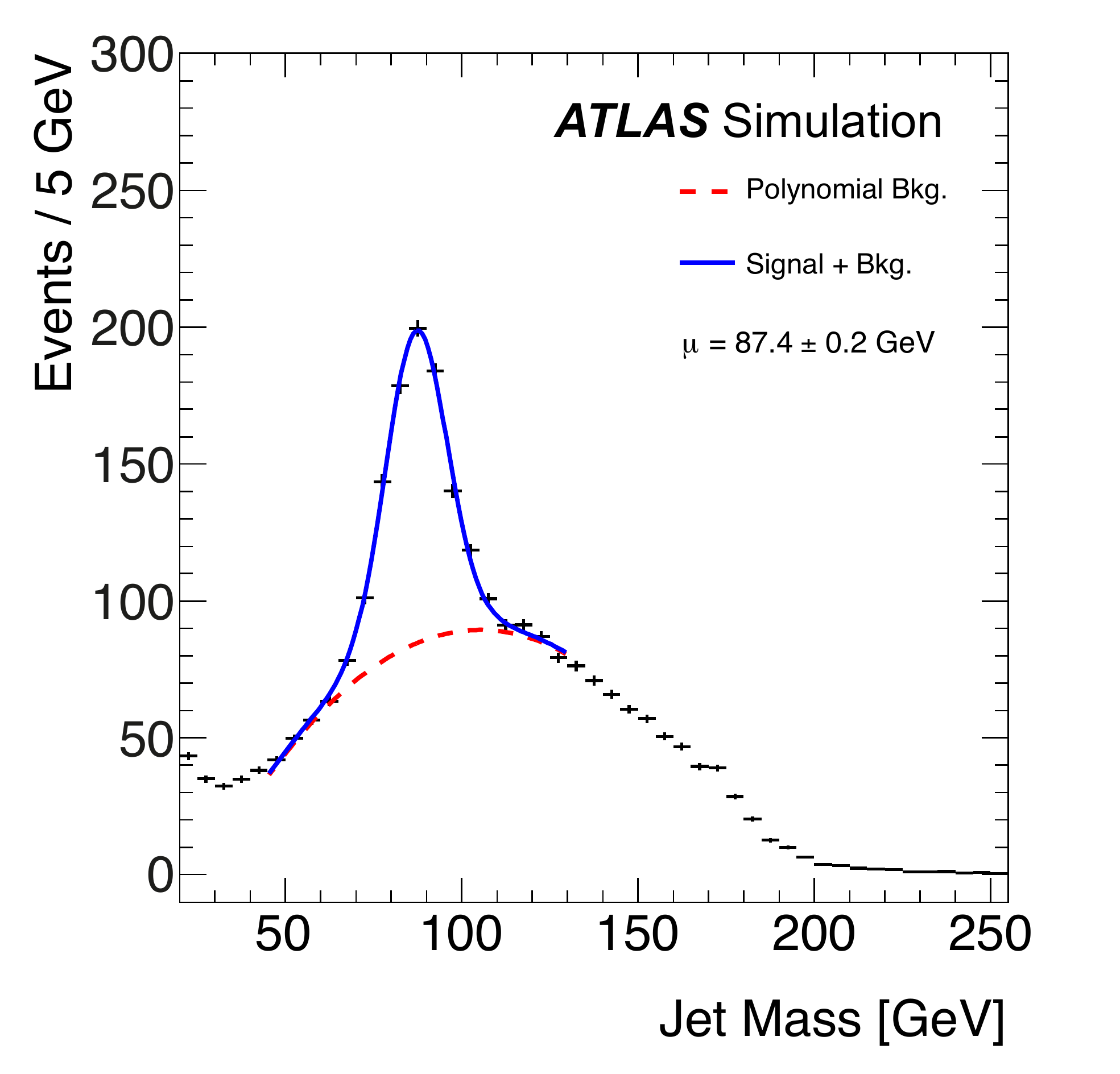}
    \label{fig:wpeakfits:mc}
  }
  \subfigure[Data]{
    \includegraphics[width=0.47\columnwidth]{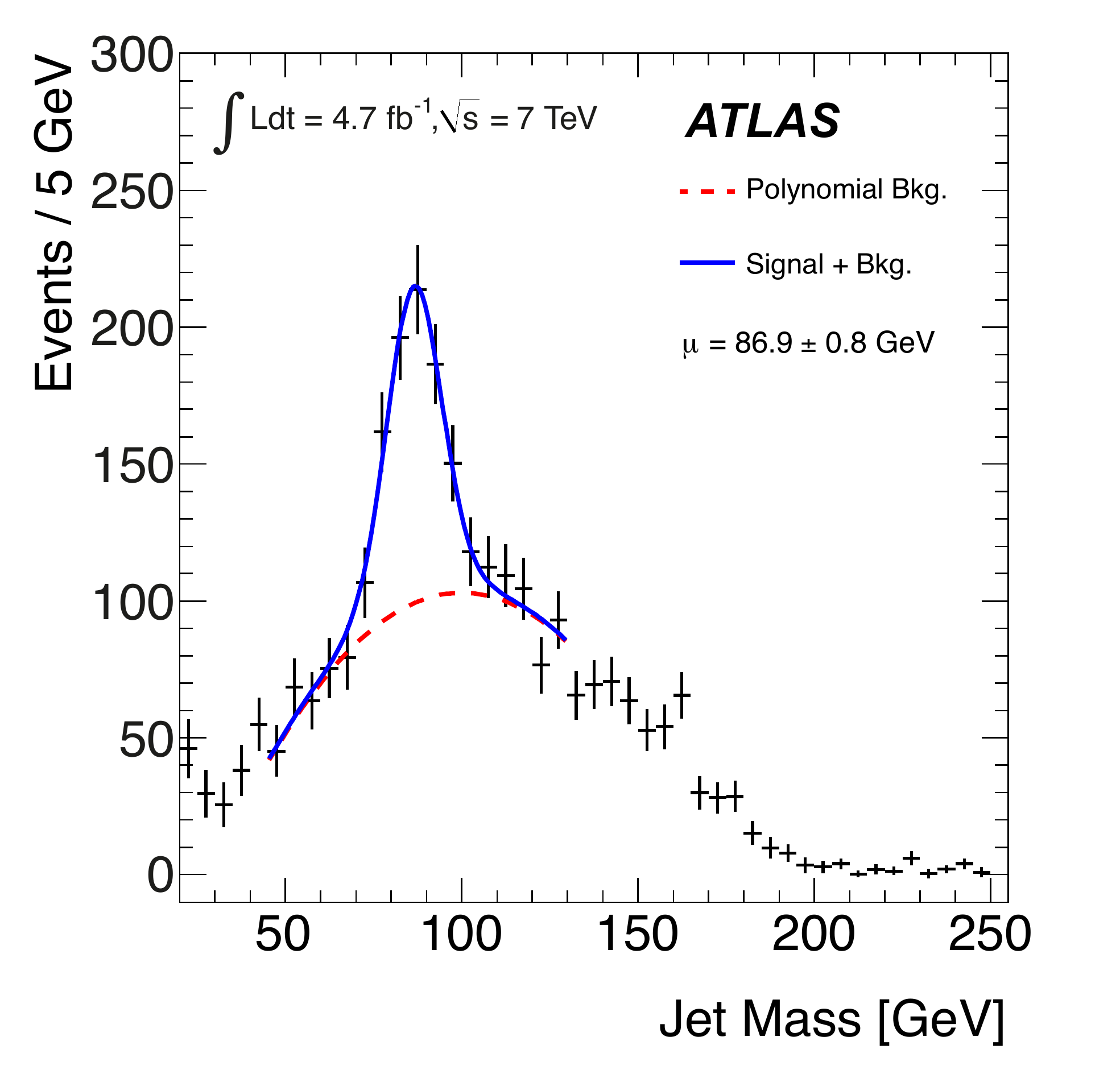}
    \label{fig:wpeakfits:data}
  }
  
  \caption{Results of fitting the mass distributions of \ca\ $R=1.2$ jets after 
           mass-drop filtering. The \Wjets and multi-jet predictions have 
           been subtracted from the distributions. The two figures show 
           \subref{fig:wpeakfits:mc} the Monte Carlo prediction and 
           \subref{fig:wpeakfits:data} data. The dashed line represents the polynomial
           fit to the background shape, while the solid line represents the combined signal
           and background fit.}
           
  \label{fig:wpeakfits}
\end{figure}

A fitting procedure is used to extract the features of the jet mass distribution. The peak produced by the hadronic decays of \W bosons is modelled by a Voigtian function, which is the convolution of Gaussian and Lorentzian functions\footnote{The Lorentzian function is defined as $L(x) = \frac{(\sigma/2\pi)}{(x-x_{0})^{2} + (\sigma/2)^{2}}$, where $x_{0}$ specifies the mean and $\sigma$ specifies the width.}. In this function the width of the Lorentzian component is fixed to the world-average \W boson width. The width of the Gaussian function is a measure of the mass resolution although this is not explored in this study. The background shape has no simple analytic form and is assumed to be modelled by a quadratic polynomial. In order to simplify the background modelling, the \Wjets MC prediction and multi-jet prediction are both subtracted from the data and the resulting distributions are compared to the sum of the \ttbar, single-top quark, and $WW$ MC samples. The results of this fit for \ca\ jets can be seen in \figref{wpeakfits}.

After fitting, the parameter controlling the mean of the Voigtian function ($\mu$) is a measure of the reconstructed mass of hadronically decaying \W bosons. As shown in \figref{wpeakfits}, this scale is observed to be $\mu_{\rm data} = 86.9\pm0.8$~GeV and $\mu_{\rm MC} = 87.4\pm0.2$~GeV. The departure from the world-average value of $\Wmass=80.385\pm0.015$~GeV~\cite{PDG2012} is due to clustering and detector effects, as well as physics effects such as contributions from the UE and hadronization that are not completely removed by the grooming procedures. The ratio ($\mu_{\rm data}/\mu_{\rm MC}$) is therefore the relevant figure of merit for any mismodelling of the reconstructed scale in Monte Carlo simulation. A relative scale difference of $(-0.6 \pm 1.0)\%$ is found for \CamKt jets after mass-drop filtering, whereas the value is $(0.5 \pm 1.2)\%$ for \akt\ jets after trimming.  The uncertainties are statistical, as extracted from the fitting procedure.

\begin{figure}[!ht]
  \centering
  \subfigure[\CAFat (mass-drop filtered)]{
    \includegraphics[width=0.47\columnwidth]{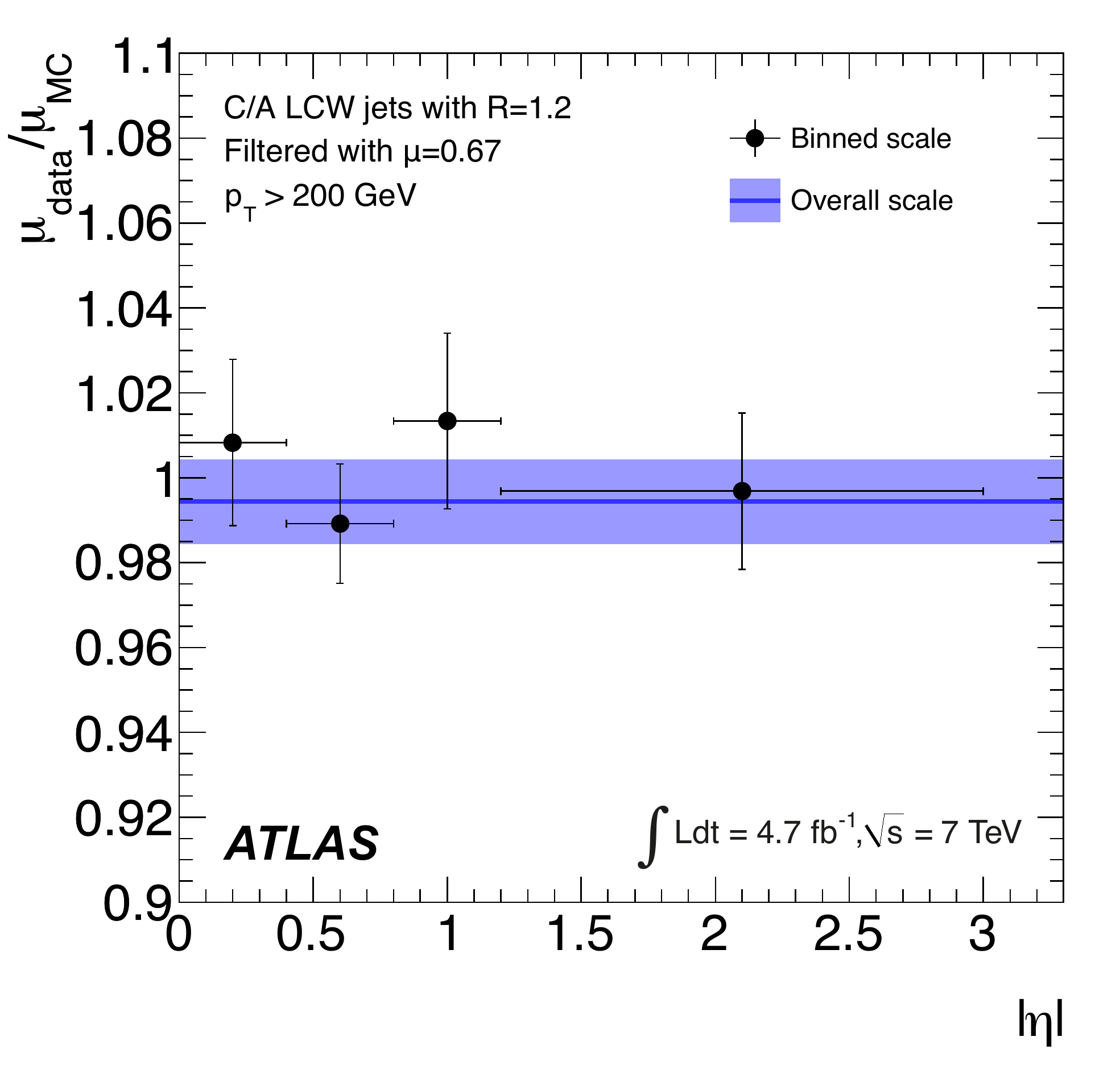}
    \label{fig:wpeaketadep:ca}
  }
  \subfigure[\AKTFat (trimmed)]{
    \includegraphics[width=0.47\columnwidth]{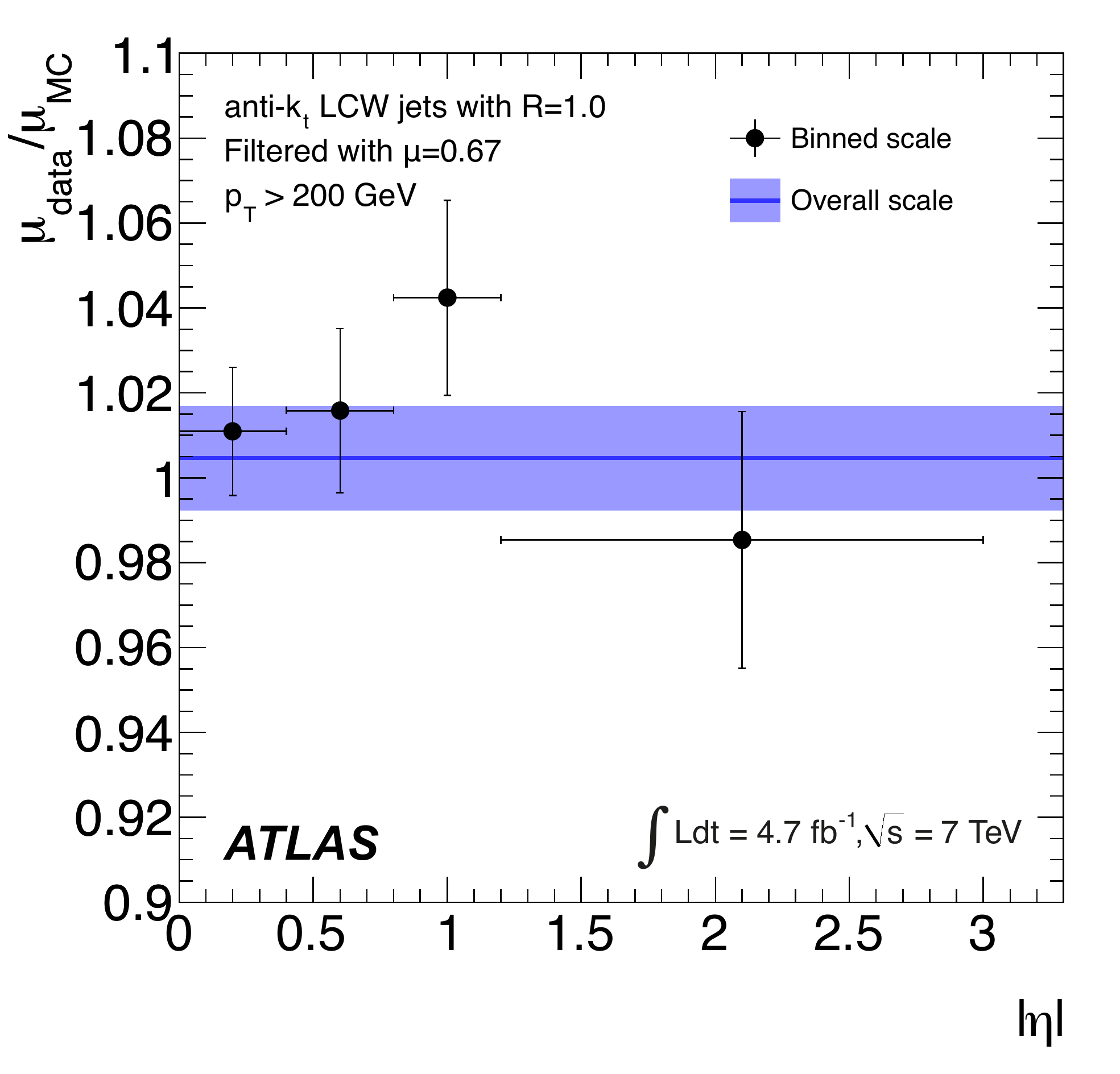}
    \label{fig:wpeaketadep:akt}
  }
  
  \caption{The ratio $\mu_{\rm data}/\mu_{\rm MC}$ as a function of jet $|\eta|$ for 
           \subref{fig:wpeaketadep:ca} \ca\ $R=1.2$ after mass-drop filtering and 
           \subref{fig:wpeaketadep:akt} \antikt\ $R=1.0$ after trimming.}
           
  \label{fig:wpeaketadep}
\end{figure}

It is also possible to repeat the fitting procedure after dividing the data and Monte Carlo simulation samples into bins of $|\eta|$. This allows a test of any potential $|\eta|$ dependence in a range outside that covered by track-jet techniques. The result of these fits can be seen in \figref{wpeaketadep}. The results extracted in individual bins show no statistically significant deviation from the average.

Furthermore, systematic uncertainties are evaluated for this technique. The Monte Carlo modelling is assessed by replacing the \ttbar\ and single-top samples with alternative samples as described in \secref{data-mc}. The impact of mismodelling of the jet resolution is assessed by applying additional smearing. The effect of mismodelling of the \pt\ scale is studied by shifting the scale in the Monte Carlo predictions. Other systematic uncertainties considered are: a possible bias in the fitting procedure; cross-section uncertainties; muon reconstruction uncertainties; energy, resolution and $b$-tagging uncertainties associated with the additional jet; \met\ reconstruction uncertainties. The most significant of these are the uncertainties associated with Monte Carlo modelling and jet resolution.

The final relative scales including all uncertainties are, for \ca\ jets $(-0.6 \pm 1.0 (\mathrm{stat}) ^{+1.9}_{-1.7} (\mathrm{syst}))\%$ and for \antikt\ jets $(+0.5 \pm 1.2 (\mathrm{stat}) ^{+2.7}_{-2.7} (\mathrm{syst}))\%$. There is therefore no observed discrepancy between data and Monte Carlo simulation for jets originating from \W bosons with $\pt \gtrsim 200$~\GeV.

\subsection{\Insitu validation of the subjet energy scale}
\label{sec:recocalib:track2calo}


The trimming and mass-drop filtering procedures both rely heavily on the energy scale of subjets in order to evaluate \fcut or \mufrac, respectively. A similar approach based on double ratios as described in \secref{recocalib:massscale} is used in order to determine the uncertainty on the energy scale of these subjets. Tracks measured by the inner detector are utilized as an independent reference with which to compare calorimeter measurements. Each subjet measured by the calorimeter has a set of tracks associated with it. The following momentum ratio is defined using the calorimeter \pt\ (\subjetpt) and the track \pt\ (\pttrk) for each subjet:

\begin{equation}
  \rtrksubjet=\frac{\sum \pttrk}{\subjetpt}
  \label{eq:ratio_pt_calo2track}
\end{equation}

\begin{figure}[ht]
      \subfigure[Mean energy fraction carried by subjets] { 
        \includegraphics[width=0.48\textwidth, keepaspectratio]{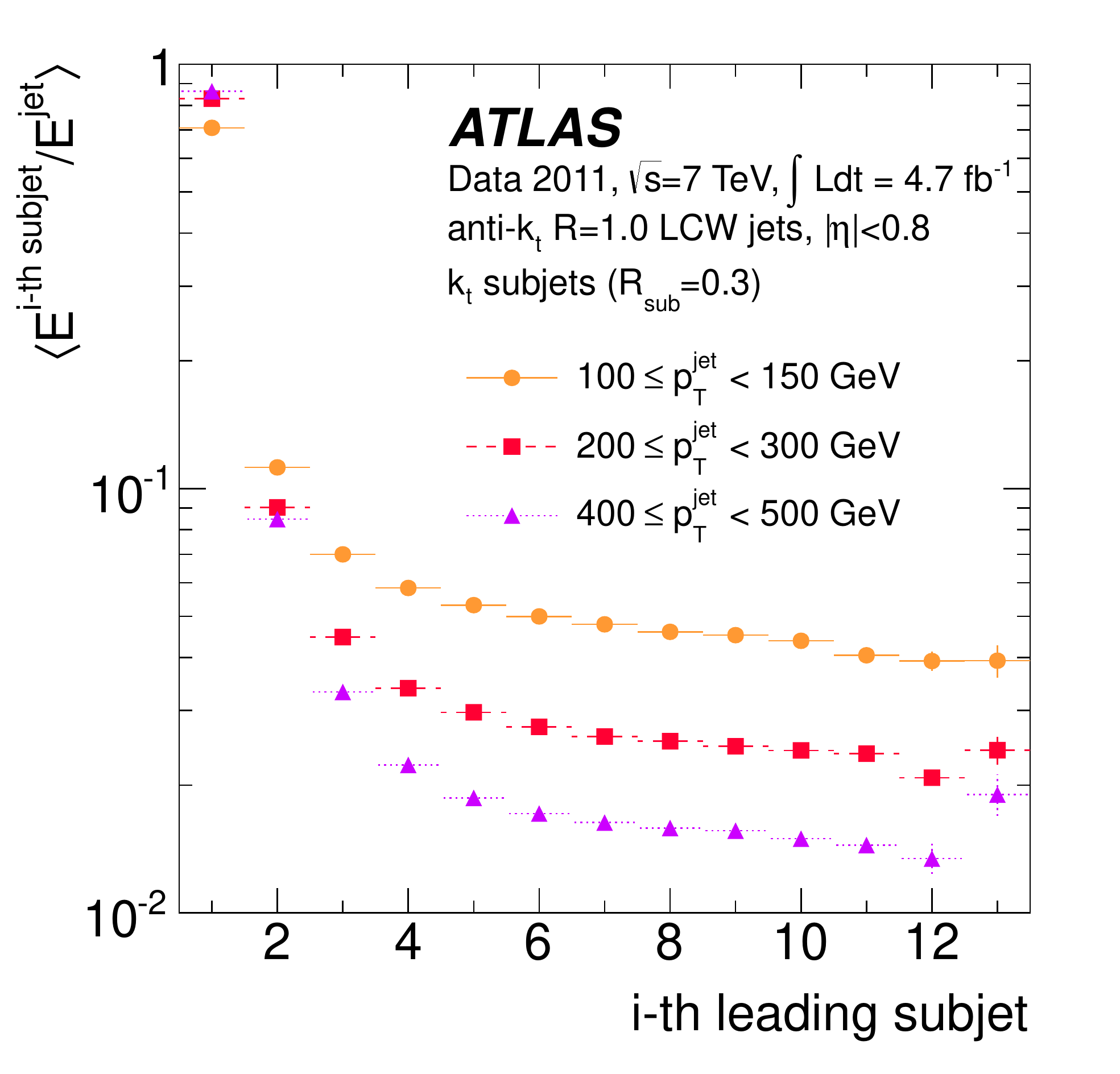} 
        \label{fig:subjets:esharing}
      }
      \subfigure[Mean radial position of subjets within the \parentjet] { 
        \includegraphics[width=0.48\textwidth, keepaspectratio]{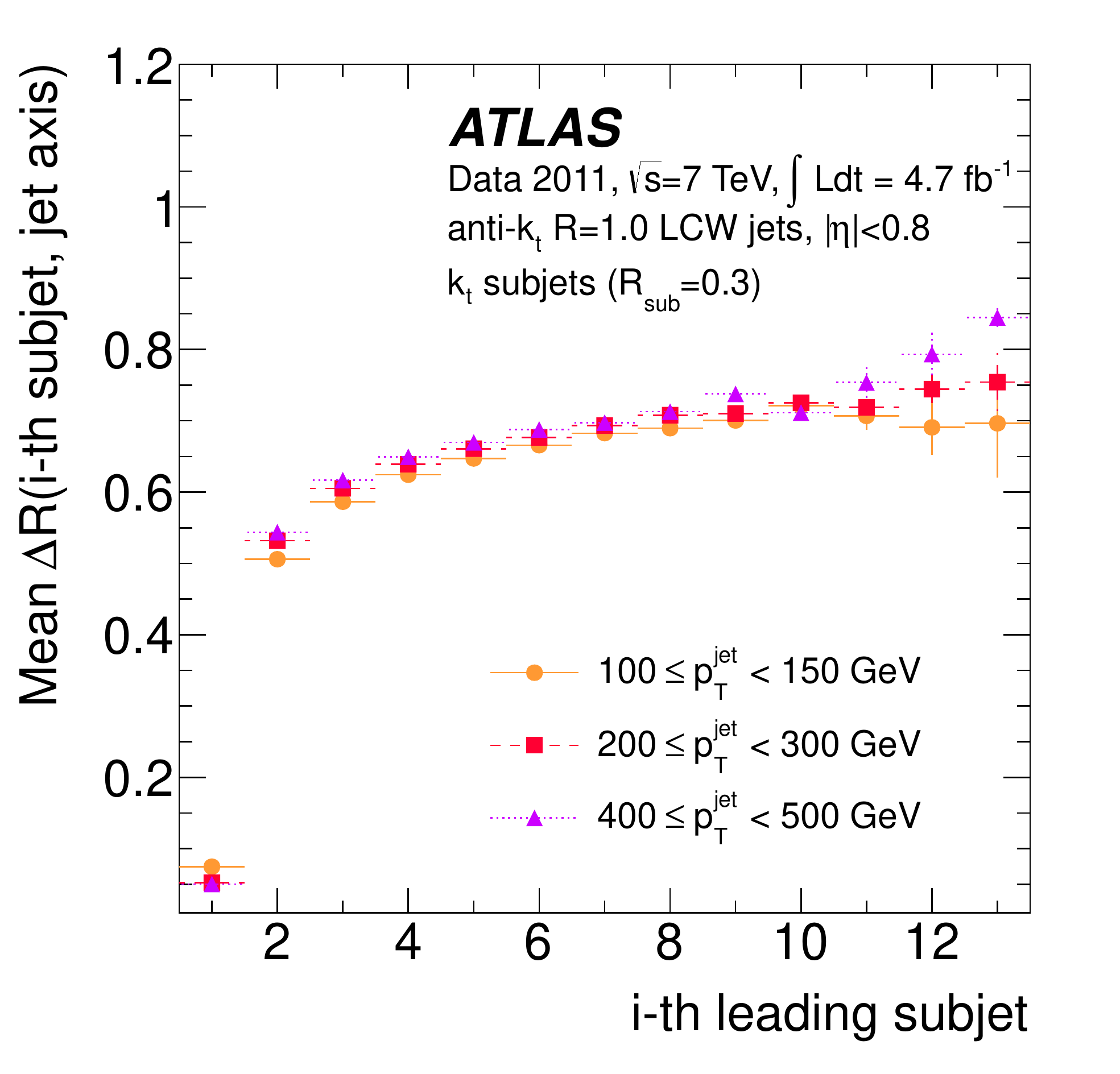} 
        \label{fig:subjets:radialpos}
      }
  
  \caption{\subref{fig:subjets:esharing} Mean subjet energy fraction and \subref{fig:subjets:radialpos} mean distance of subjets to the axis of the \parentjet for different subjets in the jet.} 
  
  \label{fig:subjets}
\end{figure}

It is useful to identify the general features of the subjet structure of \largeR jets in dijet events in order to guide the study of the kinematic properties of subjets. These jets are typically characterized by a highly energetic leading-\subjetpt subjet located close to the \parentjet axis as shown in \figref{subjets}. These leading subjets (with $\Rsub=0.3$) carry a large fraction of the \parentjet energy, and this fraction increases with the \parentjet \pt: $\subjetptl/\ptjet\approx0.71$ for $100\GeV\leq\ptjet< 150$~GeV and $\subjetptl/\ptjet\approx0.86$ for $400\GeV\leq\ptjet< 500$~GeV. The second leading-\subjetpt subjet carries approximately 10\% of the energy of the \parentjet, for jets in the range $100\GeV\leq\ptjet< 500$~GeV. The leading-\subjetpt subjet is located on average at $\DeltaR\leq 0.07$ from the axis of the \parentjet, while less energetic subjets are more distant from the axis of the \parentjet: $\DeltaR\geq 0.5$. These observations are consistent with the increased radial collimation expected in jets produced from gluons and light quarks as the jet \pt\ rises~\cite{Atlasjetshape}. Therefore, the subjet structure of jets from dijet events can be characterized by looking only at the leading and sub-leading subjets. 

Given the crowded nature of the subjet topology within a jet, many standard approaches to studying the subjet energy scale begin to fail. Associating individual tracks with calorimeter subjets may begin to suffer from the proximity of multiple such subjets to a given track, which can in turn impact the measurement of \rtrksubjet. Geometrical matching relies on assumptions that are not fulfilled in high-density environments. This association assumes a cone-like shape and an area of $\pi \Rsub^2$ for all subjets, and that all tracks within this area belong to the subjet. This assumption generally works well in the case of an isolated \antikt\ jet. Conversely, \kt, \CamKt, and even \antikt\ subjets in high-multiplicity environments often have very irregular boundaries and the question of which tracks to associate becomes more difficult to answer. Ghost-association~\cite{Cacciari:2007fd, Cacciari:2008gn} provides a much more appropriate matching of the tracks to the calorimeter subjets for this scenario. In this technique, tracks are treated as infinitesimally soft, low-\pT\ particles by setting their \pT\ to 1~eV. These tracks are then added to the list of inputs for jet finding. The low scale means the tracks do not affect the reconstruction of calorimeter-jets. However, after jet finding, it is possible to identify which tracks are clustered into which subjets. This technique shows a more stable dependence of the ratio \rtrksubjet on the angular separation between subjets. Generally, this approach facilitates the measurement of the effective area of a jet, or the so-called \emph{ghost area}. Instead of tracks, a uniform, fixed density (one per $\Delta y \times \Delta \phi=0.01\times0.01$) of infinitesimally soft particles is distributed within the event and are allowed to participate in the jet clustering algorithm. Instead of identifying tracks associated with the resulting jets, the number of such ghost particles present in the jet after  reconstruction defines the effective area of that jet.

\begin{figure}[ht]
  \begin{center}
  \begin{tabular}{cc}
      \subfigure[Cone associated tracks]
      {\includegraphics[width=0.48\textwidth]{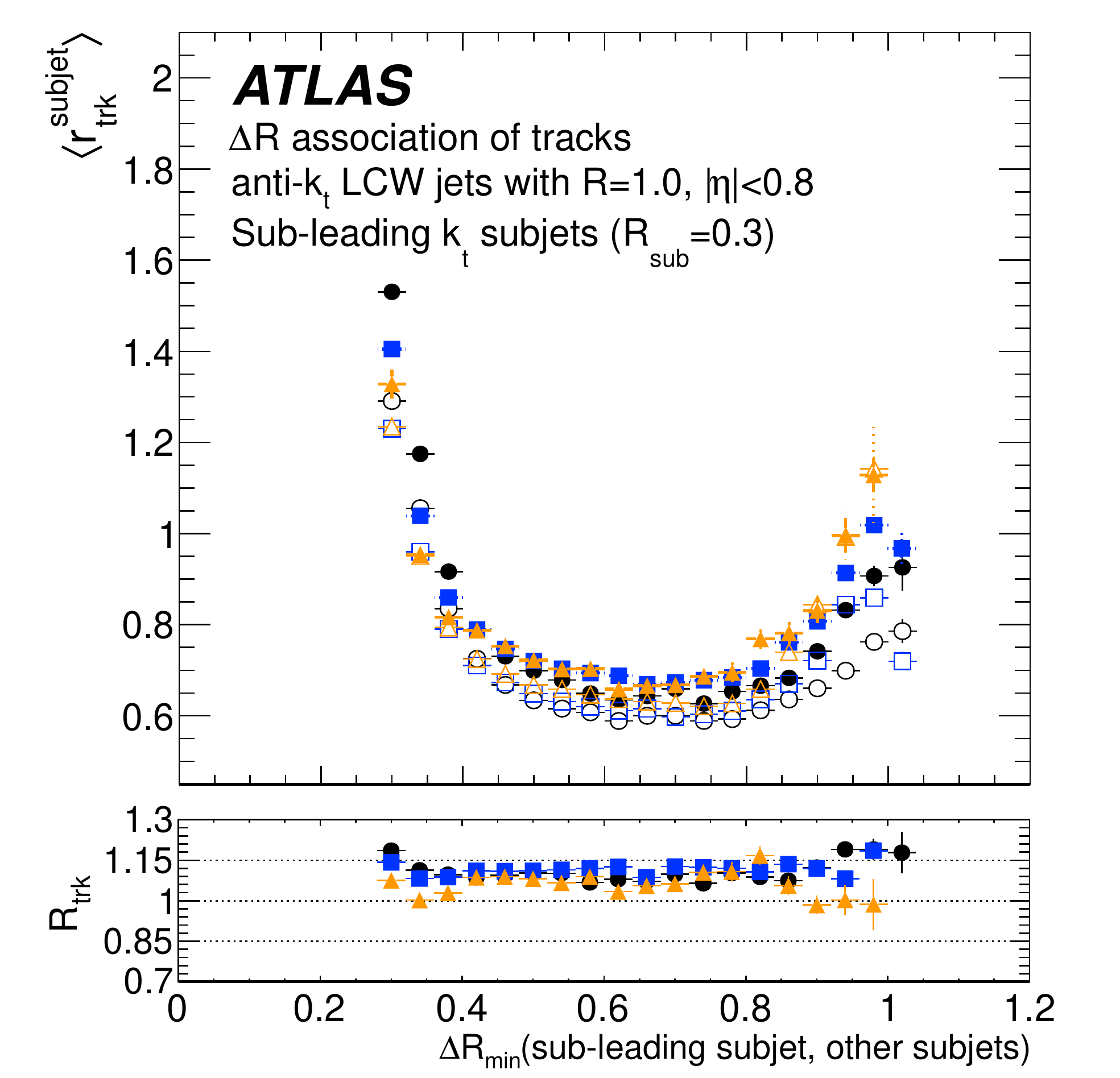}
         \label{fig:subjets-dependence-to-DRmin-cone}
       }
      \subfigure[Ghost associated tracks]
      {\includegraphics[width=0.48\textwidth]{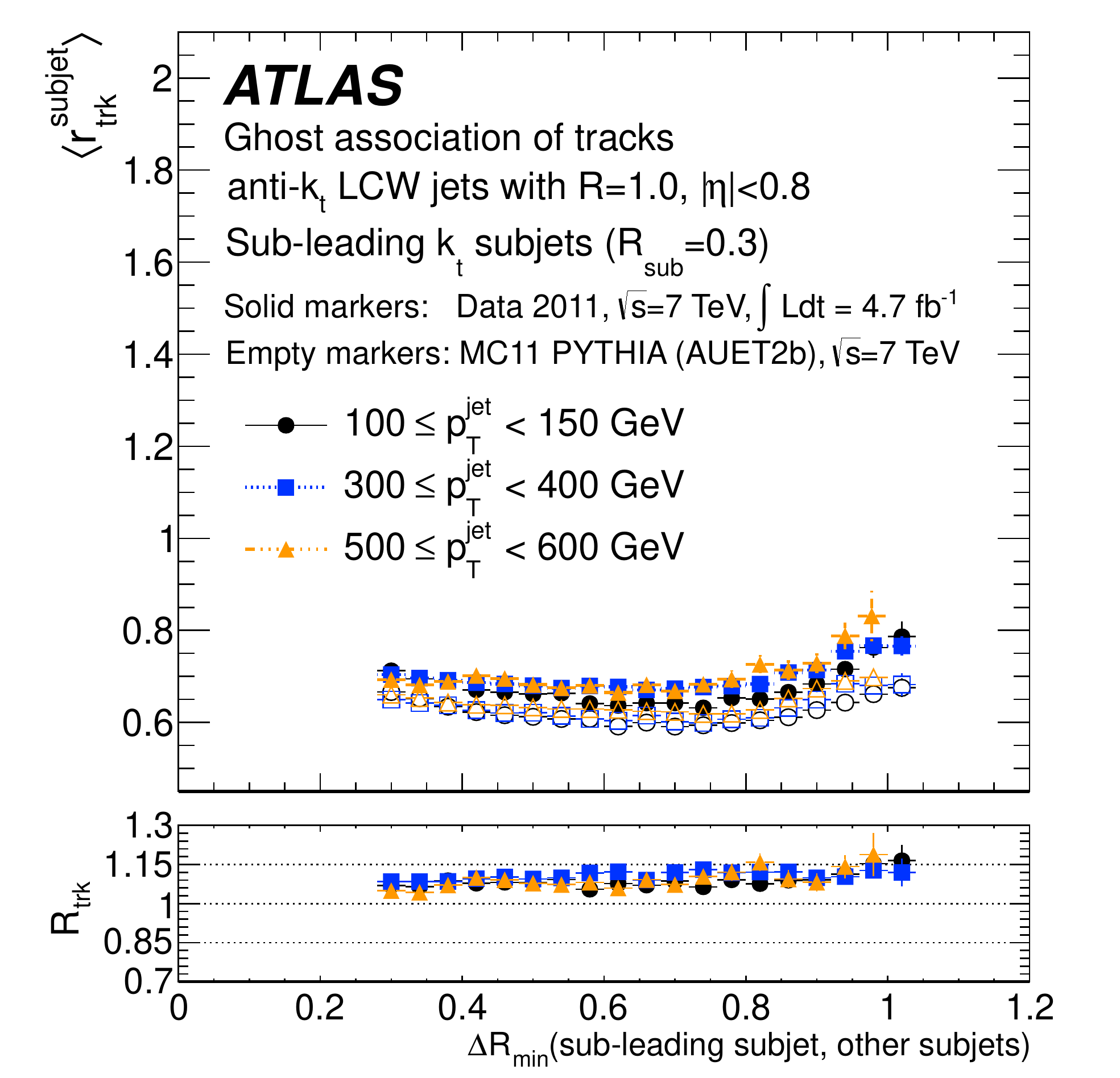}
         \label{fig:subjets-dependence-to-DRmin-ghost}
      } \\
  \end{tabular}
  \end{center}
  \caption{Mean ratio of track \pt to calorimeter \pt (\rtrksubjet) for sub-leading 
           subjets as a function of the distance of the subjet to the nearest 
           subjet within the jet for
           \subref{fig:subjets-dependence-to-DRmin-cone} geometrical association
           and \subref{fig:subjets-dependence-to-DRmin-ghost} ghost association
           techniques.
  } 
  \label{fig:subjets-dependence-to-DRmin}
\end{figure}

\Figref{subjets-dependence-to-DRmin} shows the mean ratio \rtrksubjet as a function of the distance between the subjet and its closest neighbour subjet in the $\eta-\phi$ plane (\DRmin). The numerator of \rtrksubjet corresponds to the total transverse momentum of tracks associated with the subjet using either a standard geometric association (\figref{subjets-dependence-to-DRmin-cone}) or the novel ghost-association technique (\figref{subjets-dependence-to-DRmin-ghost}). For standard geometric track association, a rise in \rtrksubjet is observed for close-by subjets with $\DRmin \leq 2 \times \Rsub$. The impact of the track association scheme is more significant for the second leading subjet, which often has a very energetic subjet nearby (i.e. the leading-\subjetpt one). For these cases geometrical association of tracks fails dramatically, and the ratio \rtrksubjet is a factor of two smaller at $\DRmin= 0.6$ than at $\DRmin= 0.3$, which is approximately the smallest separation between jets. The double ratio $\Rtrksubjet = {\langle \rtrksubjet \rangle}_{\rm data} / {\langle \rtrksubjet \rangle }_{\rm MC}$ provides an estimate of the calibration uncertainty. Any difference is well within $5\%$ for the leading-\subjetpt subjet and $20\%$ for the second leading subjet, independent of the track matching scheme.

\subsection{Calibration of subjets}
\label{sec:recocalib:calibca}


The \htt~\cite{Plehn:2010st} also relies on the energy scale of subjets to analyse the structure of $R=1.5$ jets and to reconstruct a top-quark candidate four-momentum. This section describes a dedicated calibration procedure for \CamKt subjets with a radius parameter $R=0.2-0.5$, which are used in the study of \htt performance. Both the compatibility of the structure with hadronic top-quark decay and the dependence of the reconstructed four-momentum on the calibration are discussed here. 

\CamKt jets -- reconstructed as independent jets and not as constituents of parent jets, as above -- are first calibrated using a simulation of the calorimeter response to jets by comparing the energy and pseudorapidity of a generator-level jet to that of a matched calorimeter jet. Calibration constants obtained from this procedure are then applied to the actual subjets reconstructed by the \htt. For this reason, these \CamKt jets are referred to using the same subjet notation as in the previous section. Prior to calibration, the reconstructed jet energy is lower than the particle jet's energy and is corrected as a function of \pt and in bins of \eta. For example, the correction for \CamKt jets with $R=0.4$ and $|\etajet| < 0.1$ is $+9\%$ at low \pt and up to $+2.5\%$ for $\pt > 500\GeV$. The corrected \pt  matches that of the particle jet to within $2\%$ for all energies and pseudorapidities. For \CamKt $R=0.2$ jets this closure test is $4\%$ at $\pt = 20\GeV$ and better for higher \pt.

\begin{figure}[ht]  
\begin{center}

\subfigure[]{
   \includegraphics[width=0.45\textwidth]{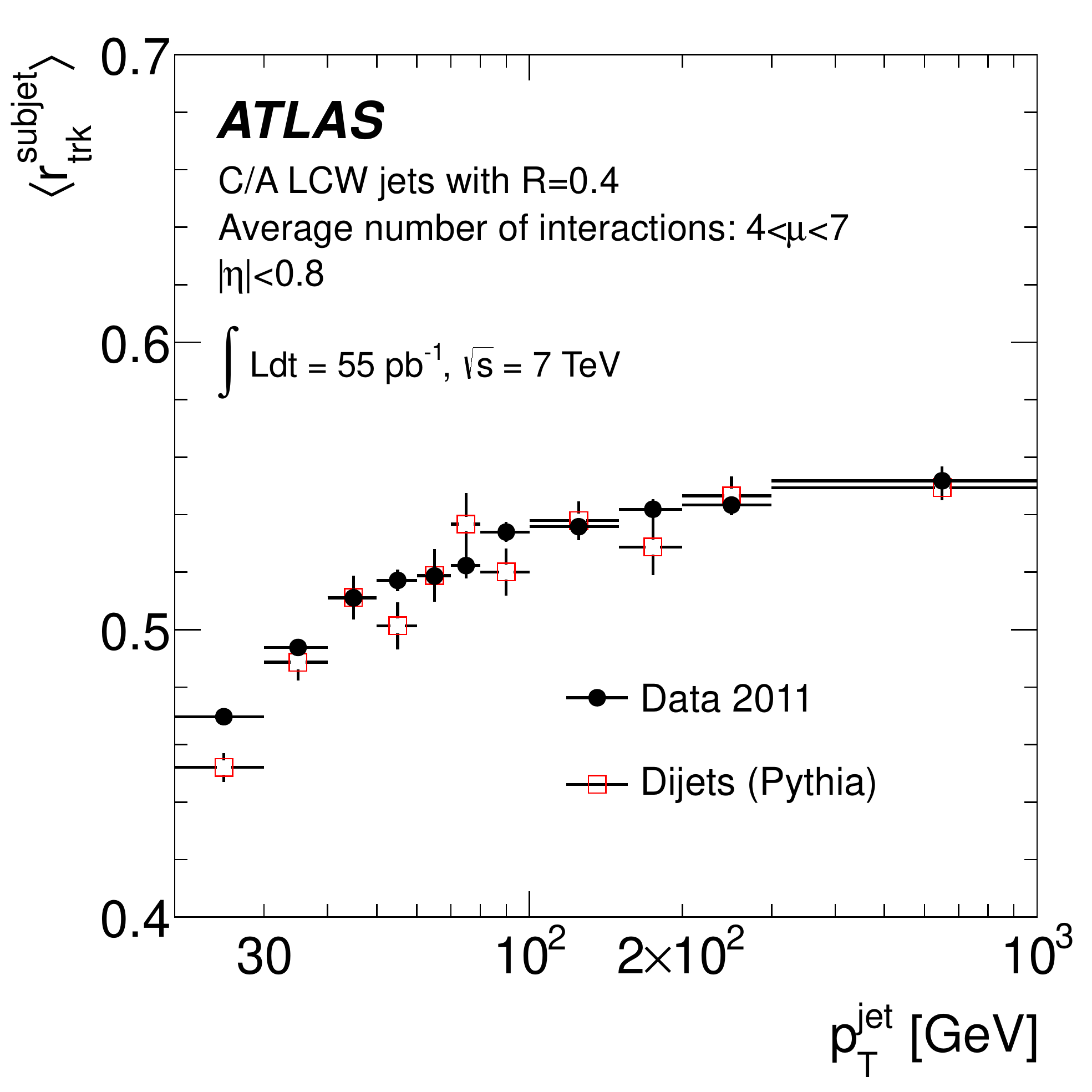} 
   \label{fig:cartrk:rtrkvspt}
}
\subfigure[]{ 
   \includegraphics[width=0.45\textwidth]{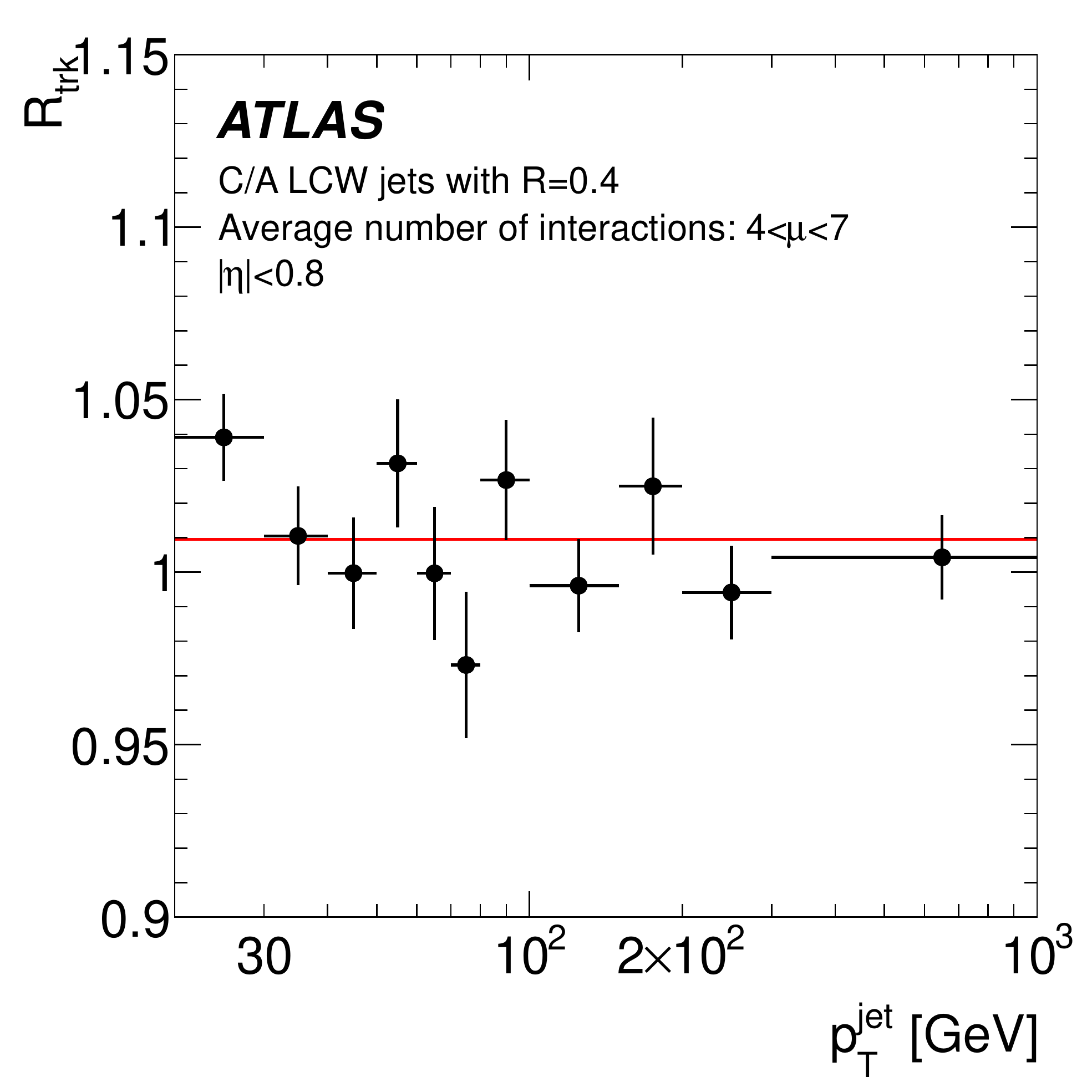} 
   \label{fig:cartrk:rtrkratiovspt}
}

\end{center}
\caption{
    Comparison of the calibrated \ca $R=0.4$ calorimeter
    jet \pt with the \pt of tracks matched to the jet for ATLAS data 
    recorded in 2011 and for dijet simulations with PYTHIA. The average 
    number $\mu$ of interactions per bunch crossing ranges from $4$ to $7$.
    \subref{fig:cartrk:rtrkvspt} The average ratio $\langle \rtrksubjet \rangle$ as a function 
    of the calibrated jet \pt.
    \subref{fig:cartrk:rtrkratiovspt} The double ratio 
    $\Rtrk = {\langle \rtrksubjet \rangle}_{\rm data} / {\langle \rtrksubjet \rangle }_{\rm MC}$.
    The horizontal line indicates the uncertainty-weighted average.}
    
    \label{fig:cartrk}

\end{figure}

Uncertainties on the jet calibration are determined from the quality of the modelling of the calorimeter-jet \pt. The direct ratio $\ptjet({\rm MC})/\ptjet({\rm data})$ is sensitive to mismodelling of jets at the hadron level. To reduce this effect, the calorimeter-jet \pt\ is normalized to the \pt\ of the tracks within the jet. This is done because the uncertainty of the track-jet \pt\ tends to be small compared to the calorimeter-jet \pt\ in the kinematic regime considered here. Tracks are matched to calorimeter-jets using ghost-track association. The jets are required to be within $|\etajet| < 2.1$ to ensure coverage of the associated tracks by the tracking detector.

The average \rtrksubjet for a subset of the data characterized by an average number of interactions per bunch crossing in the range $4 < \mu < 7$ is shown in \figref{cartrk:rtrkvspt} as a function of \ptjet for both data and simulation for \CamKt $R=0.4$ jets with $|\etajet| < 0.8$. The double ratio $\Rtrk = {\langle \rtrksubjet \rangle}_{\rm data} / {\langle \rtrksubjet \rangle }_{\rm MC}$ is shown in \figref{cartrk:rtrkratiovspt}. Deviations of \Rtrk from unity serve as an estimate of the uncertainty of the MC calibrated calorimeter-jet \pt. The largest deviation from unity is seen at low \pt and is $4\%$ with a statistical uncertainty of $1\%$. The statistical uncertainty-weighted average double ratio is indicated by the horizontal line.

Similar results are obtained when varying the jet radius parameter between $0.2$ and $0.5$ and for higher \pileup conditions (evaluated using a subset of the data characterized by $13<\mu<15$). A jet energy uncertainty of $3.5\%$ is assigned.

The imperfect knowledge of the material distribution in the tracking detector constitutes the dominating systematic uncertainty. It results in an additional uncertainty in \Rtrk of $\approx 2\%$ for $|\etajet|<1.4$ and $\approx 3\%$ for $1.4<|\etajet|<2.1$, although it does not introduce a measurable shift.

The jet \pt systematic uncertainty is taken to be the absolute deviation of the central weighted-average \Rtrk from unity, with the shifts introduced by the systematic variations added in quadrature.  The uncertainty varies between $2.3\%$ and $6.8\%$, depending on the jet \pt, \eta, and the jet radius parameter. The uncertainty has been determined independently in samples with low and high \pileup $\mu$-values and no significant difference has been found.

\begin{figure}[ht]  
\begin{center}

\subfigure[]{
   \includegraphics[width=0.45\textwidth]{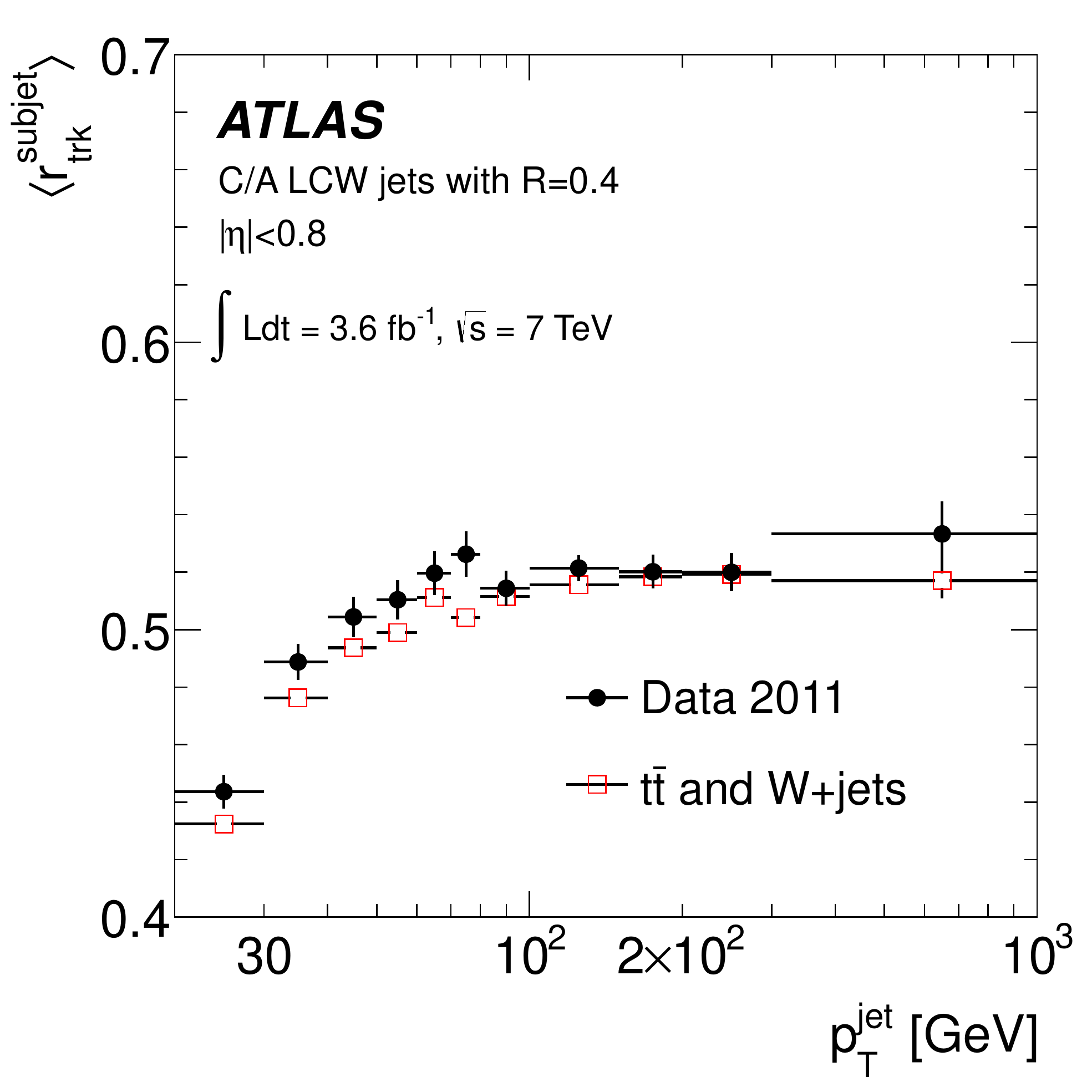} 
   \label{fig:cartrk:rtrkvspt:top}
}
\subfigure[]{ 
   \includegraphics[width=0.45\textwidth]{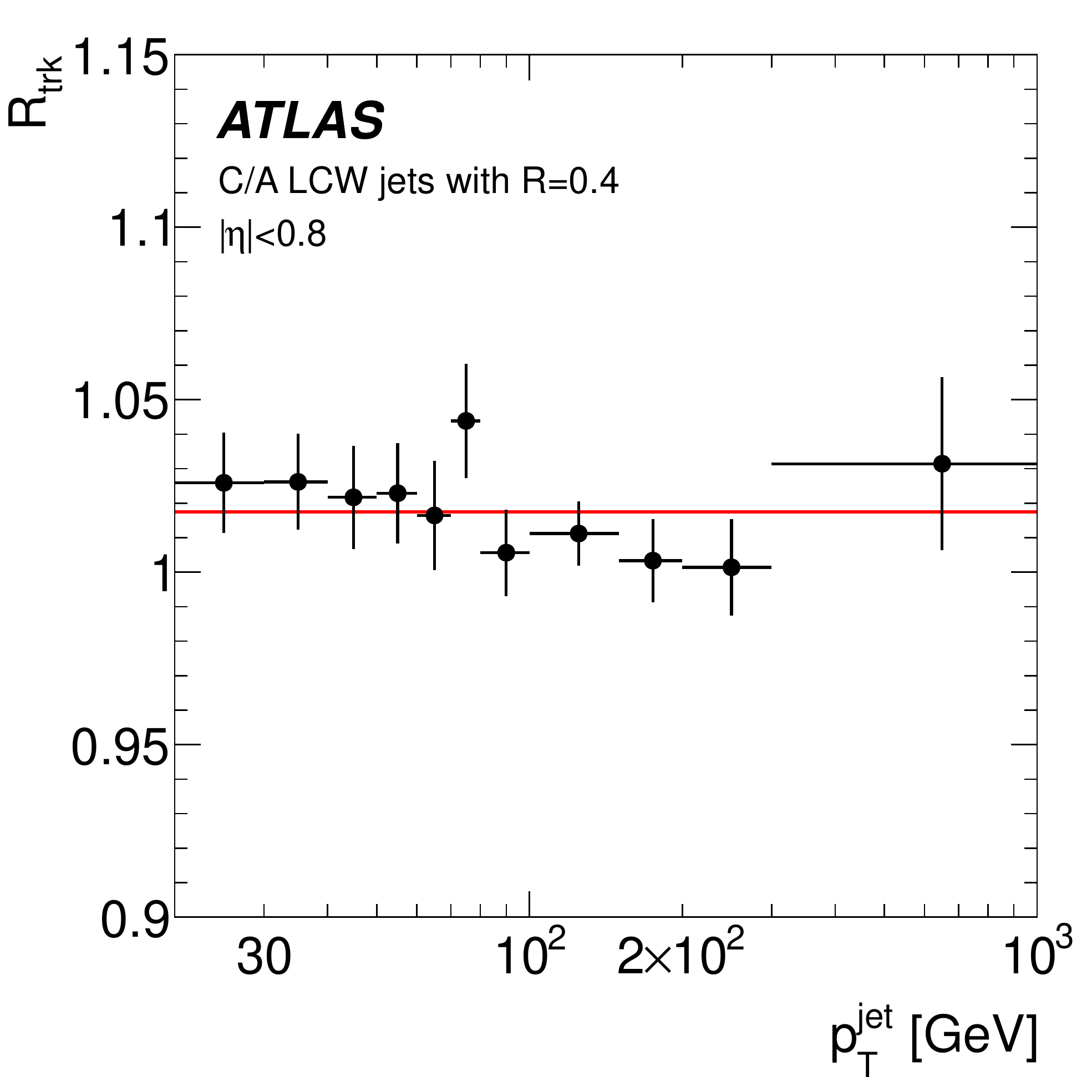} 
   \label{fig:cartrk:rtrkratiovspt:top}
}
\end{center}

\caption{
   Comparison of the MC-calibrated \ca $R=0.4$ calorimeter 
   jet \pt with the \pt of tracks matched to the jet for an event sample 
   consisting of about 50\% semileptonically decaying \ttbar{} 
   pairs with the hadronically decaying top quark having $\pt > 200$~GeV.
   \subref{fig:cartrk:rtrkvspt:top} The average ratio $\langle \rtrksubjet \rangle$ as a function 
   of the MC-calibrated jet \pt.
   \subref{fig:cartrk:rtrkratiovspt:top} The double ratio 
   $\Rtrk = {\langle \rtrksubjet \rangle}_{\rm data} / {\langle \rtrksubjet \rangle }_{\rm MC}$.}

   \label{fig:cartrkttbar}

\end{figure}

A sample of boosted top quarks is used to study the impact of heavy flavour and close-by jet topologies on the systematic uncertainties estimated above. The track-based validation is applied to a sample of events that contains $\simeq 50\%$ semileptonically decaying \ttbar\ pairs, in which the top quark with the hadronically decaying \W boson has $\toppt > 200$~\GeV. The remaining events are dominated by \Wjets production. \Figref{cartrkttbar} shows \rtrksubjet and \Rtrk for \CamKt $R=0.4$ jets in these events. The jet \pt uncertainty in this sample varies between $2.4\%$ and $5.7\%$.

\section{Jet substructure and grooming in the presence of pile-up}
\label{sec:pileup}

\subsection{Impact of \pileup on the jet energy scale and the jet mass scale}
\label{sec:pileup:jesjms}

This section elaborates on the impact of \pileup on the jet mass and other observables, and the extent to which trimming, mass-drop filtering, and pruning are able to minimize these effects. In particular, these measures of performance are used as some of the primary figures of merit in determining a subset of groomed jet algorithms on which to focus for physics analysis in \ATLAS.

\begin{figure}[h!]
  \centering
  \subfigure[Trimmed \antikt: $200\GeV\leq\ptjet<300$~GeV]{
    \includegraphics[width=0.42\columnwidth]{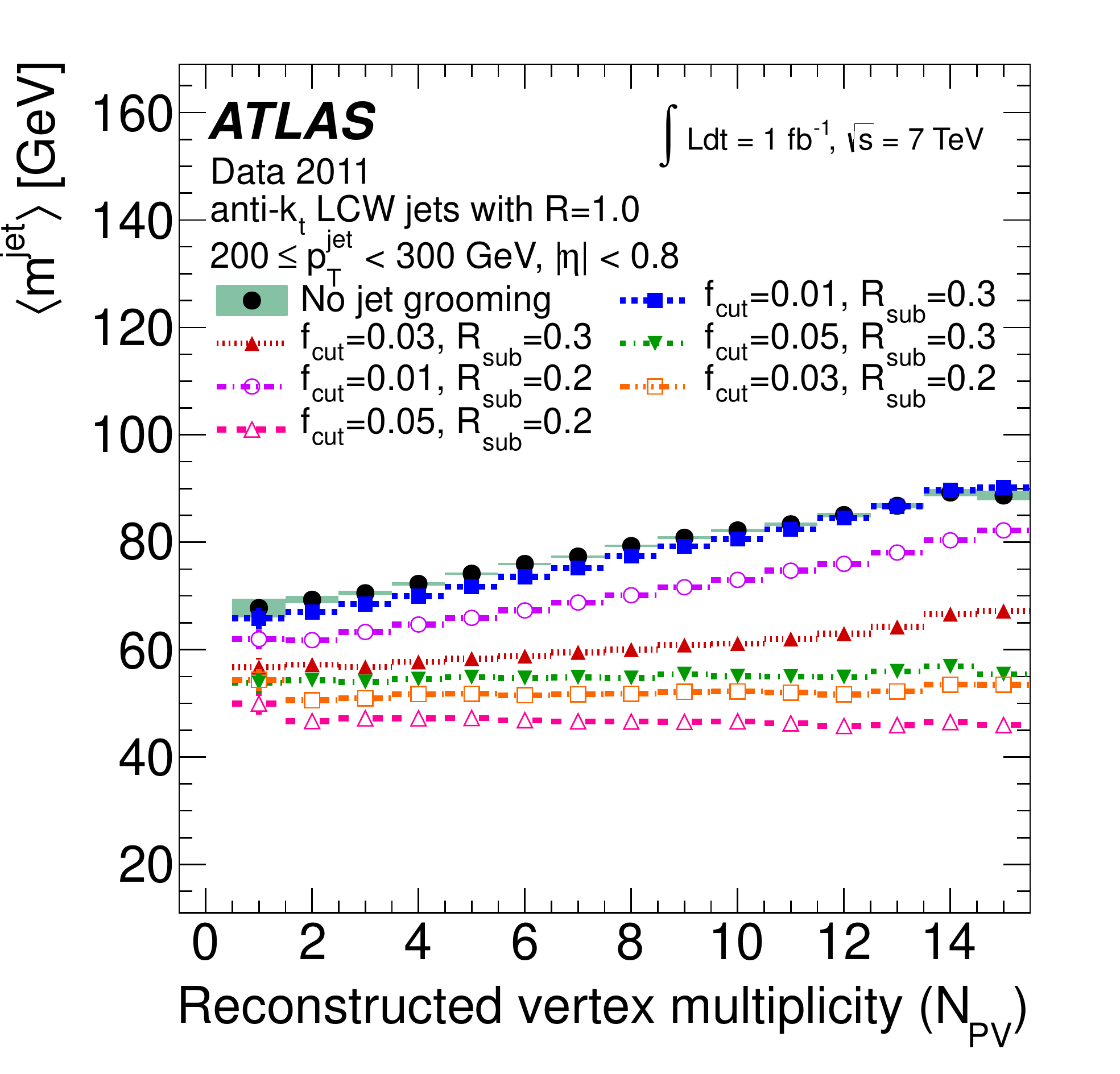}
    \label{fig:DataMC:massNpv:AKTTrim200}}
  \subfigure[Trimmed \antikt: $600\GeV\leq\ptjet<800$~GeV]{
    \includegraphics[width=0.42\columnwidth]{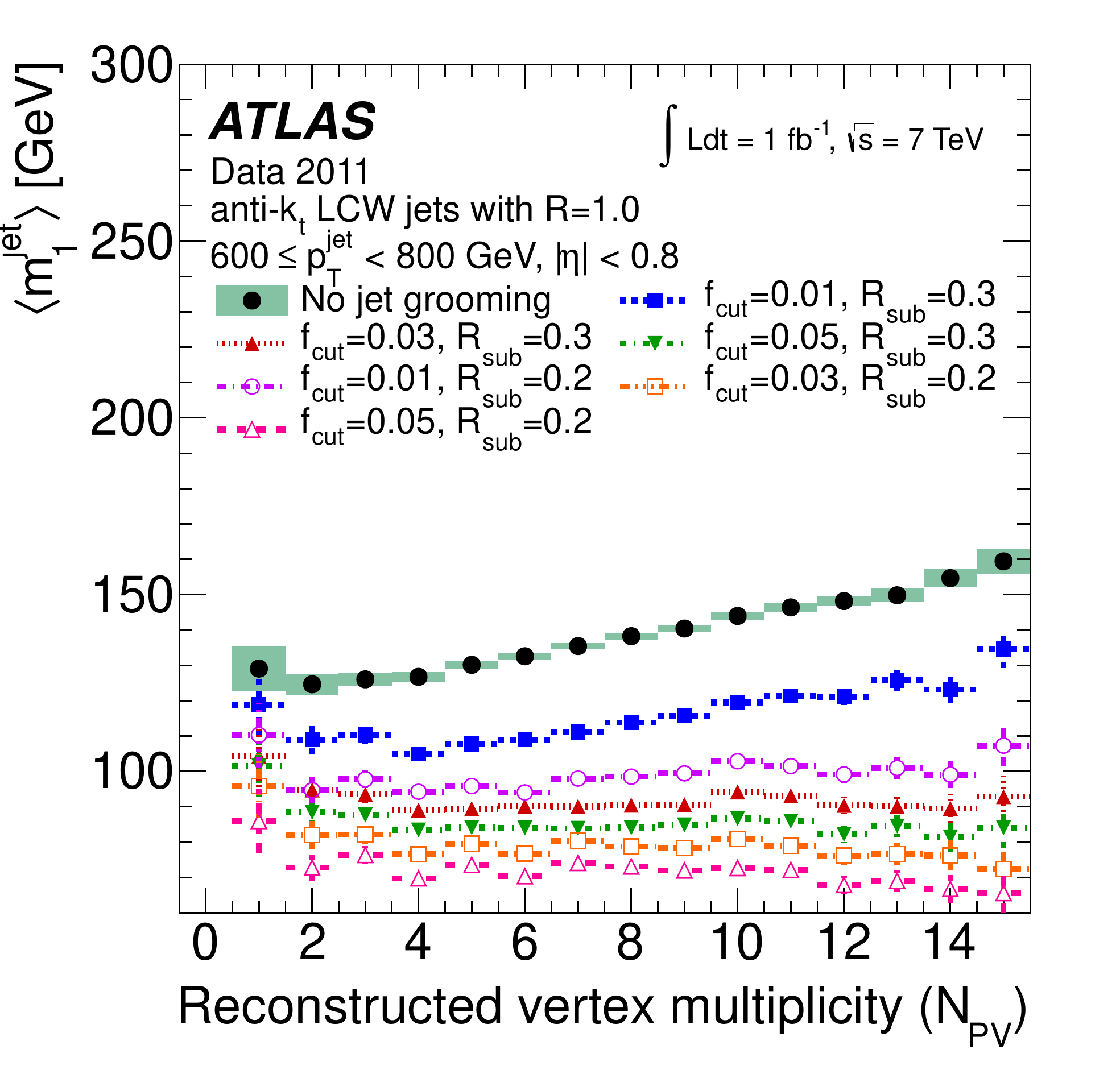}
    \label{fig:DataMC:massNpv:AKTTrim600}}
  \\ 
  
    \subfigure[Pruned \antikt: $200\GeV\leq\ptjet<300$~GeV]{
    \includegraphics[width=0.42\columnwidth]{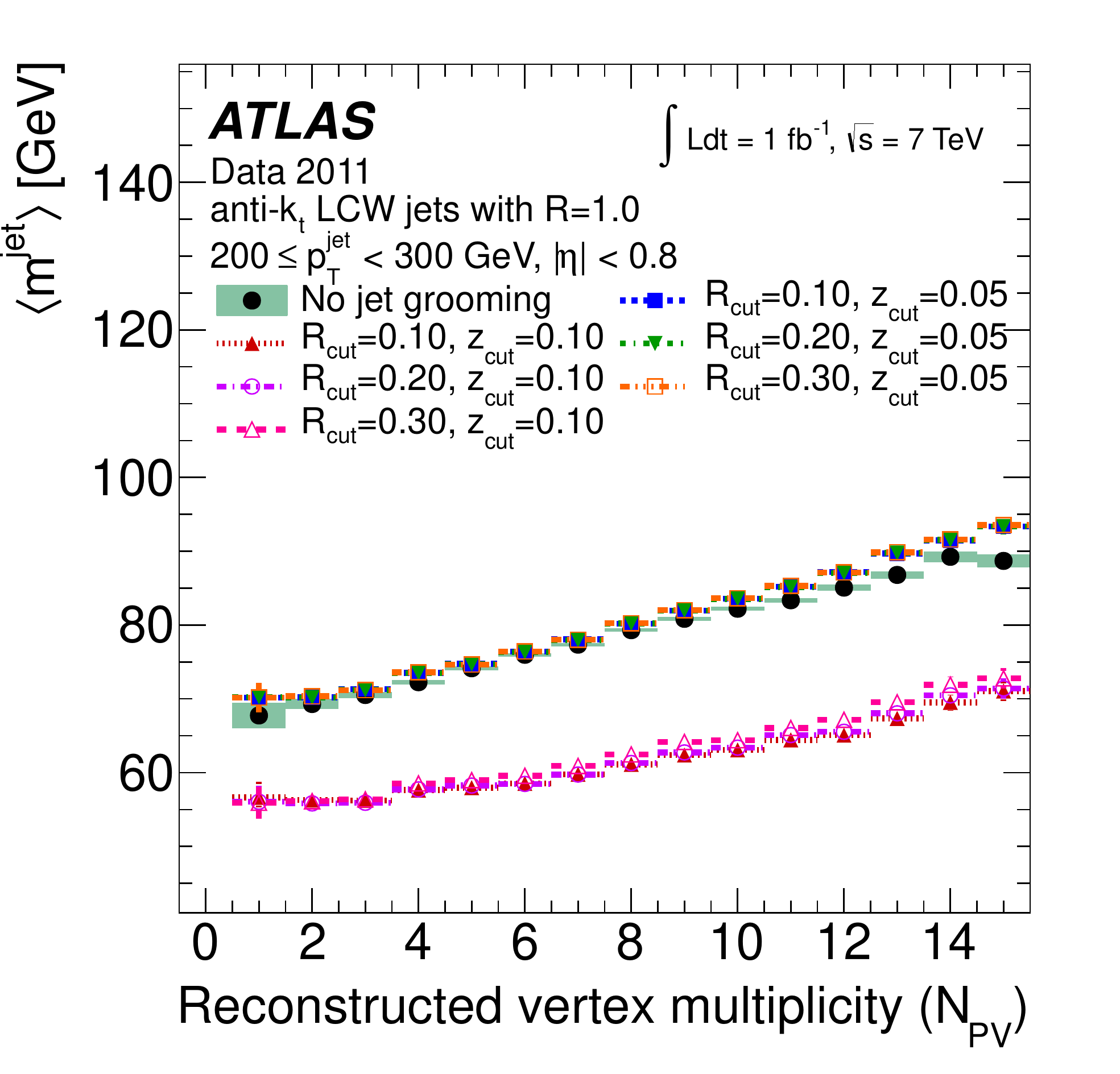}
    \label{fig:DataMC:massNpv:AKTPrune200}}
  \subfigure[Pruned \antikt: $600\GeV\leq\ptjet<800$~GeV]{
    \includegraphics[width=0.42\columnwidth]{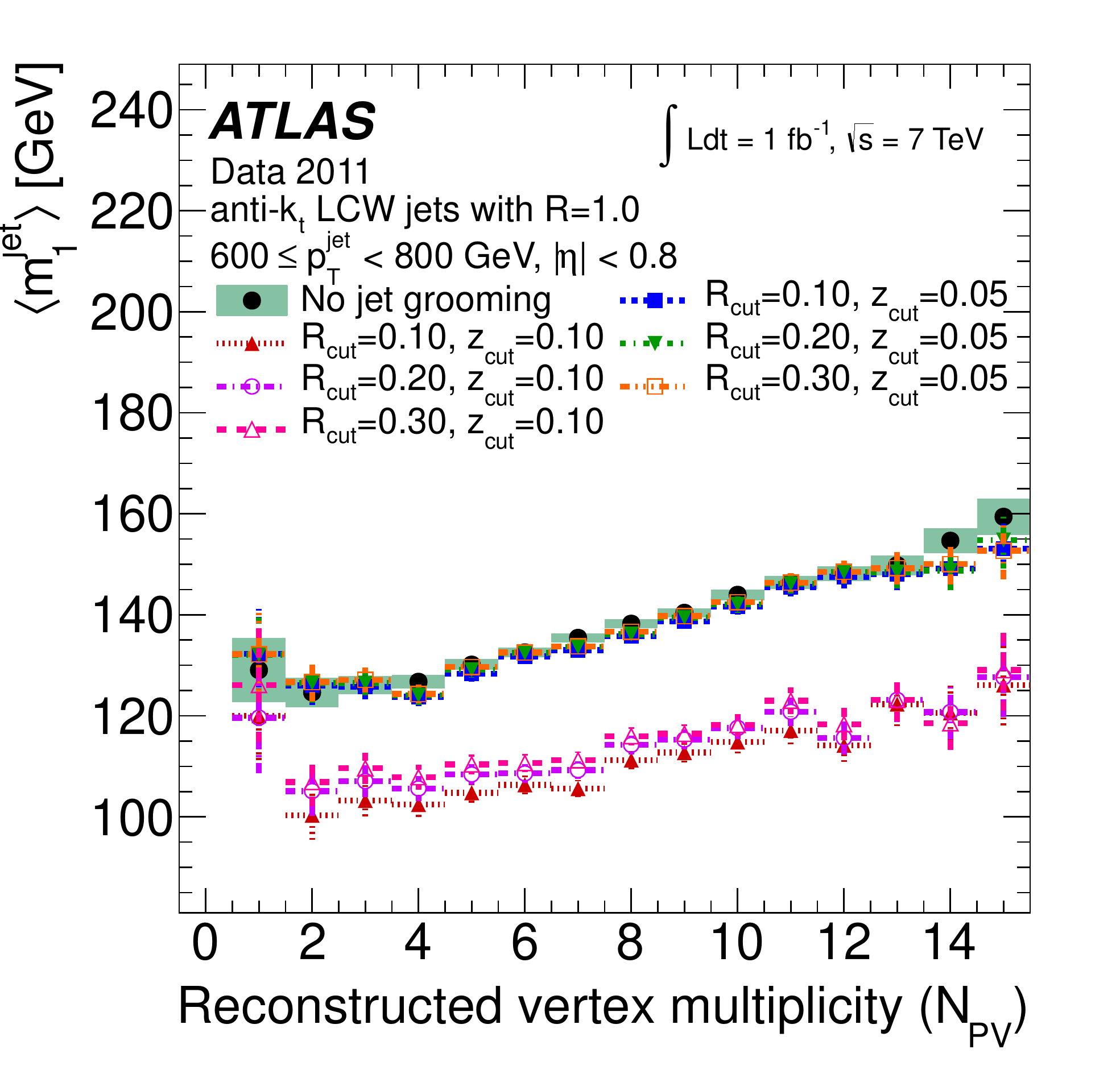}
    \label{fig:DataMC:massNpv:AKTPrune600}}
  \\ 
  
  \subfigure[Filtered \CamKt: $200\GeV\leq\ptjet<300$~GeV]{
    \includegraphics[width=0.42\columnwidth]{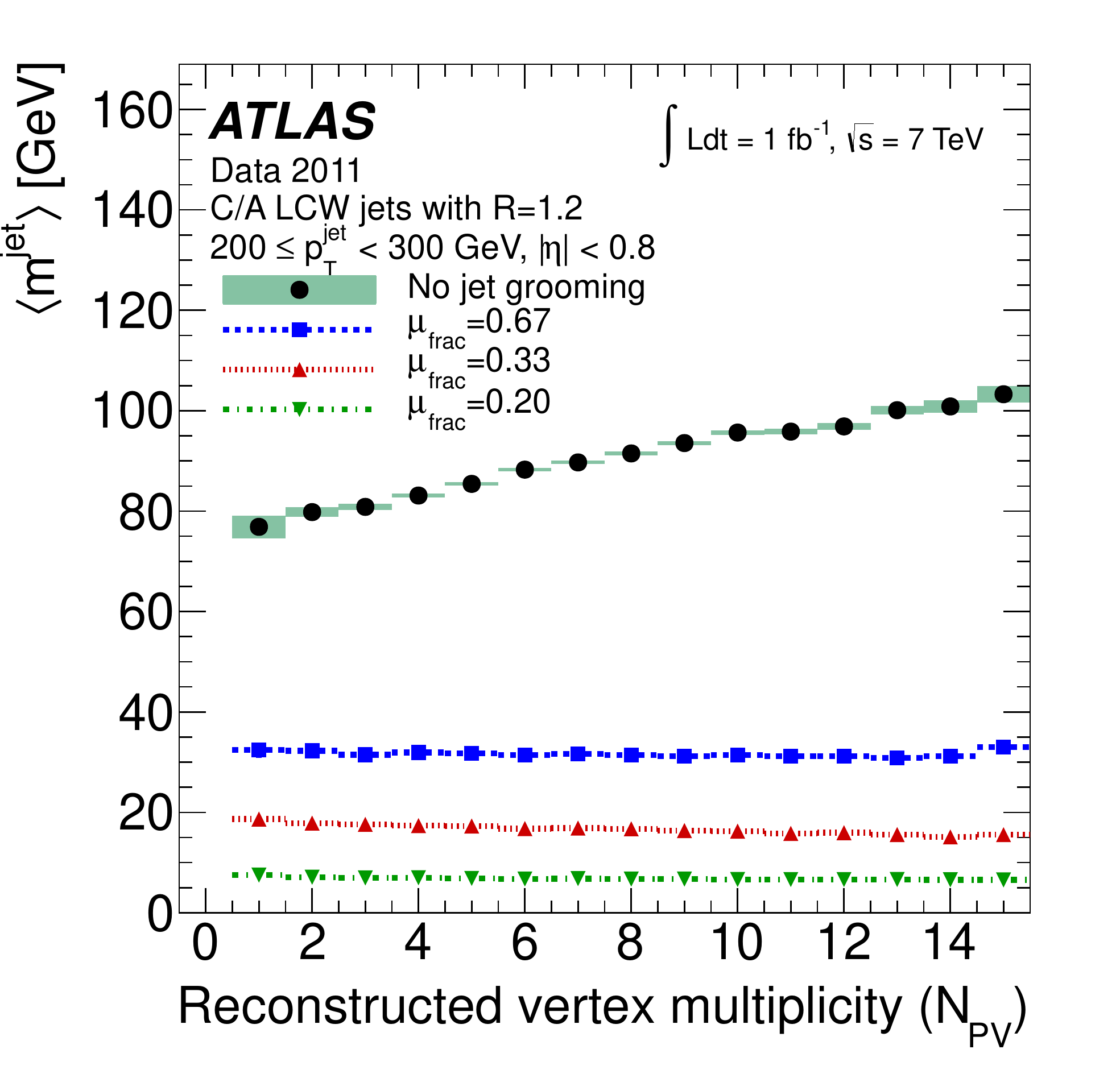}
    \label{fig:DataMC:massNpv:CAFilt200}}
  \subfigure[Filtered \CamKt: $600\GeV\leq\ptjet<800$~GeV]{
    \includegraphics[width=0.42\columnwidth]{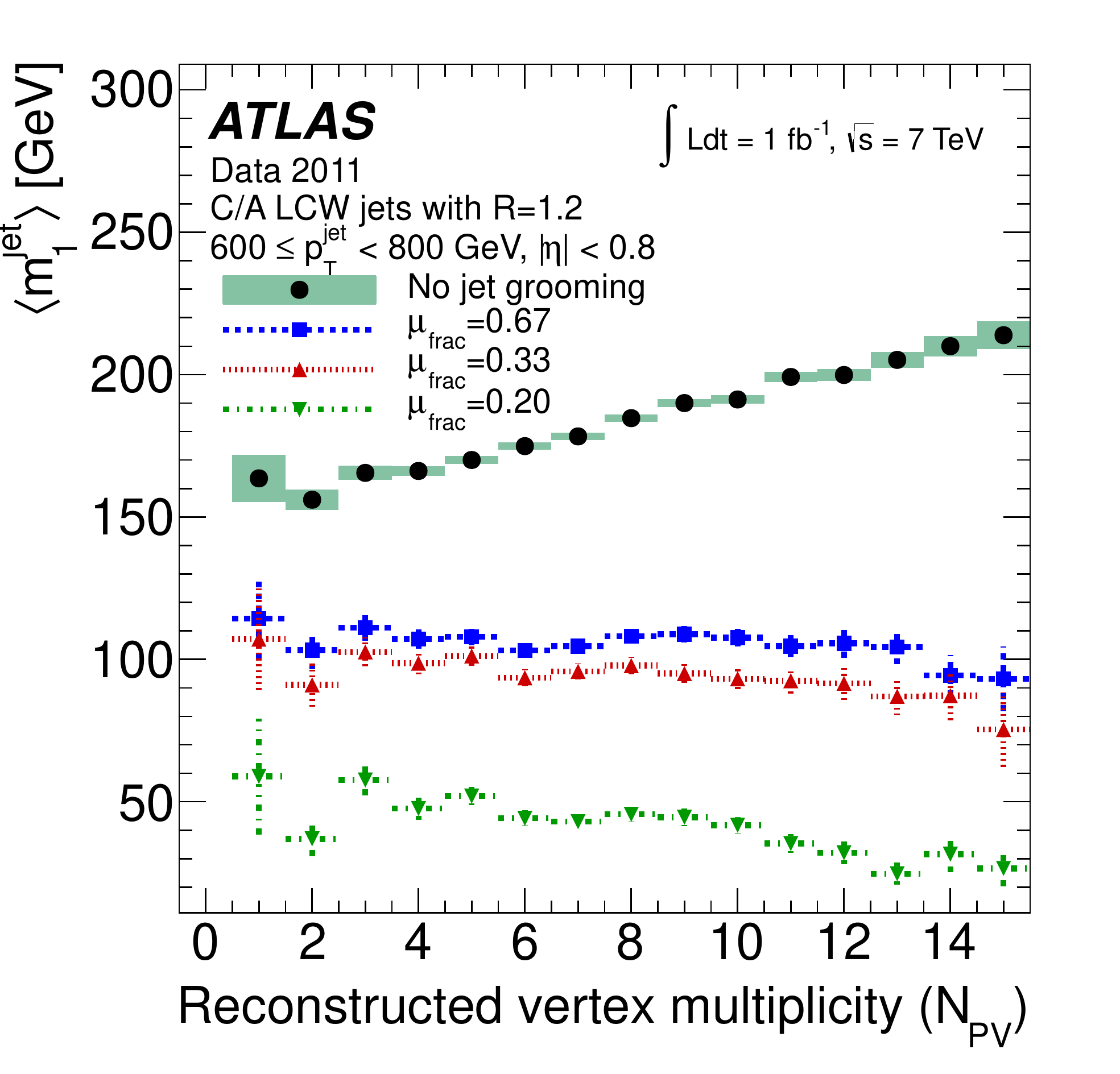}
    \label{fig:DataMC:massNpv:CAFilt600}}
    
  \caption{Evolution of the mean uncalibrated jet mass, \mjetAvg, for jets in the central region $|\eta|<0.8$ as a function of the reconstructed vertex multiplicity, \Npv\ for jets in the range $200\GeV\leq\ptjet<300$~GeV (left) and for leading-\ptjet\ jets (\mjetOneAvg) in the range $600\GeV\leq\ptjet<800$~GeV (right). \subref{fig:DataMC:massNpv:AKTTrim200}-\subref{fig:DataMC:massNpv:AKTTrim600} show trimmed \antikt\ jets with $R=1.0$, \subref{fig:DataMC:massNpv:AKTPrune200}-\subref{fig:DataMC:massNpv:AKTPrune600} show pruned \antikt\ jets with $R=1.0$, and \subref{fig:DataMC:massNpv:CAFilt200}-\subref{fig:DataMC:massNpv:CAFilt600} show mass-drop filtered \CamKt jets with $R=1.2$. The error bars indicate the statistical uncertainty on the mean value in each bin. 
           }
  \label{fig:DataMC:massNpv}
\end{figure}

\afterpage{\clearpage}

\Figref{DataMC:massNpv} shows the dependence of the mean uncalibrated jet mass, \mjetAvg, on the number of reconstructed primary vertices, \Npv, for a variety of jet algorithms in the central region $|\eta|<0.8$. The events used for these comparisons are obtained with the inclusive selection described in \secref{data-mc} and contain an admixture of light quark and gluon jets. This dependence is shown in two \ptjet\ ranges of interest for jets after trimming (\figrange{DataMC:massNpv:AKTTrim200}{DataMC:massNpv:AKTTrim600}), pruning (\figrange{DataMC:massNpv:AKTPrune200}{DataMC:massNpv:AKTPrune600}), and mass-drop filtering (\figrange{DataMC:massNpv:CAFilt200}{DataMC:massNpv:CAFilt600}). For these comparisons, only the final period of data collection from 2011 is used, which corresponds to approximately 1\invfb of integrated luminosity but is the period with the highest instantaneous luminosity recorded at \sqsseven, where $\avg{\mu}=12$, higher than the average over the whole 2011 data-taking period. The lower range, $200\GeV\leq\ptjet<300$~GeV, represents the threshold for most hadronic boosted-object measurements and searches, whereas the range $600\GeV\leq\ptjet<800$~GeV is expected to contain top quarks for which the decay products are fully merged within an $R=1.0$ jet nearly 100\% of the time. In each figure, the full set of grooming algorithm parameter settings is included for comparison. As noted in \tabref{grooming_configs}, two values of the subjet radius, \Rsub, are used for trimming, three \rcut factors for pruning are tested (using the \kt algorithm with the procedure in all cases), and three \mufrac settings are evaluated using the filtering algorithm.

Several observations can be made from \figref{DataMC:massNpv}. Of the grooming configurations tested, trimming and filtering both significantly reduce the rise with \pileup\ of \mjetAvg\ seen for ungroomed jets, whereas pruning does not. For at least one of the configurations tested, trimming and filtering are both able to essentially eliminate this dependence. Furthermore, the trimming configurations tested provide a highly tunable set of parameters that allow a relatively continuous adjustment from small to large reduction of the \pileup dependence of the jet mass. The trimming configurations with $\Rsub=0.2,\fcut=0.03$ and $\Rsub=0.3,\fcut=0.05$ exhibit good stability for both small and large \ptjet, with the $\fcut=0.05$ configuration exhibiting a slightly smaller impact from \pileup at high \Npv\ for low \ptjet (not shown). The other parameter settings either do not reduce the \pileup dependence at low \ptjet\ (e.g. $\fcut=0.01$) or result in a downward slope of \mjetAvg as a function of \pileup at high \ptjet\ (e.g. $\fcut=0.05, \Rsub=0.2$).

Pruning, on the other hand, exhibits the smallest impact on the \pileup dependence of the jet mass for these \largeR jets. Only by increasing the \zcut parameter from $\zcut=0.05$ to $\zcut=0.10$ can any reduction on the dependence of \mjetAvg on \pileup be observed. This is equivalent to reducing the low-\pt\ contributions during the jet recombination; in the language of trimming, this is analogous to raising \fcut. This change slightly reduces the magnitude of the variation of the mean jet mass as a function \Npv\ for low \ptjet. The \rcut parameter has very little impact on the performance, with nearly all of the differences observed being due to the change in \zcut. This observation holds for both small and large \ptjet.

The mass-drop filtering algorithm can be made to affect \mjetAvg significantly solely via the mass-drop criterion, \mufrac. A drastic change in \mjetAvg is observed for all configurations of the jet filtering, with the strictest $\mufrac=0.20$ setting rejecting nearly 90\% of the jets considered and resulting in a slightly negative slope in the mean jet mass versus \Npv. Nevertheless, the other two settings of \mufrac tested exhibit no significant variation as a function of the number of reconstructed vertices, and the optimum value of $\mufrac=0.67$ found previously seems to have the best stability. Studies from 2010~\cite{JetMassAndSubstructure} demonstrate that this reduction in the sensitivity to \pileup is due  primarily to the filtering step in the algorithm as opposed to the jet selection itself.\footnote{The performance of grooming also depends on the radius parameter, $R$, which was not varied in these studies for a single jet algorithm. In particular, the relative efficacy of the various grooming algorithms and configurations in mitigating the effects of \pileup can change with $R$ and will be studied in a future analysis.}

\begin{figure}[!h]
  \centering
  \subfigure[\antikt, $R=1.0$: Ungroomed]{
    \includegraphics[width=0.46\columnwidth]{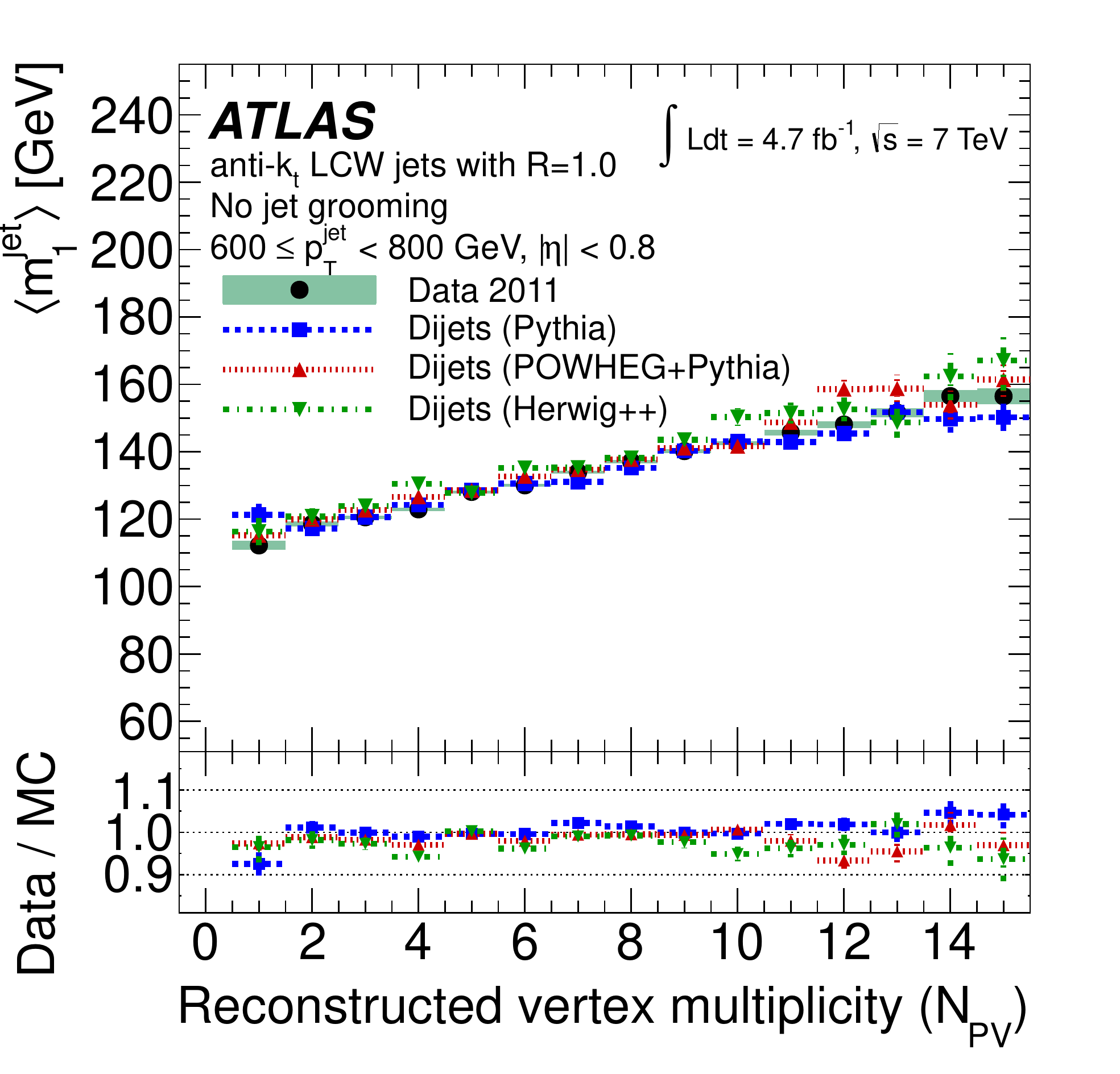}
    \label{fig:DataMC:massNpvDataMC:AKTFat600}}
  \subfigure[\antikt, $R=1.0$: Trimmed]{
    \includegraphics[width=0.46\columnwidth]{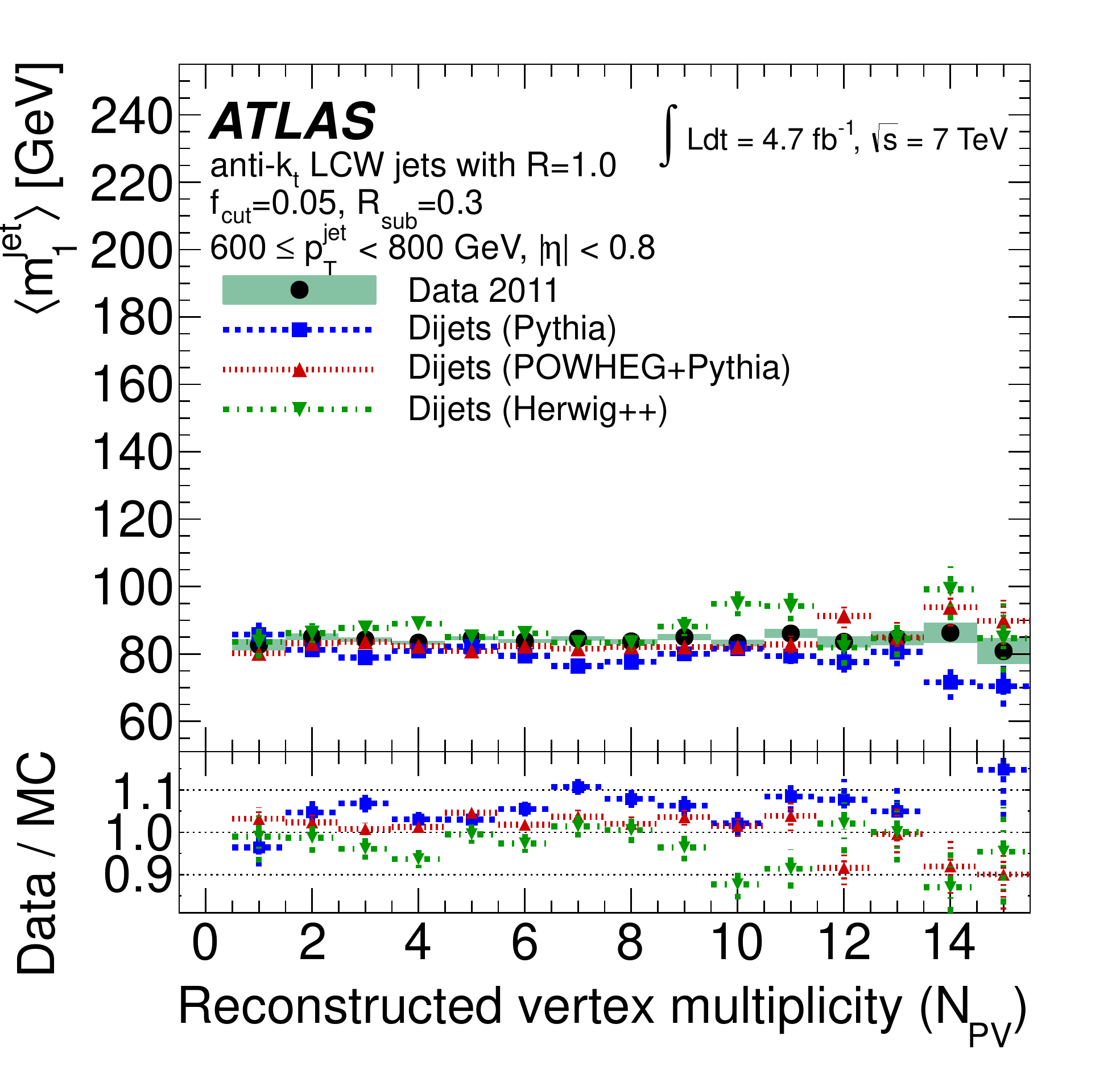}
    \label{fig:DataMC:massNpvDataMC:AKTTrim600}}
  \caption{Dependence of the mean jet mass, \mjetOneAvg, on the reconstructed vertex multiplicity for leading-\ptjet\ \antikt\ jets with $R=1.0$ in the range $600\GeV\leq\ptjet<800$~GeV in the central region $|\eta|<0.8$ for both \subref{fig:DataMC:massNpvDataMC:AKTFat600} untrimmed \antikt\ jets and \subref{fig:DataMC:massNpvDataMC:AKTTrim600} trimmed \antikt\ jets with $\fcut=0.05$. The error bars indicate the statistical uncertainty on the mean value in each bin.
           }
  \label{fig:DataMC:massNpvDataMC}
\end{figure}

\Figref{DataMC:massNpvDataMC} presents the \pileup dependence of the mean leading-\ptjet\ jet mass, \mleadavg, in data compared to the three simulations. Here, only the range $600\GeV\leq\ptjet<800$~GeV for ungroomed and trimmed \antikt\ jets is shown for brevity, but similar conclusions apply in all \ptjet\ ranges. The comparison is made using the full 2011 dataset. \Pythia, \Herwigpp, and \PowPythia all model the data fairly accurately, with a slight 5\%--10\% discrepancy appearing in the predictions from \Pythia and \Herwigpp for the trimmed jets. Most importantly, the impact of \pileup is very well modelled, with the slope of the dependence of \mleadavg on \Npv\ in data agreeing within 3\% with the \PowPythia prediction for both the ungroomed and trimmed jets.

\begin{figure}[!h]
  \centering
  
  \subfigure[Data: \antikt, $R=1.0$: Ungroomed]{
    \includegraphics[width=0.46\columnwidth]{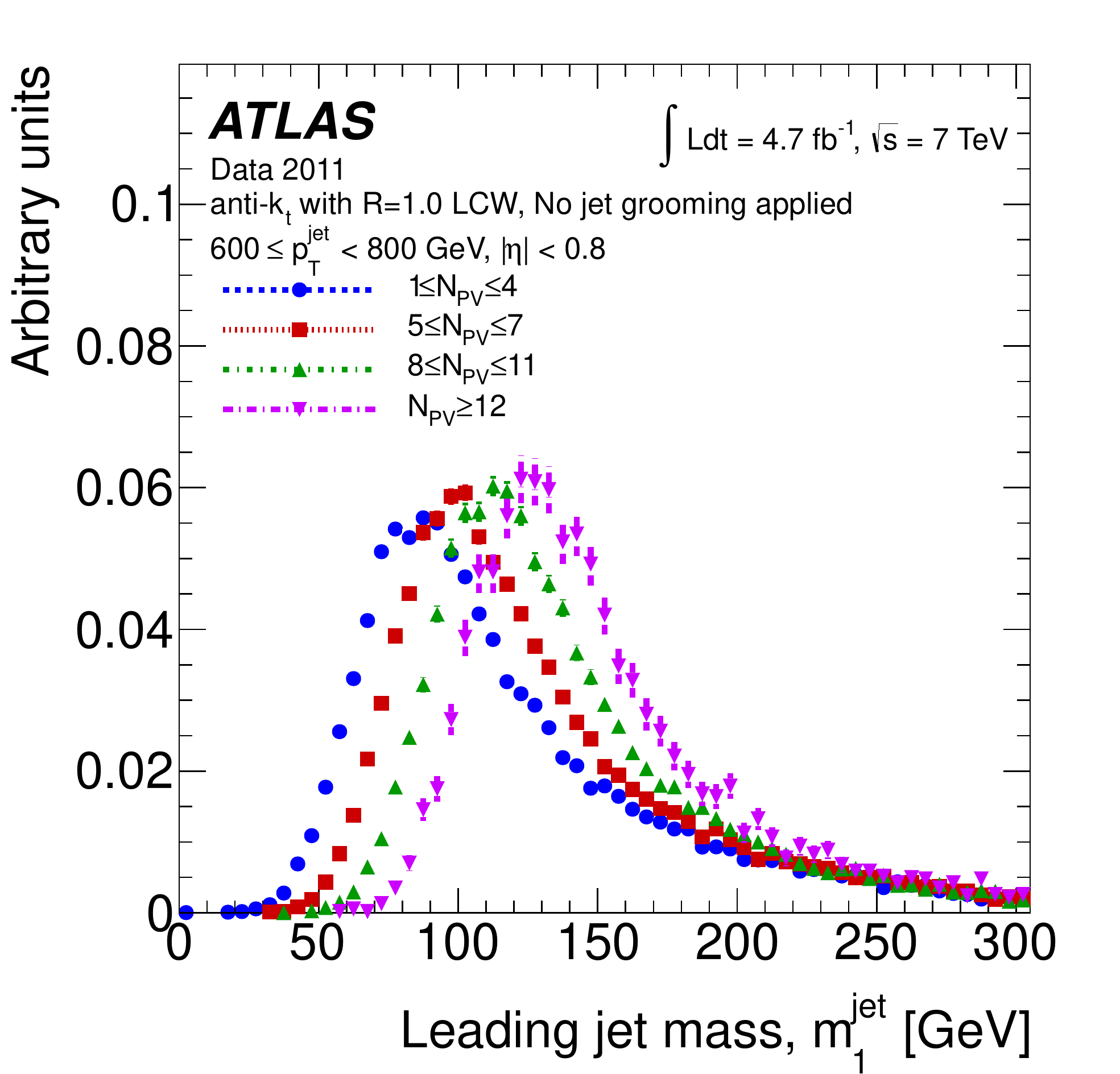}
    \label{fig:DataMC:massDistro:Data}}
  \subfigure[Data: \antikt, $R=1.0$: Trimmed]{
    \includegraphics[width=0.46\columnwidth]{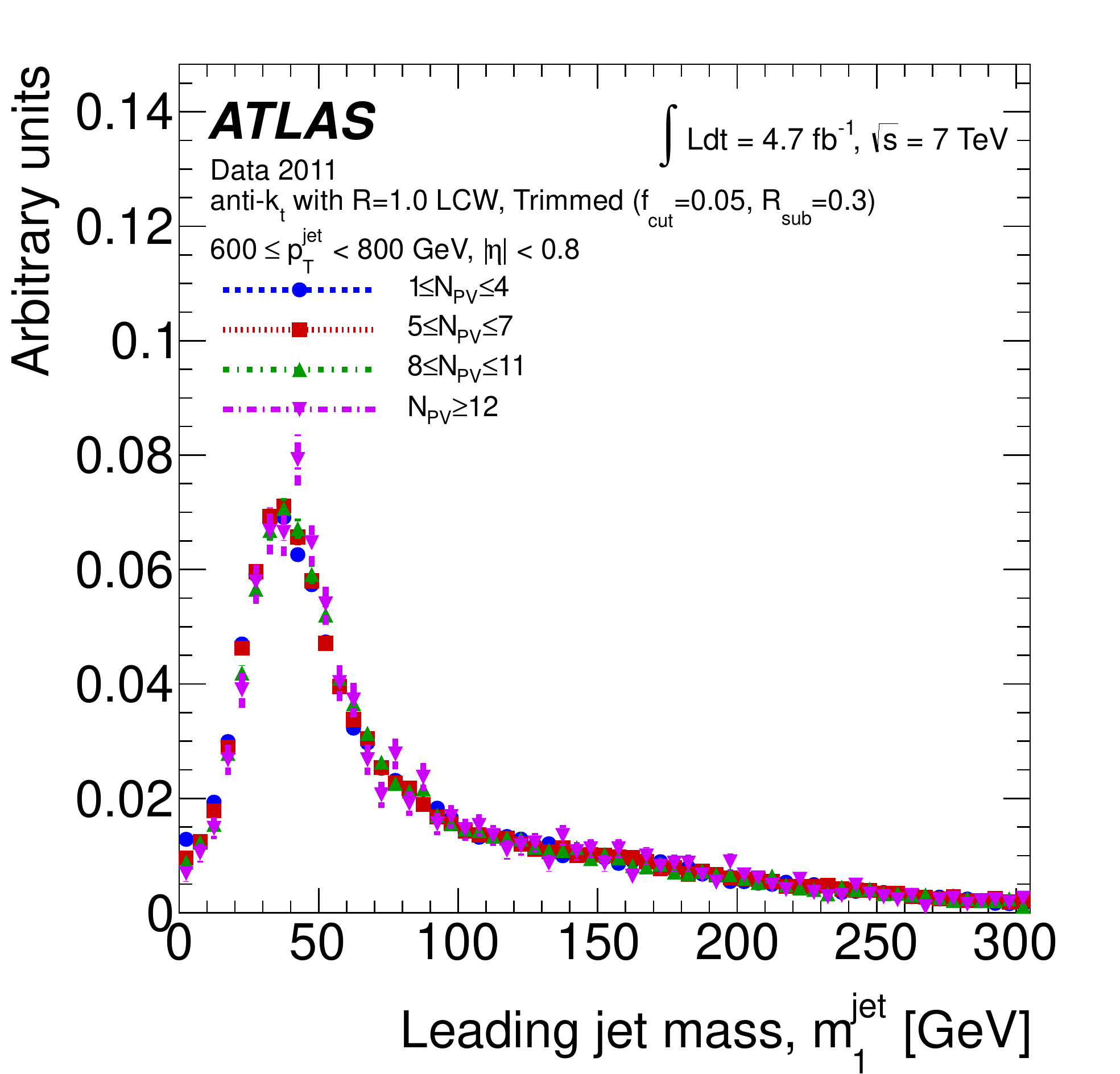}
    \label{fig:DataMC:massDistro:TrimData}} \\
    
  \subfigure[\Zprime: \antikt, $R=1.0$: Ungroomed]{
    \includegraphics[width=0.46\columnwidth]{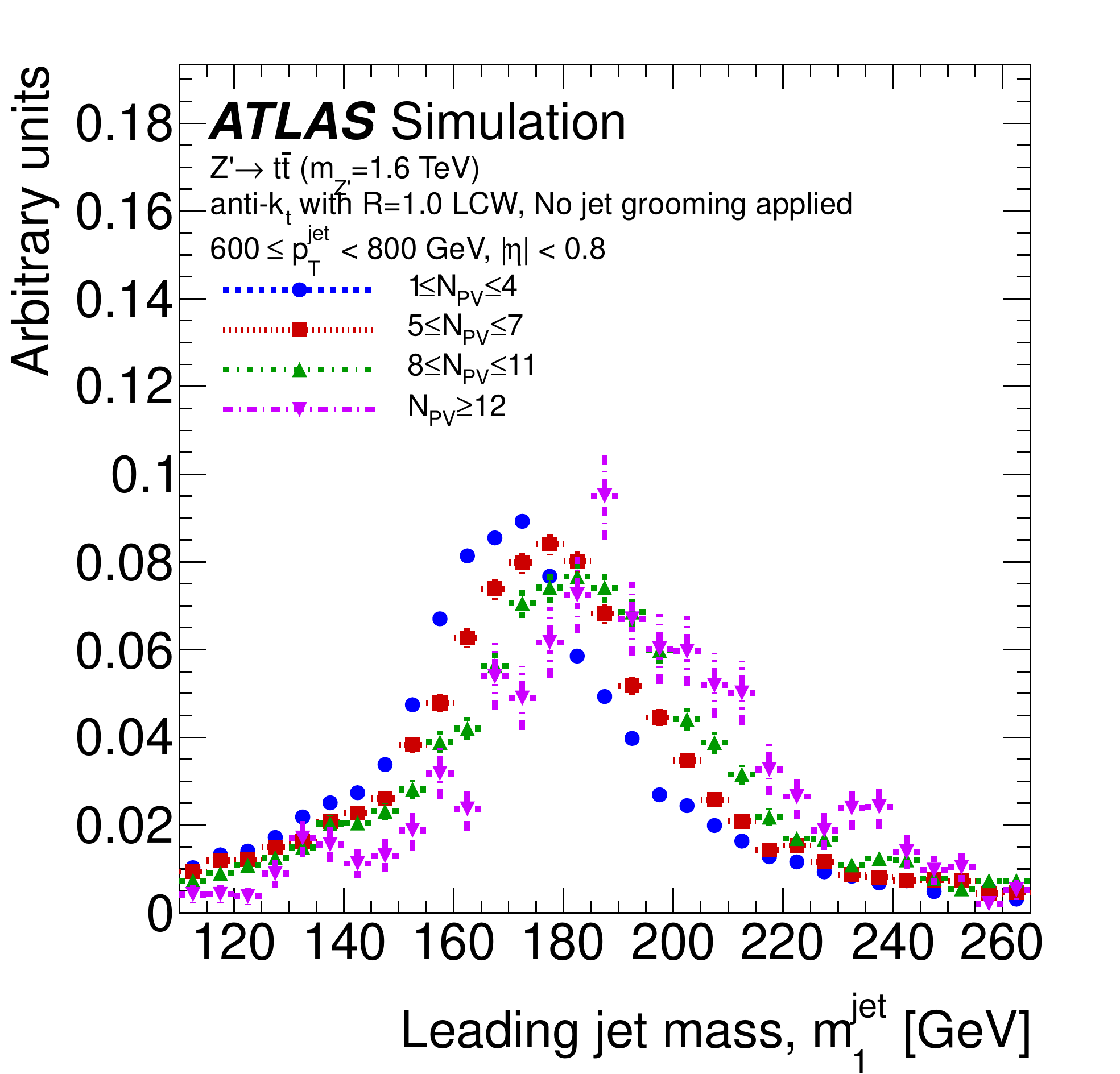}
    \label{fig:DataMC:massDistro:Zprime}}
  \subfigure[\Zprime: \antikt, $R=1.0$: Trimmed]{
    \includegraphics[width=0.46\columnwidth]{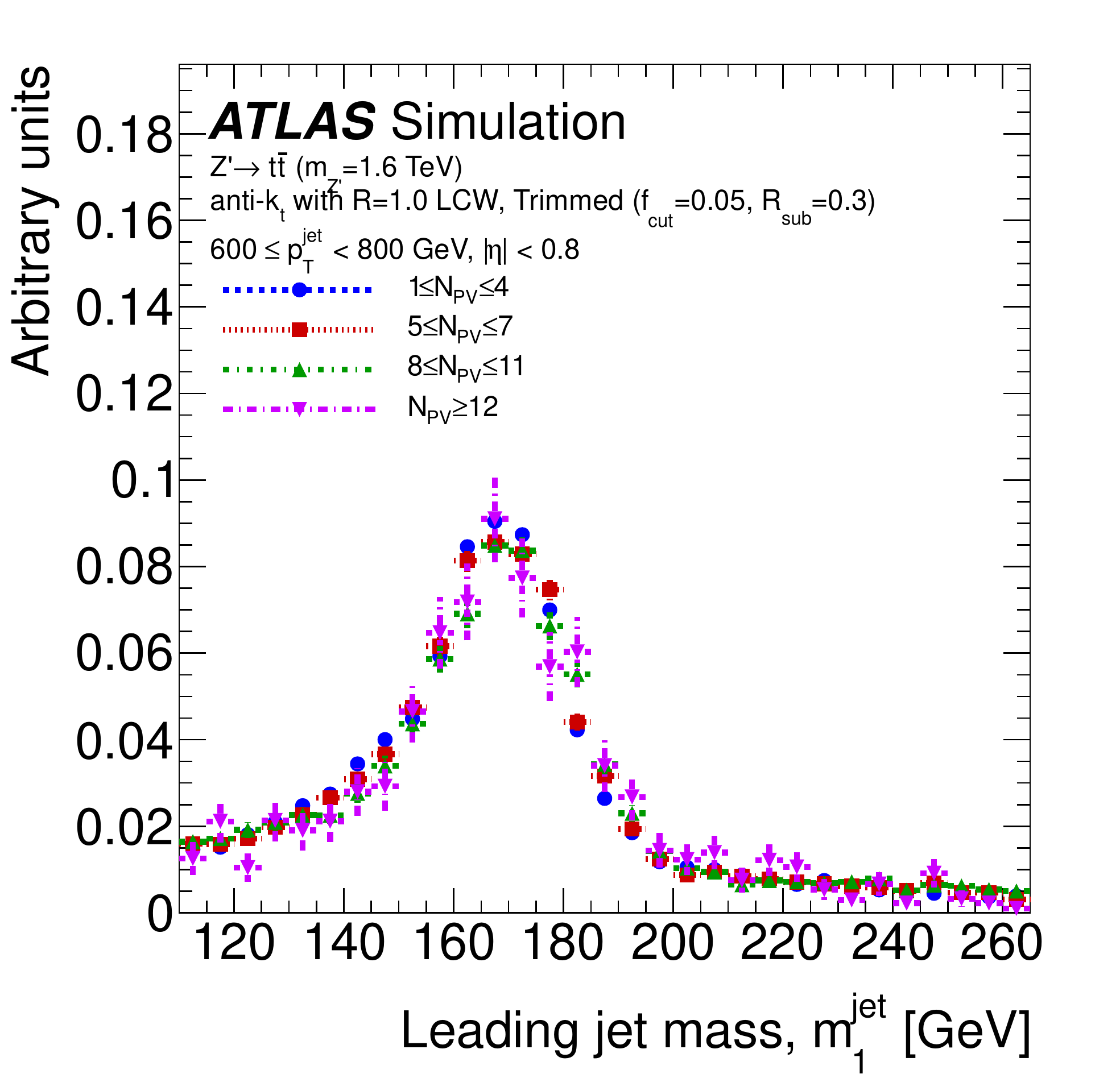}
    \label{fig:DataMC:massDistro:TrimZprime}}  
    
  \caption{Jet mass spectra for four primary vertex multiplicity ranges for \antikt\ 
           jets with $R=1.0$ in the range $600\GeV\leq\ptjet<800$~GeV. Both 
           untrimmed (left) and trimmed (right) \antikt\ jets are compared
           for the various \Npv ranges in data (top) and for a \Zprimett Monte 
           Carlo sample (bottom).
           }
           
  \label{fig:DataMC:massDistro}
\end{figure}

Beyond simply providing a \pileup-independent average jet mass, the optimal grooming configurations render the full jet mass spectrum insensitive to high instantaneous luminosity. \Figref{DataMC:massDistro} demonstrates this by comparing the jet mass spectrum for leading-\ptjet\ ungroomed and trimmed \antikt\ jets for various values of \Npv. The comparison is performed both in an inclusive data sample and using the \Zprimett MC sample, which produces a characteristic peak at the top-quark mass. The inclusive jet sample obtained from data shows that a nearly identical trimmed \massjet spectrum is obtained regardless of the level of \pileup. The peak of the leading-\ptjet\ jet mass distribution for events with $\Npv\geq12$ is shifted comparatively more due to trimming: from $\massjet\approx125$~GeV to $\massjet\approx45$~GeV compared to an initial peak position of $\massjet\approx90$~GeV for events with $1\leq\Npv\leq4$. Nonetheless, the resulting trimmed jet mass spectra exhibit no dependence on \Npv. 

Comparisons performed using the simulated \Zprimett sample demonstrate the same performance of the trimming algorithm, but in the context of the reconstruction of highly boosted top quarks. Figures~\ref{fig:DataMC:massDistro:Zprime}-\subref{fig:DataMC:massDistro:TrimZprime} indicate that the ability to render the full jet mass distribution independent of \pileup does not come at the cost of the mass resolution or scale. Prior to jet trimming, a variation in the peak position of the jet mass of nearly 15~GeV is observed between the lowest and the highest ranges of \Npv studied. After jet trimming, the resulting mass spectra for the various \Npv ranges are narrower and lie directly on top of one another, even in the case of a jet containing a highly boosted, and thus very collimated, top-quark decay. This observation, combined with that above, demonstrates that the trimming algorithm is working as expected by removing soft, wide-angle contributions to the calculation of the jet mass while retaining the relevant hard substructure of the jet.

Finally, although not shown explicitly, the mass-drop filtered jet mass is also stable with respect to \pileup.

The track-jet approach used to evaluate the jet mass uncertainty (see \secref{recocalib:massscale}) is also used to understand the effects of \pileup. It is observed that \rtrkjetM is nearly equal for the various trimming configurations in the case of little or no in-time \pileup (i.e.~$\Npv \approx 1$) whereas filtering shows a significant, although small, difference between the configurations using $\mufrac=0.67,0.33,0.20$. This shows that the filtering method does affect the magnitude of \massjet and \masstrkjet slightly differently, resulting in an approximate 12\% relative drop in  \AvgrtrkjetM after filtering. These distributions are nonetheless well modelled by the simulation, resulting in double ratios of data to simulation very close to one. The trimming configuration with $\Rsub=0.3, \fcut=0.01$ has almost no impact on the dependence of \rtrkjetM with \pileup.

\subsection{Impact of \pileup on jet substructure properties}
\label{sec:pileup:properties}

\begin{figure}[!ht]
  \centering
  \subfigure[\DOneTwo]{
    \includegraphics[width=0.46\columnwidth]{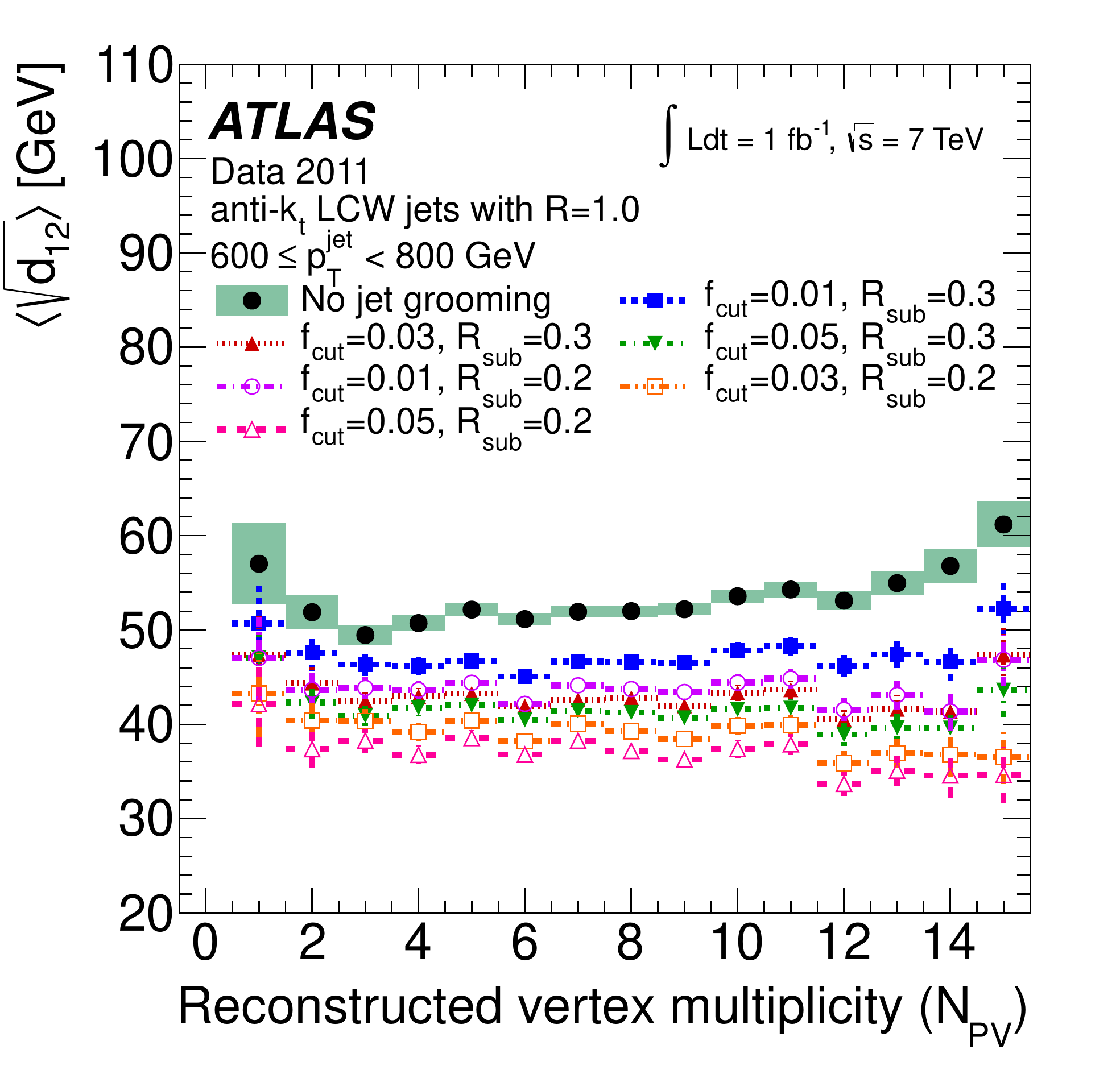}
    \label{fig:DataMC:dijNpv:d12}}
  \subfigure[\DTwoThr]{
    \includegraphics[width=0.46\columnwidth]{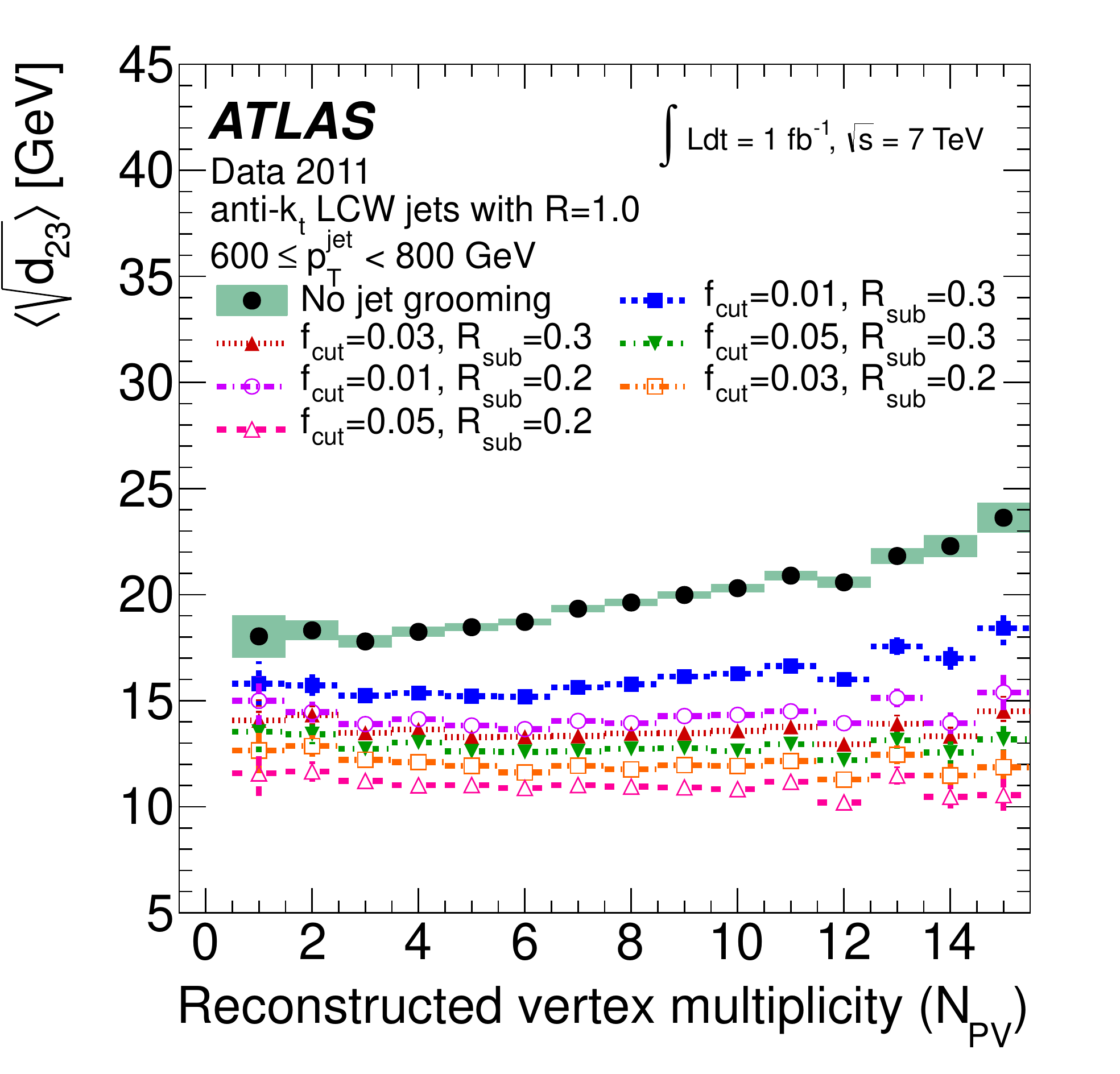}
    \label{fig:DataMC:dijNpv:d23}}
  \caption{Variation with the number of reconstructed primary vertices, \Npv, of the mean splitting scales \subref{fig:DataMC:dijNpv:d12} $\langle\DOneTwo\rangle$ and \subref{fig:DataMC:dijNpv:d23} $\langle\DTwoThr\rangle$ measured in data for \antikt\ jets with $R=1.0$ in the range $600\GeV\leq\ptjet<800$~GeV before and after trimming. The error bars indicate the statistical uncertainty on the mean value in each bin.
           }
  \label{fig:DataMC:dijNpv}
\end{figure}

\Figsref{DataMC:dijNpv}{DataMC:tauNNpv} show the variation with \pileup observed in data for the splitting scales and $N$-subjettiness observables for jets in the range $600\GeV\leq\ptjet<800$~GeV. In this case, the focus is on jet trimming since the mass-drop filtering algorithm makes a predefined choice to search for properties of a jet characteristic of a two-body decay. The constraints placed on subjet multiplicity by the filtering procedure are not appropriate for calculating generic jet shapes given the strict substructure requirements they place on a jet. Furthermore, pruning of jets with $R=1.0$ does not seem to mitigate the effects of \pileup. The trimming configurations with $\Rsub=0.3$ and $\fcut=0.03, 0.05$ yield the most stable jet substructure properties with the smallest deviation in their observed mean values at low \Npv. This conclusion holds for all other jet \ptjet\ ranges as well, with larger differences between $\fcut=0.03, 0.05$ appearing at low \ptjet.

\begin{figure}[!ht]
  \centering
  \subfigure[\tauTwoOne]{
    \includegraphics[width=0.46\columnwidth]{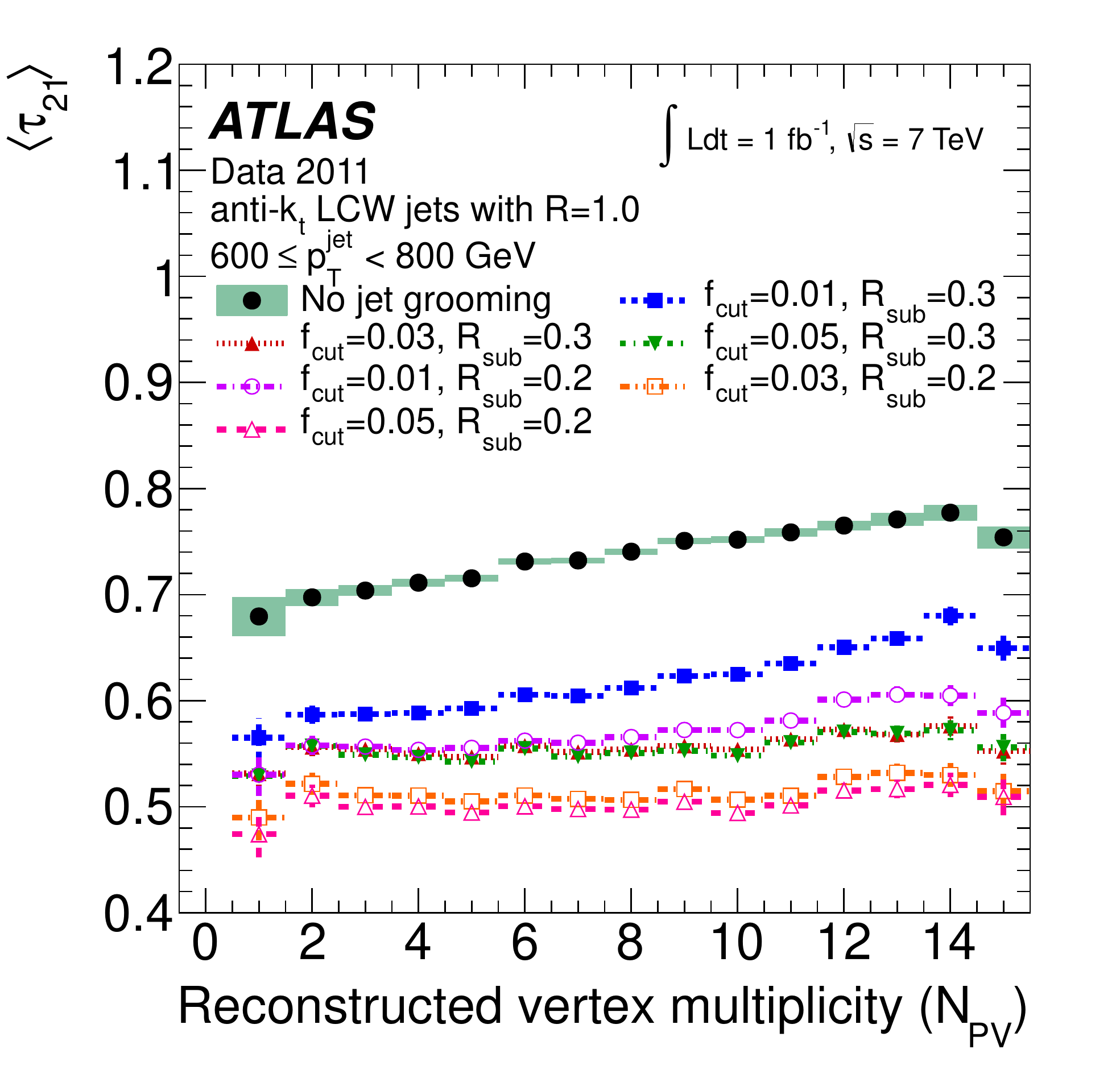}
    \label{fig:DataMC:tauNNpv:tau21}}
  \subfigure[\tauThrTwo]{
    \includegraphics[width=0.46\columnwidth]{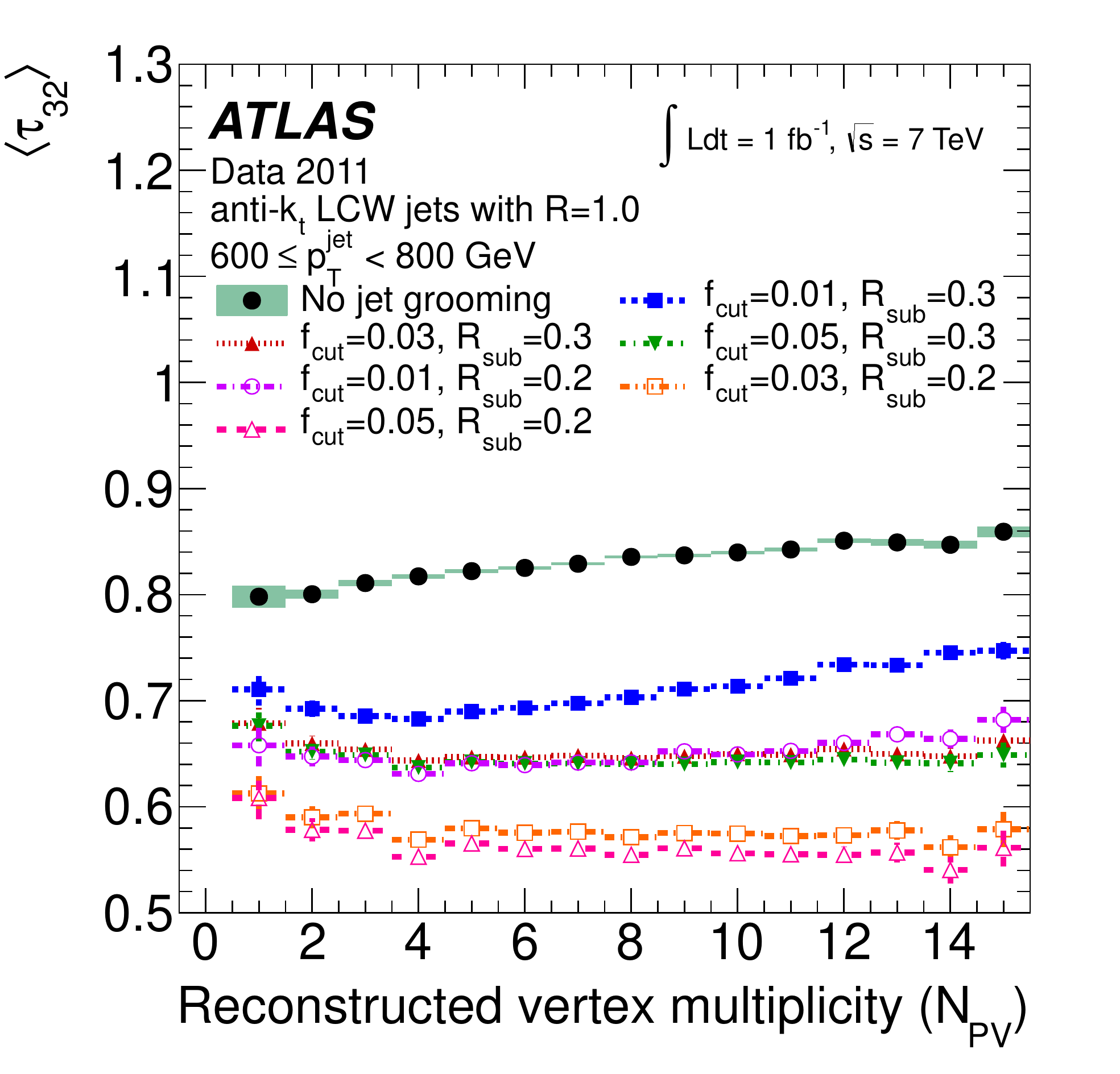}
    \label{fig:DataMC:tauNNpv:tau32}}
  \caption{Variation with the number of reconstructed primary vertices, \Npv, of the mean $N$-subjettiness ratios \subref{fig:DataMC:tauNNpv:tau21} $\langle\tauTwoOne\rangle$ and \subref{fig:DataMC:tauNNpv:tau32} $\langle\tauThrTwo\rangle$ measured in data for \antikt\ jets with $R=1.0$ in the range $600\GeV\leq\ptjet<800$~GeV before and after trimming. The error bars indicate the statistical uncertainty on the mean value in each bin.
           }
  \label{fig:DataMC:tauNNpv}
\end{figure}

\begin{figure}[!ht]
  \centering

  \subfigure[\antikt, $R=1.0$: Ungroomed]{
    \includegraphics[width=0.46\columnwidth]{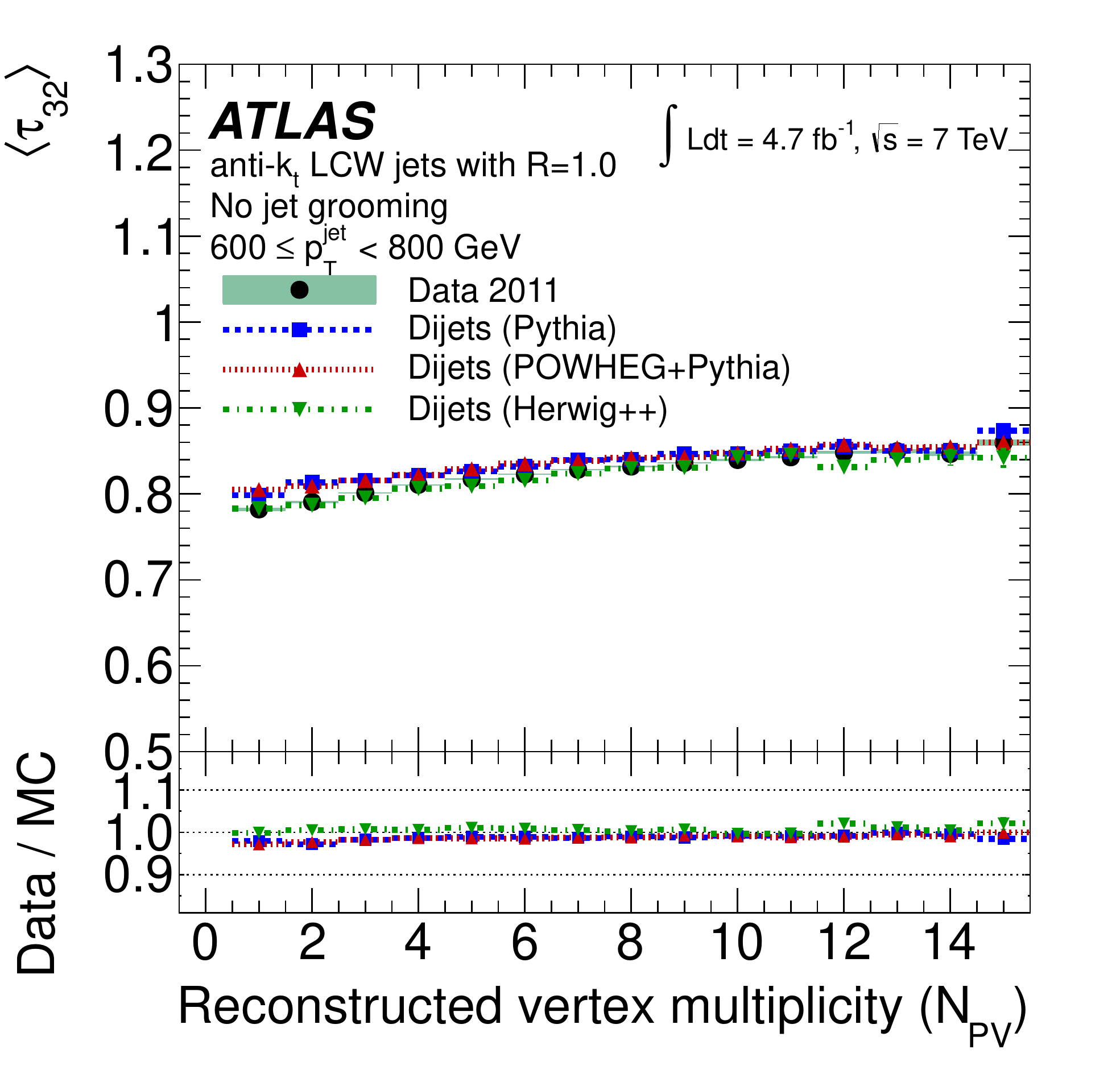}
    \label{fig:DataMC:Tau32Npv:AKTFat600}}
  \subfigure[\antikt, $R=1.0$: Trimmed]{
    \includegraphics[width=0.46\columnwidth]{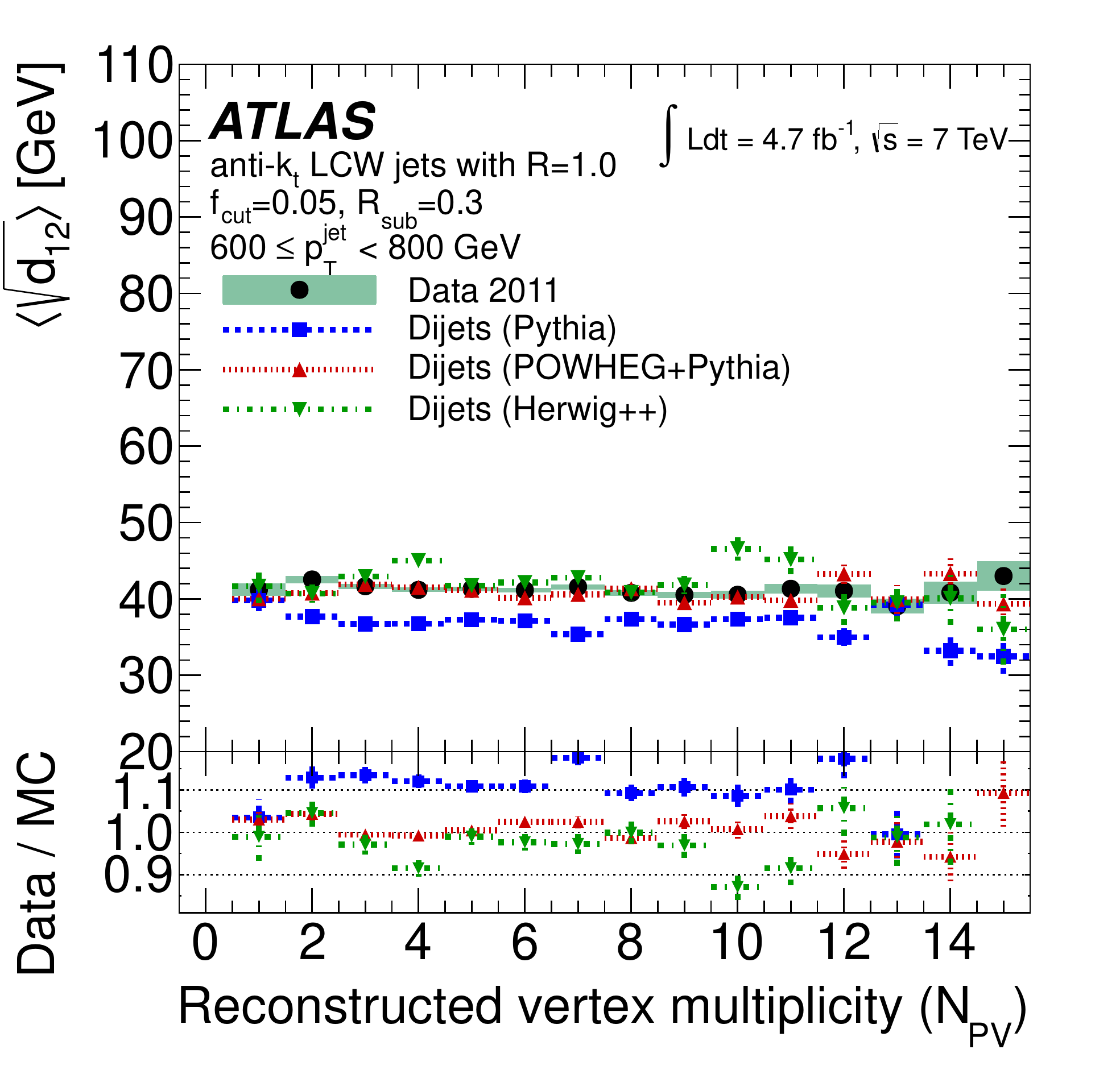}
    \label{fig:DataMC:d12NPV:AKTTrim600}}
    
  \caption{     
     Variation with the number of reconstructed primary vertices, \Npv, of 
     \subref{fig:DataMC:Tau32Npv:AKTFat600} the mean $N$-subjettiness ratio 
     ($\langle\tauThrTwo\rangle$) for ungroomed \antikt\ jets with  
     $R=1.0$ and \subref{fig:DataMC:d12NPV:AKTTrim600} the mean splitting scales 
     ($\langle\DOneTwo\rangle$) for trimmed jets ($\Rsub=0.3,\fcut=0.05$), 
     for data and simulation. 
     The error bars indicate the statistical uncertainty on the mean value in each bin. 
     The lower panels show the ratio of the mean values measured in data to simulation.
    }
  \label{fig:DataMC:d12Tau32Npv}
\end{figure}

\Figref{DataMC:d12Tau32Npv} presents a comparison of data and Monte Carlo simulation for $\langle\DOneTwo\rangle$ and $\langle\tauThrTwo\rangle$ for ungroomed and trimmed ($\Rsub=0.3,\fcut=0.05$) \antikt\ jets with $R=1.0$. The slope of these observables as a function of \Npv is well modelled by the simulation.

\subsection{Impact of \pileup on signal and background in simulation}
\label{sec:pileup:mconly}

In addition to the comparisons between data and simulation, and between the various grooming configurations, a comparison of how grooming impacts signal-like events versus background-like events in searches for resonances decaying to boosted jets is crucial.

\begin{figure}[!h]
  \centering
  \subfigure[Ungroomed]{	
    \includegraphics[width=0.46\columnwidth]{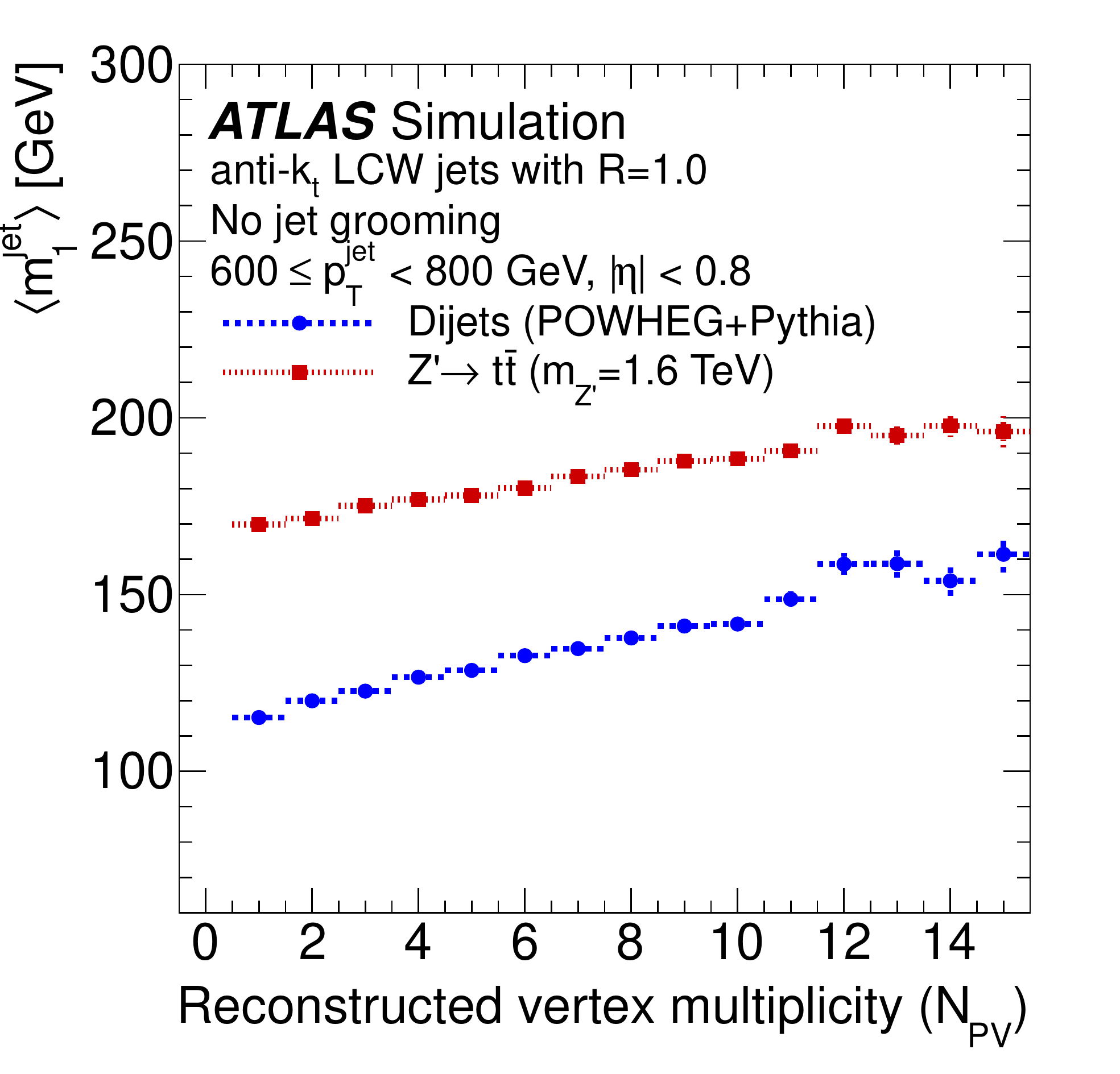}
    \label{fig:pileup:mconly:mass1:AKTFat600}}
  \subfigure[Trimmed]{
    \includegraphics[width=0.46\columnwidth]{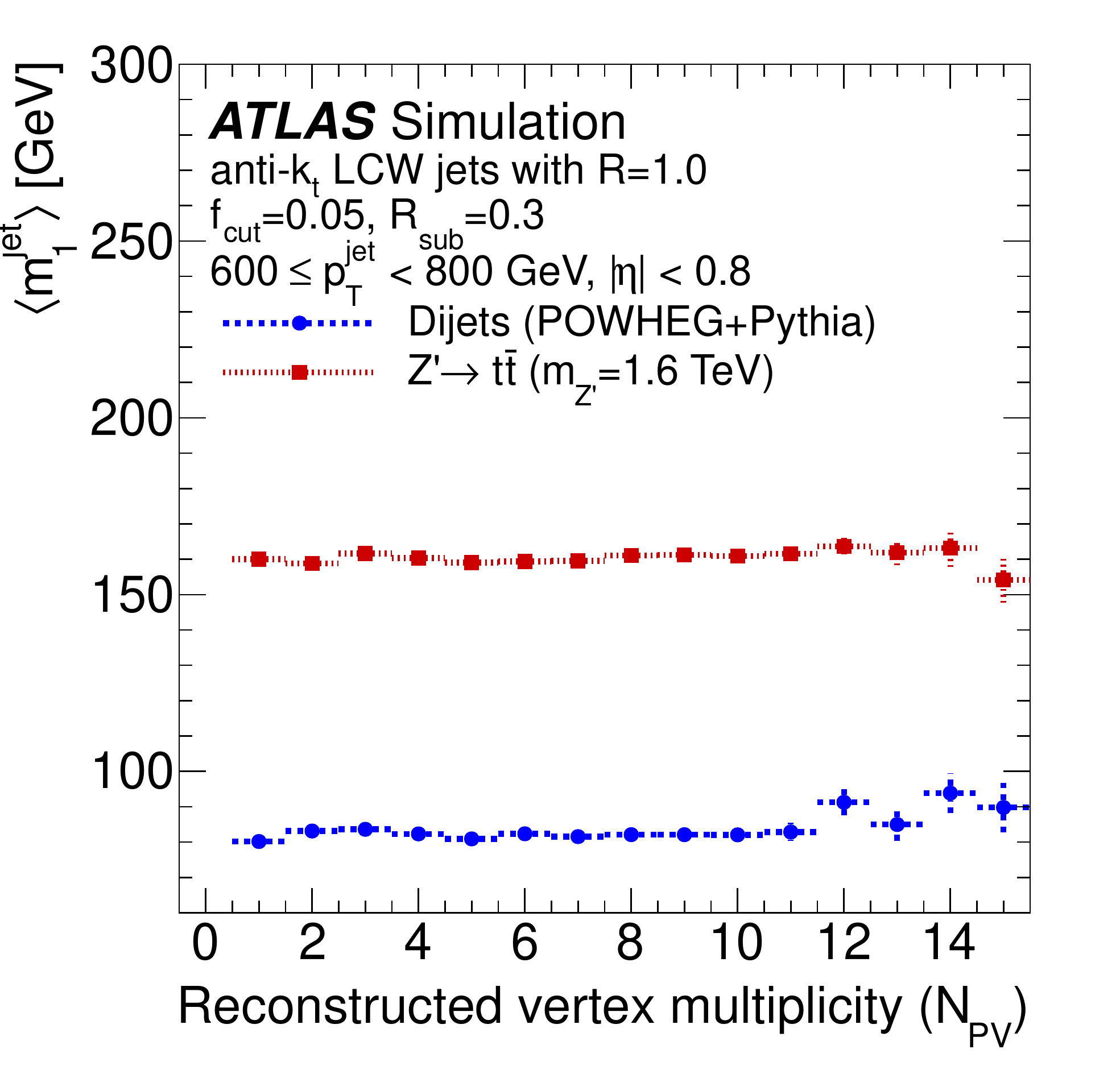}
    \label{fig:pileup:mconly:mass1:AKTTrim600}}
  \caption{Variation with the number of reconstructed primary vertices, \Npv, of the average leading-\ptjet\ jet mass, \mleadavg, in the range $600\GeV\leq\ptjet<800$~GeV for \subref{fig:pileup:mconly:mass1:AKTFat600} ungroomed and \subref{fig:pileup:mconly:mass1:AKTTrim600} trimmed jets. The error bars indicate the statistical uncertainty on the mean value in each bin.
           }
  \label{fig:pileup:mconly:mass1:AKTTrim}
\end{figure}

\Figref{pileup:mconly:mass1:AKTTrim} shows the variation of the average leading-\ptjet\ jet mass, \mleadavg, with \Npv for events with $600\GeV\leq\ptjet<800$~GeV for ungroomed and trimmed \AKTFat jets, for both the \Zprimett sample and the \PowPythia dijet sample. The average ungroomed leading-\ptjet\ jet mass in the sample of gluons and light quarks in the inclusive \PowPythia dijet events exhibits a slope of approximately $\mathrm{d}\mleadavg/\mathrm{d}\Npv\approx3$~GeV/\Npv. The leading-\ptjet\ jets in the \Zprime\ sample are typically entirely composed of fully hadronic boosted top-quark decays contained in a single jet. The mass reconstruction in this case proceeds as usual (four-momentum recombination) and the mass distribution is highly peaked near the top-quark mass of approximately 175~GeV. Jets in this peak but without grooming exhibit a slope of roughly $\mathrm{d}\mleadavg/\mathrm{d}\Npv\approx2.15$~GeV/\Npv, or about 30\% smaller than in the inclusive jet sample. In the case of trimmed jets, the slopes as a function of \Npv for both signal-like jets and jets in dijet events are consistent with zero.

Most importantly, the average separation between the mean jet mass for signal-like jets in the \Zprime\ sample and those in the \PowPythia dijet sample increases by nearly 50\% after trimming and remains stable across the full range of \Npv. This allows for much better discrimination between the two processes. The separation shown here is significant since the widths of the peaks of each of the distributions are also simultaneously narrowed by the grooming algorithm, as shown in \figref{DataMC:massDistro}. This differential impact of trimming is again due to the design of the algorithm: soft, wide-angle contributions to the jet mass that are ubiquitous in jets produced from light quarks and gluons are suppressed whereas the hard components present in a jet with true substructure -- as in the case of the top-quark jets here -- are preserved.

\section{Jet substructure and grooming with boosted objects in data and simulation}
\label{sec:boostedwtop}

Comparisons between jets containing \emph{signal-like} boosted objects and a 
light-quark or gluon jet background are presented here. Boosted objects are divided 
into two categories depending on the event topology: \emph{two-pronged}, such as hadronically 
decaying \W or \Z bosons, and \emph{three-pronged}, such as the top quark decaying into a $b$-jet 
and a hadronically decaying \W boson. Performance measures are shown for both simulated samples 
of \Zqq and top quarks (from \Zprimett), as well as for inclusive jet data and events 
enriched in boosted top quark pairs. In addition to the event and object selection 
listed in \secref{data-mc}, the \largeR ungroomed leading-\ptjet\ jet axis is required to be 
within $|\eta|<2.0$. For the signal distributions, a $\DeltaR<1.5$ match between the 
four-momentum of the hadronically decaying boosted object in the truth record and the 
reconstructed ungroomed leading-\ptjet\ jet is made to minimize the contamination from 
light-quark or gluon jets (or top quarks with a leptonically decaying \W boson in the 
\Zprime\ sample).

\subsection{Expected performance of jet substructure and grooming in simulation}
\label{sec:boostedwtop:mc}

\subsubsection{Jet mass resolution for background}
\label{sec:boostedmc:qcdjetres}

The fractional jet mass resolution is defined as the width of a Gaussian fit to the central part of the distribution that is generated by taking the difference between the generator-level jet mass and the reconstructed jet mass, divided by the same generator-level jet mass. Here, the generator-level jet is the simulated particle shower that has been groomed according to the same grooming algorithms used after jet reconstruction. \LargeR generator-level and reconstructed jets before grooming are matched if they are within $\DeltaR<0.7$. The matching is performed only once and comparisons between generator-level and reconstructed jets after grooming are made using the groomed versions of the matched ungroomed jets. The mass-drop filtering method is not applied to \antikt\ jets, as discussed in \secref{intro:definitions}.

\begin{table}[ht]
\footnotesize
\begin{center}
\begin{tabular}{|l|l||l|l|}
\hline
\hline
  Label  \B \T             & Algorithm and Parameters  & Label  \B \T             & Algorithm and Parameters \\[1ex]
\hline
  Trm PtF1 R30 \T       & Trimmed, $\fcut=1\%$, $\drsub=0.3$ &   Prn Rc10 Zc5        & Pruned, $\rcut=0.1$, $\zcut=5\%$   \\[1ex]
  Trm PtF3 R30$^\dag$ & Trimmed, $\fcut=3\%$, $\drsub=0.3$   &   Prn Rc10 Zc10       & Pruned, $\rcut=0.1$, $\zcut=10\%$    \\[1ex]
  Trm PtF5 R30$^\ddag$ & Trimmed, $\fcut=5\%$, $\drsub=0.3$  &   Prn Rc20 Zc5        & Pruned, $\rcut=0.2$, $\zcut=5\%$    \\[1ex]
  Trm PtF1 R20        & Trimmed, $\fcut=1\%$, $\drsub=0.2$   &   Prn Rc20 Zc10       & Pruned, $\rcut=0.2$, $\zcut=10\%$    \\[1ex]
  Trm PtF3 R20        & Trimmed, $\fcut=3\%$, $\drsub=0.2$   &   Prn Rc30 Zc5        & Pruned, $\rcut=0.3$, $\zcut=5\%$    \\[1ex]
  Trm PtF5 R20        & Trimmed, $\fcut=5\%$, $\drsub=0.2$   &   Prn Rc30 Zc10       & Pruned, $\rcut=0.3$, $\zcut=10\%$    \\[1ex]
  \hline
  MD Filt mf20 \T         & Mass-drop Filtered, $\mufrac=0.2$  & MD Filt mf67$^\S$\B  & Mass-drop Filtered, $\mufrac=0.67$ \\[1ex]
  MD Filt mf33         & Mass-drop Filtered, $\mufrac=0.33$ & &  \\[1ex]
\hline
\hline
\end{tabular}
\end{center}
\caption{Labels used in figures to represent the various configurations of the
grooming algorithms. $^\dag$Groomed jets have been calibrated for both
\antikt~and \CamKt jets. $^\ddag$Groomed jets have been calibrated for
\antikt~only. $^\S$Groomed jets have been calibrated for \CamKt only.}
\label{tab:groomingkey}
\end{table}

\begin{figure}[ht]  
  \begin{center}
       \subfigure[\AKTFat]{
         \includegraphics[width=0.45\textwidth]{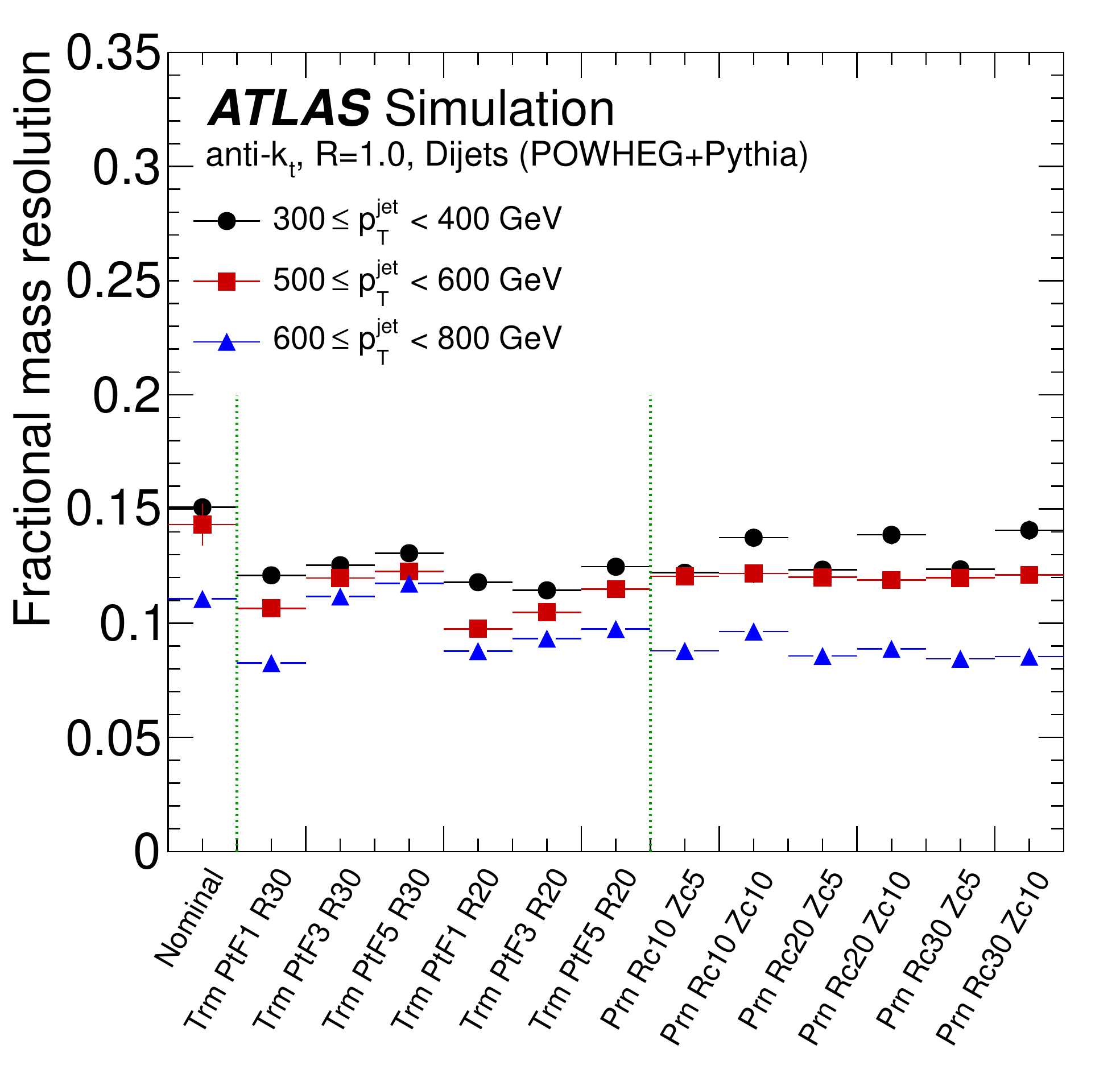} 
        }
       \subfigure[\CAFat]{
         \includegraphics[width=0.45\textwidth]{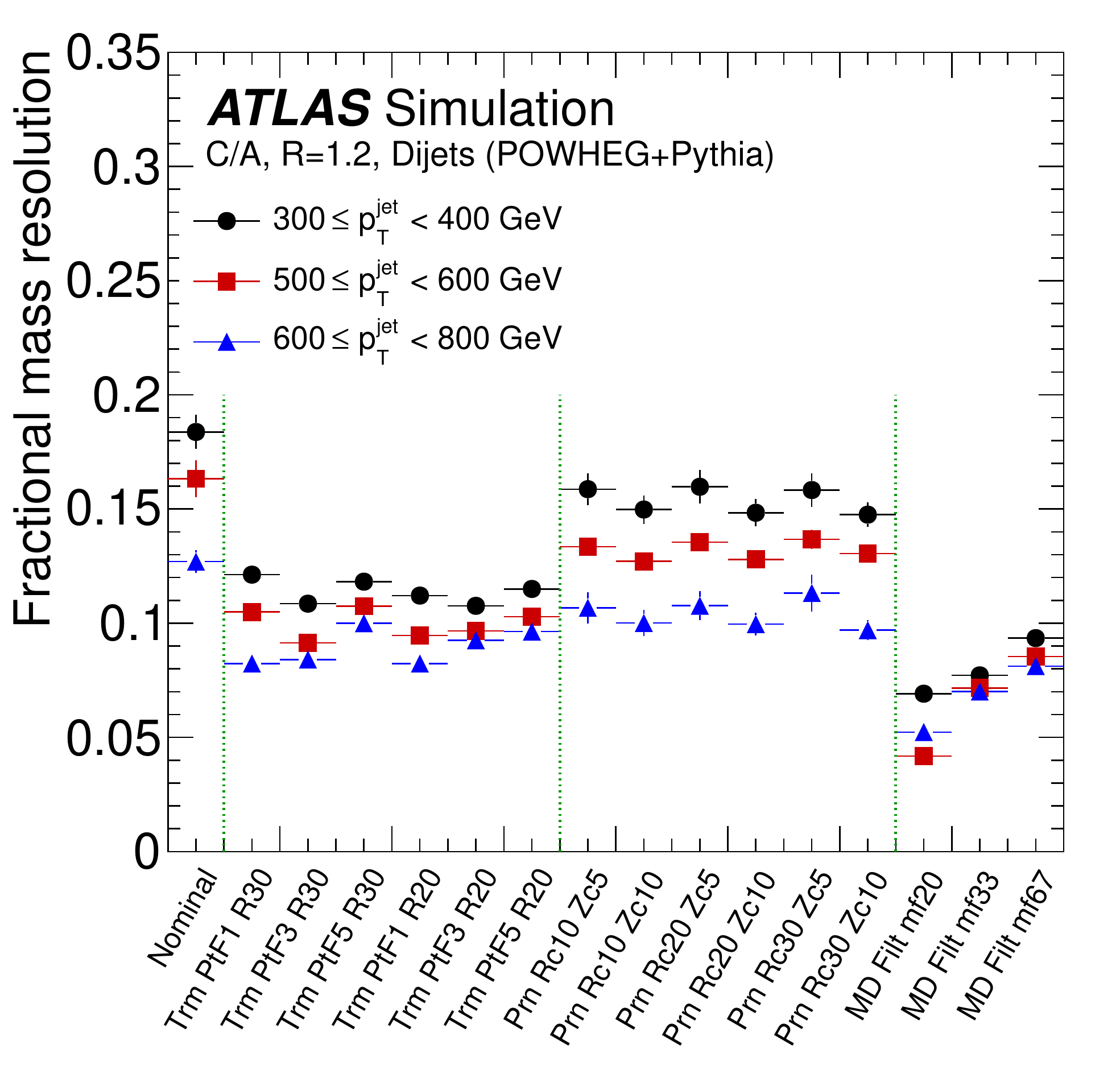} 
        }
  \end{center}
  \caption{
     Fractional mass resolution comparing the various grooming 
     algorithms (with labels defined in \tabref{groomingkey}) 
     for the leading-\ptjet\ jet in \PowPythia dijet 
     simulated events. Here, \emph{nominal} refers to jets before 
     grooming is applied. Three ranges of the nominal jet \ptjet\ 
     are shown. The uncertainty on the width of the Gaussian fit 
     is indicated by the error bars.
  \label{fig:groomed_jets_mass_resolutions_pt}} 
\end{figure}

\Figref{groomed_jets_mass_resolutions_pt} shows the fractional mass resolution for leading-\ptjet\ jets in the \PowPythia dijet sample with the same \pileup conditions as were observed in the data. Abbreviated versions of the groomed algorithm names used to label the figures are listed in \tabref{groomingkey}.  
In general, the groomed jets have better resolution than the ungroomed \largeR jets, with improvements of up to $\sim$10\% (absolute) in some cases. 
The trimmed and pruned jet resolution improves with increasing \pT, where the calibrated jets gain $\sim$$3-5\%$ over the range $300\GeV\leq\ptjet<800$~GeV.  
The pruning algorithm, especially with \CamKt jets, produces larger tails in the resolution distribution compared to the trimmed algorithm, worsening the overall fractional resolution in comparison.
The resolution is fairly stable for the mass-drop filtering algorithm over a large range of \pT.
It is important to note that the efficiency is considerably lower for jets resulting from this algorithm compared with jets produced in other grooming procedures ($\sim$30$\%$) due to the strict mass-drop requirement, which is often not met for jets without boosted object substructure.

\begin{figure}[ht]  
  \begin{center}
       \subfigure[\AKTFat]{
         \includegraphics[width=0.45\textwidth]{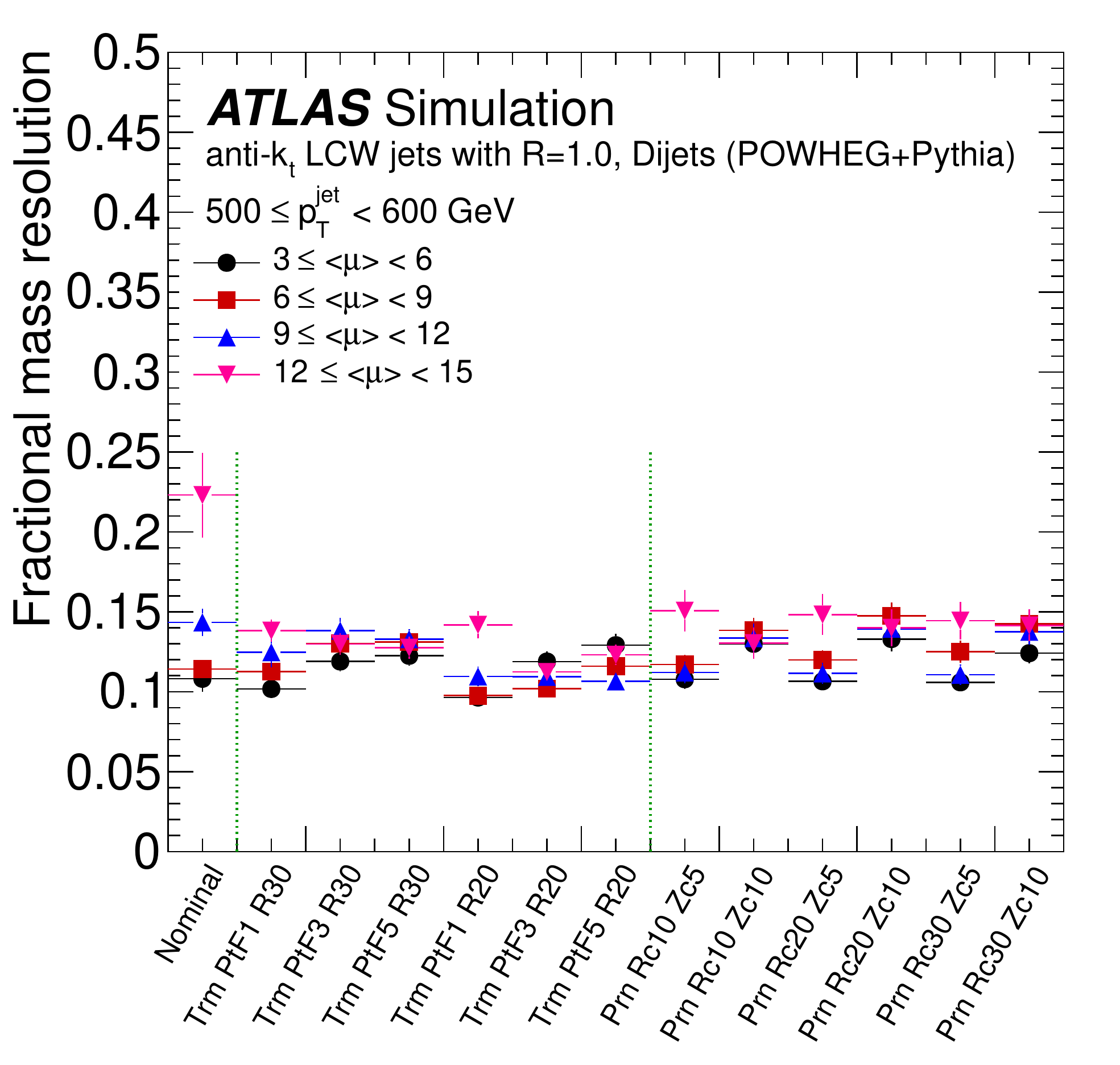} 
        }
       \subfigure[\CAFat]{
         \includegraphics[width=0.45\textwidth]{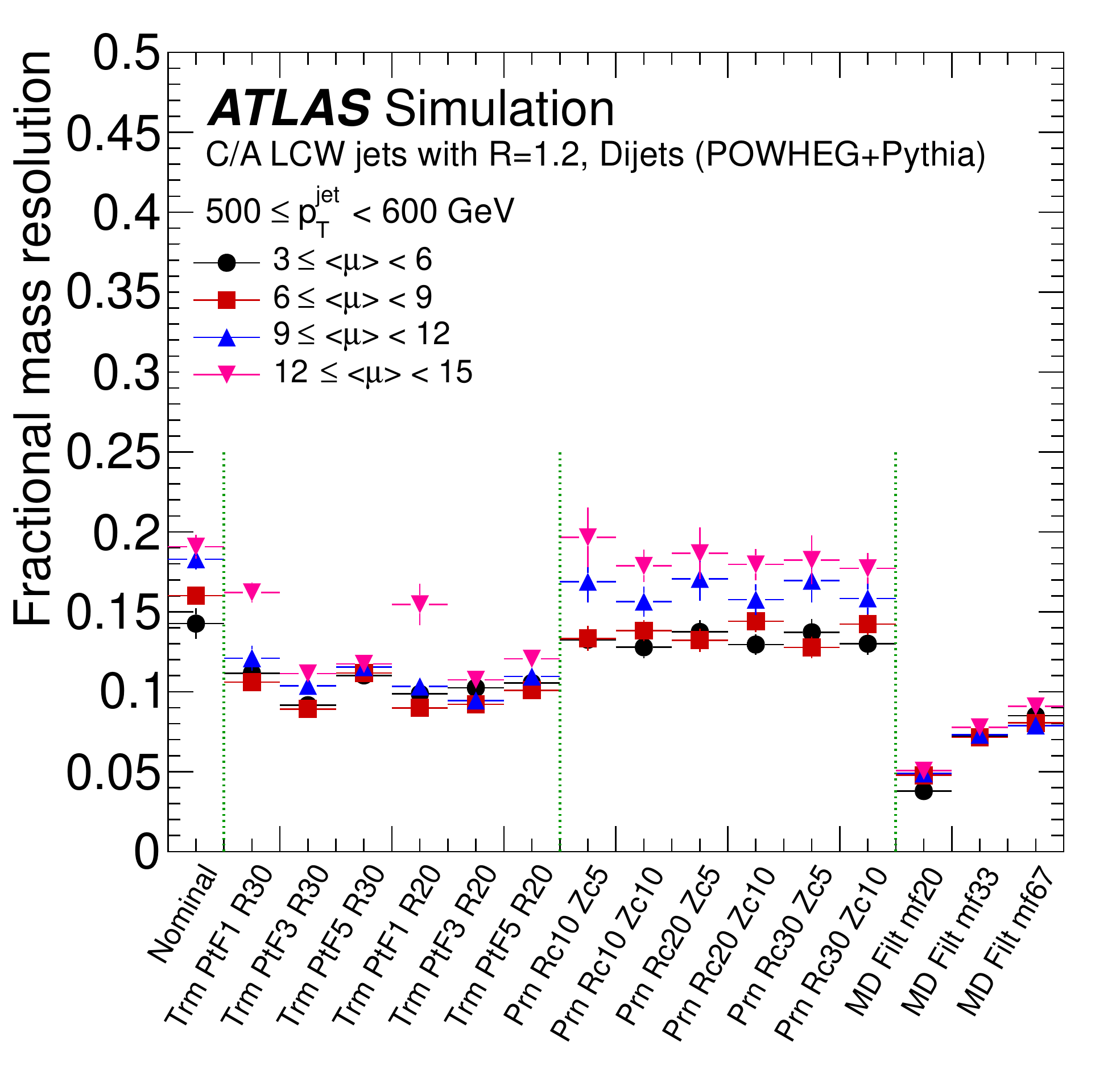} 
       }
  \end{center}
  \caption{Fractional mass resolution comparing the various grooming algorithms 
          (with labels defined in \tabref{groomingkey}) for the 
           leading-\ptjet jet in the range $500\GeV\leq\ptjet<600$~GeV 
           in dijet events, simulated with \PowPythia.
           \emph{Nominal} refers to jets before grooming is applied. Various ranges of
           the average number of interactions (\avgmu) in the events are shown. 
          The uncertainty on the width of the Gaussian fit is indicated by the error bars.
  \label{fig:groomed_jets_mass_resolutions_MU}} 
\end{figure}

A summary of the fractional mass resolution for jets before and after grooming in the presence of various \pileup conditions is shown in \figref{groomed_jets_mass_resolutions_MU}. 
Trimming in both \akt and \CamKt jets reduces the dependence of the jet mass on \pileup (spread in the points) compared to the ungroomed jet, as does the mass-drop filtering procedure in the case of \CamKt jets, while pruning has little impact. In all cases, no \pileup subtraction is applied to the ungroomed jet kinematics.
In particular, the trimming parameters $\fcut=0.03$ and 0.05 slightly outperform the looser $\fcut=0.01$ setting in events with a mean number of interactions greater than 12. They also exhibit a significantly reduced overall variation between various instantaneous luminosities.

Based on the above comparisons of mass resolution in different \ptjet\ ranges and under various \pileup conditions, two configurations, trimmed \akt\ jets ($\fcut=0.05$, $\drsub=0.3$) with $R=1.0$ and filtered \CamKt jets ($\mufrac=0.67$) with $R=1.2$, are chosen for detailed comparisons between data and simulation and are presented in \secref{boostedwtop:inclusivedatamc}.

\subsubsection{Jet mass resolution for simulated signal events}
\label{sec:boostedmc:signaljetres}

\begin{figure}[!ht]  
  \begin{center}
       \subfigure[\AKTFat]{
         \includegraphics[width=0.45\textwidth]{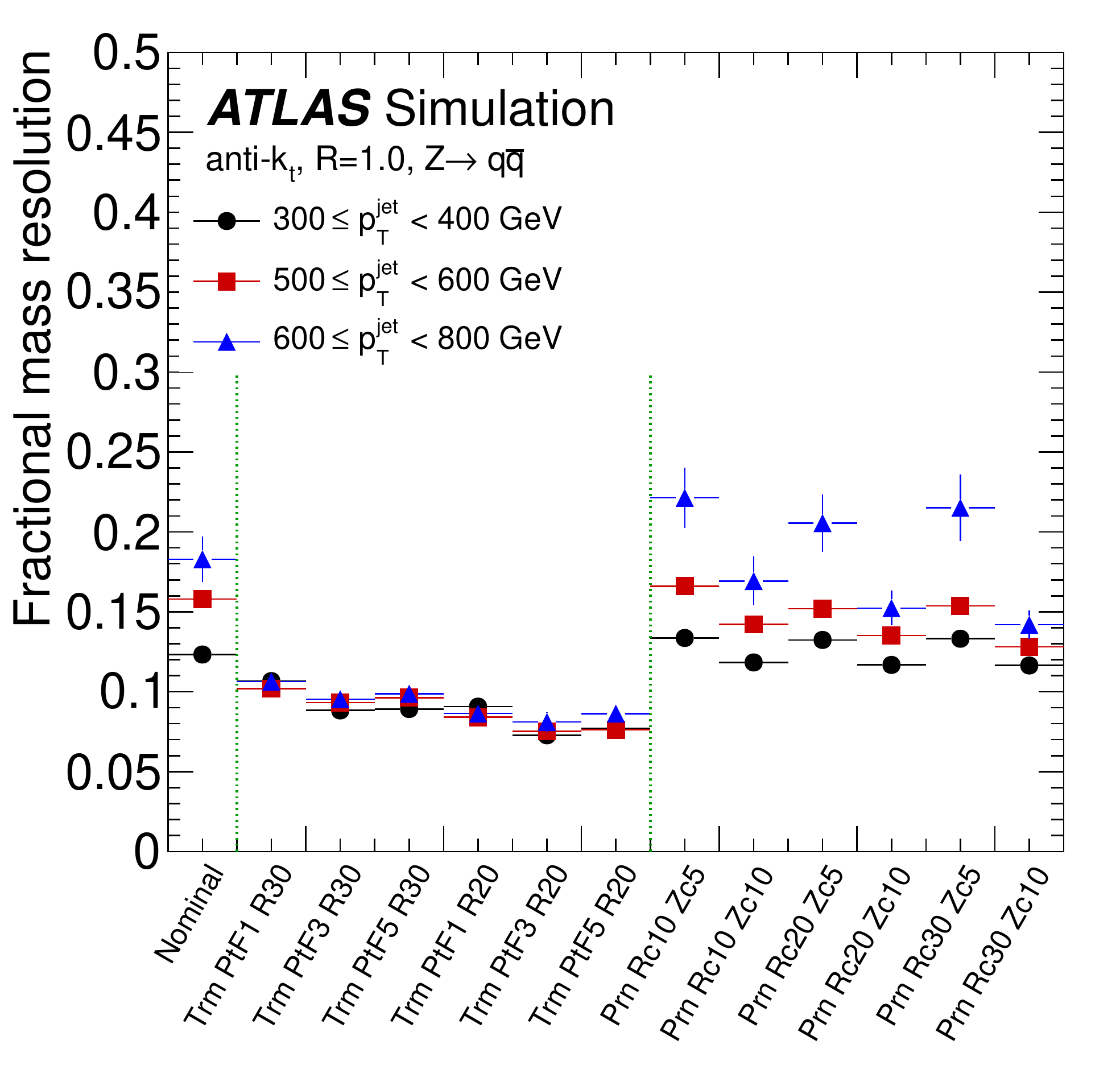} }
       \subfigure[\CAFat]{
         \includegraphics[width=0.45\textwidth]{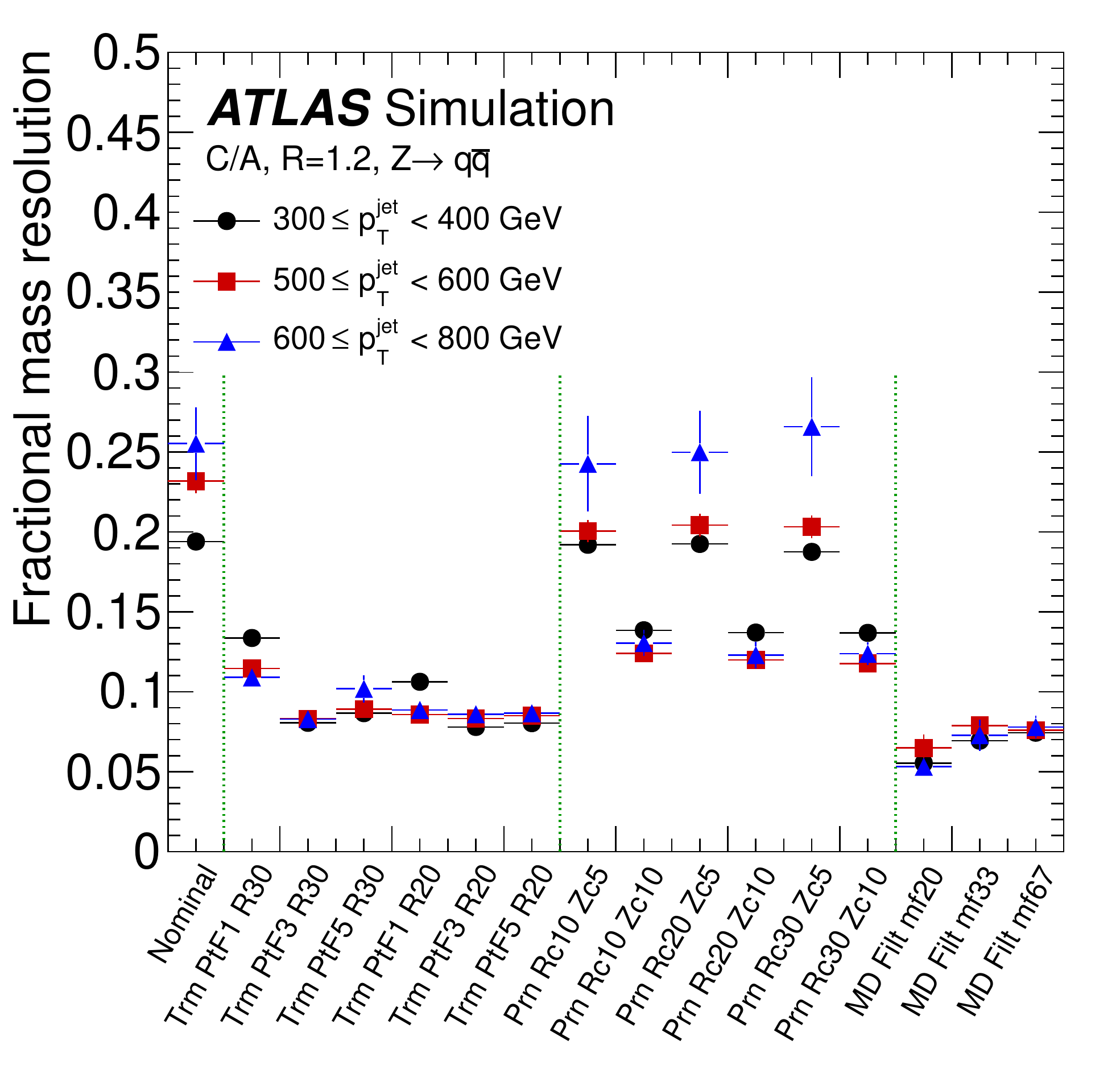} }
  \end{center}
  \caption{
     Fractional mass resolution of the leading-\ptjet\ jet in \Zqq 
     simulated events comparing the various grooming algorithms.
     Here, \emph{nominal} refers to jets before grooming is applied. 
     Three ranges of the nominal jet \ptjet\ are shown. The 
     uncertainty on the width of the Gaussian fit is indicated by the error bars.
  \label{fig:groomed_jets_mass_resolutions_pt_twoprong}} 
\end{figure}
\begin{figure}[!ht]  
  \begin{center}
      \subfigure[\AKTFat]{
        \includegraphics[width=0.45\textwidth]{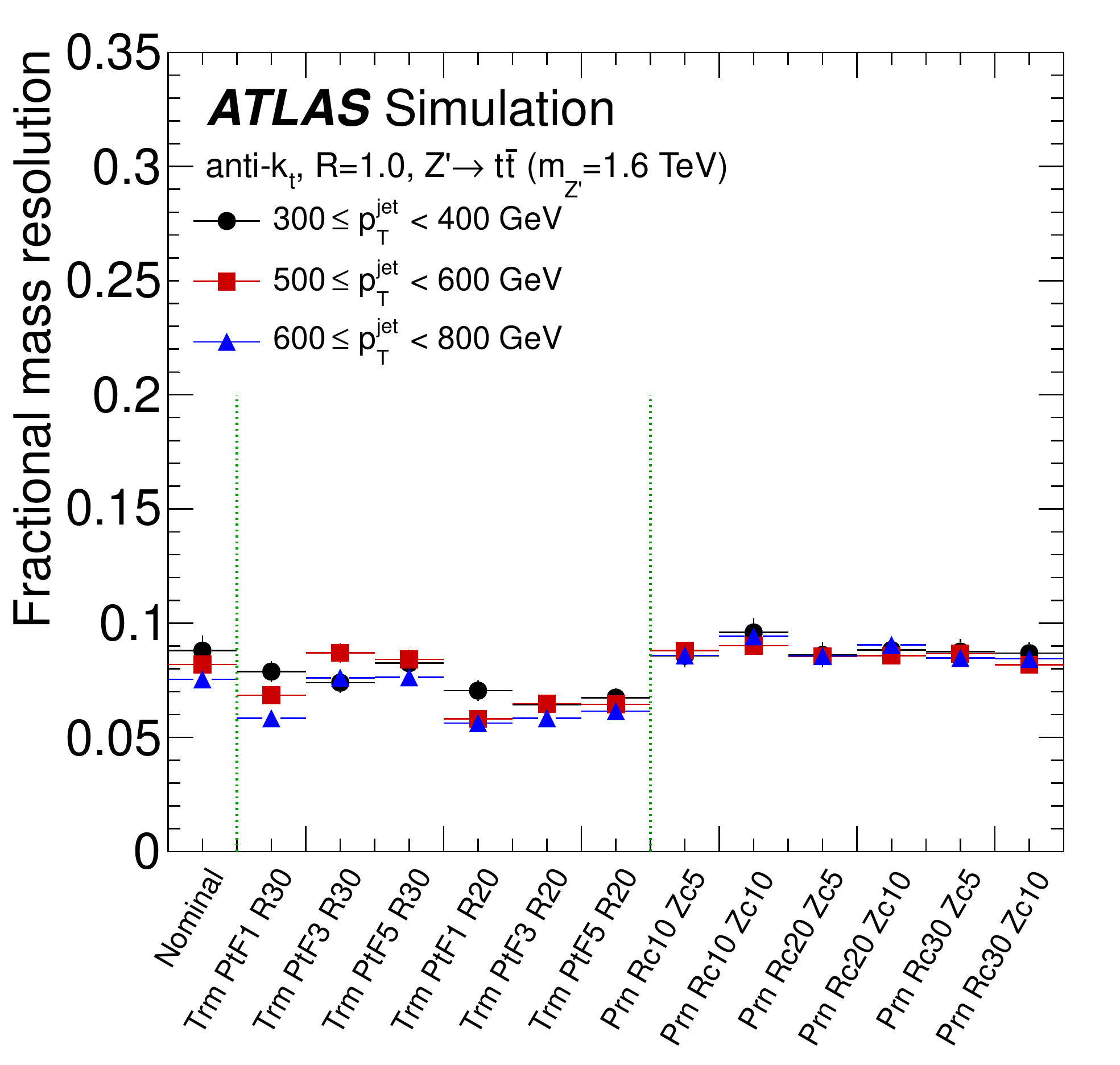} }
      \subfigure[\CAFat]{
        \includegraphics[width=0.45\textwidth]{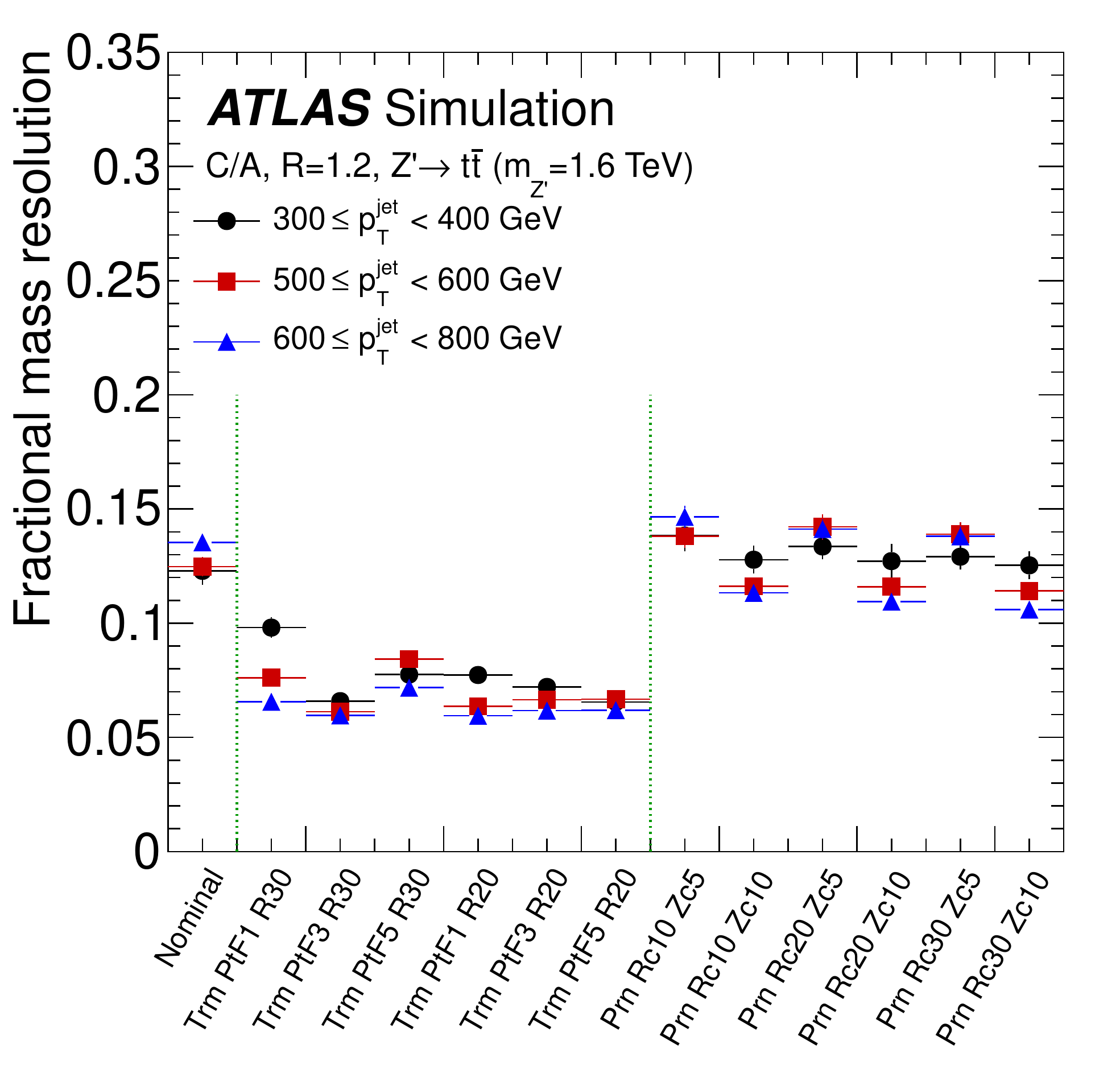} }
  \end{center}
  \caption{
     Fractional mass resolution of the leading-\ptjet\ jet in \Zprimett 
     ($\massZprime=1.6$~TeV) simulated events comparing the various 
     grooming algorithms. Here, \emph{nominal} refers to the jet 
     before grooming is applied. Three ranges of the ungroomed jet \ptjet\ are 
     shown. The uncertainty on the width of the Gaussian fit is 
     indicated by the error bars.
  \label{fig:groomed_jets_mass_resolutions_pt_threeprong}} 
\end{figure}

\Figsref{groomed_jets_mass_resolutions_pt_twoprong}{groomed_jets_mass_resolutions_pt_threeprong} show the fractional mass resolution for the two-pronged and three-pronged cases, respectively. The mass-drop filtering algorithm is shown only for the simulated two-pronged signal events with \CamKt jets.  
In the two-pronged case, as for the case of jets in the inclusive jet events shown in \figref{groomed_jets_mass_resolutions_pt}, the \CamKt mass-drop filtering algorithm performs the best, but with a signal reconstruction efficiency of $\sim45\%$ in \Zqq events (for $\mufrac=0.67$). 
In both the two-pronged and three-pronged configurations, the trimmed jets have better fractional mass resolution ($\sim5-10$\%) than the pruned jets, especially for those jets with grooming applied after the \CamKt algorithm.  
The trimmed jet mass resolution also remains fairly stable across a large $\ptjet$ range, with equivalent performance for \antikt\ and \CamKt jets.

\subsubsection{Signal and background comparisons with and without grooming}
\label{sec:boostedmc:signalbckg}

\begin{figure}[!ht]  
  \begin{center}
      \subfigure[\AKTFat]{
        \includegraphics[width=0.45\textwidth]{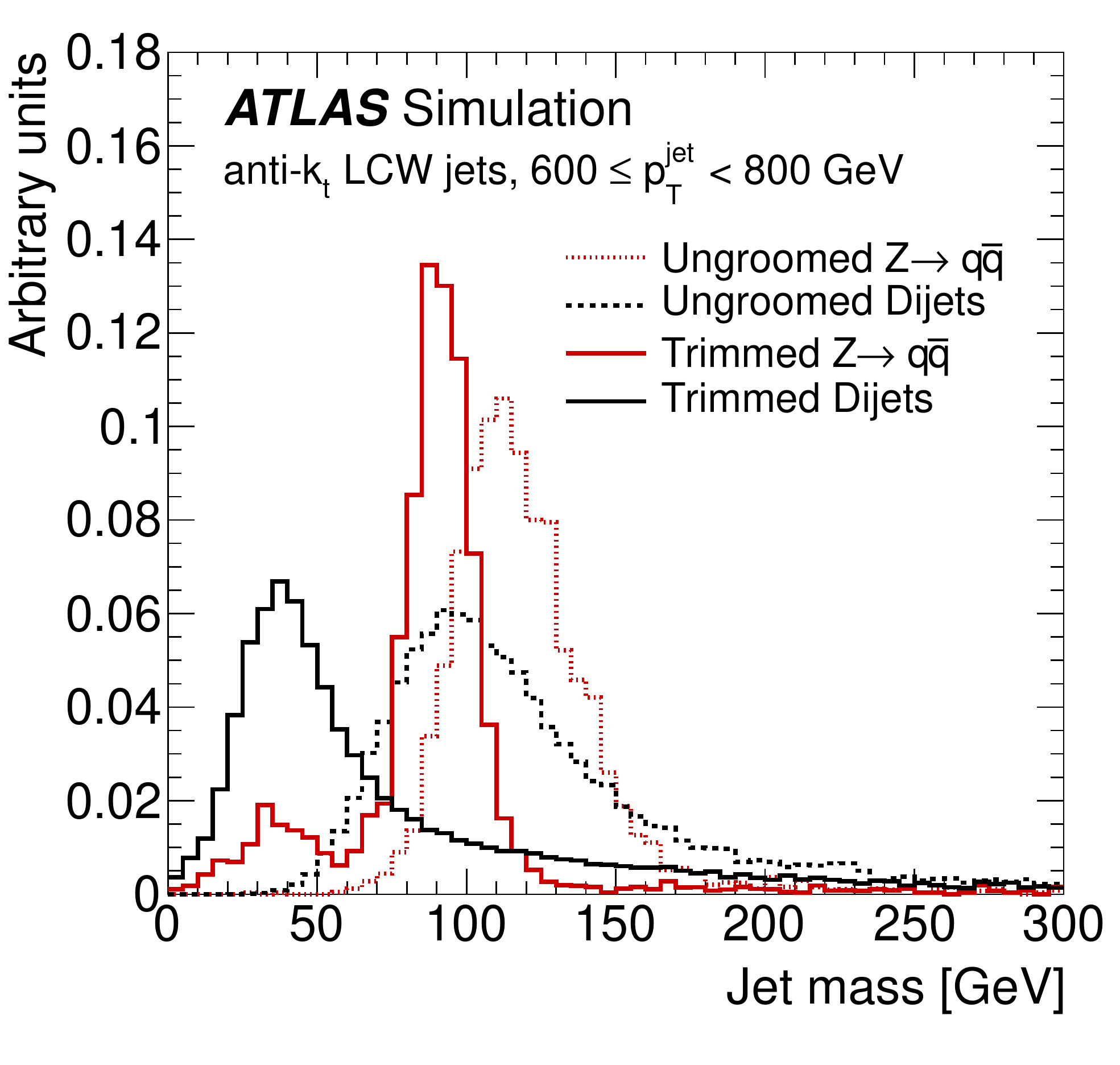} 
        \label{fig:groomed_jets_mass_compare_twoprong_AKTFat}
      }
      \subfigure[\CAFat]{ 
        \includegraphics[width=0.45\textwidth]{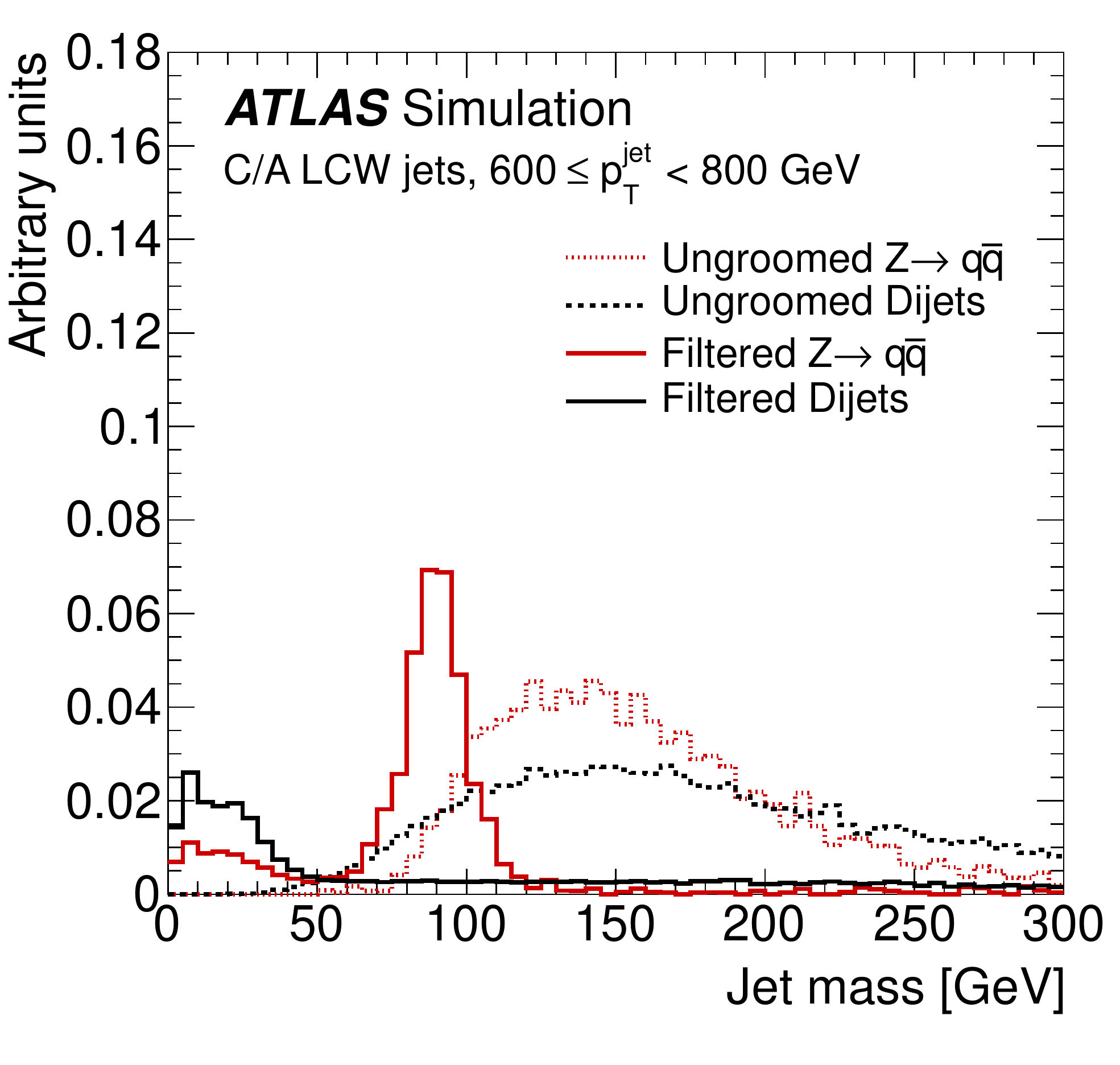} 
        \label{fig:groomed_jets_mass_compare_twoprong_CAFat}
      }
  \end{center}
  \caption{Leading-\ptjet\ jet mass for simulated \HJimmy \Zqq signal events (red) compared to
  \PowPythia dijet background events (black) for jets in the range 
  $600\GeV\leq\ptjet<800$~GeV. The dotted lines show the ungroomed jet distributions, 
  whereas the solid lines show the 
  \subref{fig:groomed_jets_mass_compare_twoprong_AKTFat} trimmed and 
  \subref{fig:groomed_jets_mass_compare_twoprong_CAFat} mass-drop filtered jet distributions. 
  The trimming parameters are $\fcut=0.05$ and $\drsub=0.3$ and the 
  mass-drop filtering parameter is $\mufrac=0.67$. The groomed distributions 
  are normalized with respect to the ungroomed distributions, which are themselves 
  normalized to unity. 
  \label{fig:groomed_jets_mass_compare_twoprong}} 
\end{figure}

\begin{figure}[!ht]  
  \begin{center}
      \subfigure[\AKTFat]{
        \includegraphics[width=0.45\textwidth]{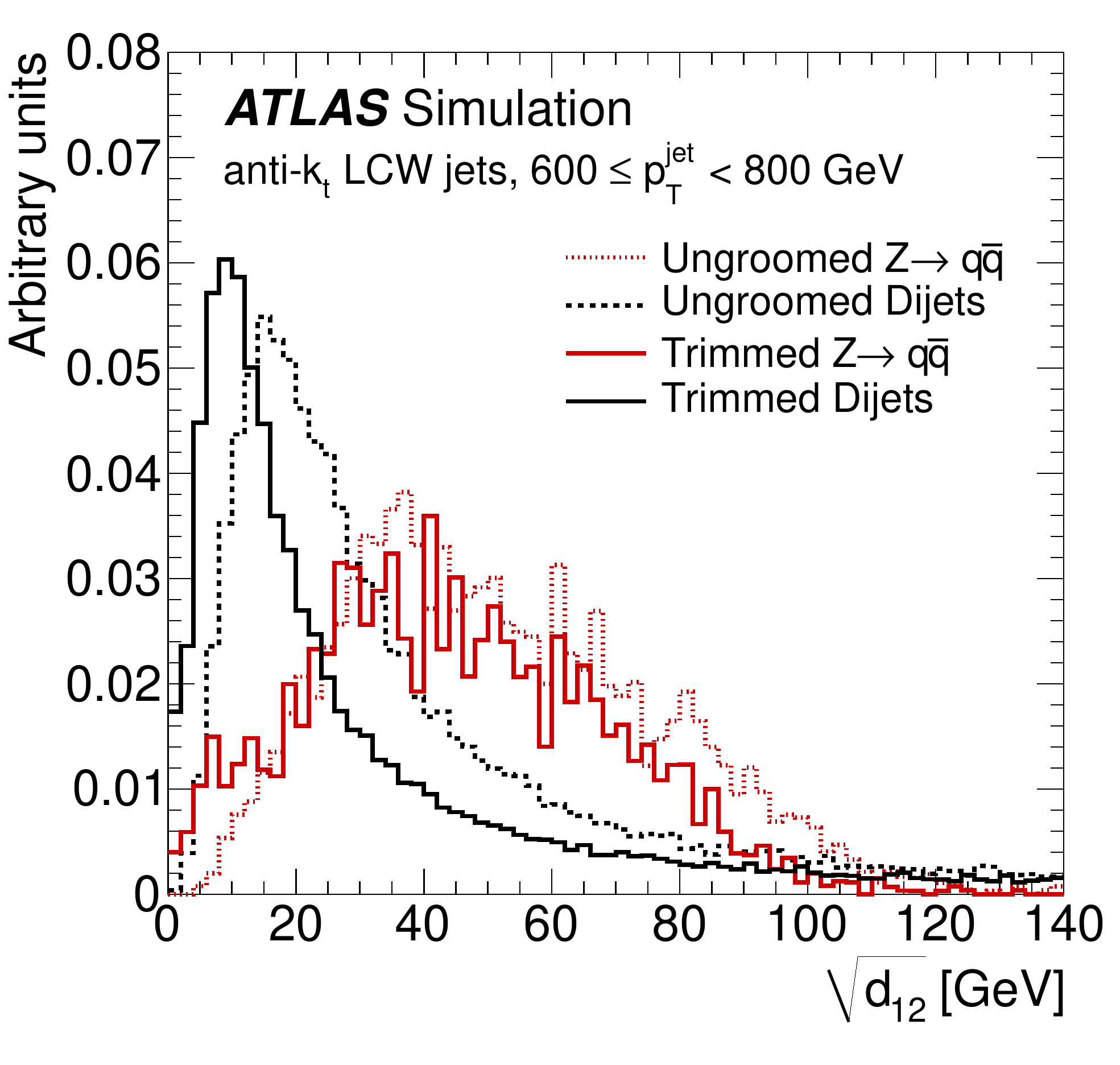} 
        \label{fig:groomed_jets_split_compare_twoprong_AKTFat}
      }
      \subfigure[\CAFat]{
        \includegraphics[width=0.45\textwidth]{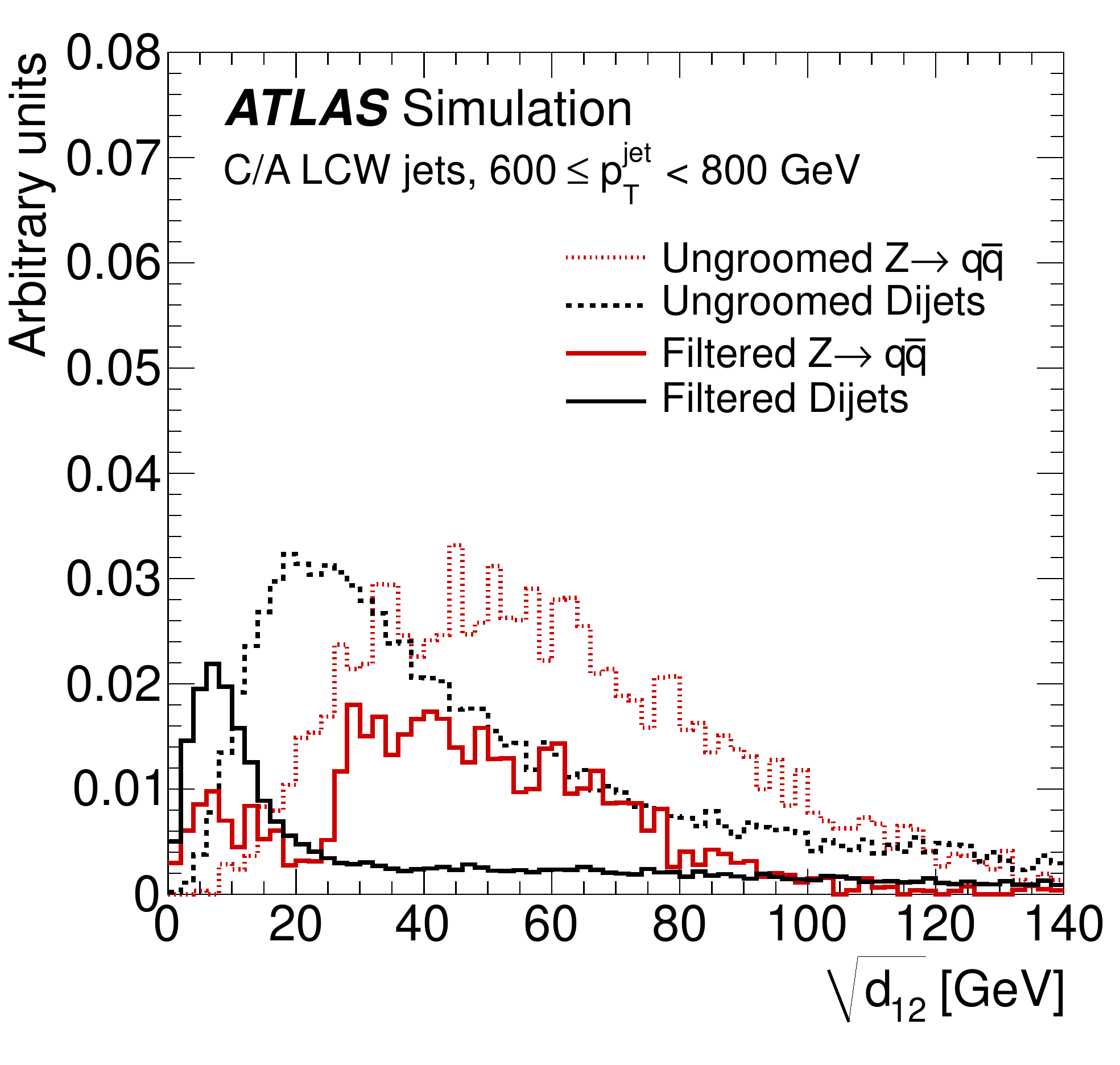} 
        \label{fig:groomed_jets_split_compare_twoprong_CAFat}
      }
  \end{center}
  \caption{Leading-\ptjet\ jet splitting scale \DOneTwo for simulated \HJimmy \Zqq signal 
  events (red) compared to \PowPythia dijet background events (black) for jets in 
  the range $600\GeV\leq\ptjet<800$~GeV. The dotted lines show the ungroomed jet 
  distributions, while the solid lines show the 
  \subref{fig:groomed_jets_split_compare_twoprong_AKTFat} trimmed and 
  \subref{fig:groomed_jets_split_compare_twoprong_CAFat} mass-drop filtered jet distributions.  
  The trimmed parameters are $\fcut=0.05$ and $\drsub=0.3$ and the mass-drop filtering 
  parameter is $\mufrac=0.67$. The groomed distributions are normalized with 
  respect to the ungroomed distributions, which are themselves normalized to unity.
  \label{fig:groomed_jets_split_compare_twoprong}} 
\end{figure}

\begin{figure}[!ht]  
  \begin{center}
      \subfigure[\AKTFat]{
        \includegraphics[width=0.45\textwidth]{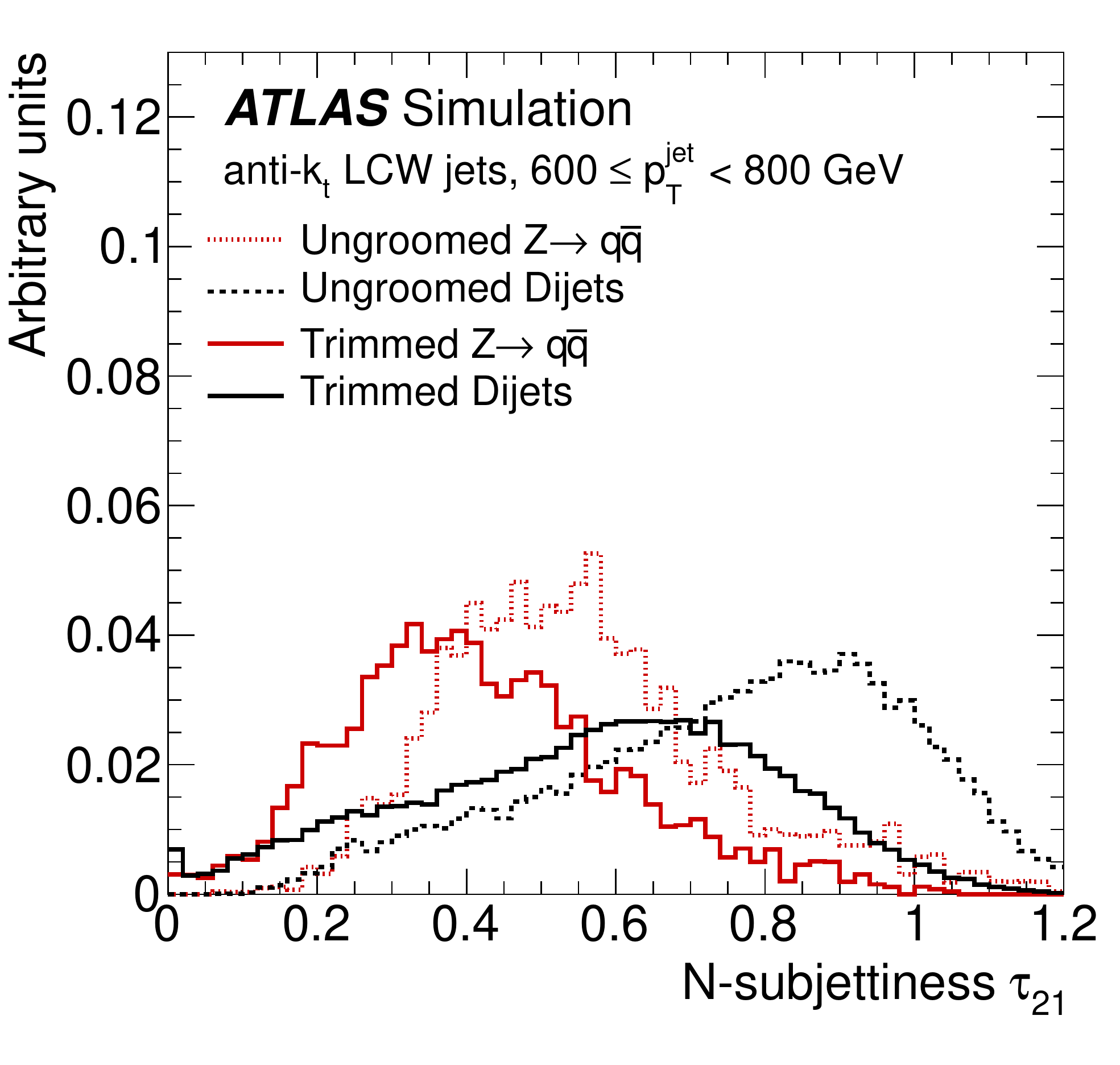} 
        \label{fig:groomed_jets_subjetiness_compare_twoprong_AKTFat}
      }
      \subfigure[\CAFat]{
        \includegraphics[width=0.45\textwidth]{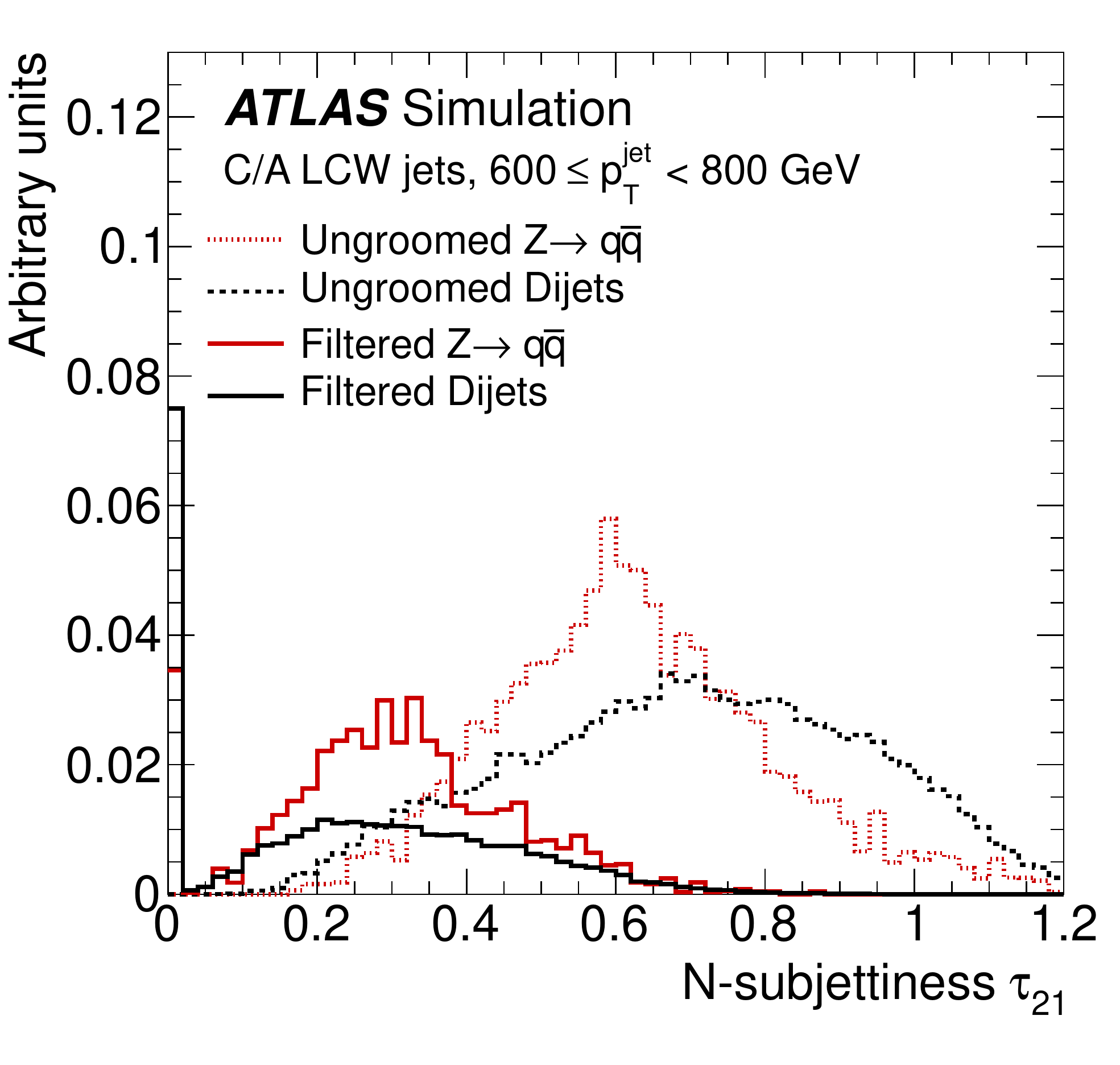} 
        \label{fig:groomed_jets_subjetiness_compare_twoprong_CAFat}
      }
  \end{center}
  \caption{Leading-\ptjet\ jet $N$-subjettiness \tauTwoOne for simulated \HJimmy \Zqq signal 
  events (red) compared to \PowPythia dijet background events (black) for jets in the range 
  $600\GeV\leq\ptjet<800$~GeV.  The dotted lines show the ungroomed jet distributions, 
  while the solid lines show the \subref{fig:groomed_jets_split_compare_twoprong_AKTFat}
  trimmed and \subref{fig:groomed_jets_split_compare_twoprong_CAFat} mass-drop filtered jet 
  distributions. The trimmed parameters are $\fcut=0.05$ and $\drsub=0.3$ and the 
  mass-drop filtering parameter is $\mufrac=0.67$. The groomed distributions 
  are normalized with respect to the ungroomed distributions, which are 
  normalized to unity.
   \label{fig:groomed_jets_subjetiness_compare_twoprong}} 
\end{figure}

\begin{figure}[!ht]  
  \begin{center}
      \subfigure[\AKTFat]{
        \includegraphics[width=0.45\textwidth]{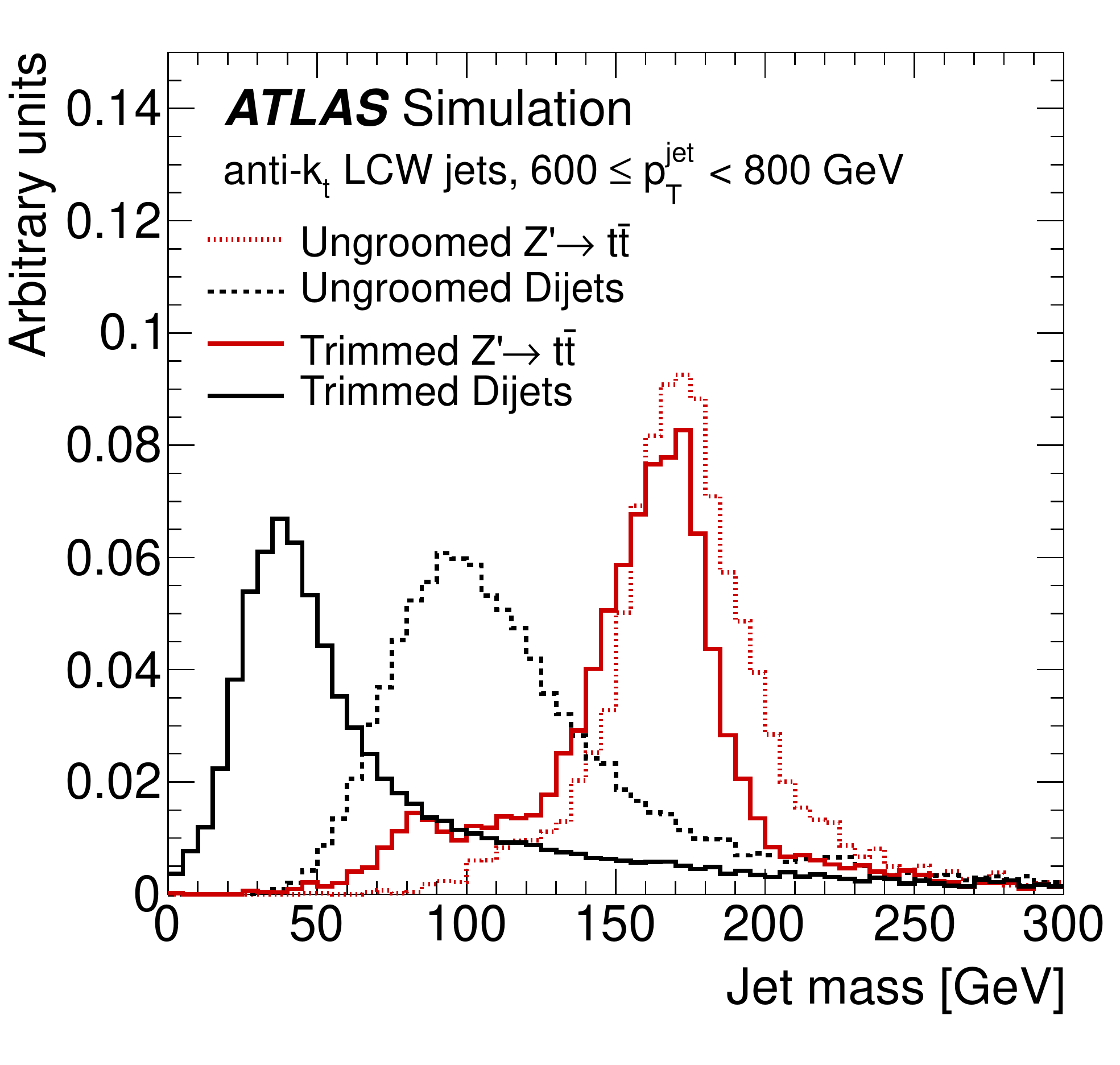} }
      \subfigure[\CAFat]{
        \includegraphics[width=0.45\textwidth]{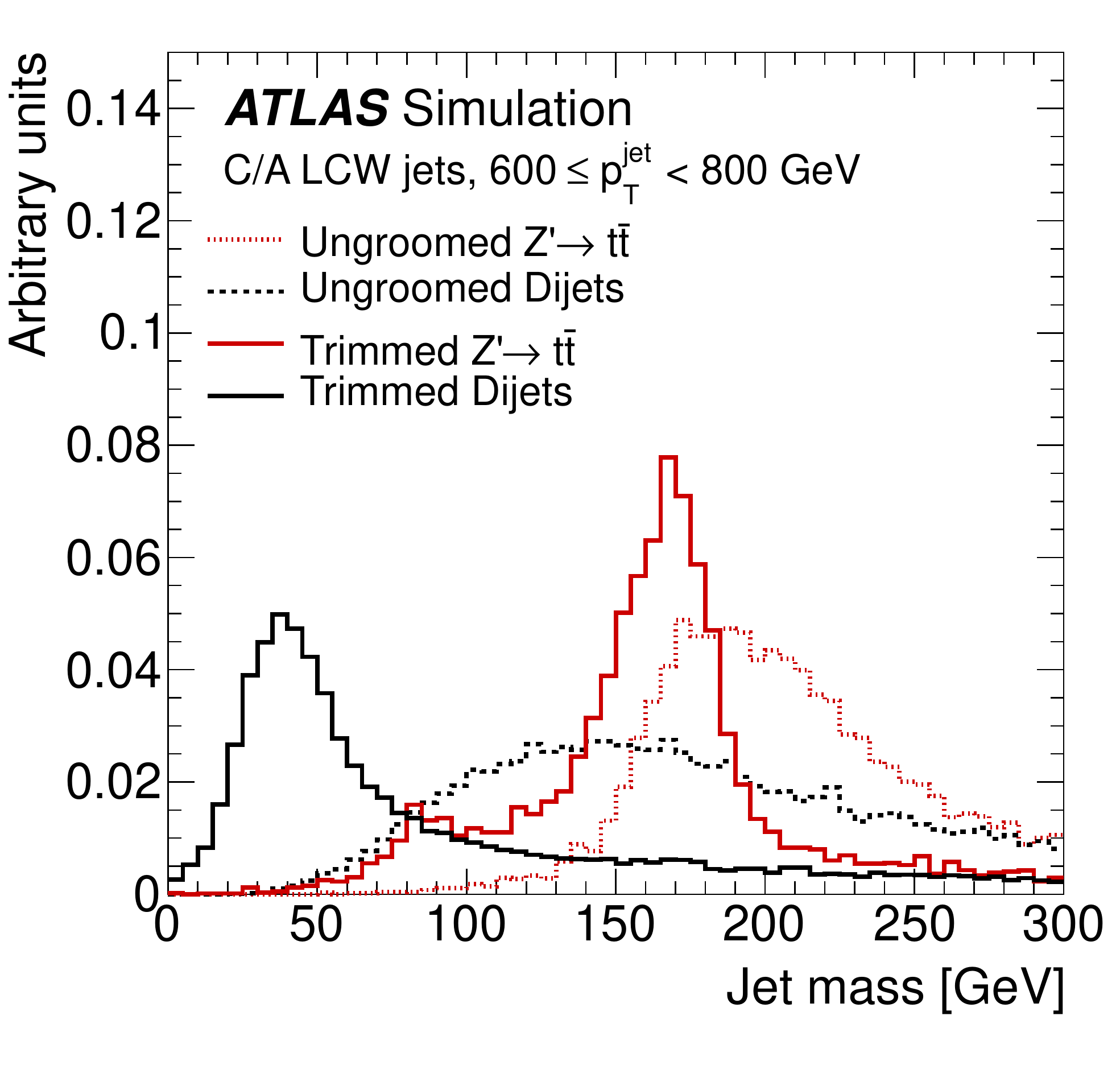} }
  \end{center}
  \caption{Leading-\ptjet\ jet mass for simulated \Pythia \Zprimett ($\massZprime=1.6$~TeV) 
  signal events (red) compared to \PowPythia dijet background events (black) for 
  jets in the range $600\GeV\leq\ptjet<800$~GeV. The dotted lines show the ungroomed 
  leading-\ptjet\ jet distribution, while the solid lines show the corresponding trimmed 
  ($\fcut=0.05$, $\drsub=0.3$) jets.  The groomed distributions are normalized with 
  respect to the ungroomed distributions, which are themselves normalized to unity.
  \label{fig:groomed_jets_mass_compare}} 
\end{figure}

\begin{figure}[!ht]  
  \begin{center}
      \subfigure[\AKTFat]{
        \includegraphics[width=0.45\textwidth]{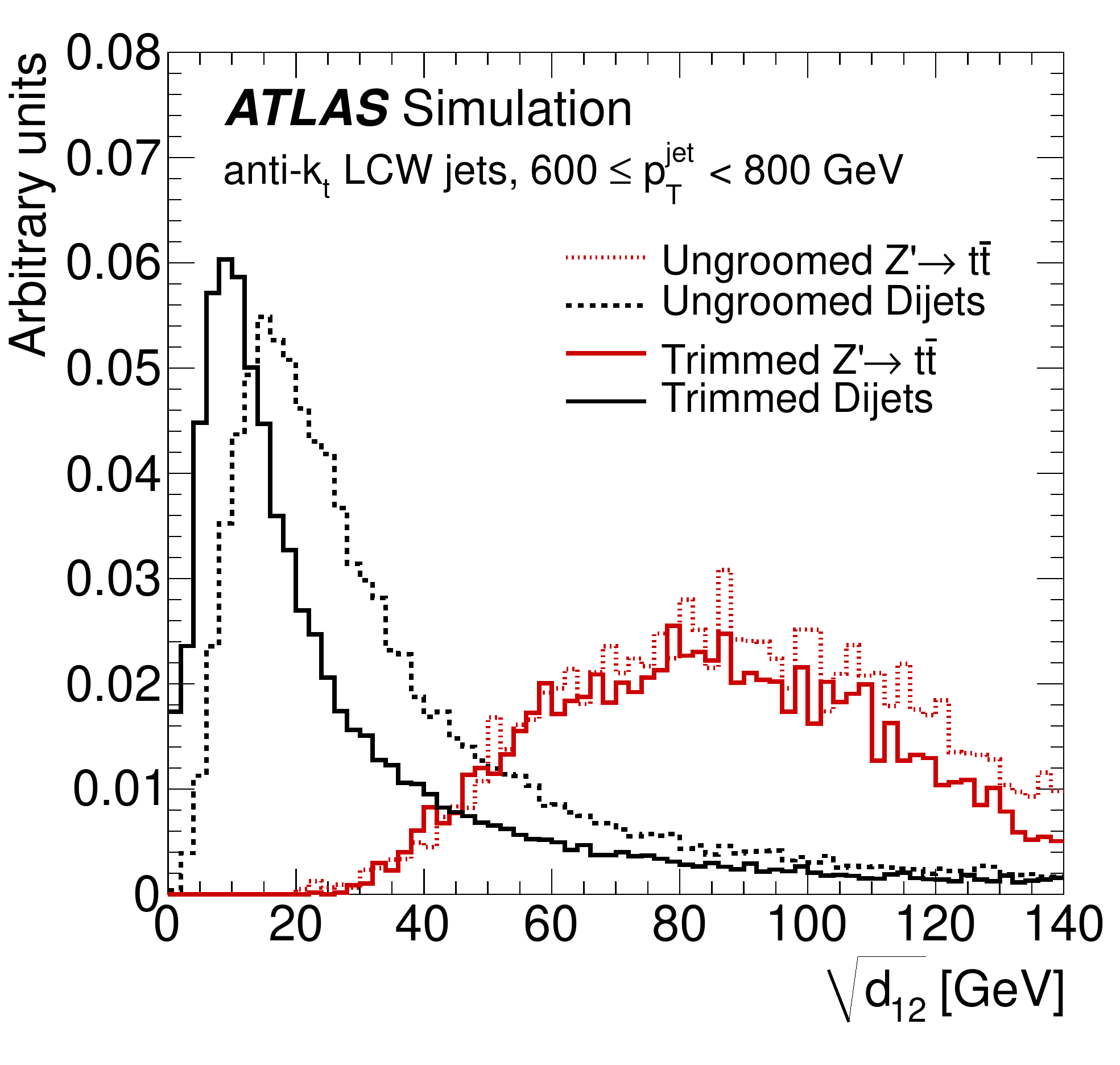} }
      \subfigure[\CAFat]{
        \includegraphics[width=0.45\textwidth]{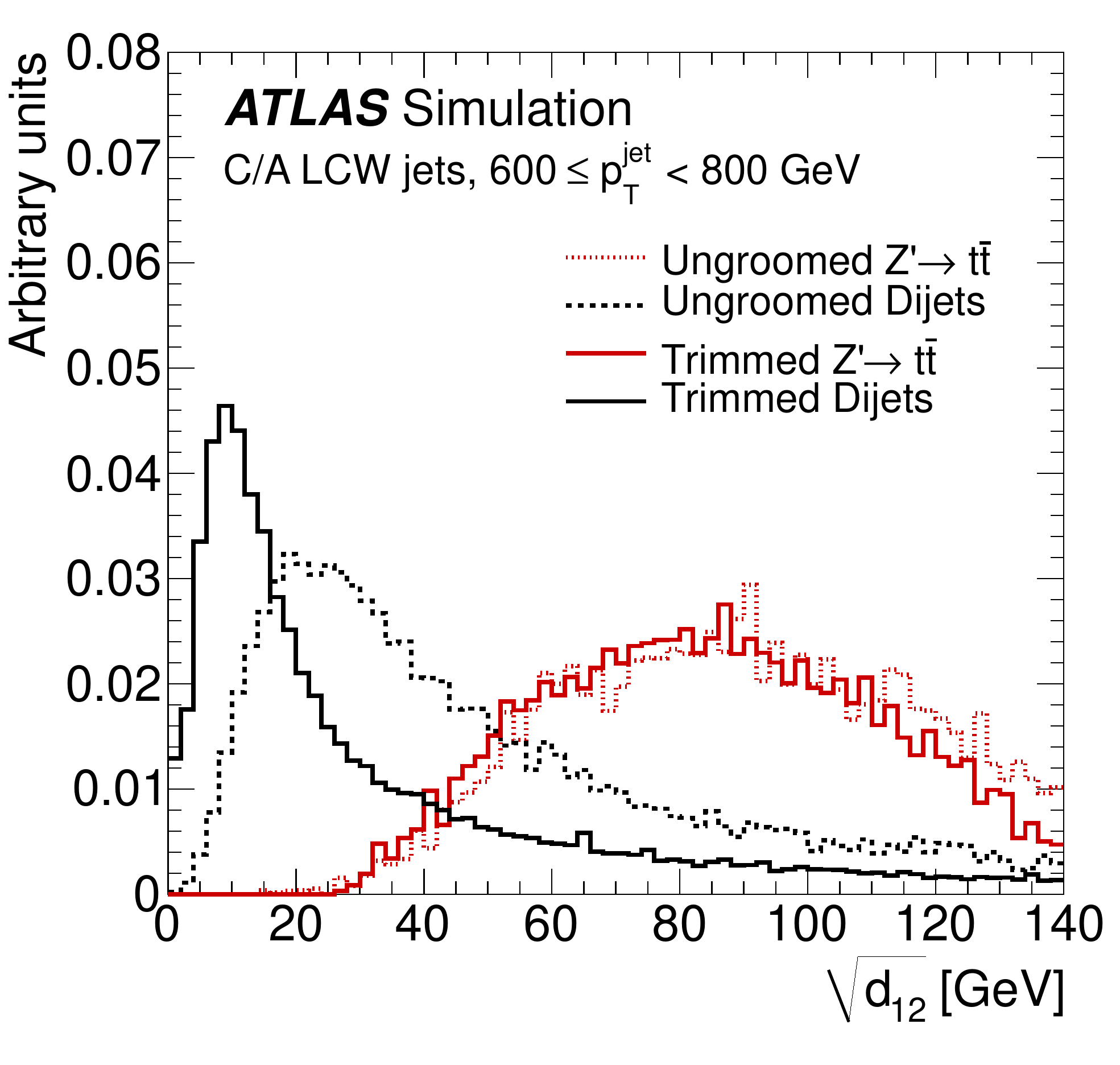} }
  \end{center}
  \caption{Leading-\ptjet\ jet splitting scale \DOneTwo for simulated \Pythia \Zprimett 
  ($\massZprime=1.6$~TeV) signal events (red) compared to \PowPythia dijet background 
  events (black) for jets in the range $600\GeV\leq\ptjet<800$~GeV. The dotted lines 
  show the ungroomed leading-\ptjet\ jet distribution, while the solid lines show the 
  corresponding trimmed ($\fcut=0.05$, $\drsub=0.3$) jets. The groomed distributions are
  normalized with respect to the ungroomed distributions, which are themselves normalized 
  to unity.
  \label{fig:groomed_jets_split12_compare}} 
\end{figure}

\begin{figure}[!ht]  
  \begin{center}
      \subfigure[\AKTFat]{
        \includegraphics[width=0.45\textwidth]{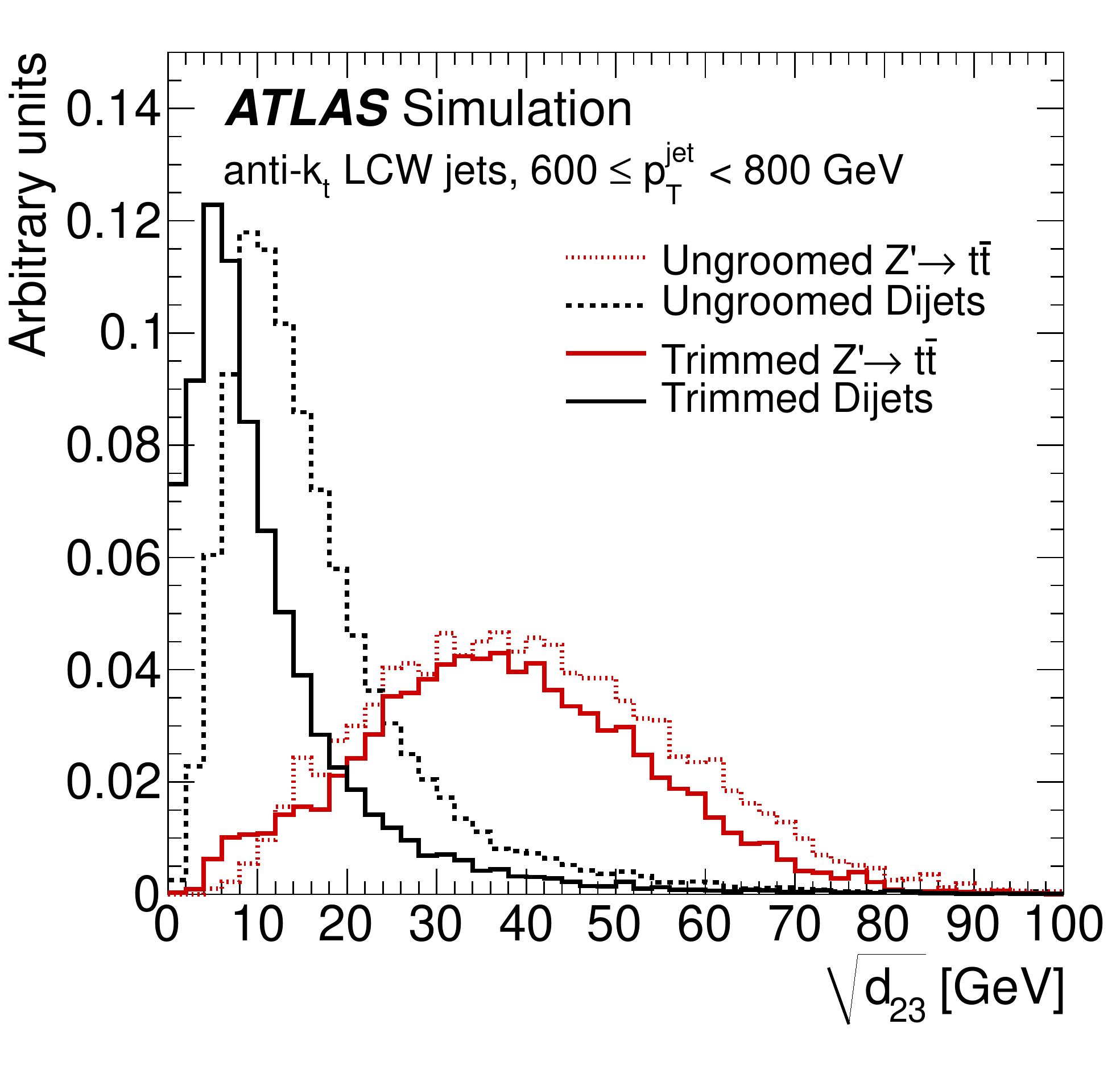} }
      \subfigure[\CAFat]{
        \includegraphics[width=0.45\textwidth]{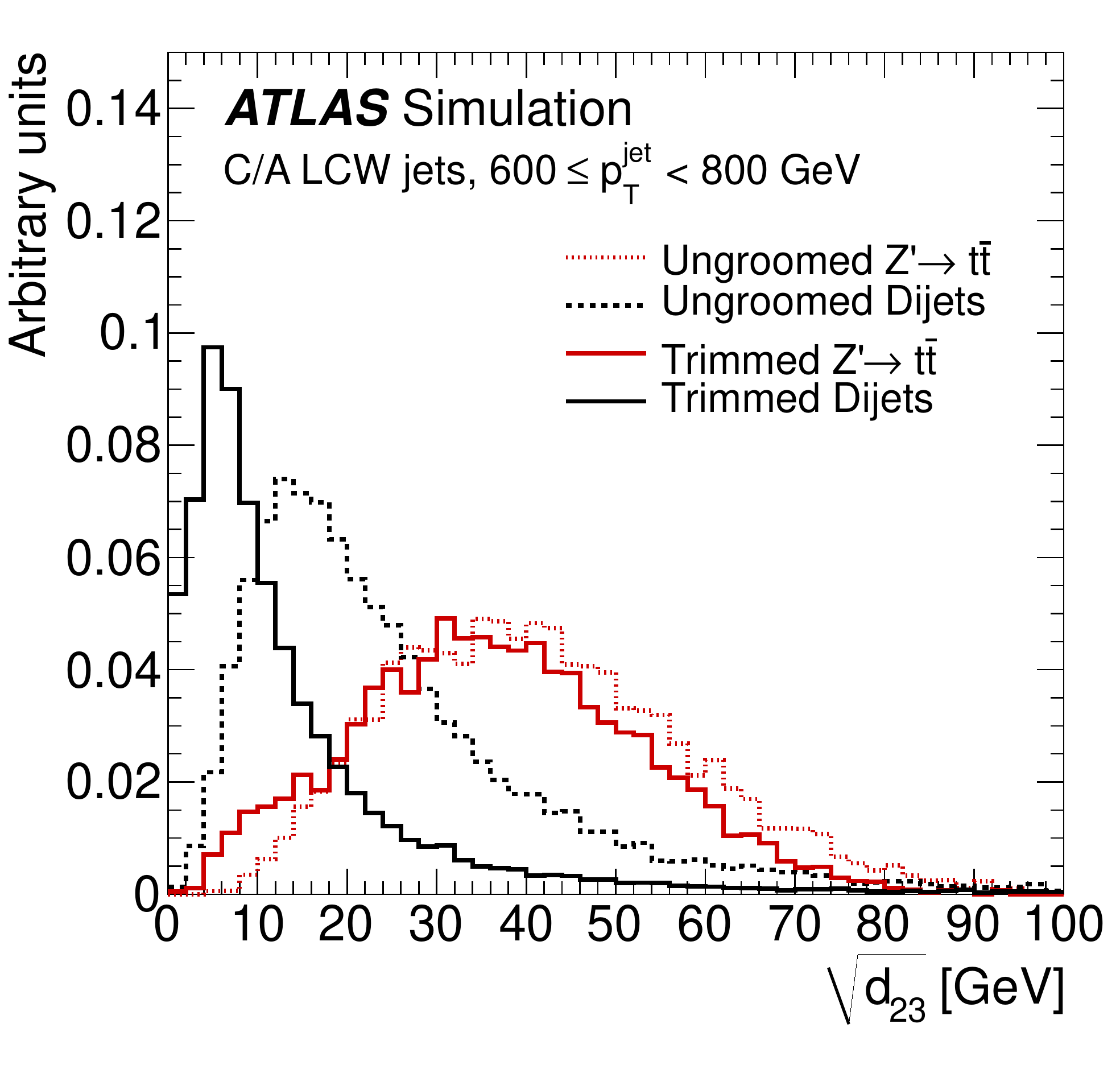} }
  \end{center}
  \caption{Leading-\ptjet\ jet splitting scale \DTwoThr for simulated \Pythia \Zprimett 
  ($\massZprime=1.6$~TeV) signal events (red) compared to \PowPythia dijet background 
  events (black) for jets in the range $600\GeV\leq\ptjet<800$~GeV. The dotted lines show 
  the ungroomed leading-\ptjet\ jet distribution, while the solid lines show the corresponding 
  trimmed ($\fcut=0.05$, $\drsub=0.3$) jets. The groomed distributions are normalized with 
  respect to the ungroomed distributions, which are themselves normalized to unity.
  \label{fig:groomed_jets_split_compare}} 
\end{figure}

\begin{figure}[!ht]  
  \begin{center}
      \subfigure[\AKTFat]{
        \includegraphics[width=0.45\textwidth]{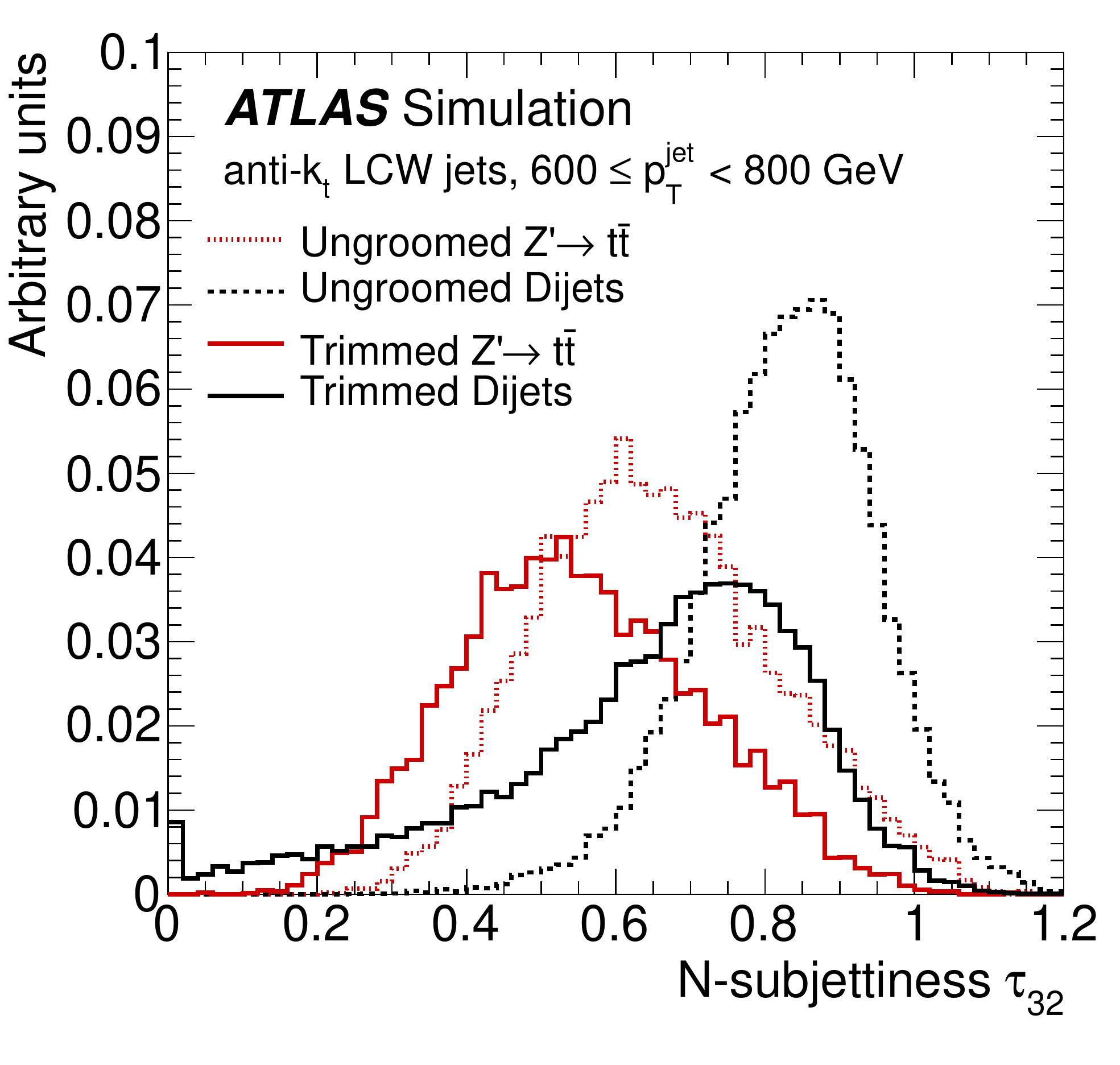} }
      \subfigure[\CAFat]{
        \includegraphics[width=0.45\textwidth]{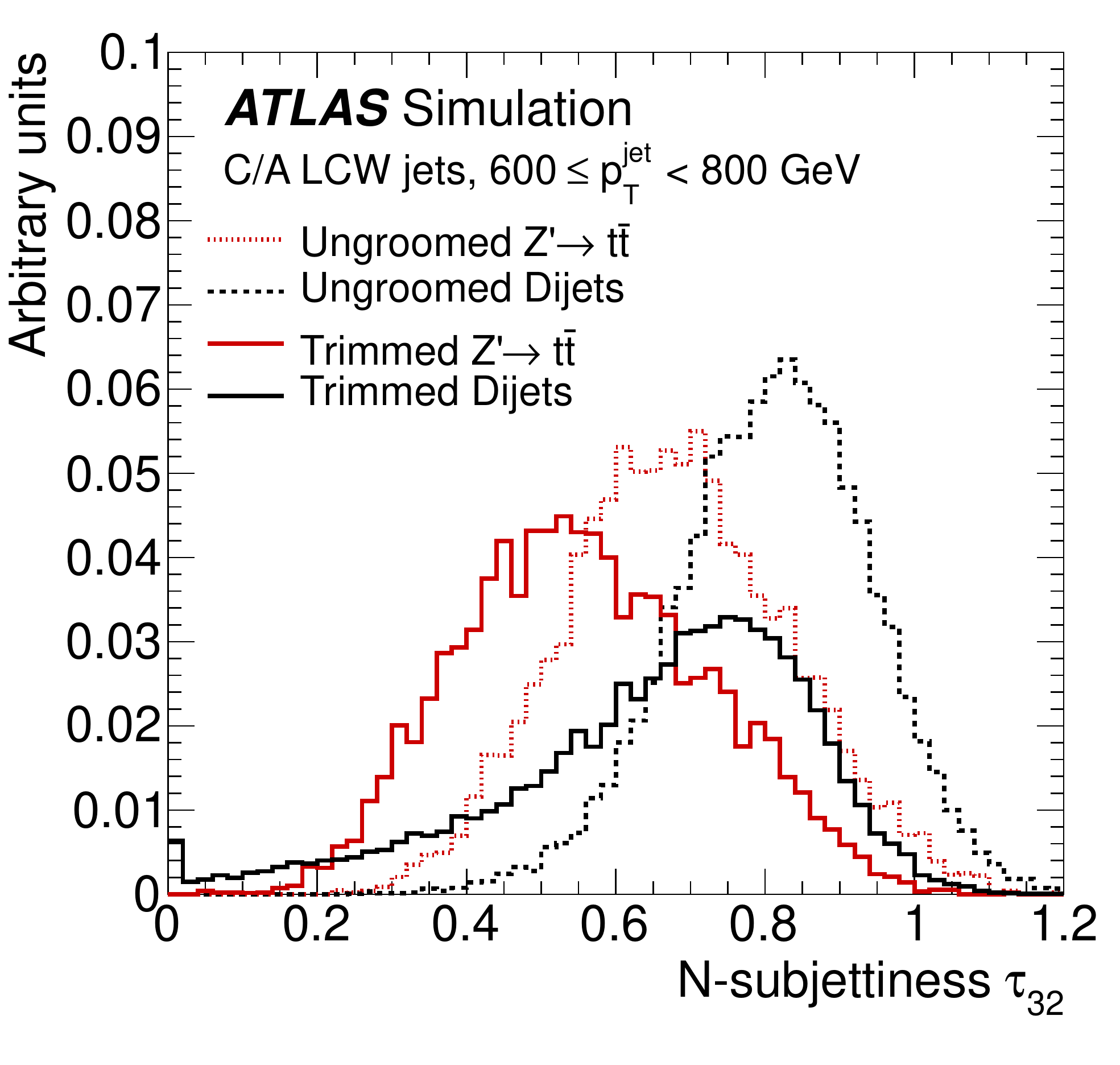} }
  \end{center}
  \caption{Leading-\ptjet\ jet $N$-subjettiness \tauThrTwo for simulated \Pythia \Zprimett 
  ($\massZprime=1.6$~TeV) signal events (red) compared to \PowPythia dijet background 
  events (black) for jets in the range $600\GeV\leq\ptjet<800$~GeV. The dotted lines show 
  the ungroomed leading-\ptjet\ jet distribution, while the solid lines show the corresponding 
  trimmed ($\fcut=0.05$, $\drsub=0.3$) jets. The groomed distributions are normalized with 
  respect to the ungroomed distributions, which are themselves normalized to unity.
  \label{fig:groomed_jets_subjetiness_compare}} 
\end{figure}

Leading-\ptjet\ jet distributions of mass, splitting scales and $N$-subjettiness are compared for jets in simulated signal and background events in the range $600\GeV\leq\ptjet<800$~GeV. 
As seen in \figrange{groomed_jets_mass_compare_twoprong}{groomed_jets_subjetiness_compare_twoprong}, showing distributions for the two-pronged decay case, and in \figrange{groomed_jets_mass_compare}{groomed_jets_subjetiness_compare} showing comparisons for the three-pronged decay case, better discrimination between signal and background is obtained after grooming. 
In these figures, the ungroomed distributions are normalized to unit area, while the groomed distributions have the efficiency with respect to the ungroomed \largeR jets folded in for comparison.  
This is especially conspicuous in the \CamKt jets with mass-drop filtering applied as mentioned previously.

The mass resolution of the simulated \Zqq signal events shown in \figref{groomed_jets_mass_compare_twoprong} dramatically improves after trimming or mass-drop filtering for \akt jets with $R=1.0$ and \CamKt jets with $R=1.2$, respectively. Mass-drop filtering has an efficiency of approximately 55\% and therefore fewer jets remain in this figure. After trimming or mass-drop filtering, the mass peak corresponding to the \Z boson is clearly seen at the correct mass. 
Note that the dijet background is pushed much lower in mass after grooming as was demonstrated in \figref{DataMC:massDistro}, while the constituents of signal jets have higher \pt and survive the grooming procedure, thus improving discrimination between signal and background. 
The small excess of signal events below 50~GeV is the result of one of the two quarks from the decay of the $Z$ boson being removed by the jet grooming, thus leaving only one quark reconstructed as the jet and making it indistinguishable from the background. \Figref{groomed_jets_split_compare_twoprong} shows the splitting scale \DOneTwo in \Zqq events.  
The signal exhibits a splitting scale roughly equal to half the mass of the jet, whereas the splitting scale distribution for jets produced in dijet events peaks at smaller values of \DOneTwo and falls more steeply. This effect is enhanced after grooming, especially in the case of \CamKt jets after mass-drop filtering.  
In \figref{groomed_jets_subjetiness_compare_twoprong}, the $N$-subjettiness variable \tauTwoOne is observed to have improved discrimination between signal and background with \antikt\ trimmed jets compared to \CamKt mass-drop filtered jets, where the discrimination is worsened after applying the mass-drop filtering criteria. The filtering step is explicitly reconstructing a fixed number of final subjets (three, in this case), thereby shaping the background and worsening the resulting separation.

The three-pronged hadronic top-jet mass distributions from \Zprimett\ events are shown in \figref{groomed_jets_mass_compare}, where the signal peak is relatively unshifted between groomed and ungroomed jets, especially with \antikt\ jets. 
Again the mass resolution for the signal improves after grooming, where the \W-mass peak can also be seen after trimming is applied. 
The enhancement of the $W$-mass peak is seen especially in jets with lower \ptjet, as the jet from the $b$-quark decay falls outside the radius of the \largeR jet.  
\Figsref{groomed_jets_split12_compare}{groomed_jets_split_compare} show the variables \DOneTwo and \DTwoThr, respectively, for \Zprimett events compared to jets produced in dijet events. 
As in the two-pronged case, signal discrimination with the splitting scales is enhanced after jet trimming.

One of the primary applications of $N$-subjettiness is as a discriminating variable in searches for highly boosted top quarks~\cite{Thaler:2010tr}. 
A common method of comparing the performance of such discriminating variables or tagging algorithms is to compare the rate at which light-quark or gluon jets are selected (the \emph{mis-tag} rate) to the efficiency for retaining jets containing the hadronic particle decay of interest~\cite{BOOST2010, BOOST2011}. This comparison is performed for both ungroomed and trimmed jets in order to assess the impact of grooming on the discrimination power of this observable.
\Figref{groomed_jets_subjetiness_compare} shows the \tauThrTwo distribution before and after trimming. Here, trimming of \antikt\ and \CamKt jets results in similar discrimination between signal and background. 
In order to understand the utility of the \tauThrTwo selection criterion and the potential impact of jet grooming, trimmed \AKTFat jets are compared to their ungroomed counterparts in a boosted top sample for two jet momentum ranges. 
The signal mass range is defined as that which contains a large fraction of the boosted top signal. For ungroomed jets this fraction is set to 90\%, and the mass range that satisfies this requirement is $100\GeV\leq\Mjet<250$~GeV. 
A slightly lower signal fraction of 80\% for the same mass range is required for groomed jets; this is motivated by the tendency for trimmed jets to populate an additional small peak around the \W mass, as shown in \figref{groomed_jets_mass_compare}.

\begin{figure}[!t]
  \centering
  \subfigure[$600\GeV\leq\ptjet<800$~\GeV]{
    \includegraphics[width=0.46\columnwidth]{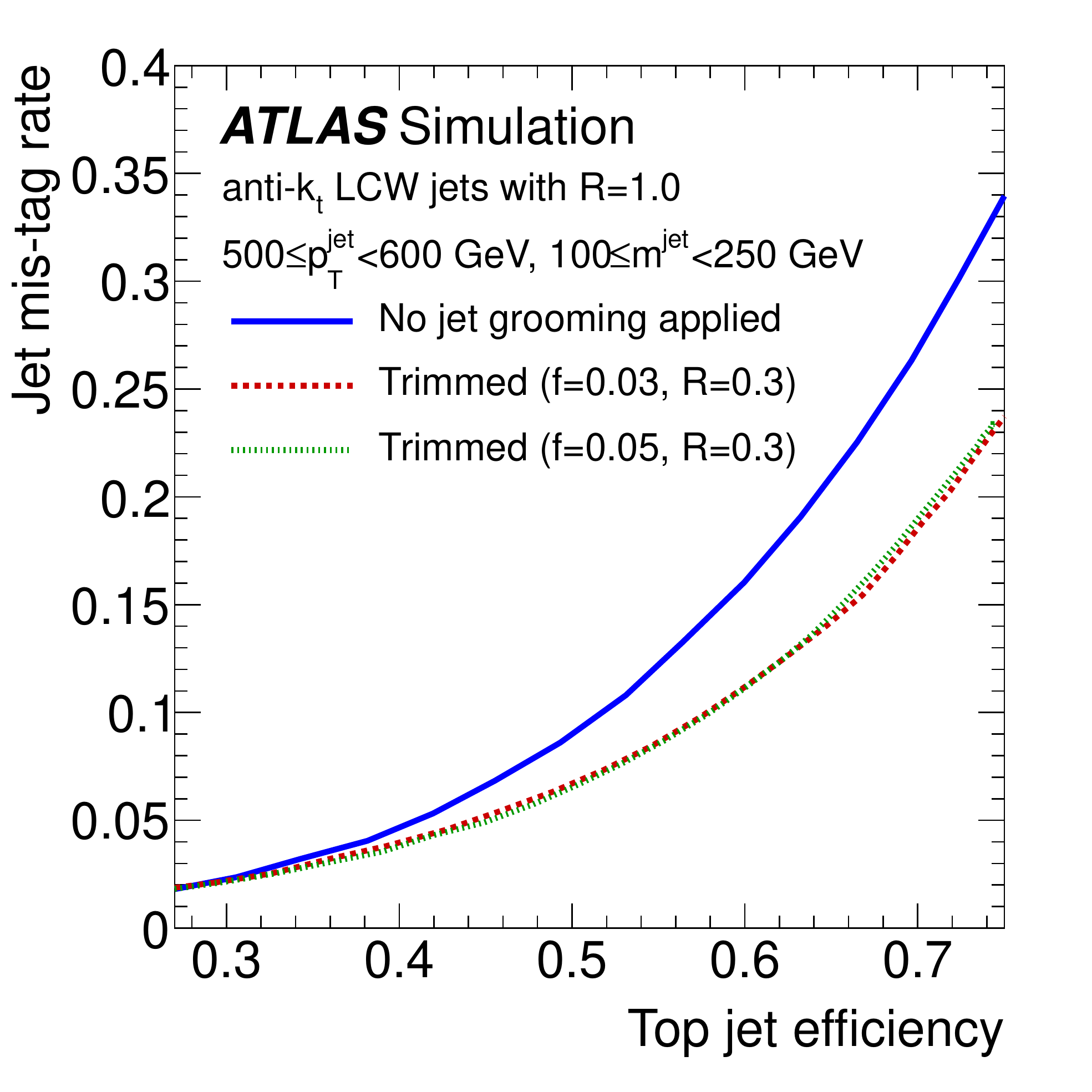}
    \label{fig:topTag:tau32:lowPt}}
  \subfigure[$800\GeV\leq\ptjet<1000$~GeV]{
    \includegraphics[width=0.46\columnwidth]{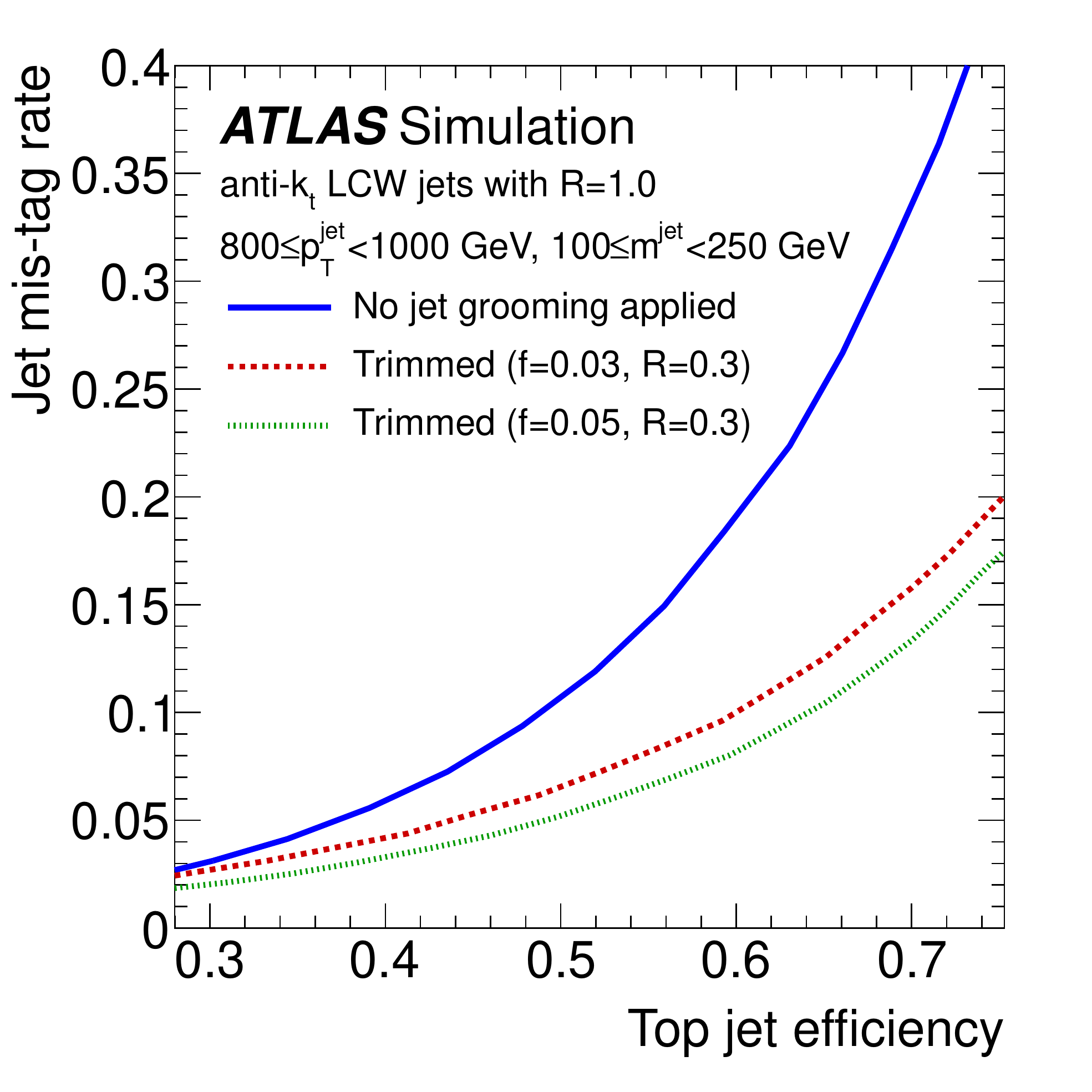}
    \label{fig:topTag:tau32:hiPt}}
  \caption{Jet mis-tag rate for dijet events vs. top-jet efficiency curves using \tauThrTwo as a top tagger for \subref{fig:topTag:tau32:lowPt} $600\GeV\leq\ptjet<800$~\GeV~and \subref{fig:topTag:tau32:hiPt} $800\GeV\leq\ptjet<1000$~GeV with masses in the range $100\GeV\leq\Mjet<250$~GeV.
           }
  \label{fig:topTag:tau32}
\end{figure}

The mis-tag rate is defined as the fraction of the \Powheg dijet sample that remains in the mass window after a simple selection based on \tauThrTwo. 
The signal \emph{top jet} efficiency is defined as the fraction of top jets selected in the \Zprime\ sample with the same \tauThrTwo selection. 
\Figref{topTag:tau32} shows the performance of the $N$-subjettiness tagger for jets with $600\GeV\leq\ptjet<800$~GeV and $800\GeV\leq\ptjet<1000$~GeV. 
In both cases, for a fixed top-jet efficiency, the reduction in high-invariant-mass jets due to trimming results in a relative reduction of several percent in the mis-tag rate.
Moreover, in the case of very high \ptjet, as in \figref{topTag:tau32:hiPt}, the slightly more aggressive trimming configuration results in a slight performance gain as well.

\subsection{Inclusive jet data compared to simulation with and without grooming}
\label{sec:boostedwtop:inclusivedatamc}

Previous studies conducted by ATLAS~\cite{JetMassAndSubstructure} and CMS~\cite{CMSHadronicTTResonance2011} suggest that even complex jet-substructure observables are fairly well modelled by the MC simulations used by the LHC experiments. 
This section reviews the description provided by \Pythia, \Herwigpp, and \PowPythia of the jet grooming techniques introduced above, and of the substructure of the ungroomed and groomed jets themselves.


\begin{figure}[!ht]
  \centering
  \subfigure[\AKTFat (ungroomed)]{
    \includegraphics[width=0.46\columnwidth]{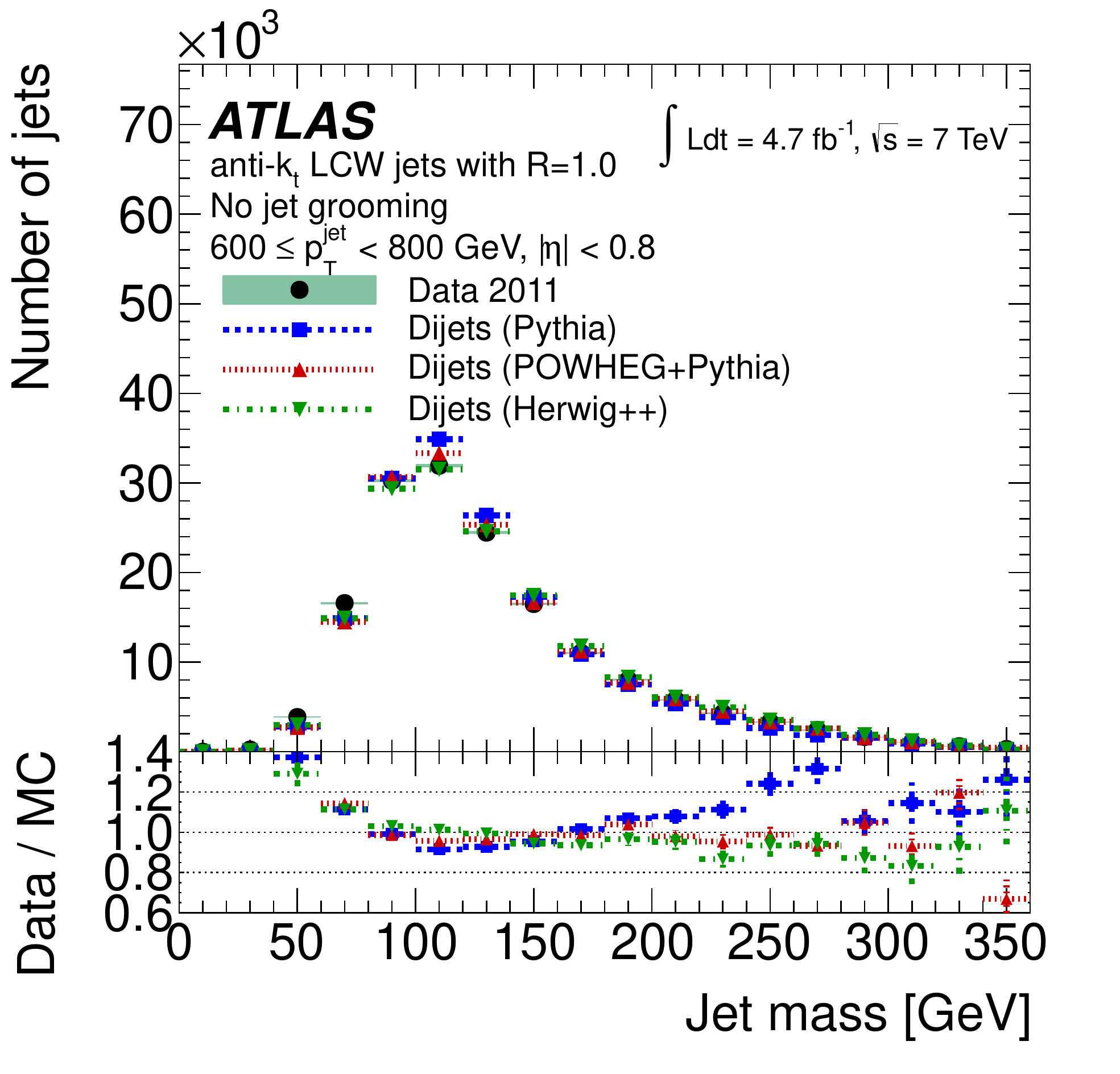}
    \label{fig:DataMC:mass:AKTFat}}
  \subfigure[\AKTFat (trimmed)]{
    \includegraphics[width=0.46\columnwidth]{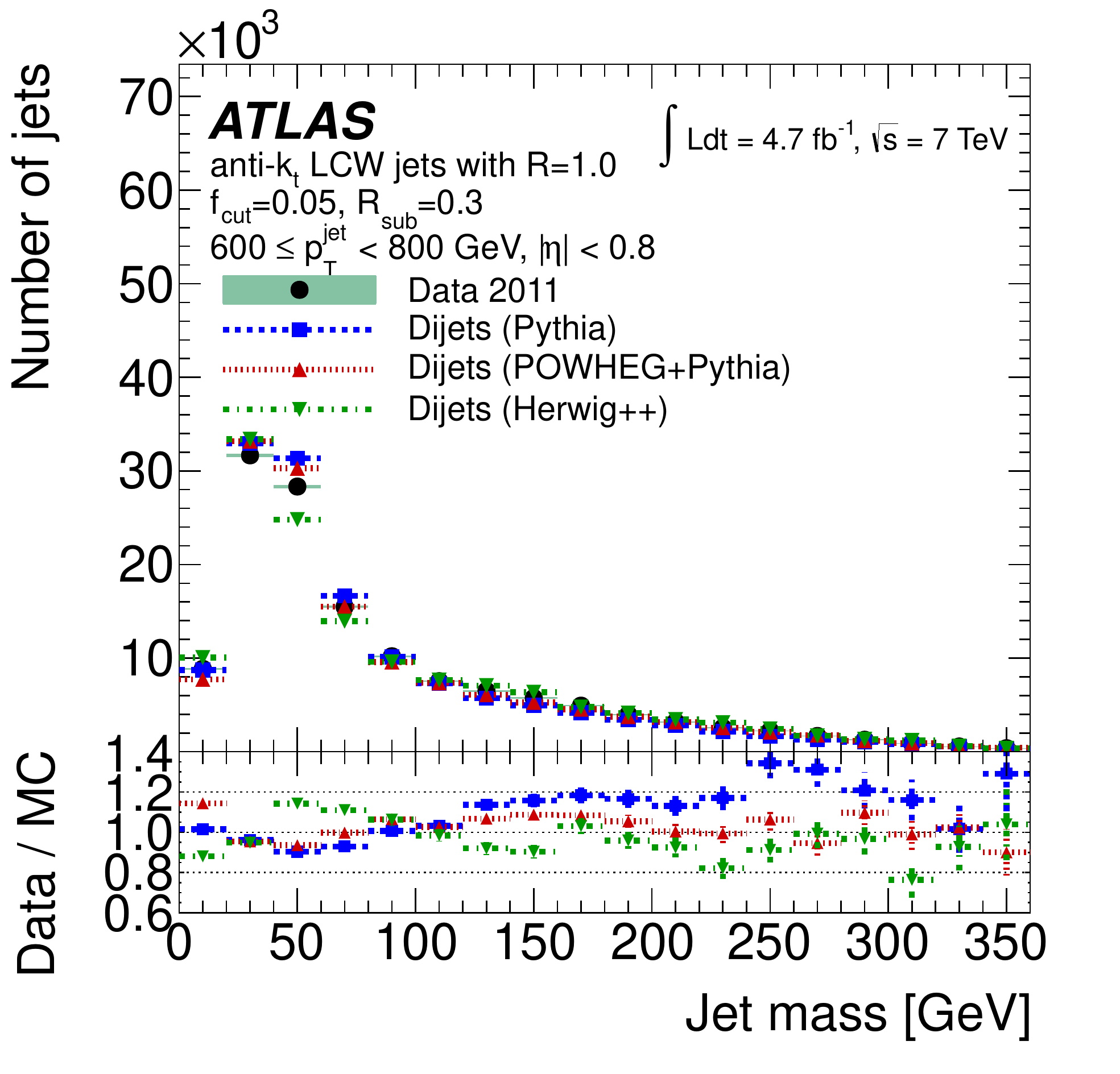}
    \label{fig:DataMC:mass:AKTTrim}}
  \\
  \subfigure[\CAFat (ungroomed)]{
    \includegraphics[width=0.46\columnwidth]{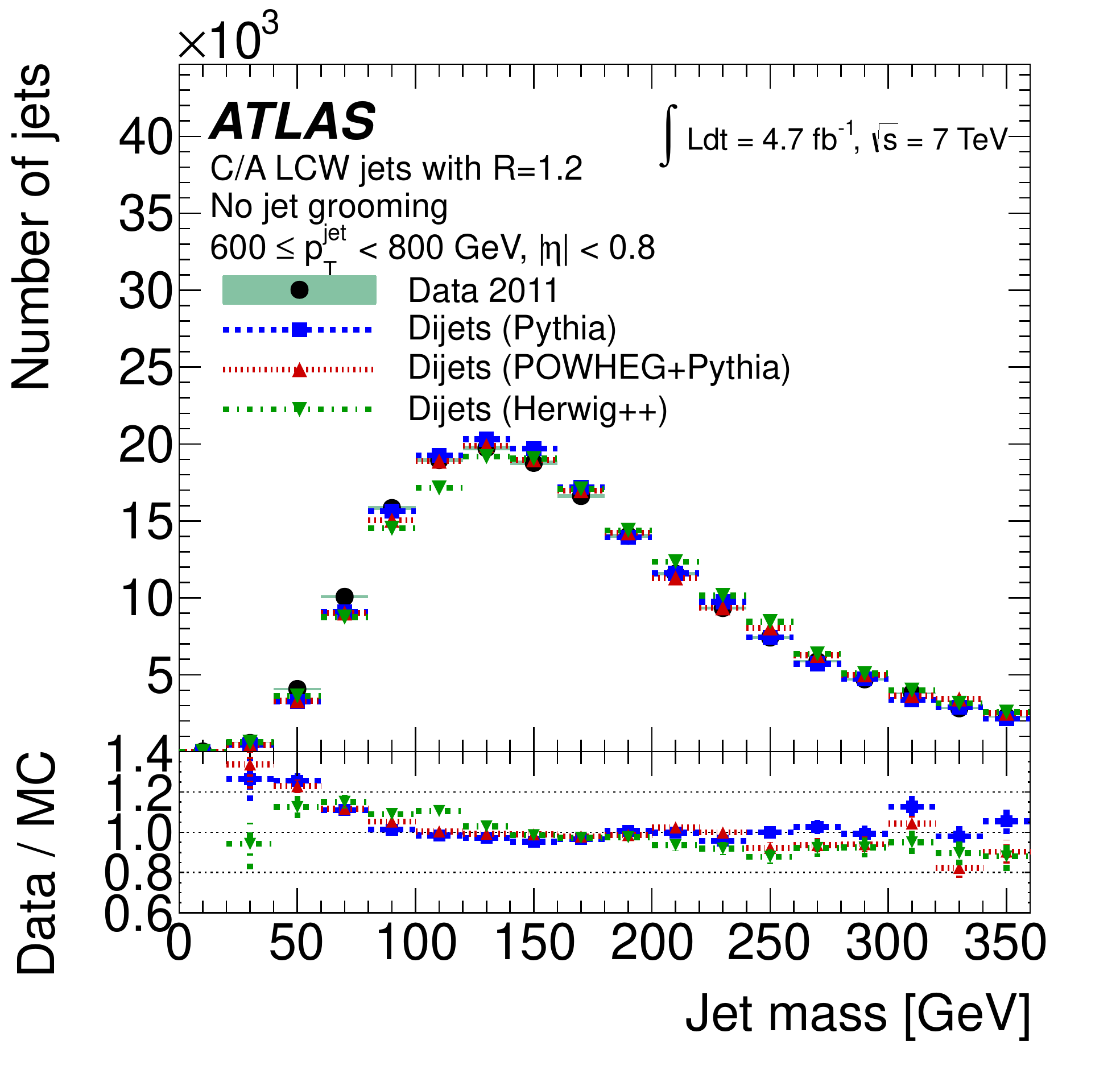}
    \label{fig:DataMC:mass:CAFat}}
  \subfigure[\CAFat (filtered)]{
    \includegraphics[width=0.46\columnwidth]{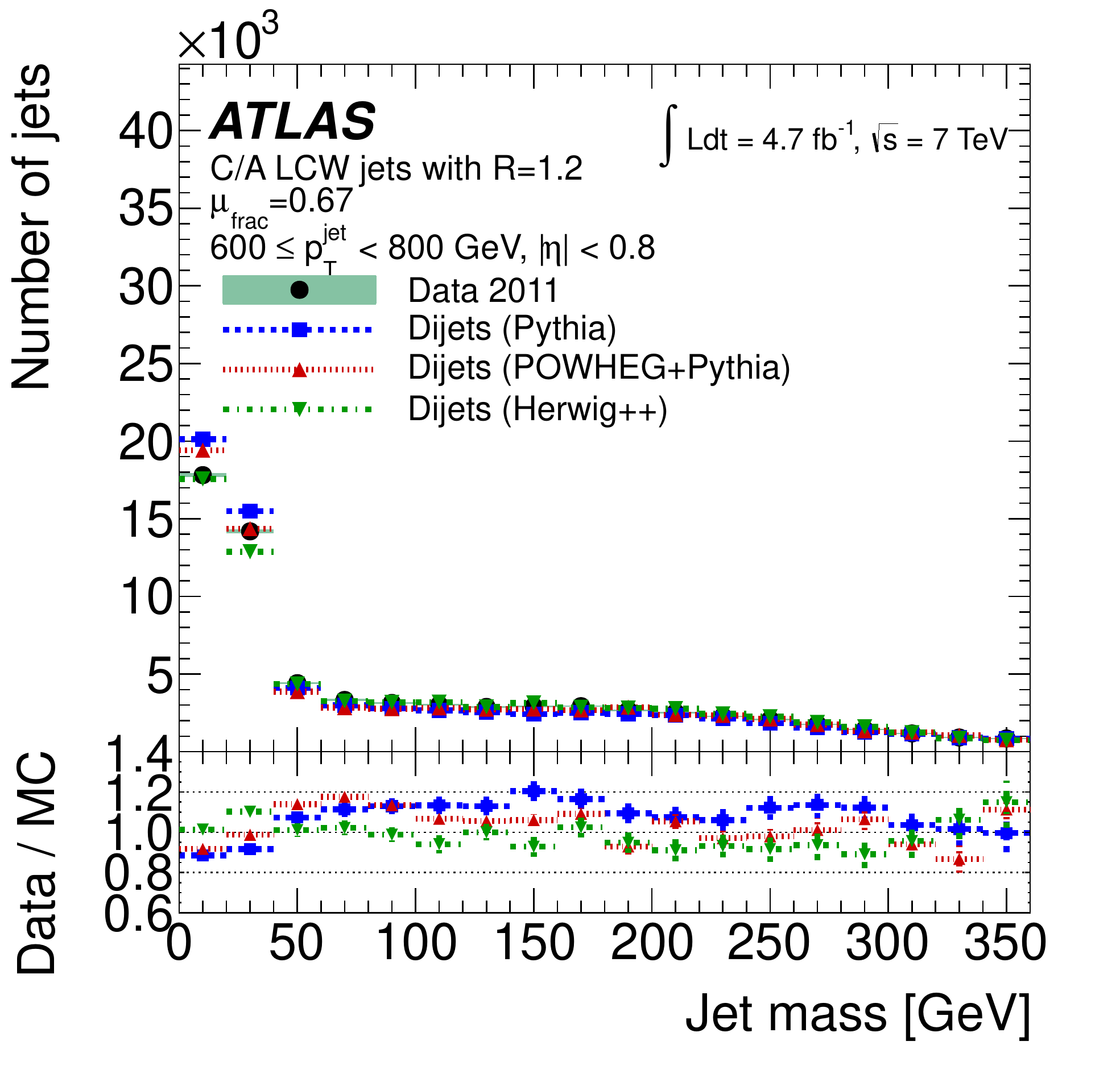}
    \label{fig:DataMC:mass:CAFilt}}
  \caption{Mass of jets in the range $600\GeV\leq\ptjet<800$~GeV and in the 
  central calorimeter ($|\eta|<0.8$). Shown are \subref{fig:DataMC:mass:AKTFat} 
  ungroomed and \subref{fig:DataMC:mass:AKTTrim} trimmed ($\fcut=0.05$, 
  $\drsub=0.3$) \akt\ jets with $R=1.0$; and \subref{fig:DataMC:mass:CAFat} 
  ungroomed and \subref{fig:DataMC:mass:CAFilt} filtered ($\mufrac=0.67$) C/A 
  jets with $R=1.2$. The ratios between data and MC distributions are shown 
  in the lower section of each figure.
           }
  \label{fig:DataMC:mass}
\end{figure}

\Figref{DataMC:mass} presents a comparison of the jet invariant mass for ungroomed,  trimmed, and filtered jets in the range $600\GeV\leq\ptjet<800$~GeV and in the central calorimeter, $|\eta|<0.8$. 
Similar performance is observed in all other \pt regions in the range $\pt>300$~GeV, and $|\eta|<2.1$. 
The description of both the ungroomed and trimmed \antikt\ jets with $R=1.0$ provided by \Pythia is poor for large masses. The descriptions provided by \Herwigpp as well as for the \NLO generator  \PowPythia are more accurate. 
\Pythia tends to underestimate the fraction of high-mass \largeR \antikt\ jets, whereas \Herwigpp and \PowPythia are accurate to within a few percent, even for very massive jets. The ungroomed \AKTFat jets are poorly described by all three MC simulations at low mass; this could be due to non-perturbative and detector effects which increase the jet mass.
This generally soft contribution is removed by grooming.

A similarly poor description of the low-mass region is observed for \CamKt jets with $R=1.2$. In this case however, \Pythia, 
in addition to both \Herwigpp and \Powheg$+$\Pythia, provides a fairly good description of the high-mass regime of the jet mass spectrum.
This suggests that there is a slight angular scale dependence, and the slightly smaller radius used for the \largeR \antikt\ jets in these studies could play a role in the observed discrepancy with \Pythia. 
\Figref{DataMC:mass} also shows that the shape of the jet mass distribution is significantly affected by the mass-drop filtering technique. 
This change is well described by all of the MC simulations, although the accuracy of the \Herwigpp and \PowPythia predictions is again observed to be slightly better.

\begin{figure}[!ht]
  \centering
  \subfigure[\AKTFat]{
    \includegraphics[width=0.46\columnwidth]{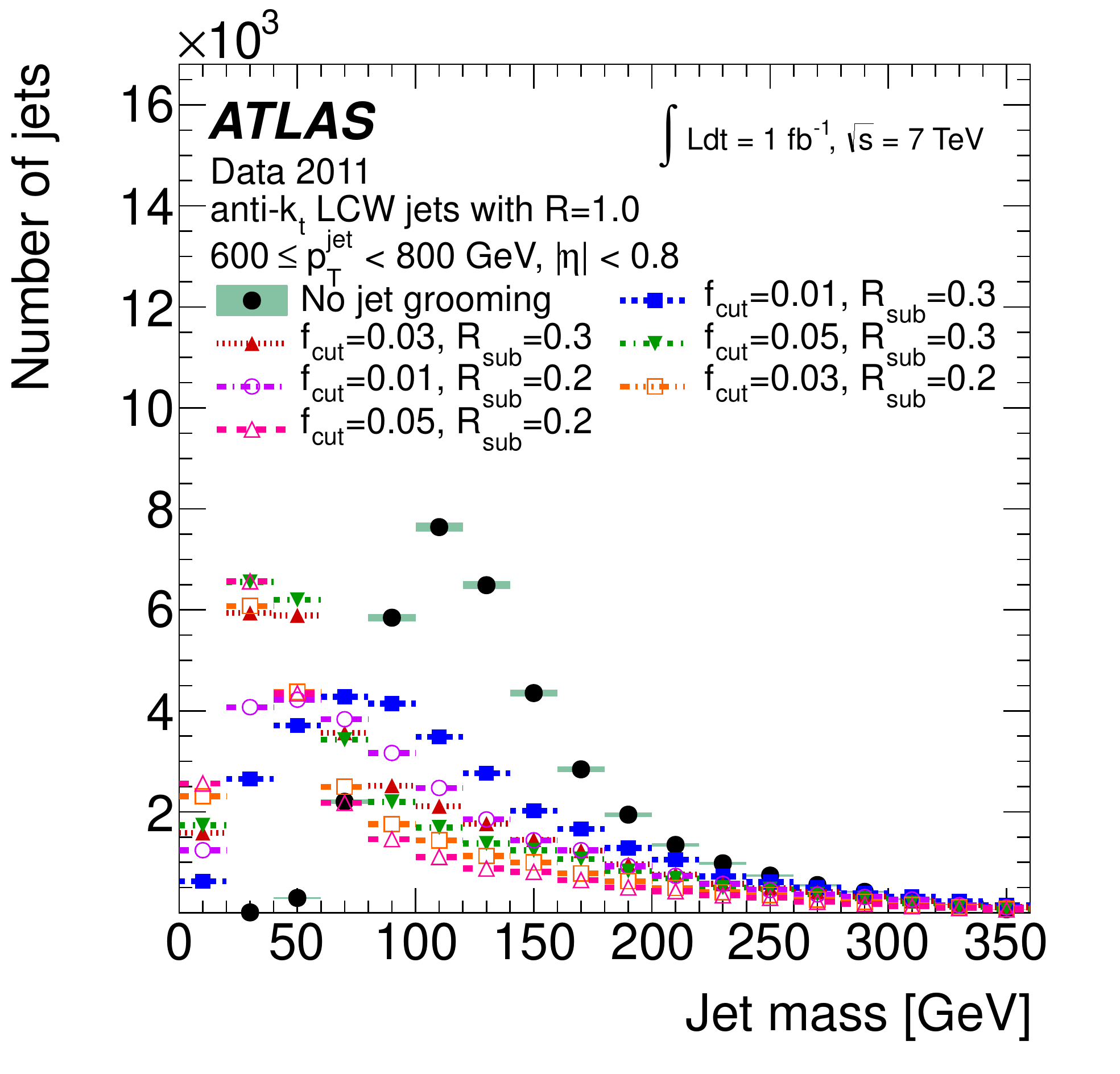}
    \label{fig:DataMC:grooming:mass:AKTFat}}
  \subfigure[\CAFat]{
    \includegraphics[width=0.46\columnwidth]{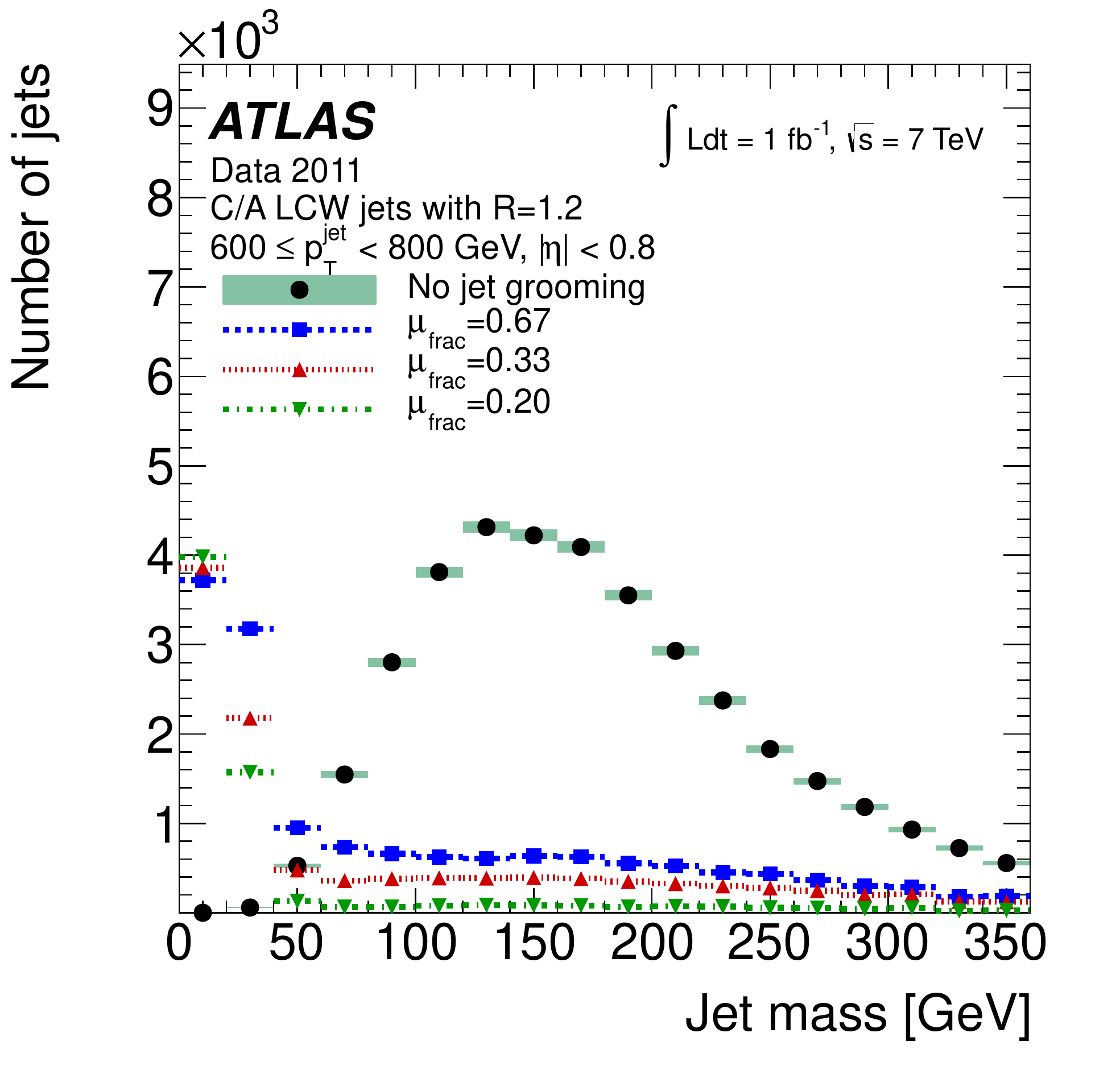}
    \label{fig:DataMC:grooming:mass:CAFat}}
  \caption{Mass of jets in the range $600\GeV\leq\ptjet<800$~GeV and in the
central calorimeter ($|\eta|<0.8$). In \subref{fig:DataMC:grooming:mass:AKTFat} \akt\
jets with $R=1.0$ are compared before (ungroomed) and after trimming
with several configurations of \pt fraction ($\fcut$) and subjet size ($\drsub$). 
In \subref{fig:DataMC:grooming:mass:CAFat} C/A jets with $R=1.2$ are
compared before and after filtering with three different values of mass-drop fraction ($\mufrac$). 
}
  \label{fig:DataMC:grooming:mass}
\end{figure}

\Figref{DataMC:grooming:mass} presents an overview of the shape of the jet mass spectrum for several configurations of the jet trimming algorithm for \akt jets and for \CamKt jets with mass-drop filtering applied. 
These spectra are measured using approximately 1\invfb of data from the last data-taking period of 2011 where $\avg{\mu}=12$, higher than the average over the whole 2011 data-taking period. 
The significant spectral shift and shape difference compared to the original jet is apparent for both grooming algorithms shown (and also for pruning, which is not shown here). 
Significant variation is also observed among the configurations tested, with the large \fcut, small \drsub setting for trimming and the small \mufrac setting for mass-drop filtering exhibiting the most dramatic changes.
For jet masses in the range of $50\GeV\leq\massjet<300$~GeV, which is expected to be the most relevant in searches for new physics, the trimming configurations exhibit efficiencies in the range of 30\%--70\%, defined as the ratio of the yield after grooming to that prior to grooming. In particular, the trimming configuration with $\fcut=0.05$ and $\drsub=0.3$ yields an approximate 47\% efficiency in this mass range. Mass-drop filtering provides a more stringent selection, yielding efficiencies in the same $50\GeV\leq\massjet<300$~GeV mass range of 20\%, 12\%, and 3\% for $\mufrac=0.67,0.33,0.30$, respectively.


The significant change observed in the jet mass distribution is due primarily to a reduction in the effective area of each jet (see \secref{recocalib:track2calo} a detailed description of the jet area).
Soft and wide-angle jet constituents are removed from the jet, thereby reducing the overall catchment area. 
This has the desirable effect of also reducing the impact of \pileup on the jet properties.

\begin{figure}[!ht]
  \centering
  \subfigure[\AKTFat (ungroomed)]{
    \includegraphics[width=0.46\columnwidth]{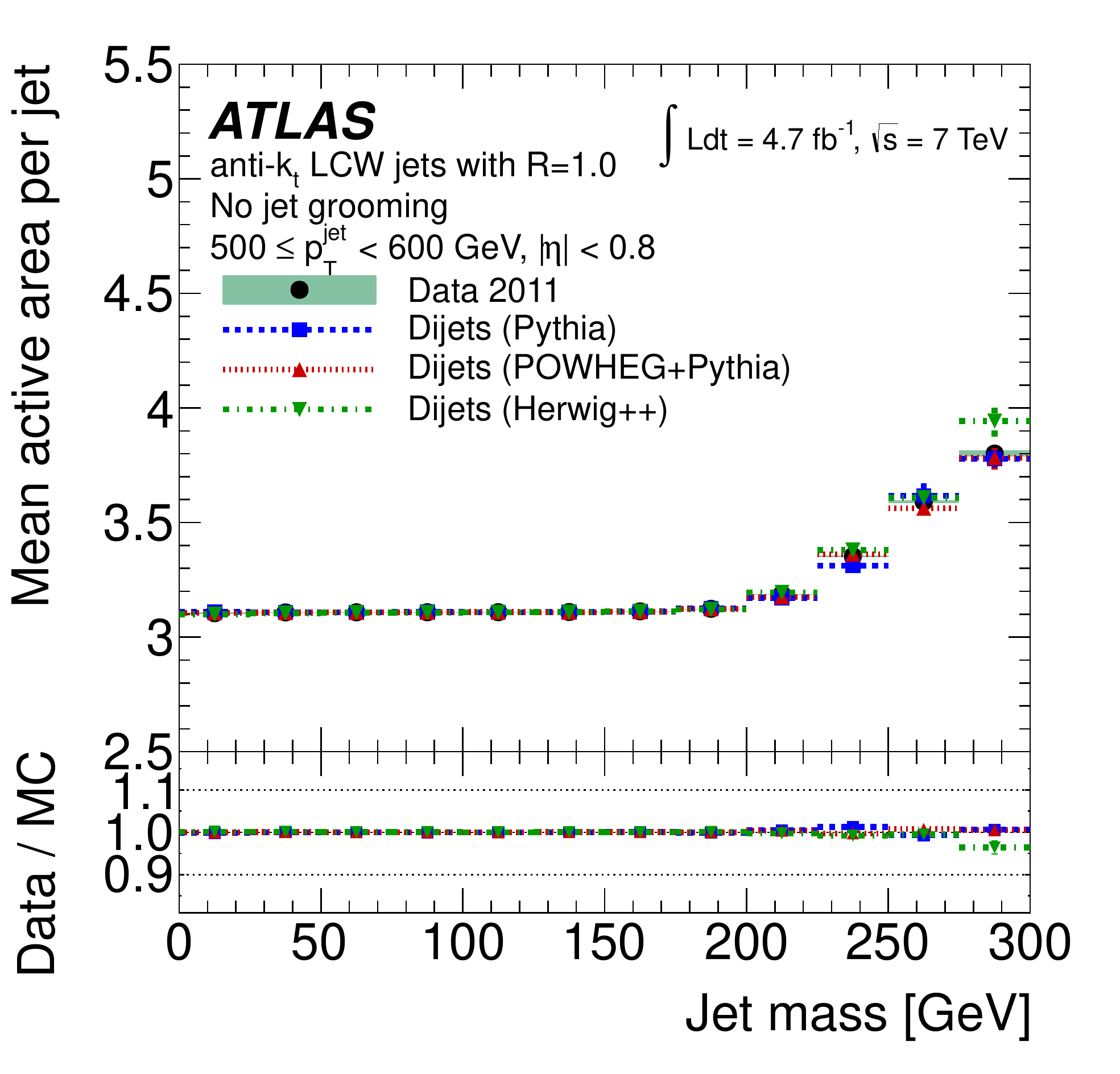}
    \label{fig:DataMC:area:AKTFat}}
  \subfigure[\AKTFat (trimmed)]{
    \includegraphics[width=0.46\columnwidth]{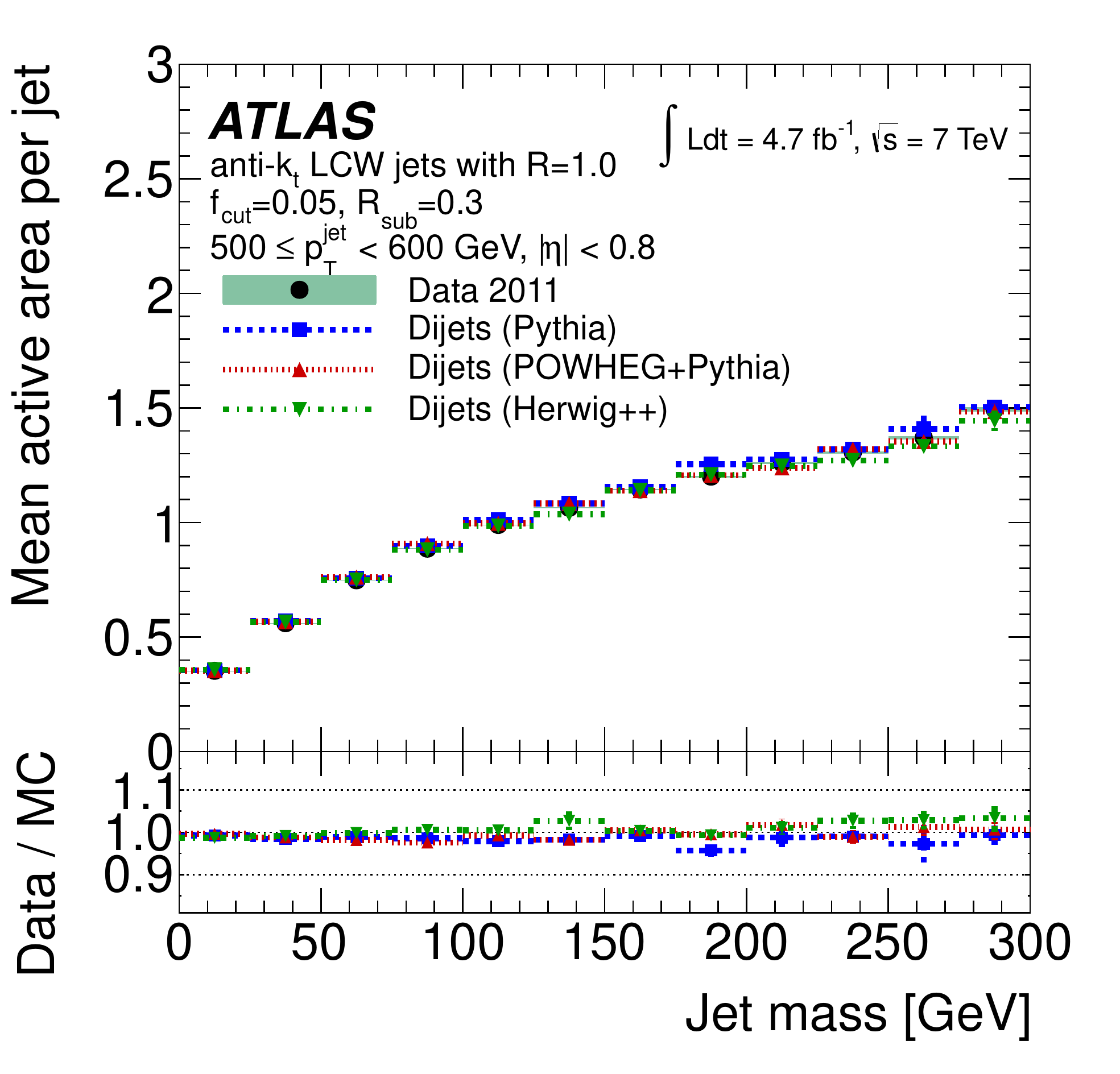}
    \label{fig:DataMC:area:AKTTrim}}
  \caption{Average area of jets in the range
           $500\GeV\leq\ptjet<600$~GeV as a function of jet mass. Shown are 
           \subref{fig:DataMC:d12:AKTFat} ungroomed and 
           \subref{fig:DataMC:d12:AKTTrim} trimmed 
           ($\fcut=0.05$, $\drsub=0.3$) \akt\ jets with $R=1.0$.
The ratios between data and MC distributions are shown
    in the lower section of each figure.
           }
  \label{fig:DataMC:area}
\end{figure}

\Figref{DataMC:area} shows the effect of grooming on the average jet area as a function of the jet mass for both the ungroomed and trimmed \AKTFat jets. 
Prior to jet trimming, the \AKTFat area is very close to $\pi$ (i.e.~$\pi R^2$, with $R=1.0$). A small rise in the ungroomed jet area is observed for jets with very large mass, characterized by small additional clusters near the edge of the jet. 
For trimmed jets at low mass, the average jet area is reduced by a factor of 3--5 and continuously rises as a function of the jet mass to a maximum of approximately half of the original jet area for high-mass ungroomed jets. 
These features are very well described by the MC simulations across the entire spectrum of jet mass, both before and after trimming. 
Similar observations are made with respect to lower and higher \ptjet\ bins, as well as for pruning and filtering.

\begin{figure}[!ht]
  \centering
  \subfigure[\AKTFat (ungroomed)]{
    \includegraphics[width=0.46\columnwidth]{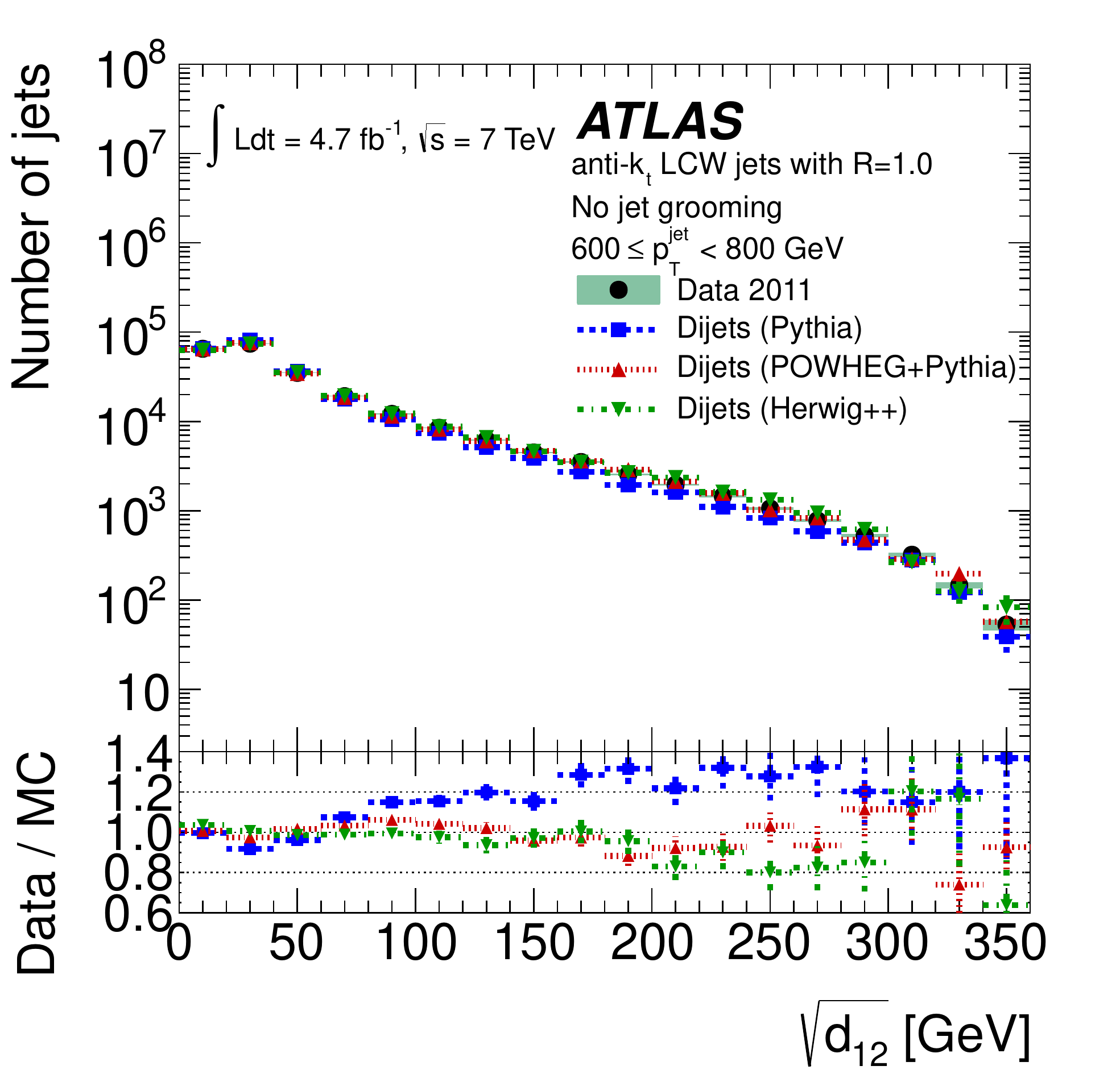}
    \label{fig:DataMC:d12:AKTFat}}
  \subfigure[\AKTFat (trimmed)]{
    \includegraphics[width=0.46\columnwidth]{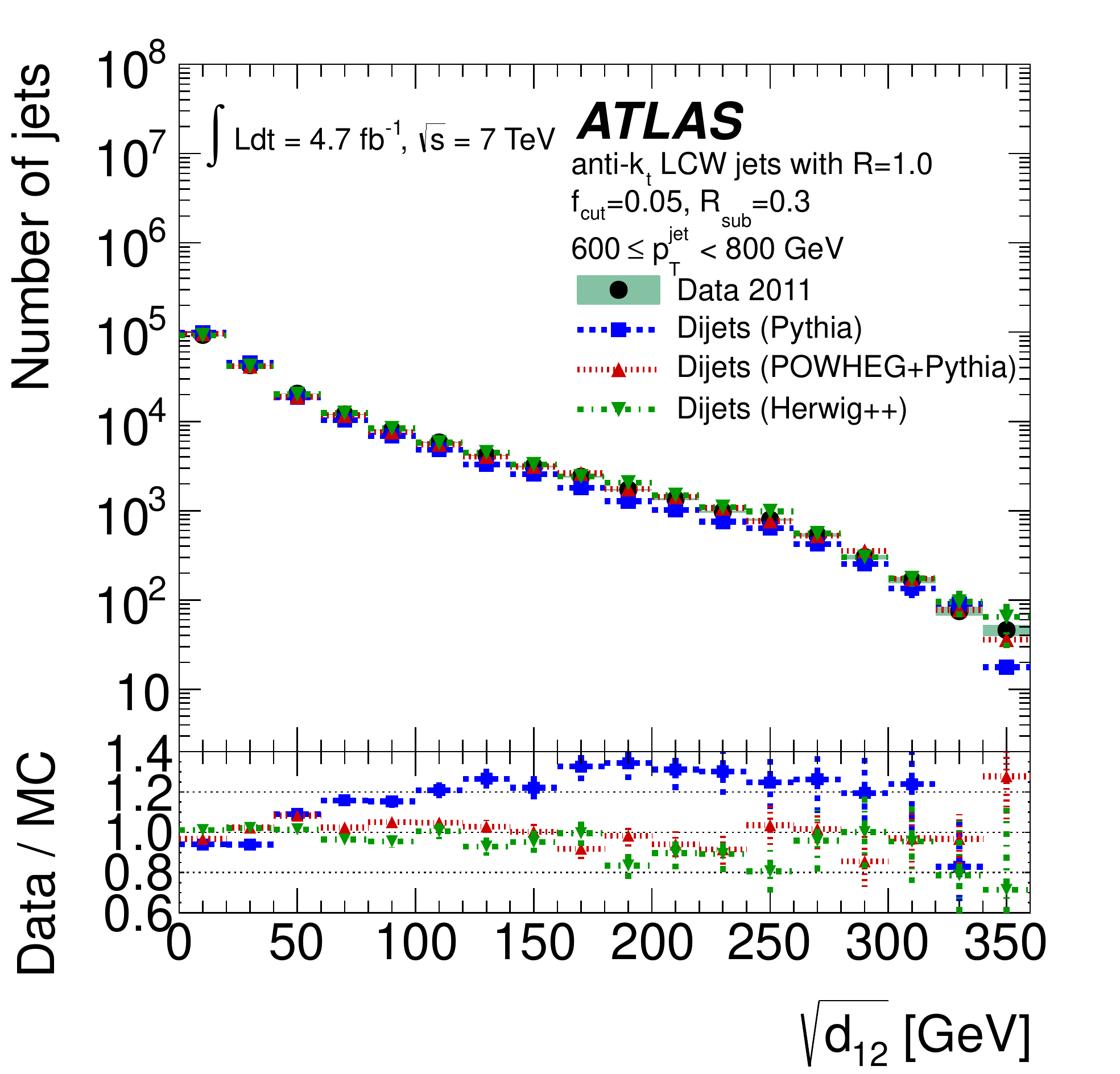}
    \label{fig:DataMC:d12:AKTTrim}}
  \caption{Splitting scale \DOneTwo of jets in the range
$600\GeV\leq\ptjet<800$~GeV.
Shown are \subref{fig:DataMC:d12:AKTFat} ungroomed and 
\subref{fig:DataMC:d12:AKTTrim} trimmed ($\fcut=0.05$, $\drsub=0.3$) \akt\ jets with $R=1.0$.
The ratios between data and MC distributions are shown
    in the lower section of each figure. 
          }
  \label{fig:DataMC:d12}
\end{figure}
\begin{figure}[!ht]
  \centering
  \subfigure[\AKTFat (ungroomed)]{
    \includegraphics[width=0.46\columnwidth]{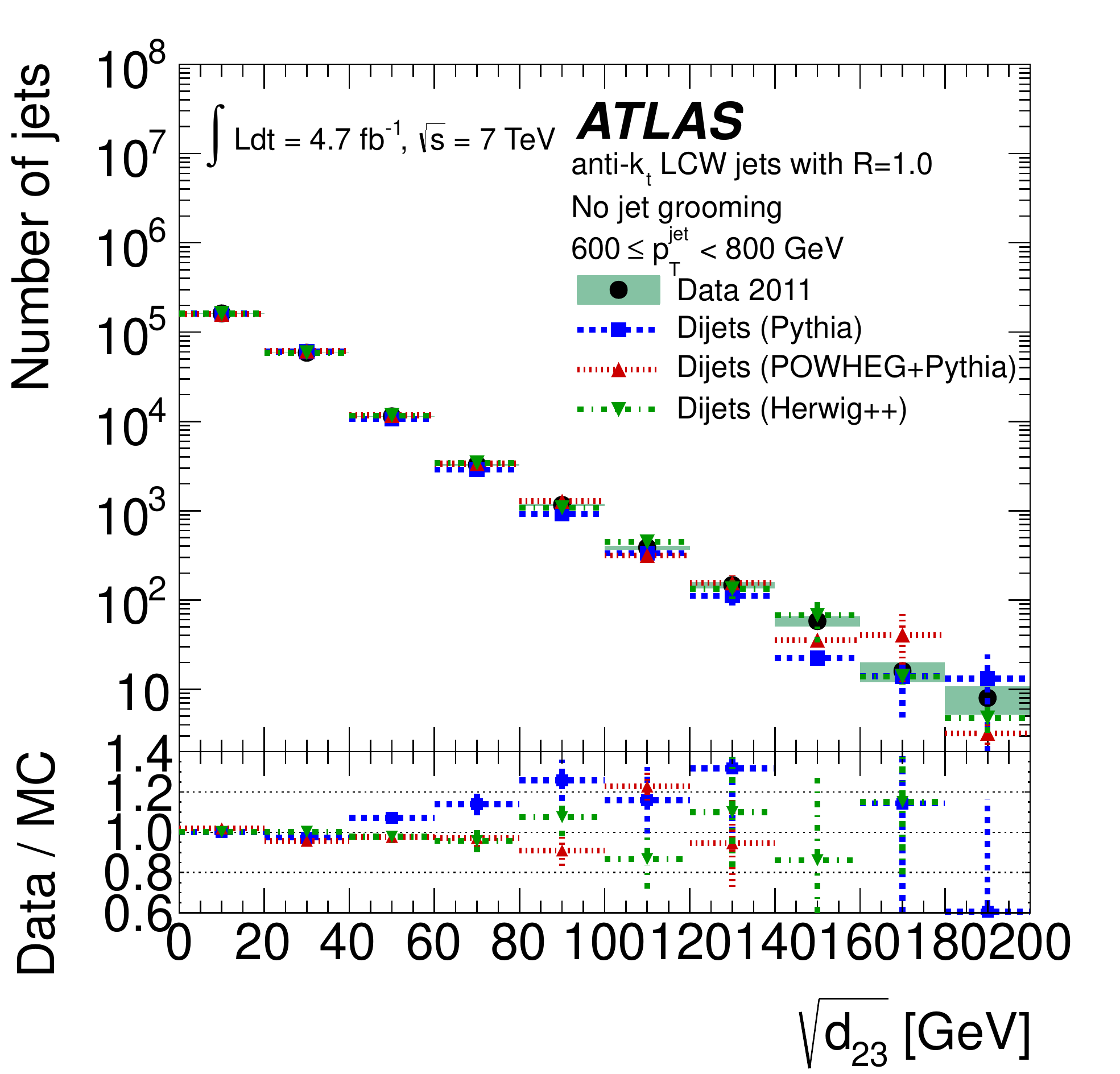}
    \label{fig:DataMC:d23:AKTFat}}
  \subfigure[\AKTFat (trimmed)]{
    \includegraphics[width=0.46\columnwidth]{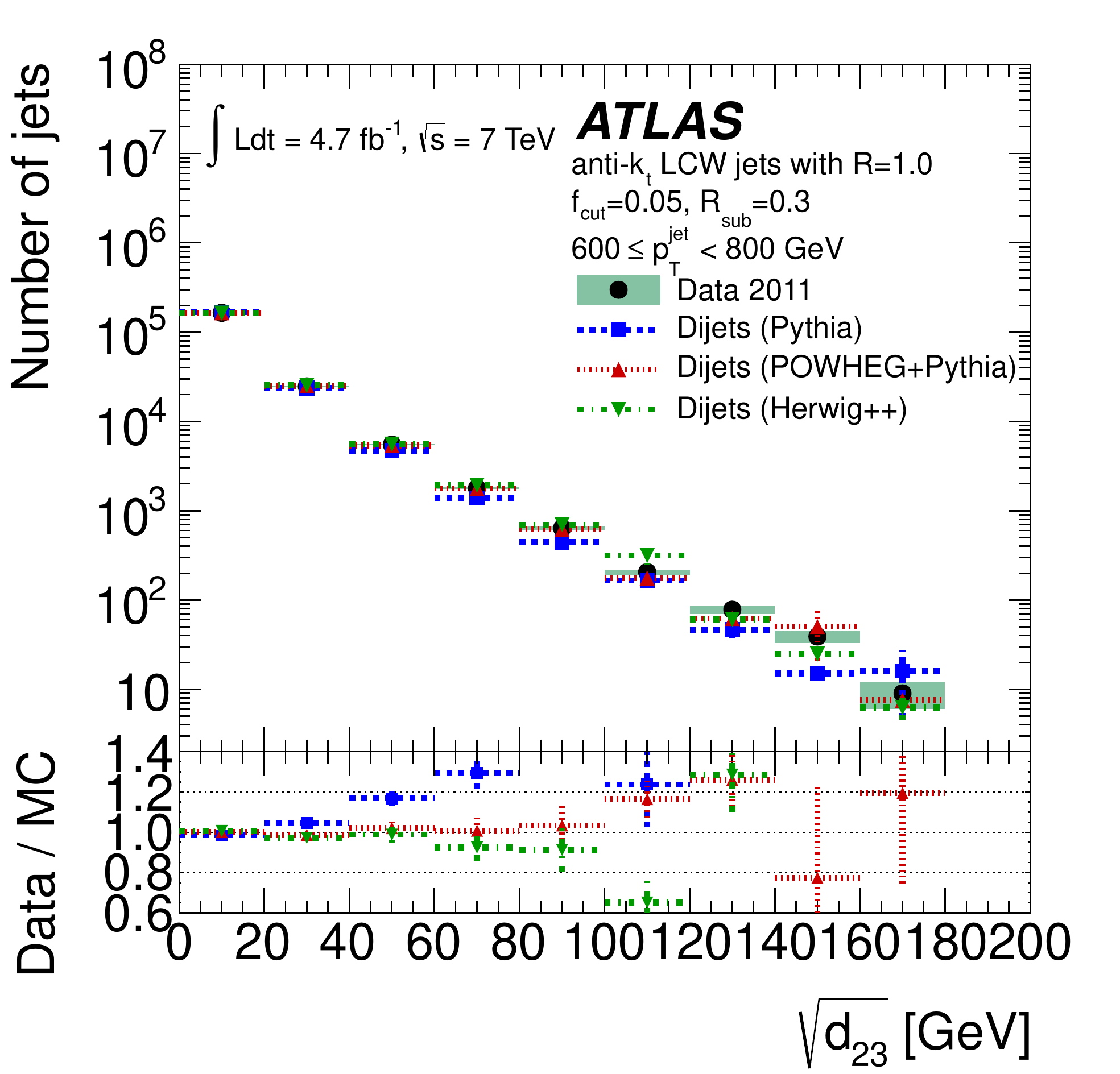}
    \label{fig:DataMC:d23:AKTTrim}}
  \caption{Splitting scale \DTwoThr of jets in the range
$600\GeV\leq\ptjet<800$~GeV.
Shown are \subref{fig:DataMC:d23:AKTFat} ungroomed and 
\subref{fig:DataMC:d23:AKTTrim} trimmed ($\fcut=0.05$, $\drsub=0.3$) \akt\ jets with $R=1.0$.
  The ratios between data and MC distributions are shown
    in the lower section of each figure.
         }
  \label{fig:DataMC:d23}
\end{figure}

As discussed in \secref{intro:definitions}, the splitting scales \DOneTwo and \DTwoThr are designed to give approximate measures of the relative mass of the leading, sub-leading, and third leading subjets, ordered by \subjetpt. 
These observables are therefore dominated by contributions originating from energetic partons, either from the parton shower or from massive particle decay.  
It is therefore not surprising that the jet grooming, which removes low-\pt\ components of the jet, does not significantly affect \DOneTwo, shown in \figref{DataMC:d12}. 
It is also not surprising that \PowPythia describes this variable better than \Pythia, especially at large values. Interestingly, \Herwigpp also models this substructure characteristic well, despite providing only a LO description of the hard process.
In the case of events selected only for the presence of a single high-\pt\ jet, a second hard splitting is not highly probable, as also evidenced by the spectrum of \DTwoThr, shown in \figref{DataMC:d23}, falling more steeply than \DOneTwo. 
This demonstrates that the trimming tends to affect this splitting scale slightly more when a third -\subjetpt subjet within the parent jet is modified during the trimming procedure.

\begin{figure}
  \centering
  \subfigure[\AKTFat (ungroomed)]{
    \includegraphics[width=0.46\columnwidth]{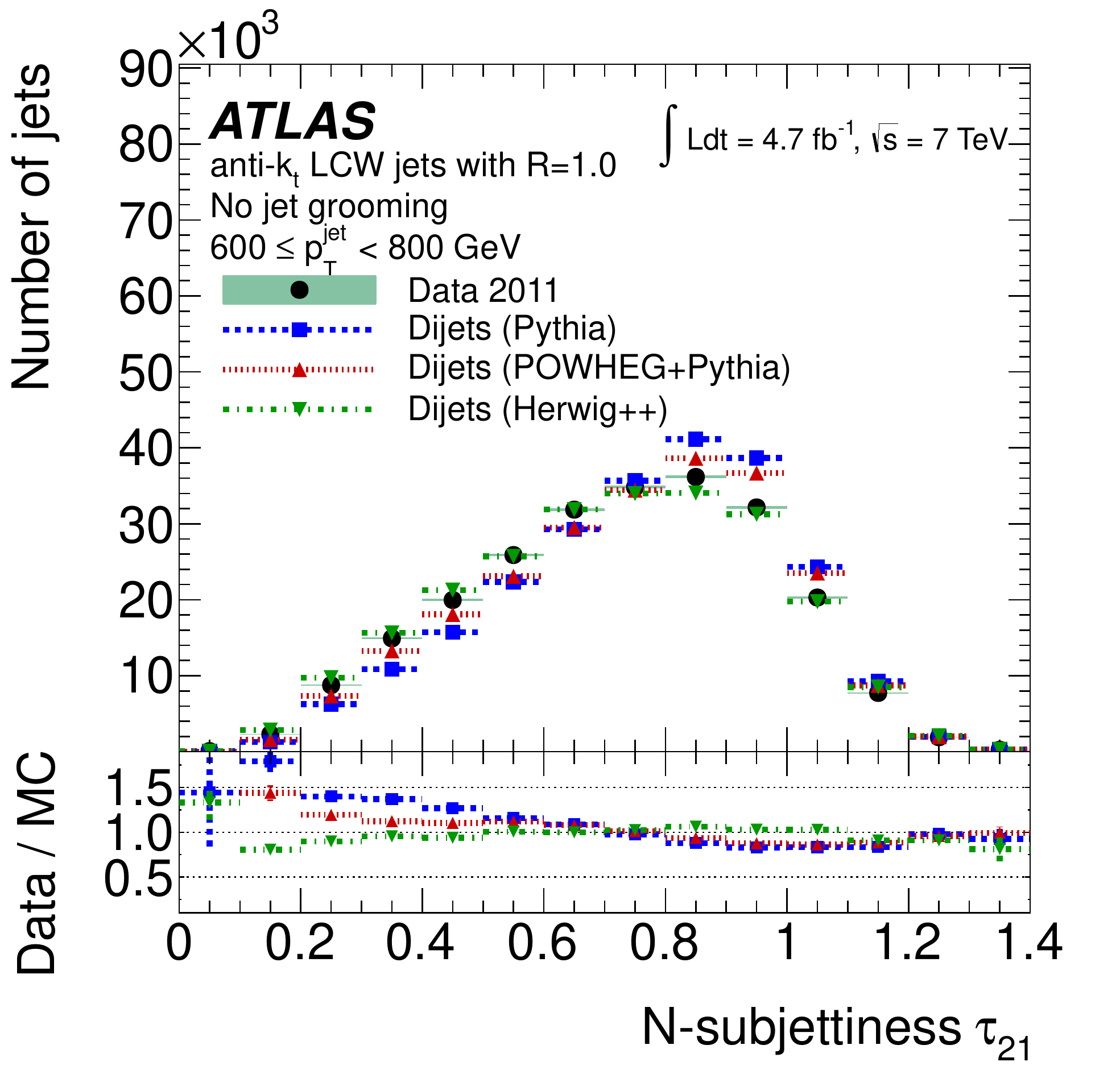}
    \label{fig:DataMC:tau21:AKTFat}}
  \subfigure[\AKTFat (trimmed)]{
    \includegraphics[width=0.46\columnwidth]{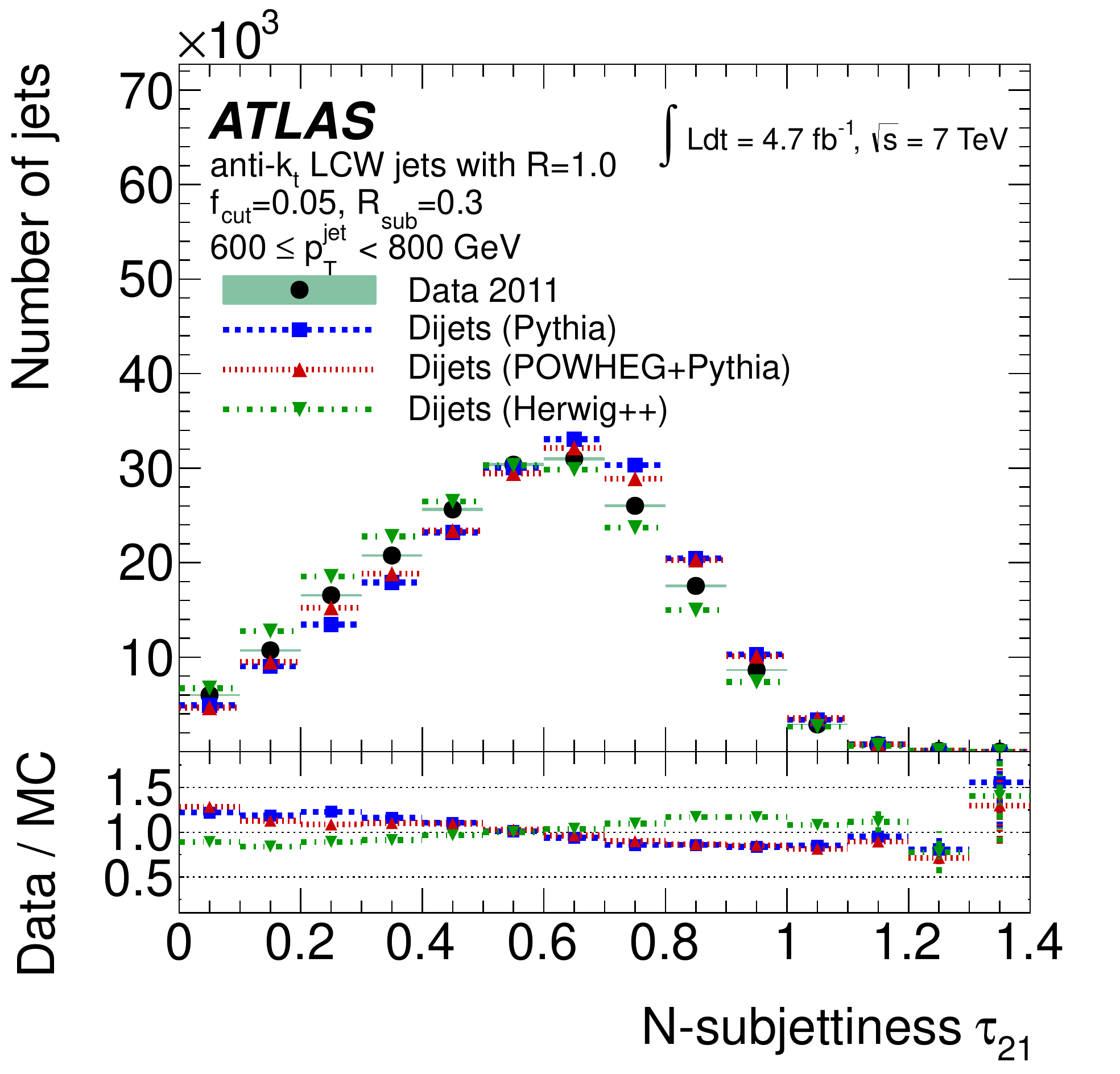}
    \label{fig:DataMC:tau21:AKTTrim}}
  \caption{$N$-subjettiness \tauTwoOne for jets in the range 
           $600\GeV\leq\ptjet<800$~GeV. Shown are \subref{fig:DataMC:tau21:AKTFat} 
           ungroomed and \subref{fig:DataMC:tau21:AKTTrim} trimmed 
           ($\fcut=0.05$, $\drsub=0.3$) \akt\ jets with $R=1.0$. The ratios 
           between data and MC distributions are shown in the lower section 
           of each figure.
           }
  \label{fig:DataMC:tau21}
\end{figure}

$N$-subjettiness helps to discriminate between jets that have well-formed substructure and those that do not. 
\Figsref{DataMC:tau21}{DataMC:tau32} demonstrate that the MC simulations model the distributions of \tauTwoOne and \tauThrTwo observed in the data within about 20\%. 
The jets in this case are selected to have $600\GeV\leq\ptjet<800$~GeV. Both \Pythia and \PowPythia exhibit a shift in the \tauTwoOne distribution towards larger values of \tauTwoOne with respect to the data for both the ungroomed and trimmed jets. \Herwigpp, however, provides a much better description of \tauTwoOne for ungroomed jets, and shows a slight shift towards smaller values of \tauTwoOne compared to the data.
The fraction of events with a small value of $\tau_{ij}$ is predicted to be slightly smaller than in data, while it is the opposite for high values of $\tau_{ij}$. 
In addition, the distributions of both \tauTwoOne and \tauThrTwo are broadened for trimmed jets (\figsref{DataMC:tau21:AKTTrim}{DataMC:tau32:AKTTrim}) and shifted slightly towards smaller values of $\tau_{ij}$ and into the region expected to be populated by boosted hadronic particle decays. 
These observations suggest that the use of shape observables for ungroomed jets, and of jet mass and substructure observables (like \DOneTwo) may lead to better discrimination between signal and background.

\begin{figure}
  \centering
  \subfigure[\AKTFat (ungroomed)]{
    \includegraphics[width=0.46\columnwidth]{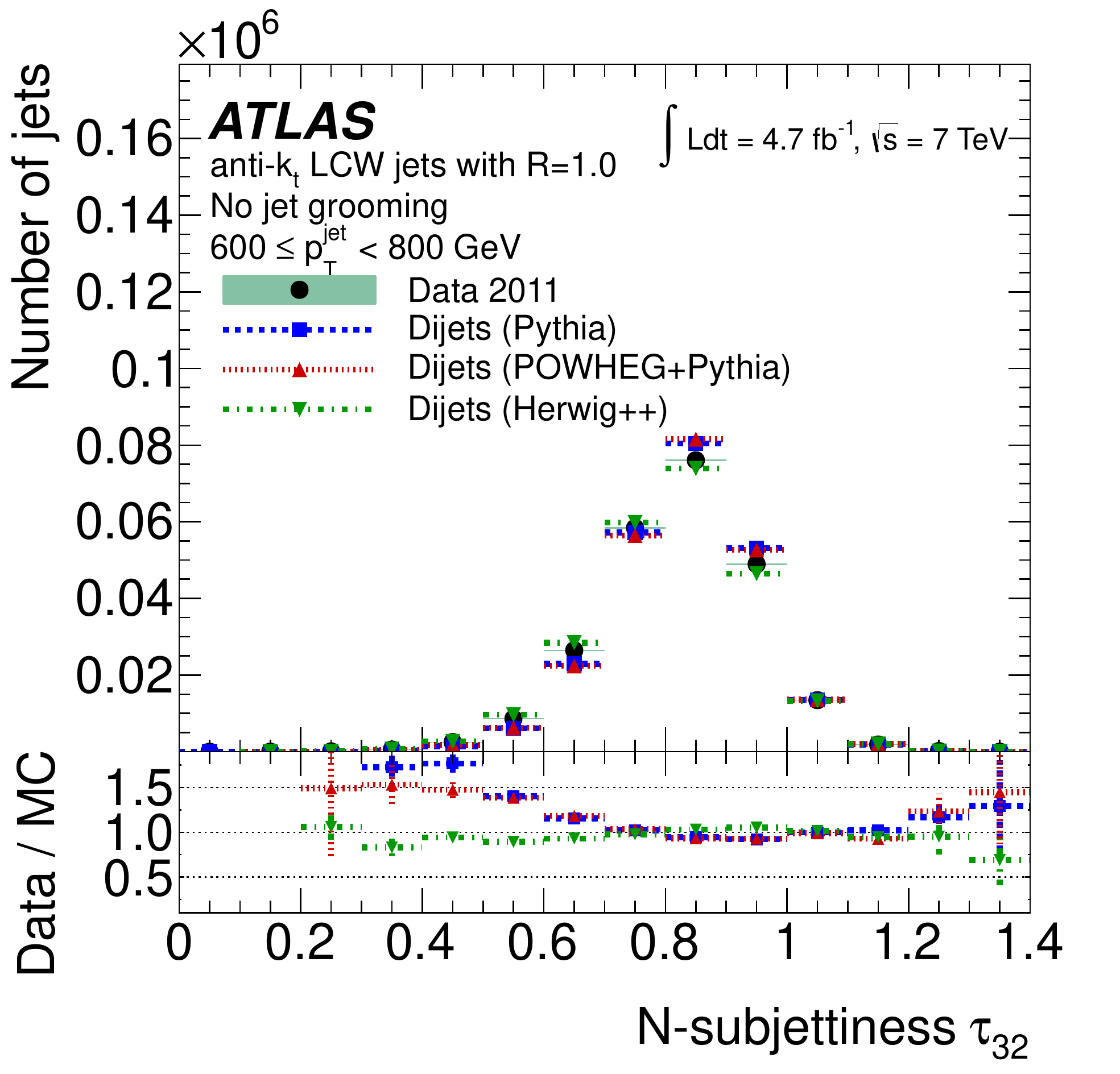}
    \label{fig:DataMC:tau32:AKTFat}}
  \subfigure[\AKTFat (trimmed)]{
    \includegraphics[width=0.46\columnwidth]{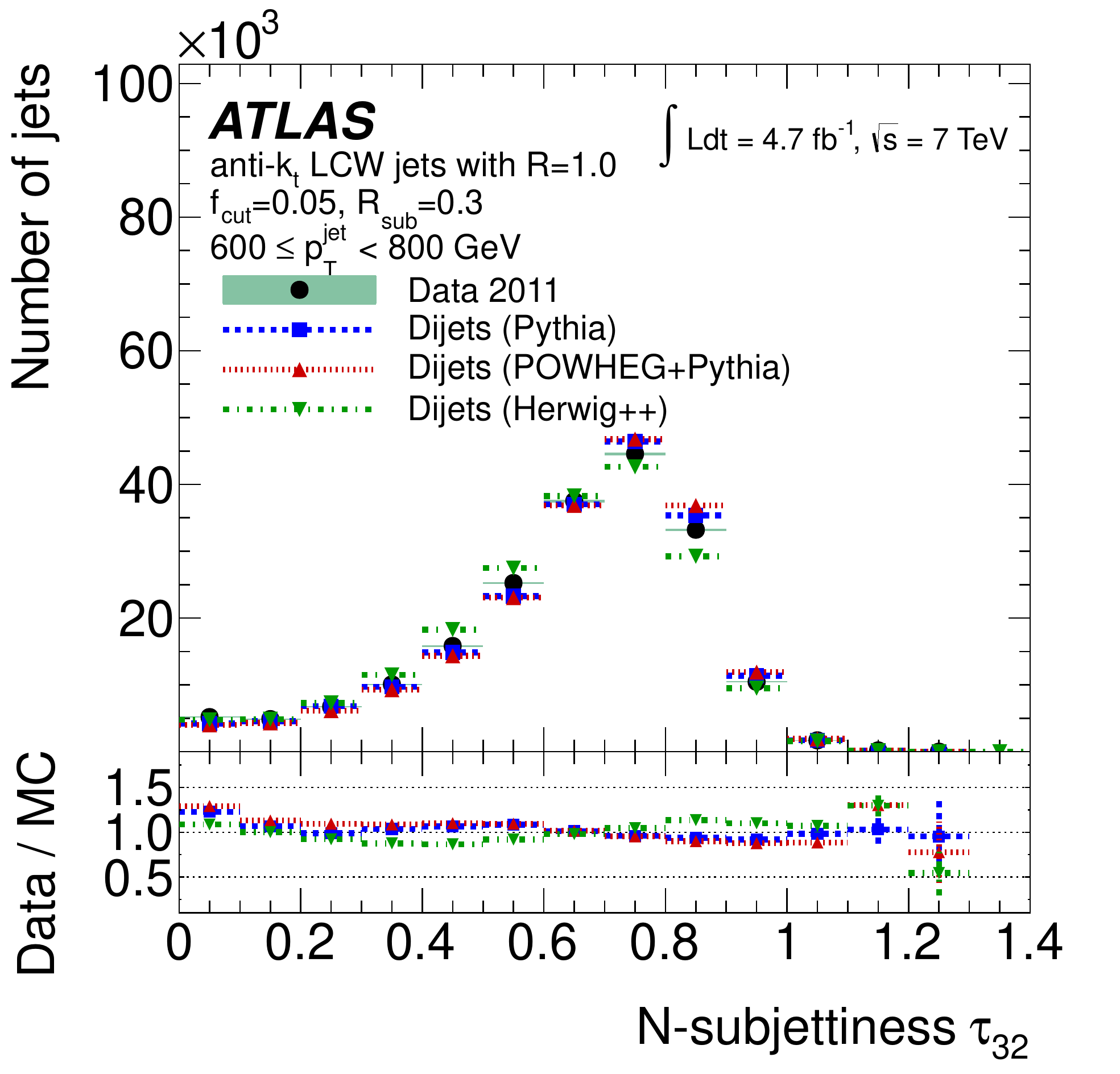}
    \label{fig:DataMC:tau32:AKTTrim}}
  \caption{$N$-subjettiness \tauThrTwo for jets in the range 
           $600\GeV\leq\ptjet<800$~GeV. Shown are 
           \subref{fig:DataMC:tau32:AKTFat} ungroomed and 
           \subref{fig:DataMC:tau32:AKTTrim} trimmed ($\fcut=0.05$, $\drsub=0.3$) 
           \akt\ jets with $R=1.0$. The ratios between data and MC 
           distributions are shown in the lower section of each figure.
           }
  \label{fig:DataMC:tau32}
\end{figure}

For \tauThrTwo, the MC distributions slightly underestimate the fraction of jets with $0.3< \tauThrTwo < 0.7$, which is the signal range for boosted top-quark candidates. 
Since these observables are intended to be used as discriminants between boosted object signal events and the inclusive jet background, such differences are important for the  resulting estimation of signal efficiency compared to background rejection. 
However, the variations observed in the distribution of each observable translate into much smaller differences in efficiency and rejection.

\begin{figure}
  \centering
  \subfigure[\AKTFat (ungroomed)]{
    \includegraphics[width=0.46\columnwidth]{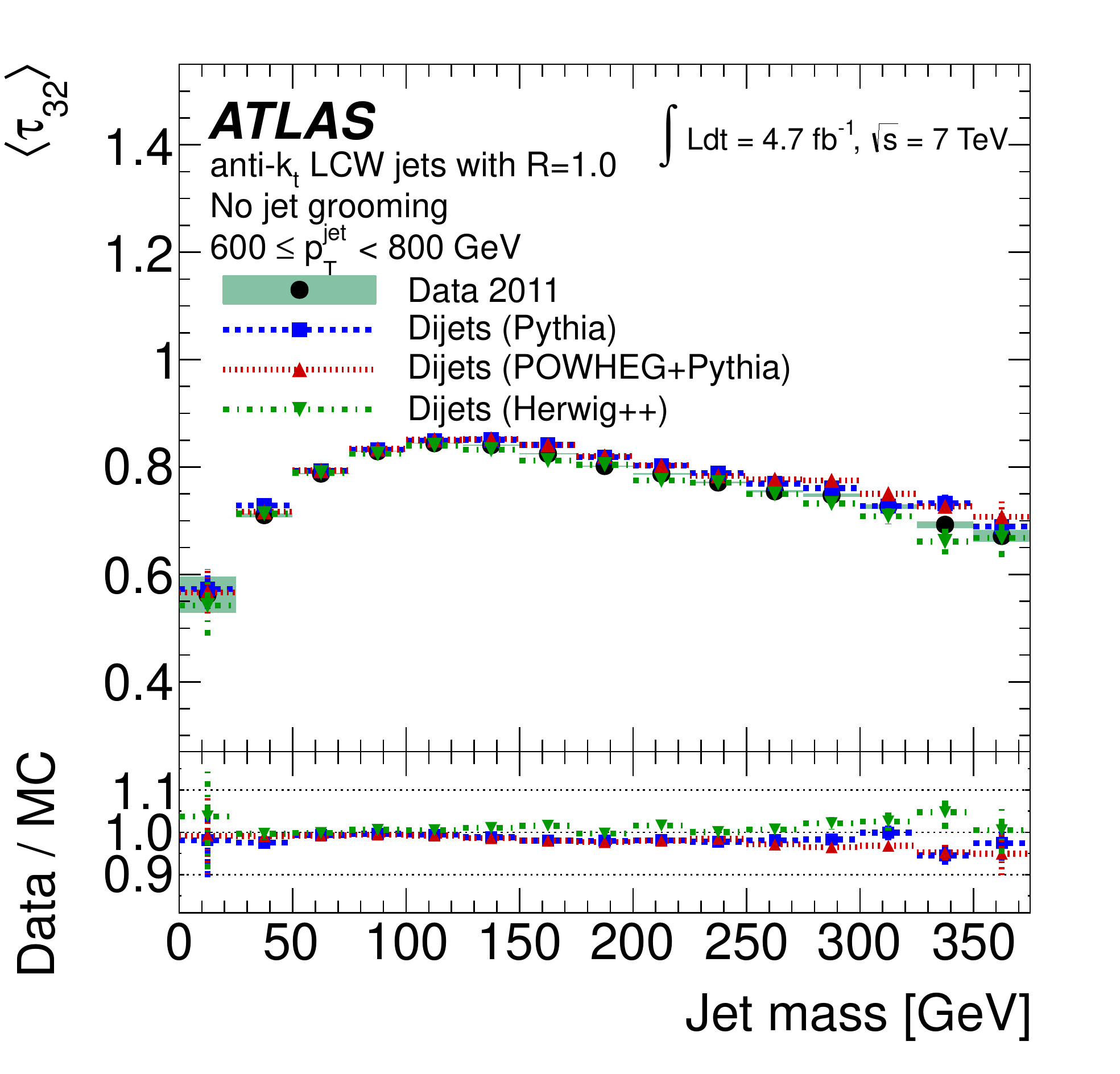}
    \label{fig:DataMC:tau32vsM:AKTFat}}
  \subfigure[\AKTFat (trimmed)]{
    \includegraphics[width=0.46\columnwidth]{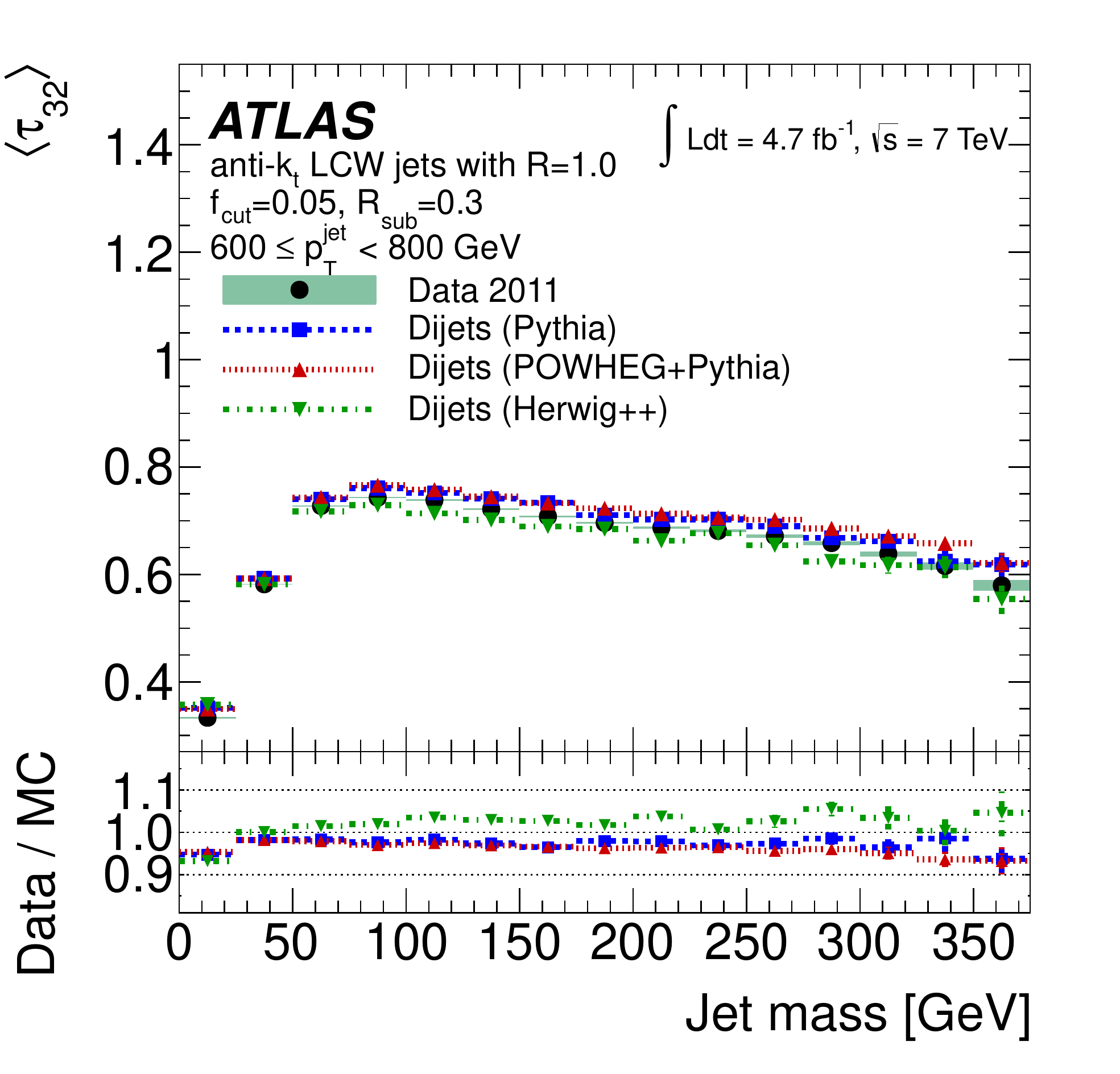}
    \label{fig:DataMC:tau32vsM:AKTTrim}}
  \caption{Mean \tauThrTwo as a function of jet mass for
\subref{fig:DataMC:tau32vsM:AKTFat} ungroomed and
\subref{fig:DataMC:tau32vsM:AKTTrim} trimmed ($\fcut=0.05$, $\drsub=0.3$)
 \akt\ jets with $R=1.0$ in the range $600\GeV\leq\ptjet<800$~GeV.
           }
  \label{fig:DataMC:tau32vsM}
\end{figure}

The modelling of the background with respect to massive boosted objects can be tested by evaluating, for example, the evolution of the mean of substructure variables as a function of jet mass. 
This is shown for $\langle\tauThrTwo\rangle$ in \figref{DataMC:tau32vsM} for the same $600\GeV\leq\ptjet<800$~GeV range as used above. Three important observations can be made. 
Values of $\langle\tauThrTwo\rangle$ are slightly lower in data than those predicted by the MC simulations, and the trimmed values are lower compared to ungroomed jets. 
Furthermore, $\langle\tauThrTwo\rangle$ is a slowly varying function of the jet mass for both the ungroomed and trimmed jets. This variation is slightly reduced for trimmed jets.

\subsection{Performance of jet grooming in boosted top-quark events}
\label{sec:boostedwtop:ttbar}

\subsubsection{Semi-leptonic \ttbar\ selection}
\label{sec:ttbar:selection}

A selection of \ttbar\nobreak$\rightarrow\nobreak(Wb)\nobreak(W\bar{b})\rightarrow\nobreak(\mu\nu b)(q\bar{q}\bar{b})$ events is used to demonstrate in data the effect of grooming on \largeR jets with substructure.
The semileptonic \ttbar\ decay mode in which one \W boson decays into a neutrino and a muon is chosen in order to tag the \ttbar\ event and reduce the overwhelming multi-jet background so that the top-quark signal is visible.
This provides a relatively pure sample of top quarks and is also very close to the selection used in searches for resonances that decay to pairs of boosted top quarks~\cite{Aad:2012dpa, Aad:2013nca}.
The following event-level and physics object selection criteria are applied to data and simulation:

\begin{description}

  \item[Event-level trigger and data quality selection:]{ 
The standard data quality and vertex requirements described in \secref{data-mc} are applied. Events are selected if they satisfy the single-muon Event Filter trigger with muon $\pt>18$~GeV.
  }

  \item[Event-level jet selection:]{
Events are required to have at least four \akt jets with $R=0.4$ having $\ptjet>25$~GeV and jet vertex fraction $|\JVF|>$0.75. The jet vertex fraction is a discriminant that contains information regarding the probability that a jet originated from the selected primary vertex in an event~\cite{ATLASWJetspaper2010}.
   }

  \item[Lepton selection:]{
Muons must be reconstructed in both the inner detector and the muon spectrometer and have $\pT>20$~GeV and $|\eta|<2.5$. The opening angle between the muon and any $R=0.4$ jet with $\pT > 25$~GeV and jet vertex fraction $|\JVF| > 0.75$ must be greater than $\DeltaR=0.4$ to be well isolated. Events with one or more electrons passing standard criteria as described in ref.~\cite{Aad:2013nca} are rejected.
  }

  \item[Event-level neutrino and leptonic \W-decay requirement:]{
To tag events with a leptonically decaying \W boson from a top-quark decay, events are required to have missing transverse momentum $\Etmiss>20$~GeV. Additionally, the scalar sum of \Etmiss\ and the transverse mass of the leptonic \W boson candidate must satisfy $\met+\WmassT>60$~GeV, where $\WmassT=\sqrt{2\pT\met(1-\cos\Dphi)}$ is calculated from the muon \pt\ and \met\ in the event, and \Dphi is the azimuthal angle between the charged lepton and the \Etmiss, which is assumed to be due to the neutrino.
  }

\end{description}

After these selection requirements, the \Wjets process constitutes the  largest background, with smaller contributions from \Zjets and single-top-quark processes.

\subsubsection{Performance of trimming in \ttbar\ events}
\label{sec:ttbar:grooming}

\begin{figure}[!ht]
  \centering
  \subfigure[\AKTFat, ungroomed]{
      \includegraphics[width=0.47\columnwidth]{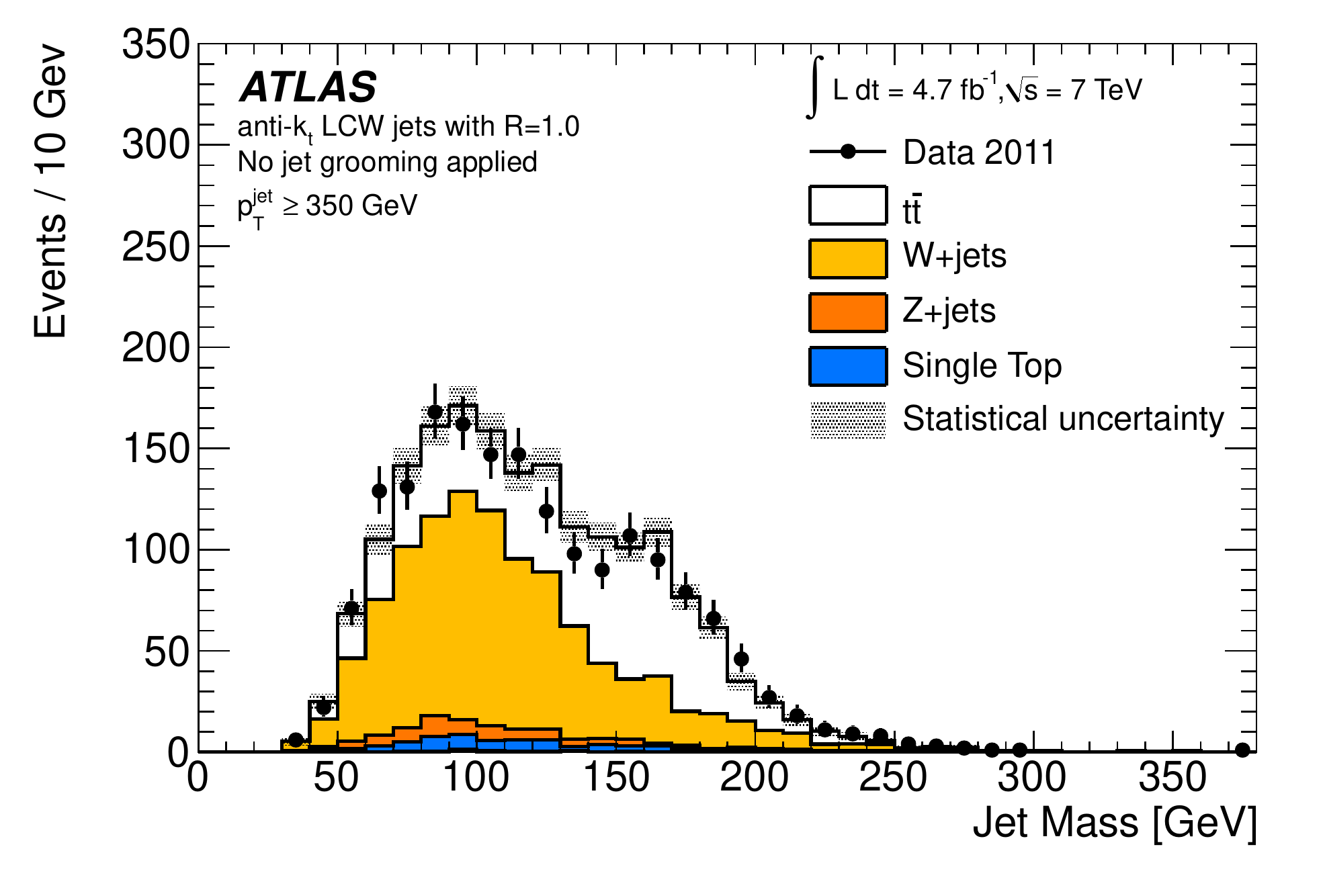}
    \label{fig:top_mass_ungroomed}}
  \subfigure[\AKTFat, trimmed]{
    \includegraphics[width=0.47\columnwidth]{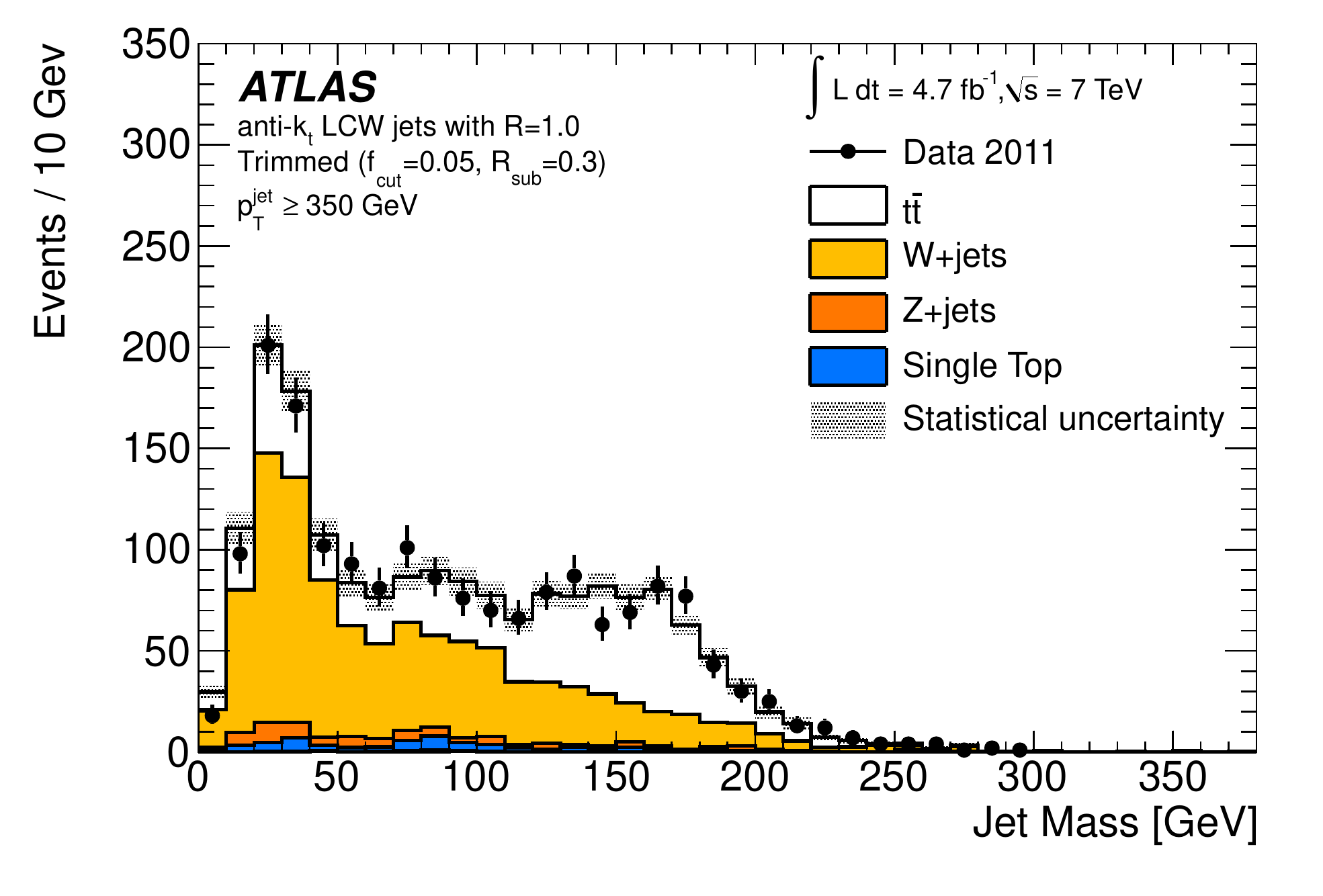}
    \label{fig:top_mass_groomed}}
  \caption{Jet mass for leading-\ptjet\ \akt jets with $R=1.0$ for 
      \subref{fig:top_mass_ungroomed} ungroomed jets and 
      \subref{fig:top_mass_groomed} trimmed jets ($\fcut=0.05$ and $\drsub=0.3$).
  The shaded band represents the bin-by-bin statistical uncertainty in simulation.
           }
  \label{fig:top_mass_akt}
\end{figure}

\Figref{top_mass_akt} shows the leading-\ptjet\ jet mass of \akt\ jets with $R=1.0$ having $\pt>350$~GeV before and after trimming ($\fcut=0.05$, $\drsub=0.3$) after the above selection criteria are applied.
The data and simulation agree within statistical uncertainty.  The \Wmunu events produced in association with jets form the largest background.
Since \largeR jets in \W events are formed from one or more random light-quark or gluon jets, trimming causes the mass spectrum to fall more steeply and the peak of the distribution to lie at smaller masses, similar to the multi-jet background in \figref{DataMC:grooming:mass:AKTFat}.
However, trimming does not alter the signal mass spectrum drastically, and any signal loss near the top-mass peak is due to events in which the top quark is not boosted enough to have all three hadronic decay products fall within $R=1.0$.

\begin{figure}[!ht]
  \centering
  \subfigure[\AKTFat, ungroomed]{
    \includegraphics[width=0.47\columnwidth]{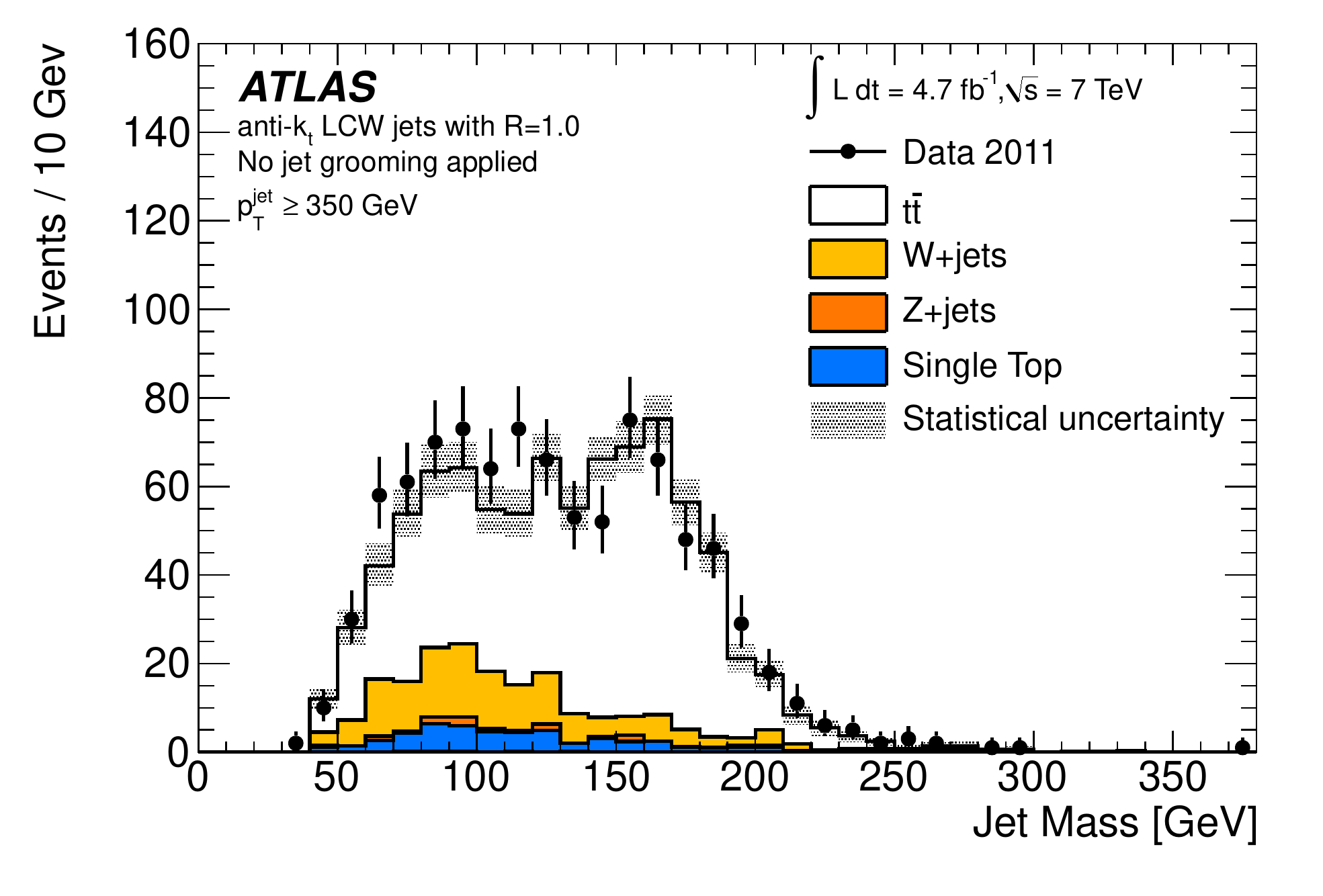}
    \label{fig:top_btag_ungroomed}}
  \subfigure[\AKTFat, trimmed]{
    \includegraphics[width=0.47\columnwidth]{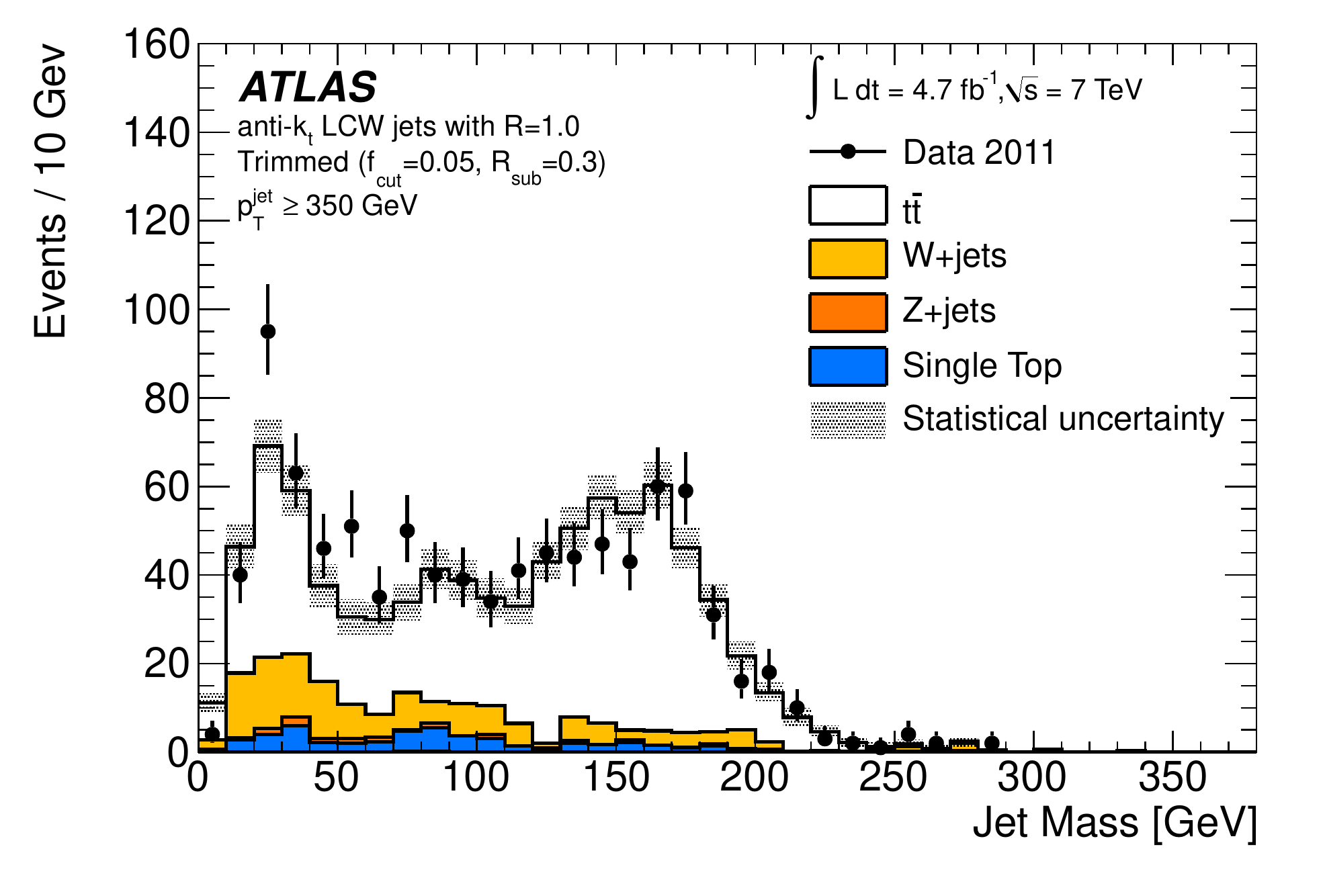}
    \label{fig:top_btag_groomed}}
  \caption{Jet mass for leading-\ptjet\ \akt jets with $R=1.0$ for 
      \subref{fig:top_btag_ungroomed} ungroomed jets and
      \subref{fig:top_btag_groomed} trimmed jets ($\fcut=0.05$ and $\drsub=0.3$), 
      where one \akt jet with $R=0.4$ was tagged as a $b$-jet.
           }
  \label{fig:top_btag_akt}
\end{figure}

In order to look at events with a reduced \W + jets background, a $b$-tagging requirement on at least one \akt\ jet in the event with $R=0.4$ is applied in addition to the selection criteria described in \secref{ttbar:selection}. \Figref{top_btag_akt} shows the effect of trimming on the mass of the leading-\ptjet\ \akt jet with $R=1.0$ in a sample of nearly pure \ttbar\ events. 
Trimming clearly enhances the mass discrimination compared to the ungroomed case, with a peak at low mass corresponding to \largeR jets containing one quark or gluon (probably from a fully leptonic \ttbar\ event) and a peak around the top mass, where all three top decay products, the $b$-jet and hadronic \W-decay daughters, fall inside the \largeR jet radius.

\begin{figure}[!ht]
  \centering
  \subfigure[\AKTFat, ungroomed]{
    \includegraphics[width=0.47\columnwidth]{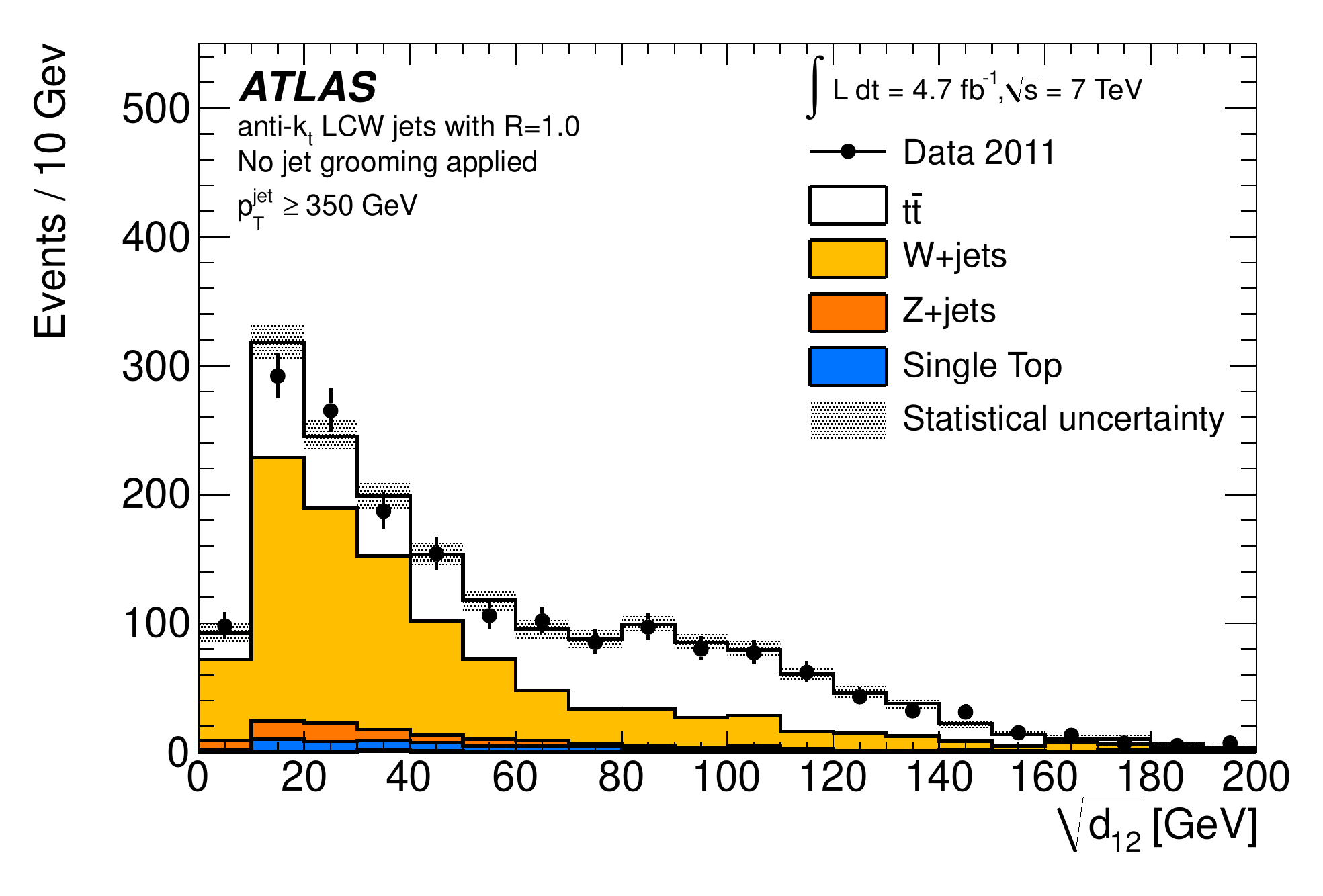}
    \label{fig:top_split12_ungroomed}}
  \subfigure[\AKTFat, trimmed]{
    \includegraphics[width=0.47\columnwidth]{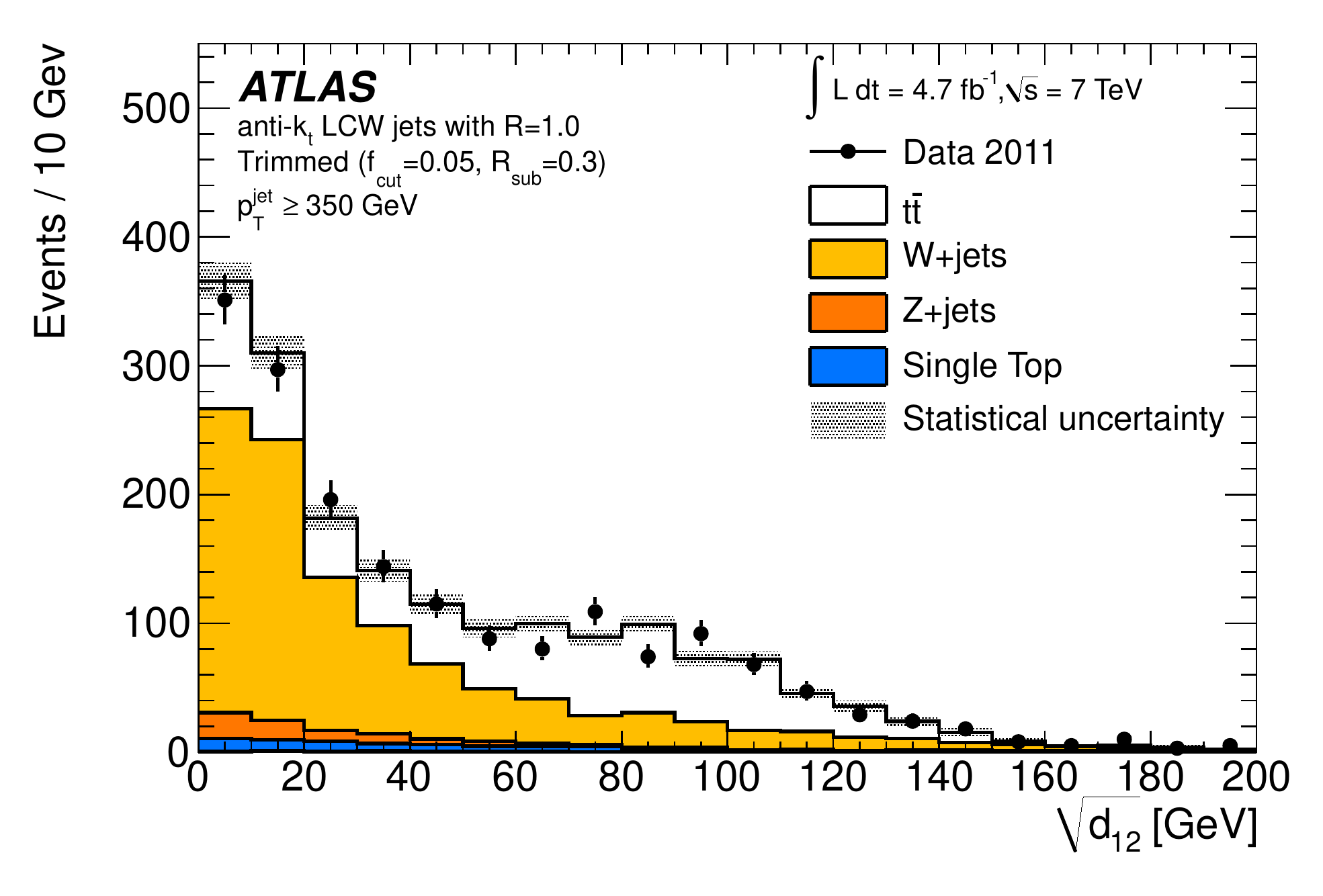}
    \label{fig:top_split12_groomed}} \\
  \subfigure[\AKTFat, ungroomed]{
    \includegraphics[width=0.47\columnwidth]{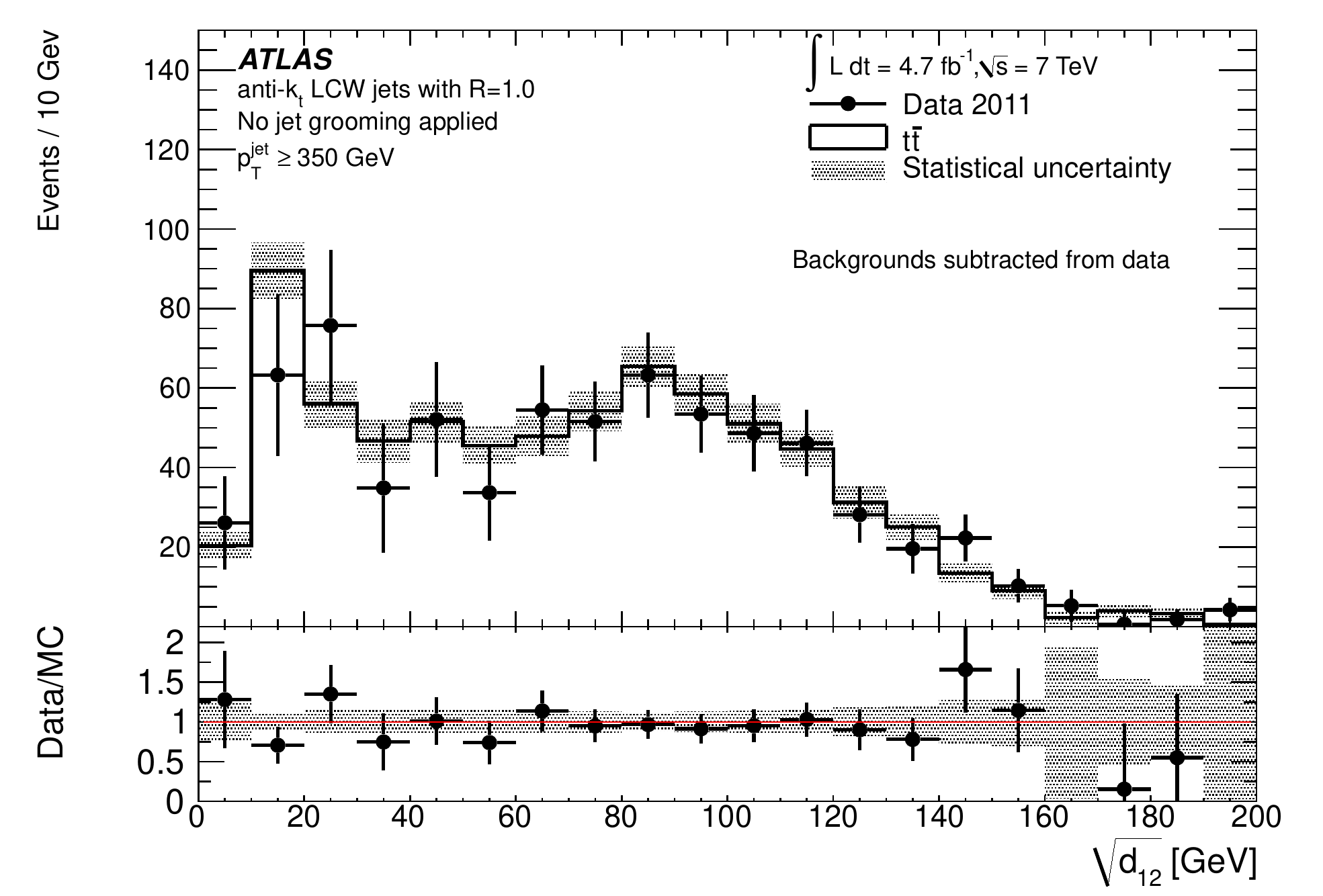}
    \label{fig:top_nobkgd_d12_ungroomed}}
  \subfigure[\AKTFat, trimmed]{
    \includegraphics[width=0.47\columnwidth]{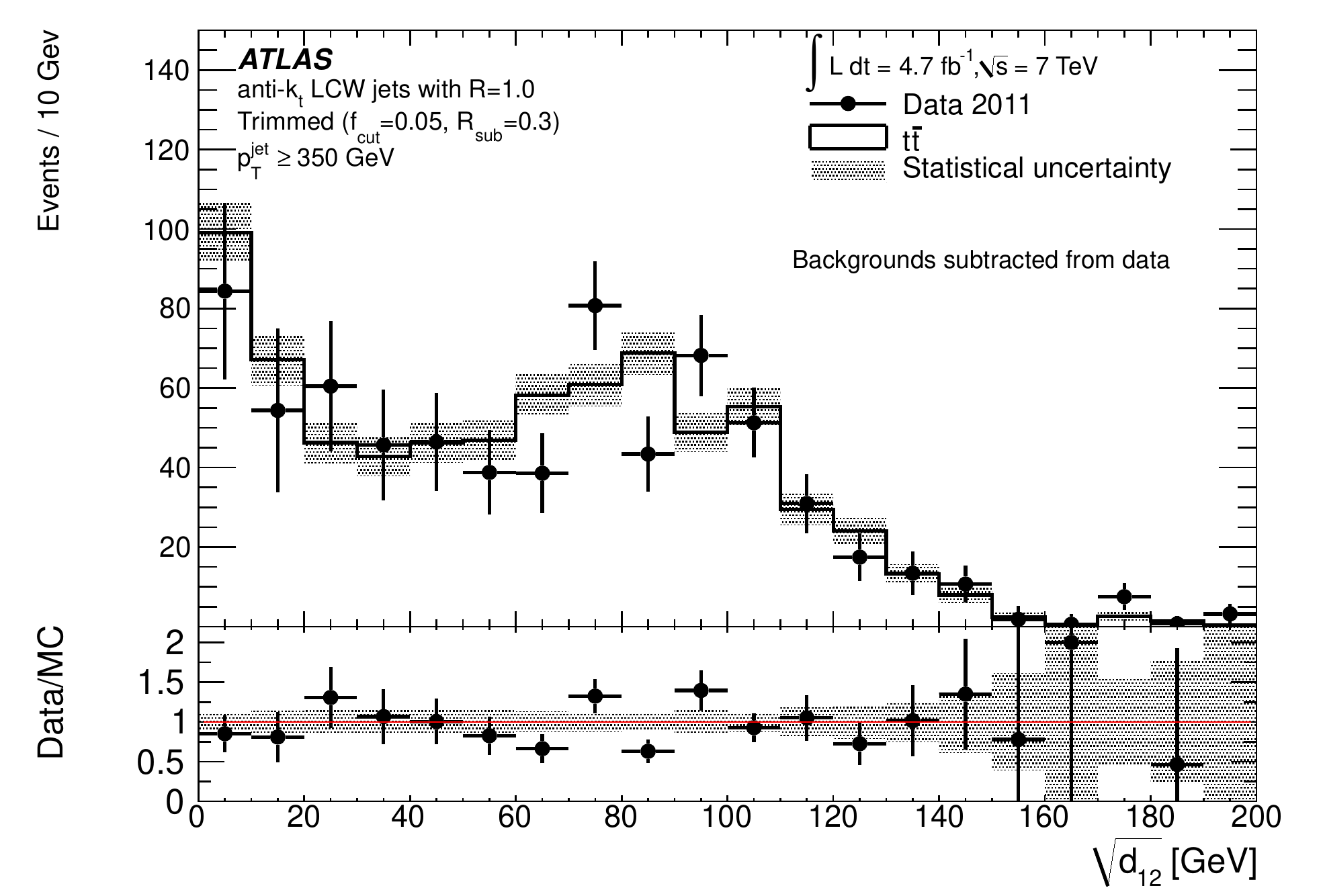}
    \label{fig:top_nobkgd_d12_groomed}}
  \caption{Splitting scale \DOneTwo for leading-\ptjet\ \akt jets with $R=1.0$ for 
     \subref{fig:top_split12_ungroomed} ungroomed jets and 
     \subref{fig:top_split12_groomed} trimmed jets ($\fcut=0.05$ and $\drsub=0.3$).
     The background-subtracted distributions for ungroomed jets and trimmed jets 
     are also shown in \subref{fig:top_nobkgd_d12_ungroomed} and 
     \subref{fig:top_nobkgd_d12_groomed}, respectively. 
           }
  \label{fig:top_split12_akt}
\end{figure}

\begin{figure}[!ht]
  \centering
  \subfigure[\AKTFat, ungroomed]{
    \includegraphics[width=0.47\columnwidth]{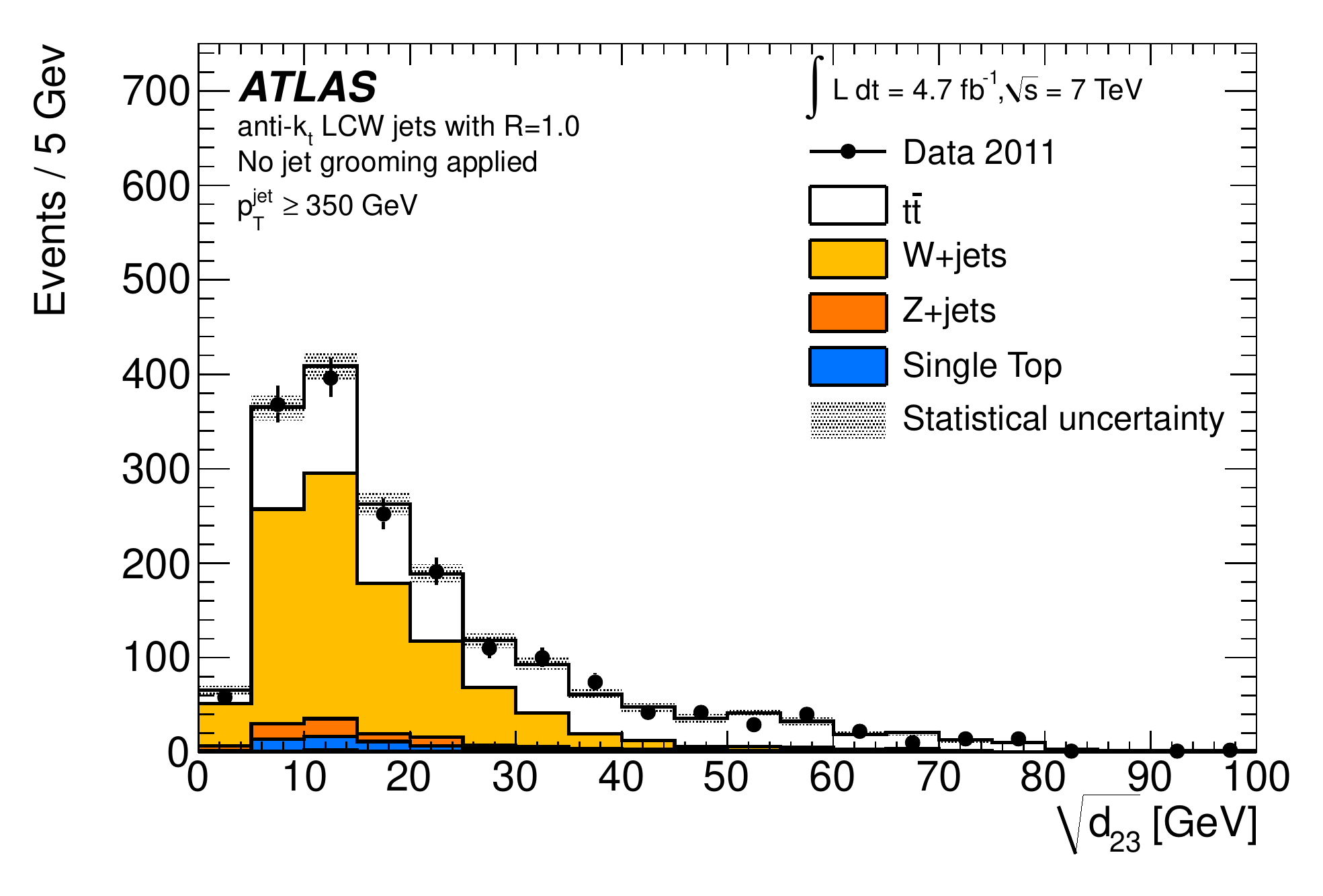}
    \label{fig:top_split23_ungroomed}}
  \subfigure[\AKTFat, trimmed]{
    \includegraphics[width=0.47\columnwidth]{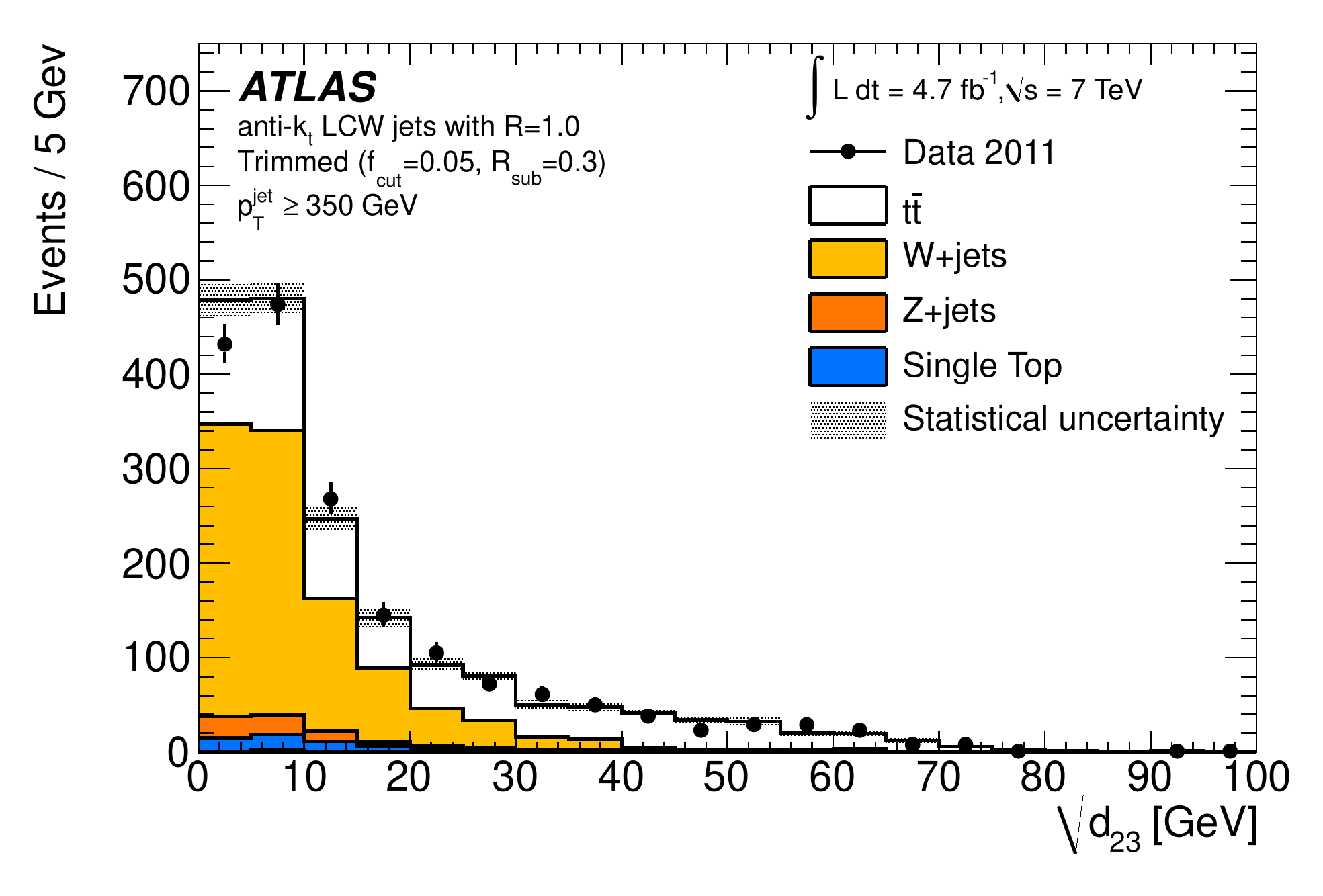}
    \label{fig:top_split23_groomed}} \\
   \subfigure[\AKTFat, ungroomed]{
    \includegraphics[width=0.47\columnwidth]{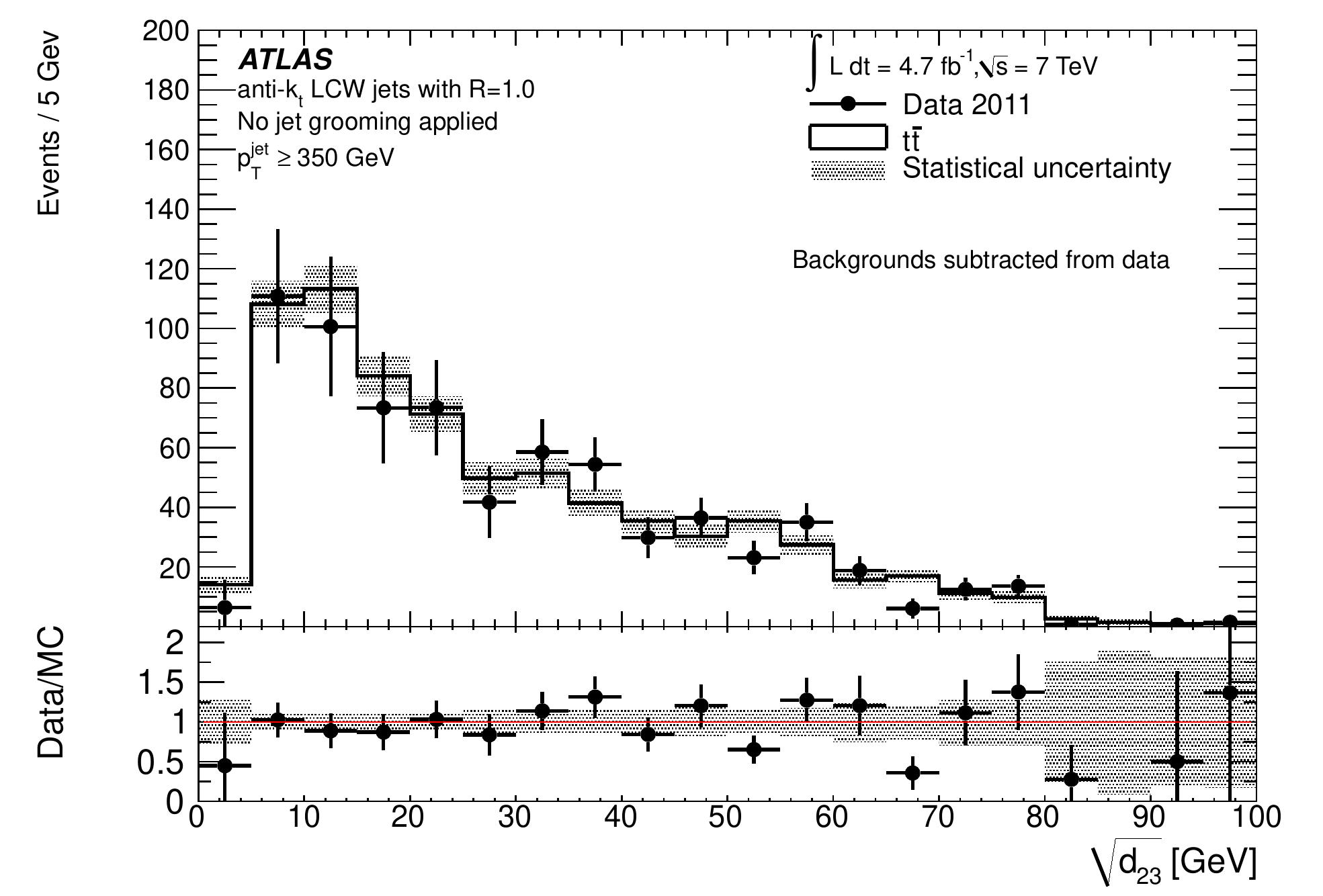}
    \label{fig:top_nobkgd_d23_ungroomed}}
  \subfigure[\AKTFat, trimmed]{
    \includegraphics[width=0.47\columnwidth]{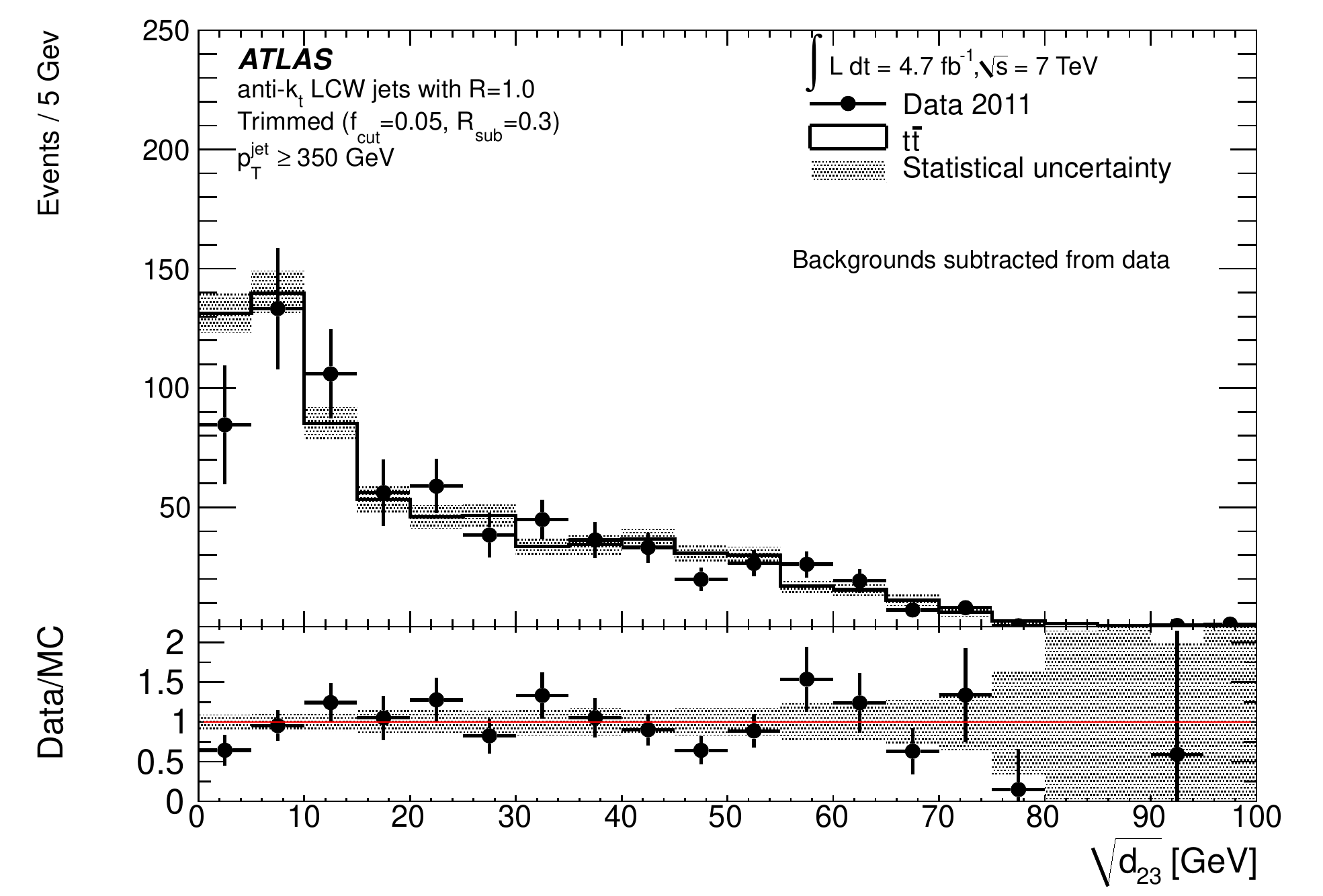}
    \label{fig:top_nobkgd_d23_groomed}}
  \caption{Splitting scale \DTwoThr for leading-\ptjet\ \akt jets with $R=1.0$ for 
      \subref{fig:top_split23_ungroomed} ungroomed jets with $R=1.0$ and
      \subref{fig:top_split23_groomed} trimmed jets ($\fcut=0.05$ and $\drsub=0.3$). 
      The background-subtracted distributions for ungroomed jets and trimmed jets 
     are also shown in \subref{fig:top_nobkgd_d23_ungroomed} and 
     \subref{fig:top_nobkgd_d23_groomed}, respectively. 
           }
  \label{fig:top_split23_akt}
\end{figure}

\Figsref{top_split12_akt}{top_split23_akt} show the distributions of \DOneTwo and \DTwoThr, respectively, before and after trimming.
Again, the top-quark distribution remains relatively unaffected by trimming, while the \Wjets background is pushed to lower values; however, the effect is smaller for these variables than for the mass. 
Also shown are the distributions with the simulated background subtracted. Comparing the shape before and after grooming with the signal distributions shown in \figsref{groomed_jets_split12_compare}{groomed_jets_split_compare} for very high-\pT\ top quarks, it is apparent that not all top quarks in this data sample were sufficiently boosted to have their decay products fall within a single $R=1.0$ jet.

\begin{figure}[!ht]
  \centering
  \subfigure[\AKTFat, ungroomed]{
    \includegraphics[width=0.47\columnwidth]{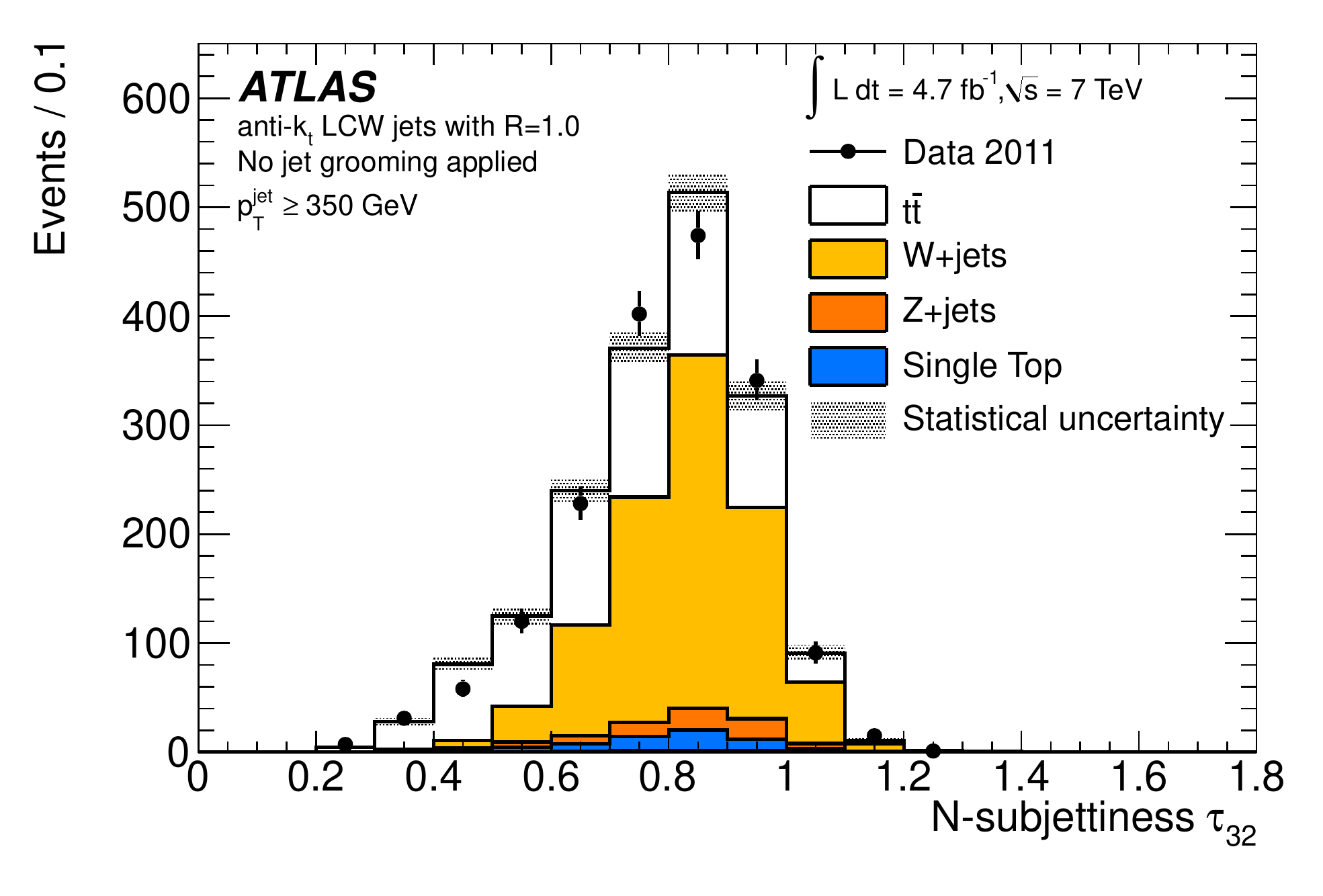}
    \label{fig:top_tau32_ungroomed}}
  \subfigure[\AKTFat, trimmed]{
    \includegraphics[width=0.47\columnwidth]{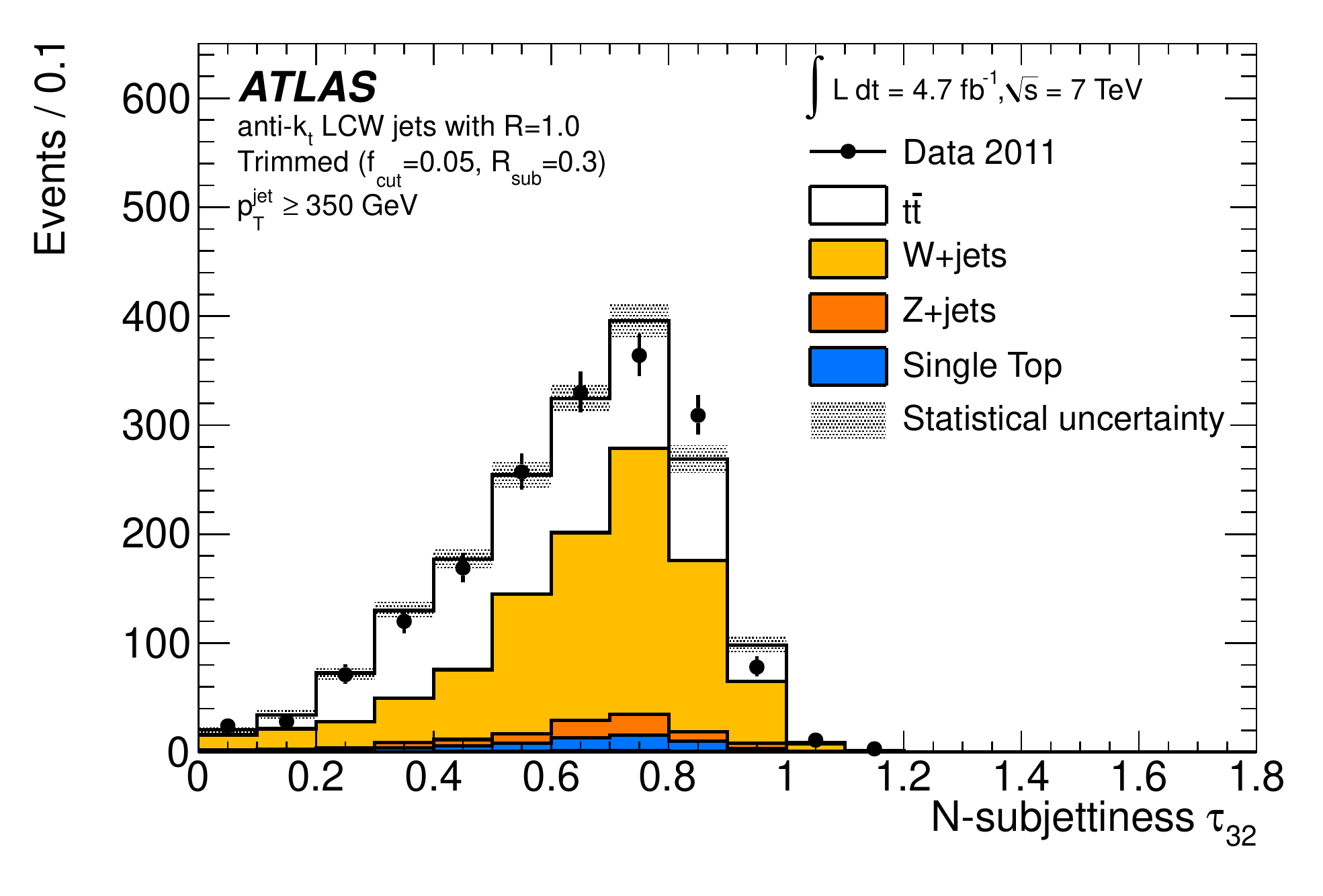}
    \label{fig:top_tau32_groomed}}\\
  \subfigure[\AKTFat, ungroomed]{
    \includegraphics[width=0.47\columnwidth]{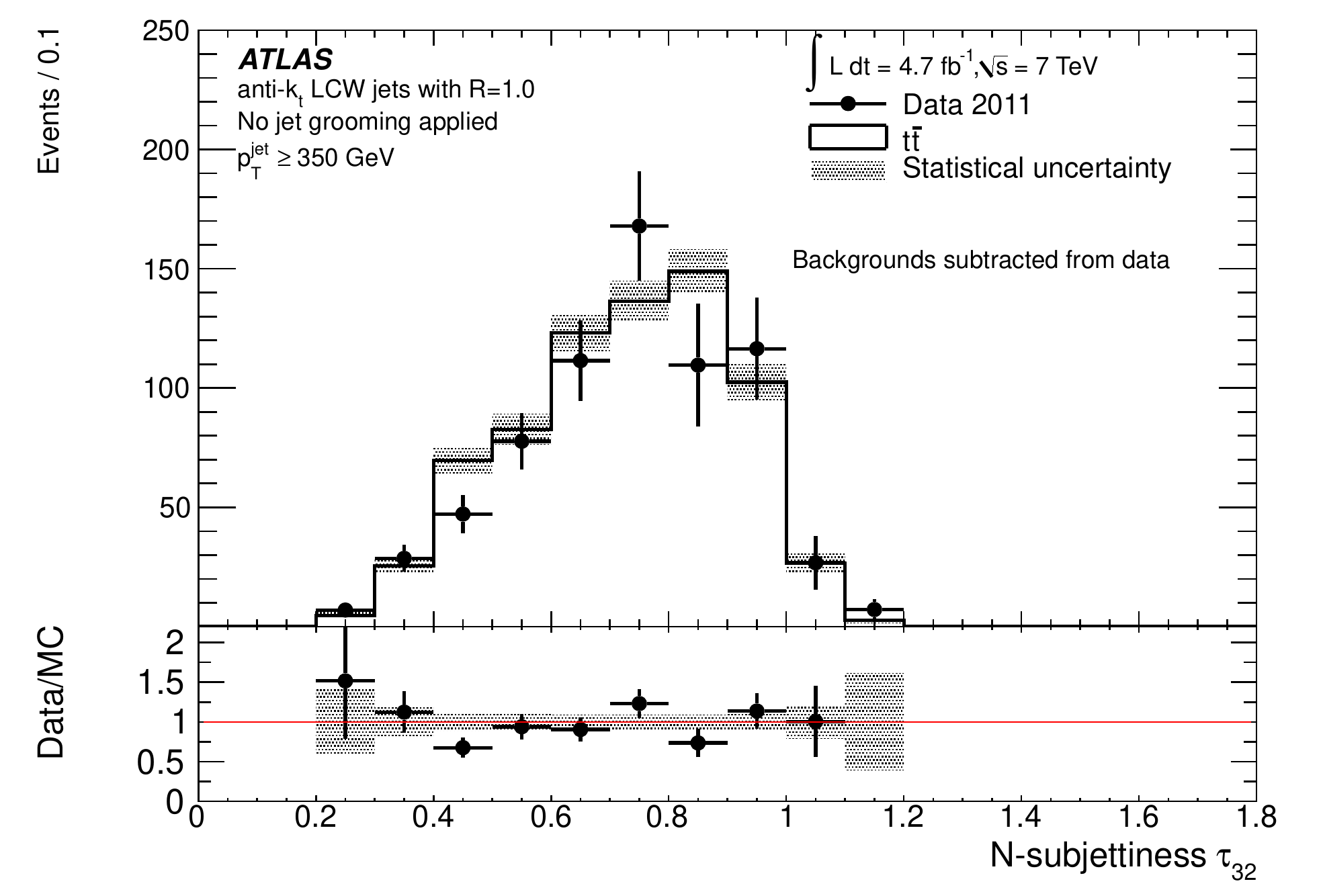}
    \label{fig:top_nobkgd_tau32_ungroomed}}
  \subfigure[\AKTFat, trimmed]{
    \includegraphics[width=0.47\columnwidth]{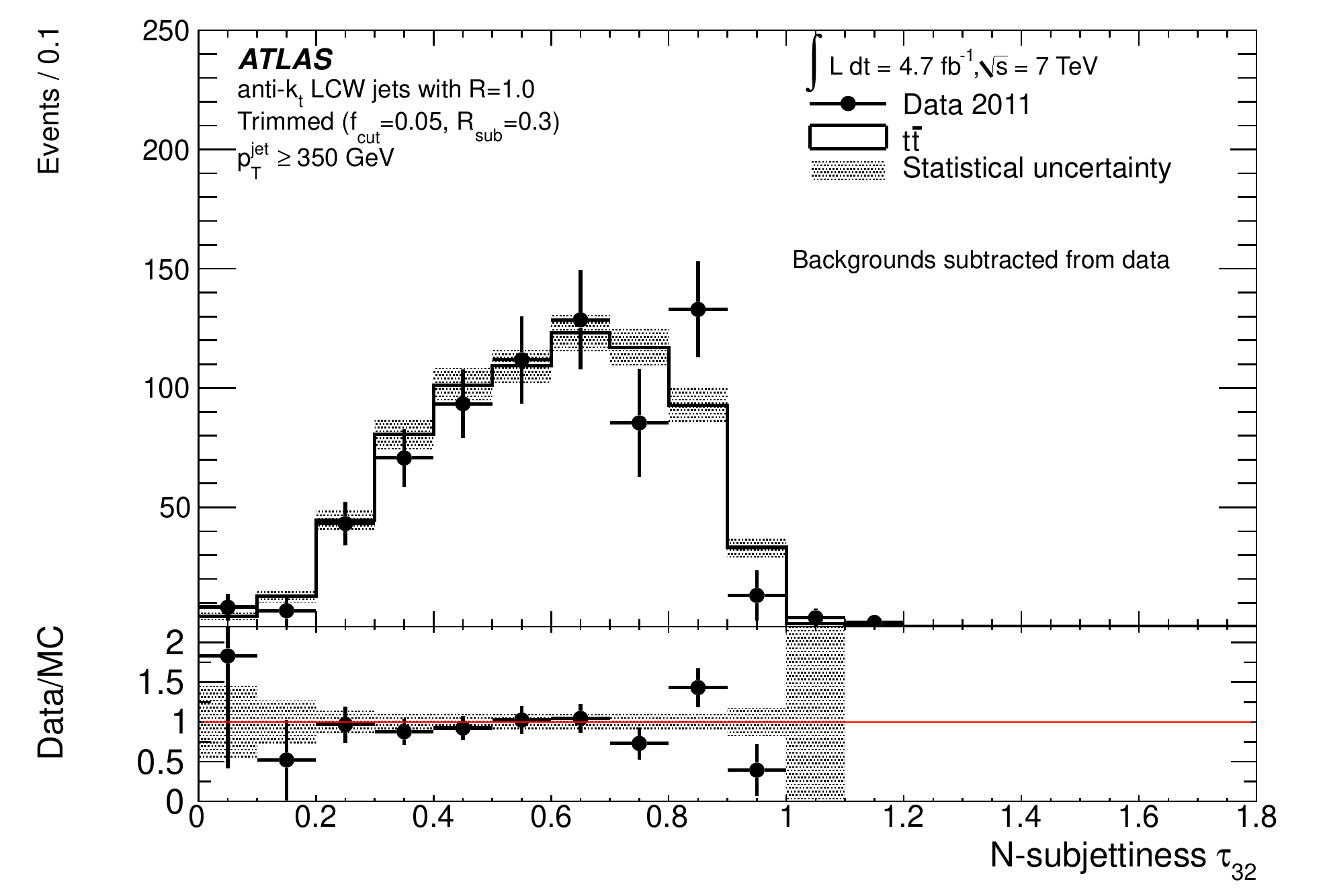}
    \label{fig:top_nobkgd_tau32_groomed}}
  \caption{$N$-subjettiness \tauThrTwo for leading-\ptjet\ \akt jets with $R=1.0$ for 
      \subref{fig:top_tau32_ungroomed} ungroomed jets with $R=1.0$ and
      \subref{fig:top_tau32_groomed} trimmed jets ($\fcut=0.05$ and $\drsub=0.3$). 
      The background-subtracted distributions for ungroomed jets and trimmed jets 
     are also shown in \subref{fig:top_nobkgd_tau32_ungroomed} and 
     \subref{fig:top_nobkgd_tau32_groomed}, respectively. 
           }
  \label{fig:top_tau32_akt}
\end{figure}

\Figref{top_tau32_akt} shows the signal-enriched distributions of \tauThrTwo before and after trimming. Here, there is less discrimination between the \Wjets background and the top-quark signal. 
This is due to the fact that multiple quark and gluon jets in \W events can be reconstructed as one $R=1.0$ \akt jet and mimic the subjettiness signature of the \largeR jet containing the hadronic decay products of the top quark. 
For shape comparisons with \figref{groomed_jets_subjetiness_compare}, the background-subtracted plots are also shown.

\subsubsection{Application of the \htt in $\mathbf{\ttbar}$ events}
\label{sec:ttbar:htt}

\begin{figure}[!ht]
  \begin{center}
    \subfigure[$R=1.5$]{ 
      \includegraphics[width=0.46\columnwidth]{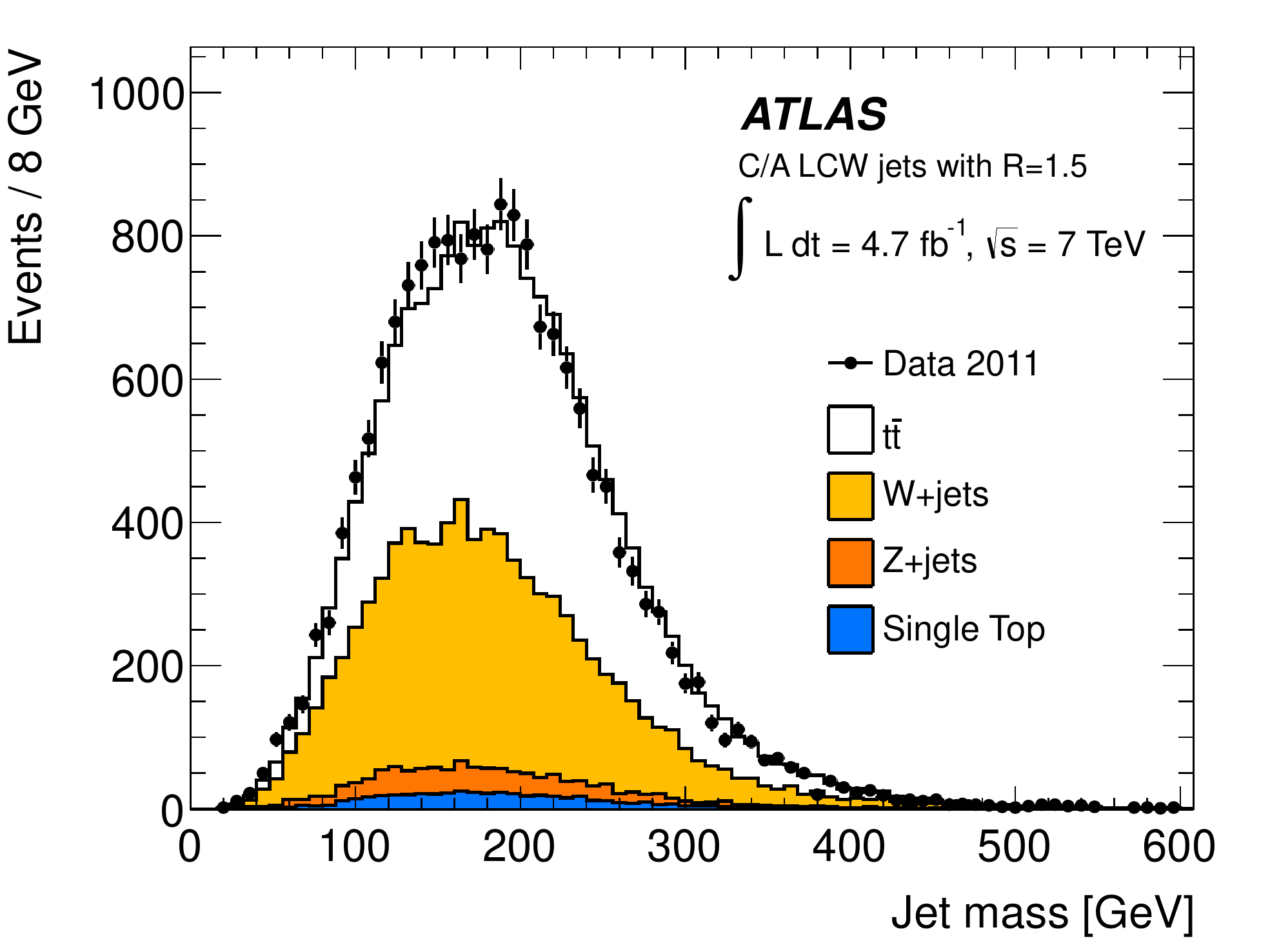} 
      \label{fig:HTT:kinematics:fatjet:m15}
    }
    \subfigure[$R=1.8$]{ 
      \includegraphics[width=0.46\columnwidth]{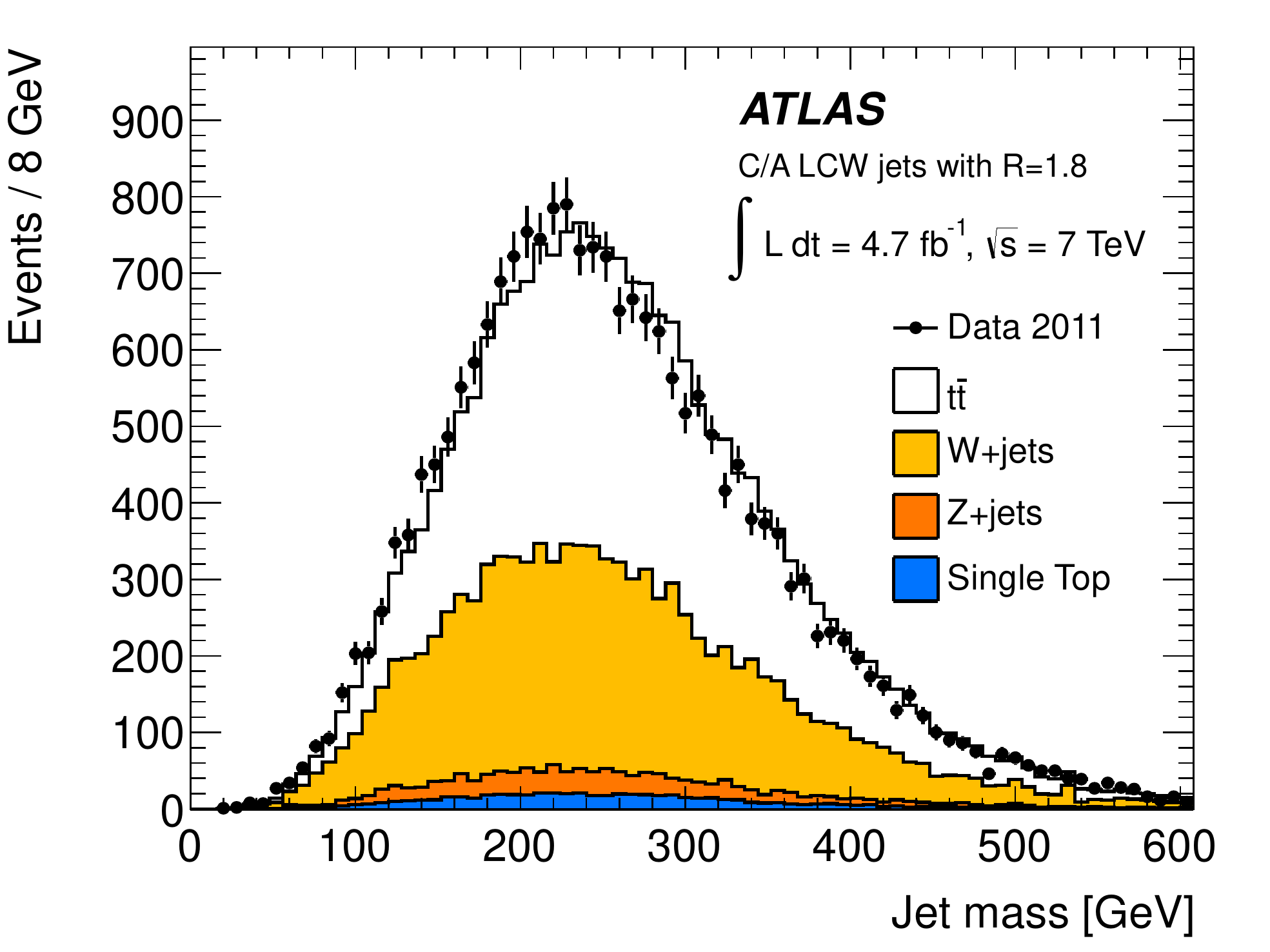} 
      \label{fig:HTT:kinematics:fatjet:m18}
    }\\
  \end{center}
  \caption{Mass distribution for \CamKt jets with 
     \subref{fig:HTT:kinematics:fatjet:m15} $R=1.5$ and
     \subref{fig:HTT:kinematics:fatjet:m18} $R=1.8$ before running the \htt.
  }
  \label{fig:HTT:kinematics:fatjet:m}
\end{figure}

Basic kinematic distributions for \largeR jets before and after applying the \htt algorithm, but without any $b$-tagging requirement, are shown in this section. 
Top-quark candidates in data and simulation are compared after selecting events according to the criteria listed in \secref{ttbar:selection}. 
\Figref{HTT:kinematics:fatjet:m} shows the jet mass distributions of C/A jets with $R=1.5$ and $R=1.8$ before applying the \htt, for jets having $\pt>200$~GeV. 
The sample consists of approximately 50\% \ttbar-pair events, with other contributions coming mainly from \Wjets and \Zjets events. 
A larger contribution from multi-jet events compared to that observed in \secref{ttbar:grooming} is expected, due to the larger jet radius and lower \pt\ threshold.
The \largeR jet mass is generally well described by the simulation.

\begin{figure}[!ht]
  \begin{center}
    \subfigure[$R=1.5$, default filtering]{ 
      \includegraphics[width=0.46\columnwidth]{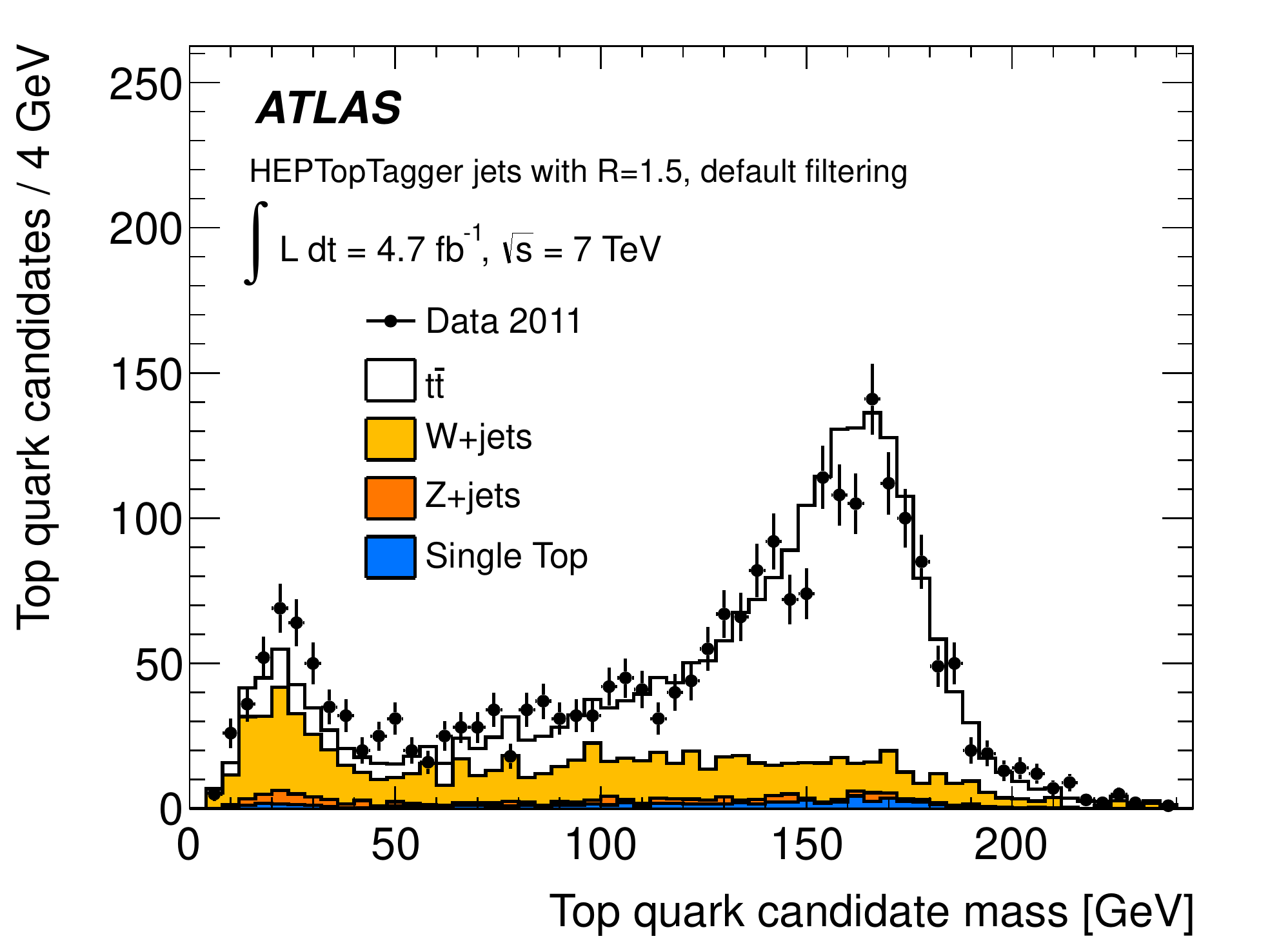} 
      \label{fig:HTT:kinematics:cand:m15def}
    }
    \subfigure[$R=1.8$, default filtering]{ 
      \includegraphics[width=0.46\columnwidth]{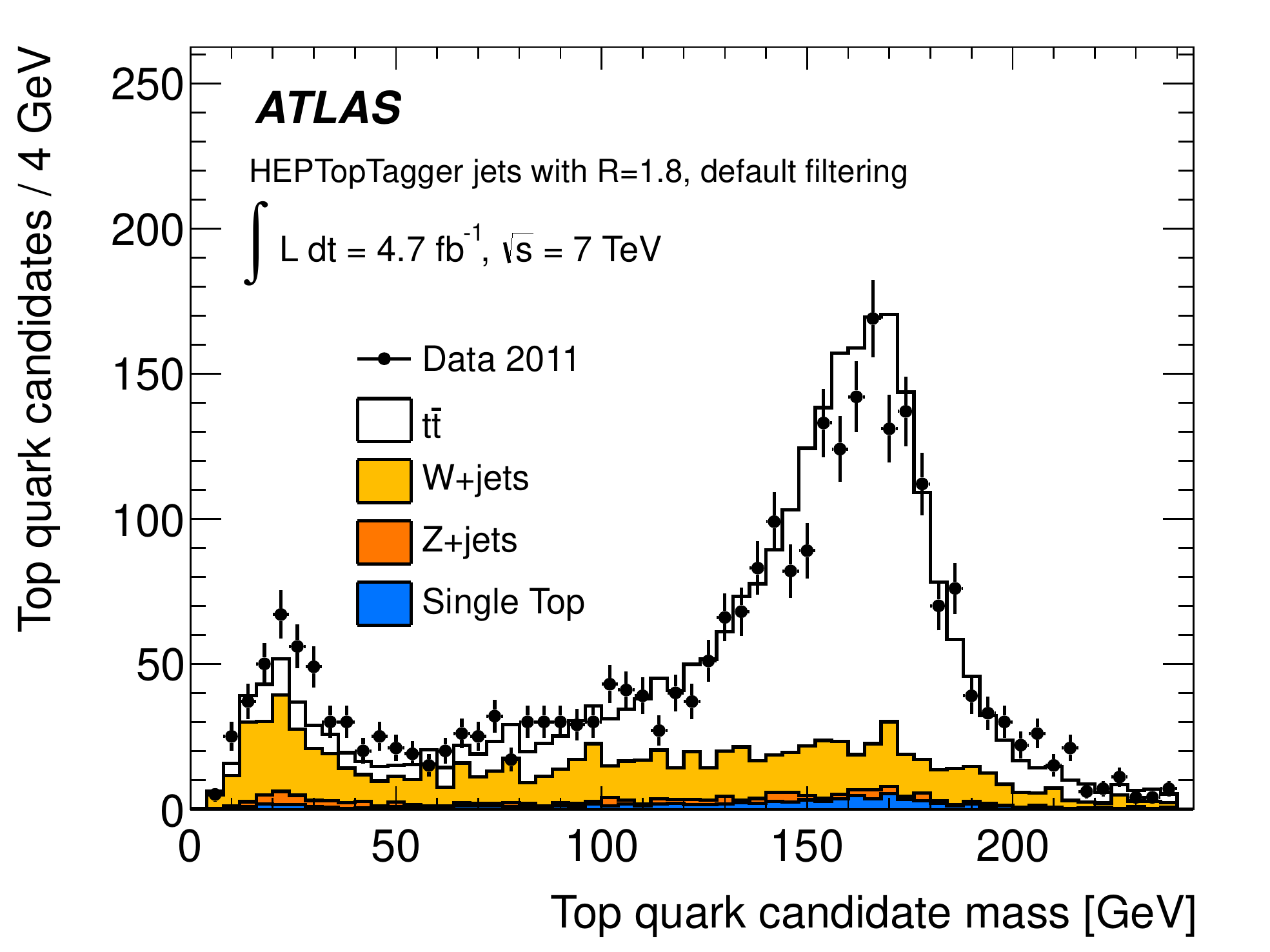}
      \label{fig:HTT:kinematics:cand:m18def}
    }\\
    \subfigure[$R=1.5$, tight filtering]{ 
      \includegraphics[width=0.46\columnwidth]{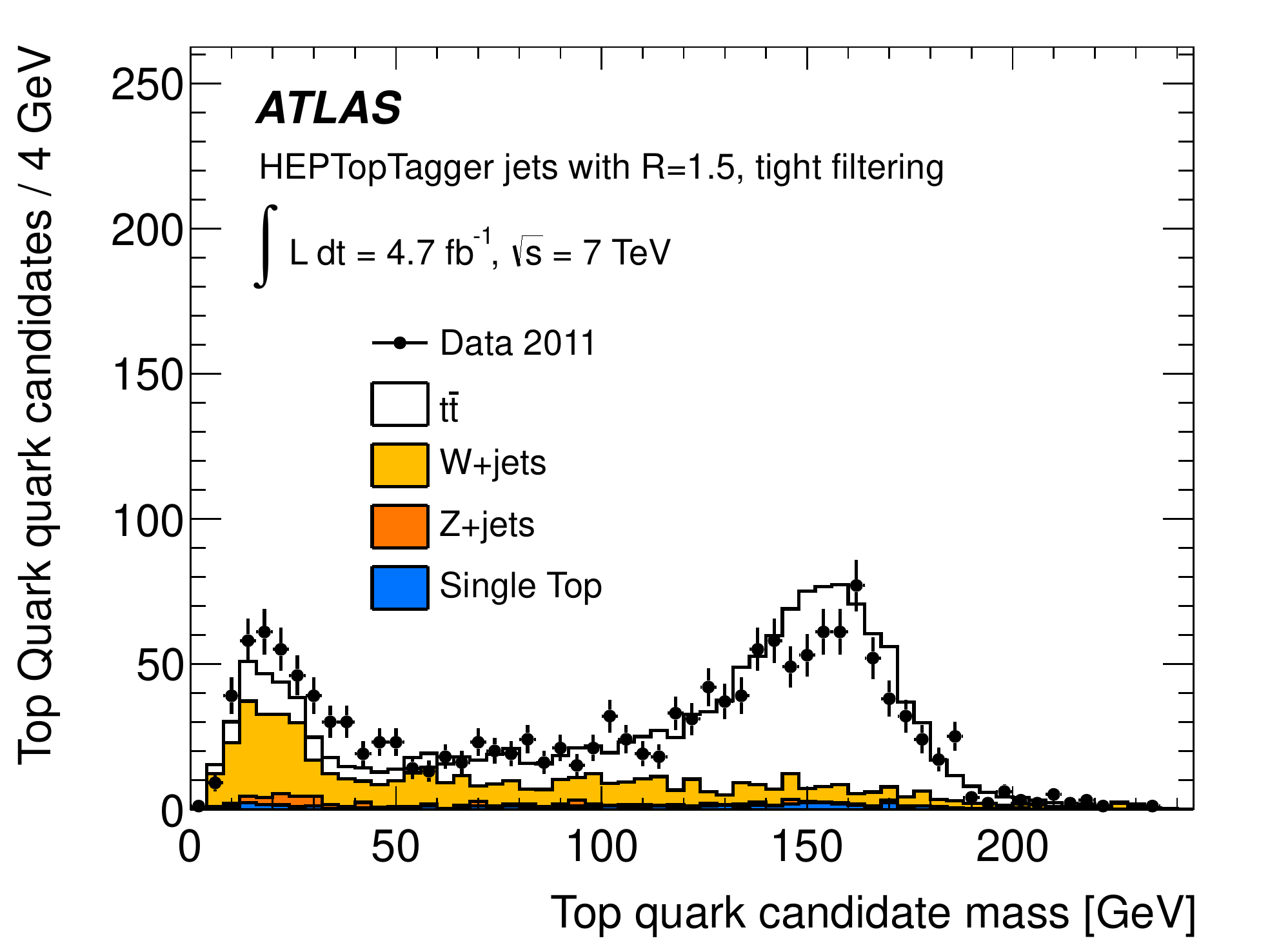} 
      \label{fig:HTT:kinematics:cand:m15t}
    }
    \subfigure[$R=1.5$, loose filtering]{ 
      \includegraphics[width=.46\textwidth]{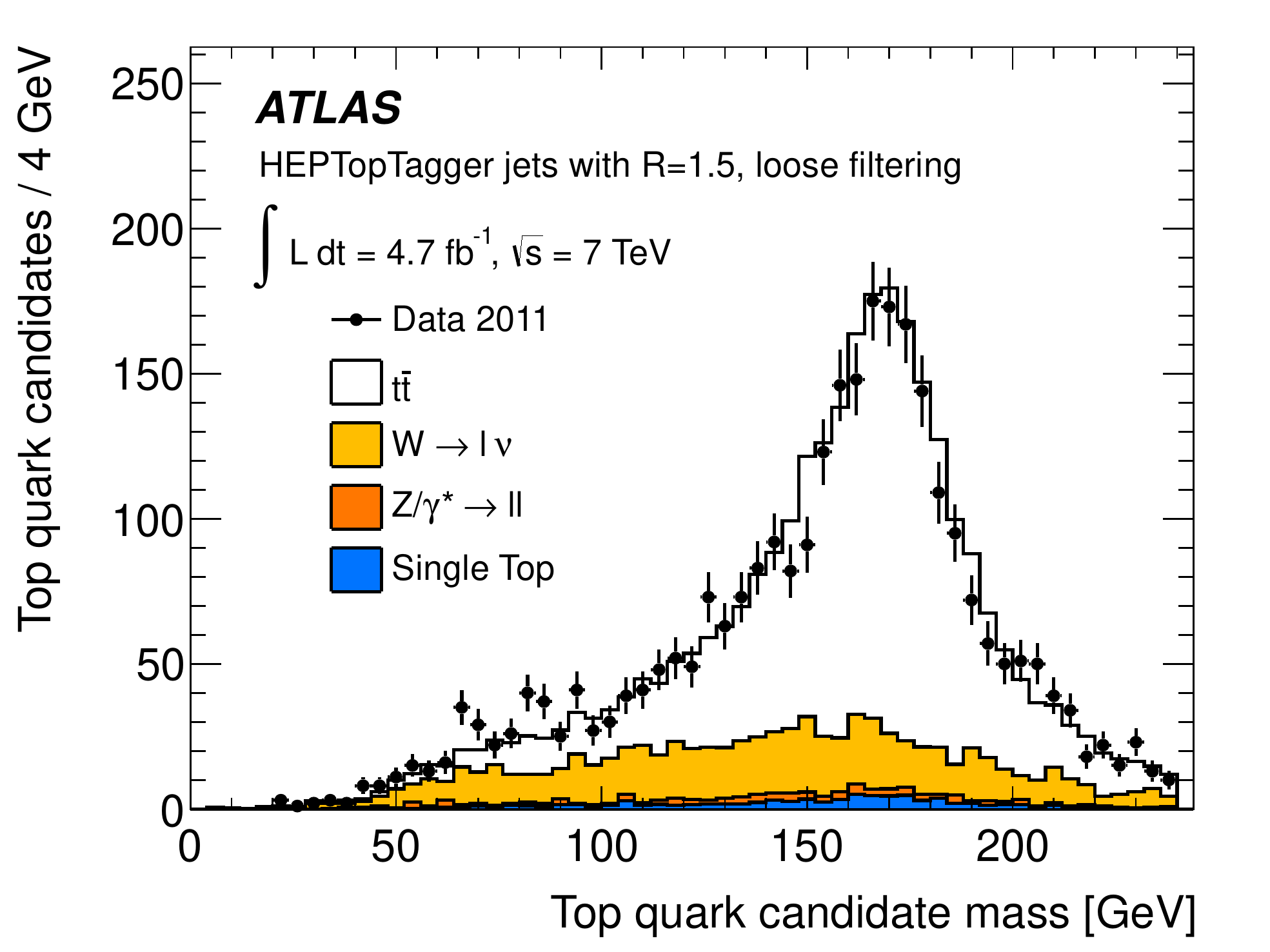} 
      \label{fig:HTT:kinematics:cand:m15l}
    }

  \end{center}
  \caption{Top candidate mass distribution after the \htt
    procedure (before applying a \topmass\ window). 
    The distance parameter for \largeR jet finding is $R=1.5$ in 
    \subref{fig:HTT:kinematics:cand:m15def}, \subref{fig:HTT:kinematics:cand:m15t}, 
    and \subref{fig:HTT:kinematics:cand:m15l}, and $R=1.8$ in
    \subref{fig:HTT:kinematics:cand:m18def}. 
    Distributions \subref{fig:HTT:kinematics:cand:m15def} and 
    \subref{fig:HTT:kinematics:cand:m18def} use the default filtering settings, 
    whereas \subref{fig:HTT:kinematics:cand:m15t} and \subref{fig:HTT:kinematics:cand:m15l}
    have been optimized for high purity and high signal efficiency, respectively.
  }
  \label{fig:HTT:kinematics:cand:m}
\end{figure}

\Figref{HTT:kinematics:cand:m} shows the top-candidate mass distribution after applying the \htt with four different filtering and \largeR jet configurations (see \tabref{htt631} for details). 
For all settings, the top-mass peak shape is generally well described by MC simulation and a relatively pure \ttbar\ selection is obtained for top-quark candidate masses above 120~GeV.

The good agreement between data and simulation both before and after applying the \htt shows that the exploited substructure is modelled well, even for jets with a very large radius.

\begin{figure}[!ht]
  \begin{center}
    \subfigure[$\arctan(m_{13}/m_{12})$]{ 
      \includegraphics[width=0.46\columnwidth]{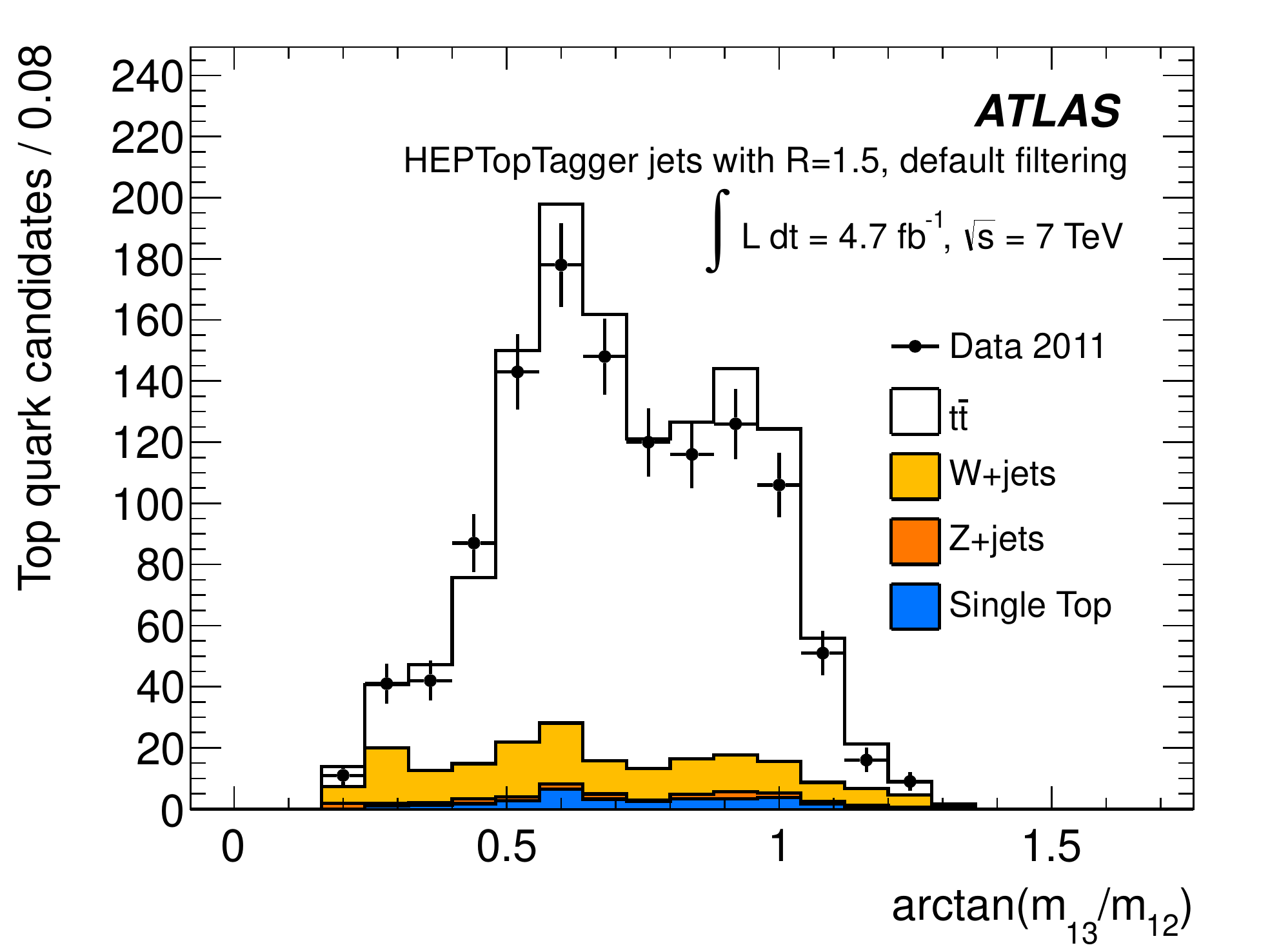} 
      \label{fig:HTT:substructureA}
    }
    \subfigure[$m_{23}/m_{123}$]{ 
      \includegraphics[width=0.46\columnwidth]{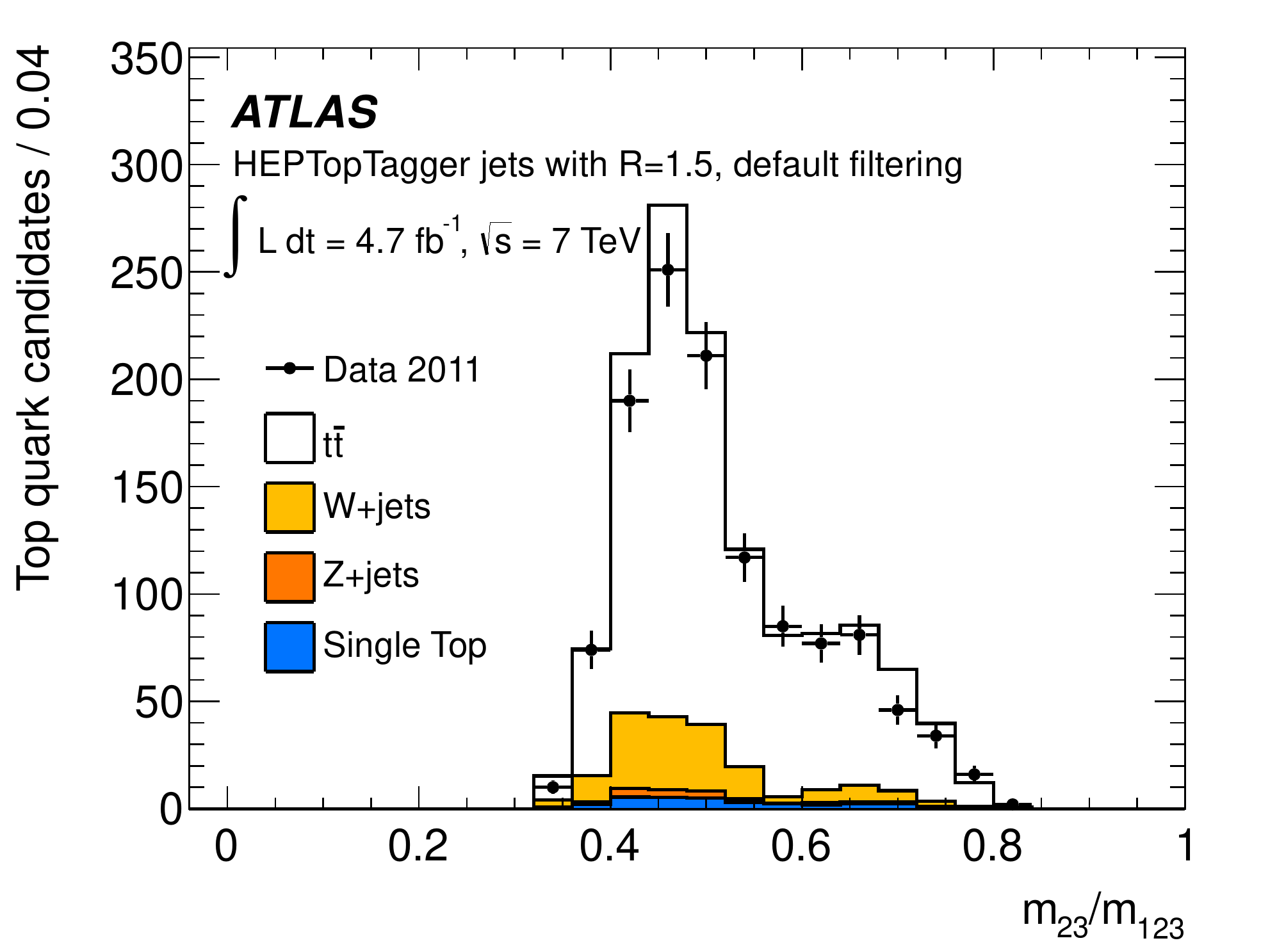} 
      \label{fig:HTT:substructureB}
    }\\
    \subfigure[\Wmass]{ 
      \includegraphics[width=.48\textwidth]{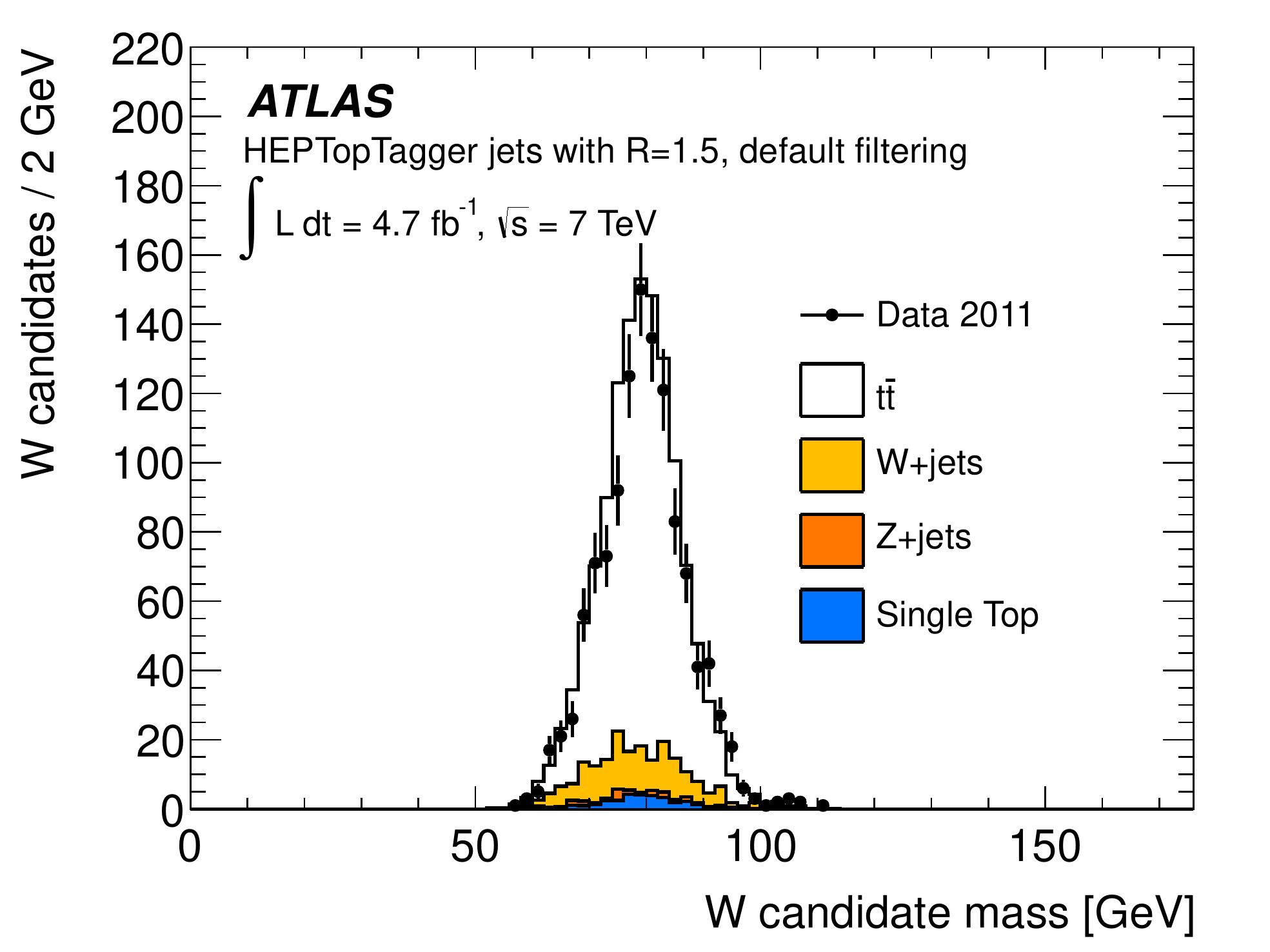} 
      \label{fig:HTT:substructureC}
    }
  \end{center}

  \caption{Substructure variables \subref{fig:HTT:substructureA} $\arctan(m_{13}/m_{12})$, 
     \subref{fig:HTT:substructureB} $m_{23}/m_{123}$, and 
     \subref{fig:HTT:substructureC} \Wmass for \htt-tagged top candidates 
     using the default filtering parameters and a jet size of $R=1.5$.   
  }
  \label{fig:HTT:substructure}
\end{figure}

\Figref{HTT:substructure} shows the variables used in the requirements imposed by the \htt on the subjet mass ratios, defined in \equref{htt_masscut} and \equref{htt_masssym}. 
For example in \figref{HTT:substructureB}, if the sub-leading \pT\ and the sub-sub-leading \pT\ subjets are the decay products of a $W$ boson then $m_{23}/m_{123}$ peaks at $\Wmass/\topmass\sim 0.46$. 
The \Wmass distribution in \figref{HTT:substructureC} is obtained by taking the subjet pair with the invariant mass closest to the true \Wmass, which also influences the shapes of the background distributions. 
All distributions are well modelled by the simulation.


The efficiency of the \htt is measured as a function of the transverse momentum of the generated top quark, and is the product of the \largeR jet finding efficiency and the efficiency to tag the jet correctly: $\varepsilon (\text{total}) = \varepsilon(\text{\largeR jet}) \cdot \varepsilon(\text{tag})$.
%
\begin{figure}[!ht]
  \begin{center}
      \includegraphics[width=0.46\columnwidth]{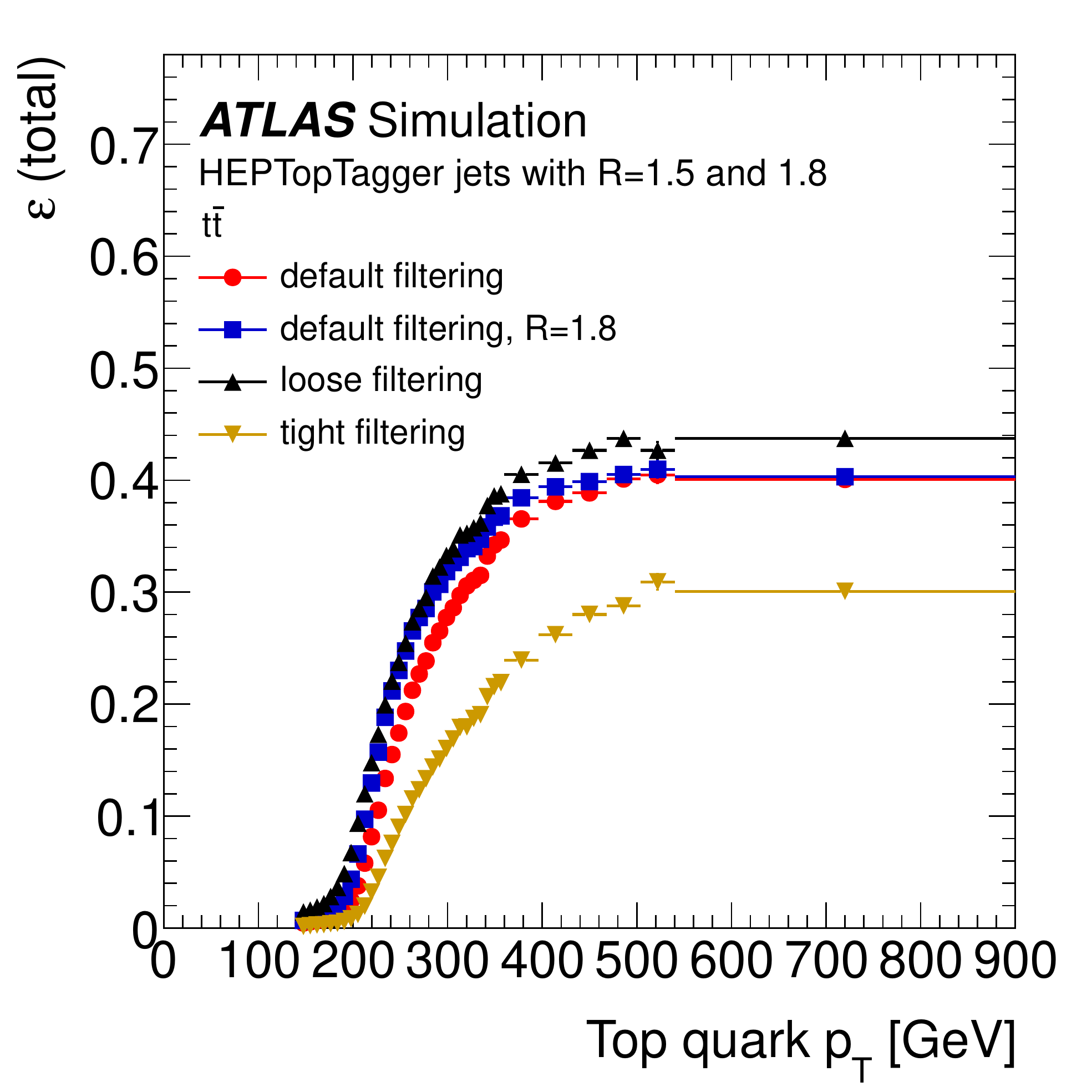}
  \end{center}
  \caption{Tagging efficiency per top quark as a function of the 
    generator-level top-quark \pT\ for various filtering settings 
    of the \htt, evaluated using the semileptonic \ttbar\ MC 
    sample.
    \label{fig:HTT:eff:signal}
  }
\end{figure}
%
\Figref{HTT:eff:signal} shows $\varepsilon (\text{total})$ for four different filtering configurations of the \htt as a function of the generator-level true top-quark \pT\ for the \ttbar\ MC sample.  
The efficiency for the default settings is 20\% at 250~GeV and reaches a plateau of 40\% at 500~GeV. 
Below 400~GeV the efficiency can be improved by 5\% by using a larger radius parameter of $R=1.8$. 
The maximum efficiency for the tight filtering settings is 30\%.

\begin{figure}[!ht]
  \begin{center}
    \subfigure[Efficiency as a function of \aktfour jet \pt with various taggers.]{ 
      \includegraphics[width=0.46\columnwidth]{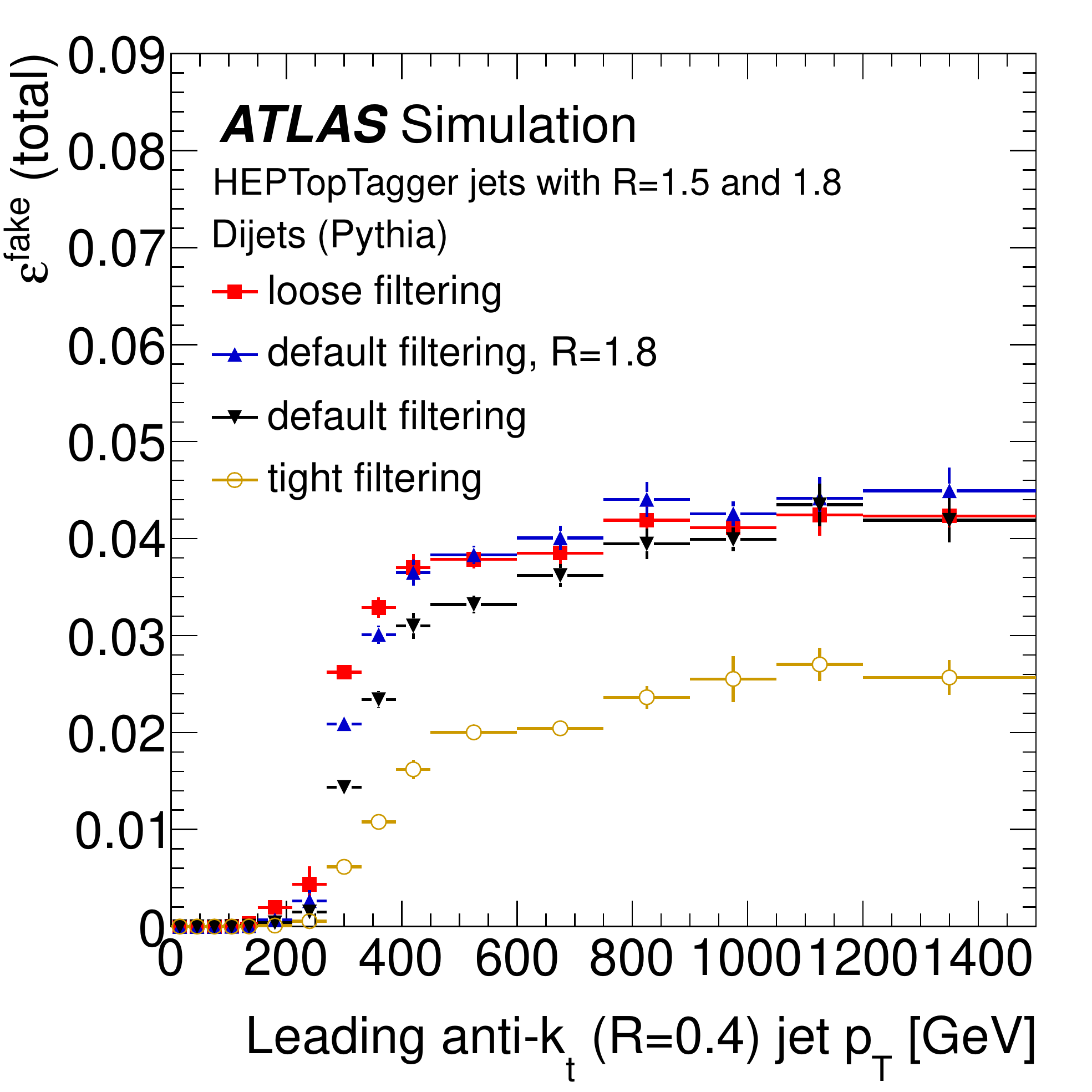}
      \label{fig:HTT:eff:bgA}
    }~~~~~
    \subfigure[Efficiency as a function \pt for top-jet candidates.]{ 
      \includegraphics[width=0.46\columnwidth]{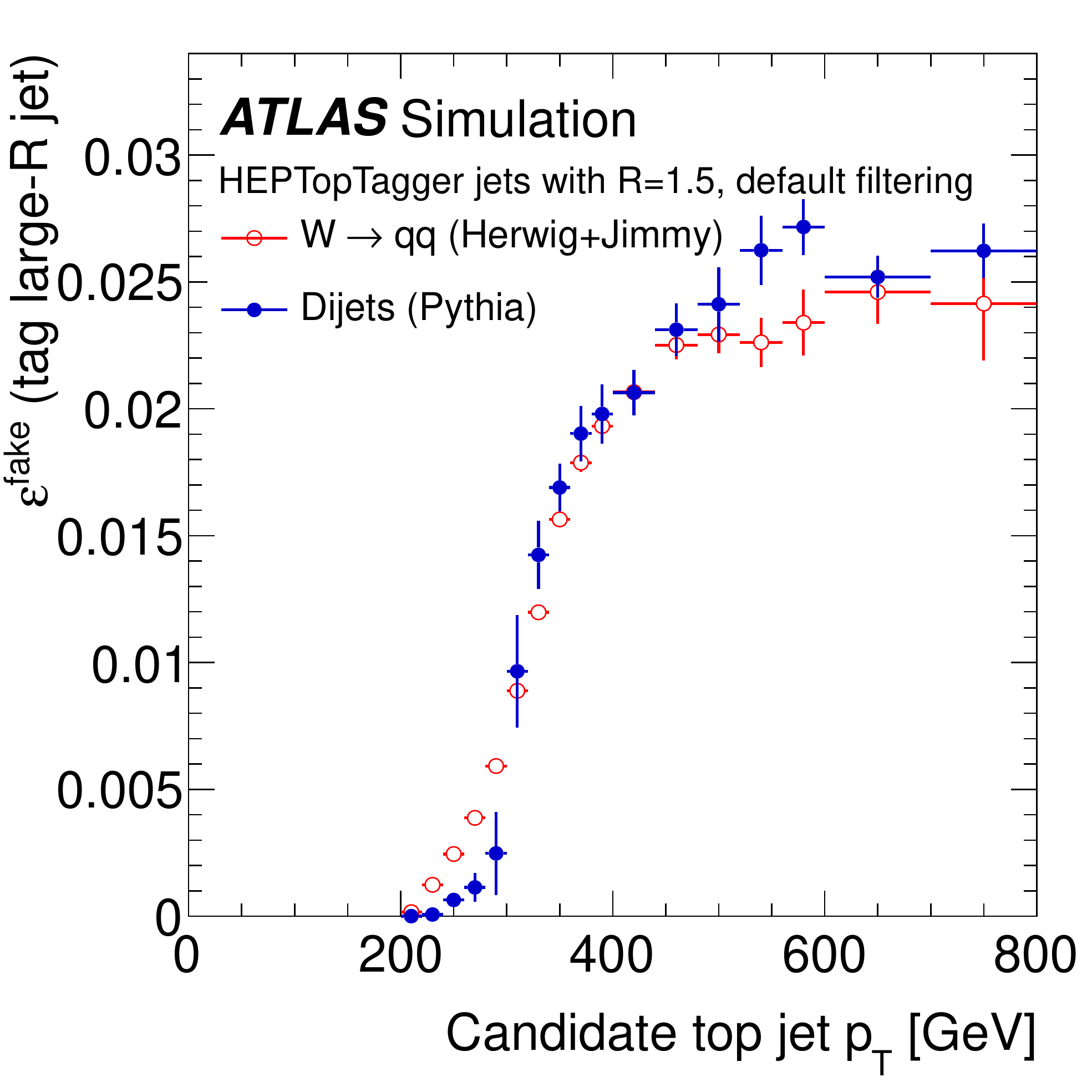}
      \label{fig:HTT:eff:bgB}
    }
  \end{center}

  \caption{\subref{fig:HTT:eff:bgA} Per-event fake efficiency as a function of the 
           leading-\ptjet\ \akt $R=0.4$ jet \pT\ in the event for different filtering 
           settings, measured using the dijet MC sample and
           \subref{fig:HTT:eff:bgB} a comparison of the per-jet fake efficiency
           between the dijet sample (\Pythia) and the \Wqq\ sample (\HJimmy) taken from 
           \ttbar\ events, both using the default filtering.   
    \label{fig:HTT:eff:bg}
  }
\end{figure}

The fake efficiency, shown in \figref{HTT:eff:bgA}, is defined in exactly the same way but is evaluated using the \Pythia inclusive jet sample. 
The \pT\ of the leading-\ptjet\ \akt jet with $R=0.4$ has been chosen to compute the efficiency as it provides a measure for the energy available in the event and is easily comparable between different tagging approaches.
The fake efficiency shows a sharp turn-on around 200~GeV with efficiencies below 0.5\% below and a plateau of 4\% (2.5\%) for the default and loose (tight) filtering settings.

\Figref{HTT:eff:bgB} shows the fake tagging efficiency as a function of the top-jet \pT\ in a multi-jet background sample and for events with a hadronically decaying \W boson. The fake efficiency rises sharply at 300~GeV and reaches a plateau of 2.5\% at a \largeR jet \pT\ of 400~GeV.

The efficiency of the \htt to select jets from boosted top quarks can be increased by varying the filtering parameters. Since this also increases the fake efficiency, the optimal working-point  depends on the analysis in question.

\section{Conclusions}
\label{sec:Conclusions}
It has been demonstrated experimentally in this paper that jet grooming algorithms can improve the identification of Lorentz-boosted physics objects that decay to jets, as well as increase sensitivity to several new physics processes. The performance of \largeR jets is improved overall, and the dependence on \pileup and the underlying event is reduced.

Jet mass calibrations have been derived in simulation for various \largeR jet algorithms, subjets, as well as for jets with grooming applied. These have been validated for boosted \W bosons in \ttbar\ events and using calorimeter-jet versus track-jet double ratios. Uncertainties on the jet energy scale and the jet mass scale have been provided over a wide range of \largeR jet momentum.

The mass distributions observed in data for \largeR jets in the inclusive jet sample, before and after grooming, are well reproduced by the ATLAS simulation, especially using the \Powheg \NLO generator. The substructure variables presented here also show good agreement between data and simulation, typically within 5\% for key observables for modern \NLO plus parton shower Monte Carlo programs such as \PowPythia, as well as for the \LO MC program \Herwigpp.

The parameters of trimmed, pruned and mass-drop filtered jet algorithms have been optimized for searches and precision Standard Model measurements using various performance measures. Among the configurations tested here, the trimming algorithm exhibits better performance than the pruning algorithm, with superior mass resolution and reduced dependence on \pileup. In particular, the \akt algorithm with $R=1.0$ and trimming parameters $\fcut=0.05$ and $\drsub=0.3$ is recommended for boosted top physics analyses, where a minimum \pT\ requirement of 200 GeV is typical. It is important to note, though, that only the \kt-pruning for \largeR ($R=1.0$) jets has been tested in this work, and that future studies should expand the comparisons to include the \CamKt-pruning as well. Additionally, \CamKt jets with $R=1.2$ using the mass-drop filtering parameter $\mufrac=0.67$ are recommended for boosted two-pronged analyses such as \Hbb\ or searches using \Wqq.

The benefit of using these grooming algorithms along with substructure variables has been demonstrated in top-tagging studies, where the efficiency of finding a boosted top quark for a given background jet mis-tag rate is greatly increased after grooming is applied due to the improved mass resolution. Grooming has been shown to leave the boosted signal mass peak relatively unaffected while systematically shifting the light-quark and gluon jet background lower in mass, thus increasing the discrimination of signal from background.

The \htt has been demonstrated to be a robust and versatile tool to reconstruct hadronically-decaying top quarks in the presence of the underlying event and \pileup, using jet grooming and substructure techniques. A comparison to data shows that the algorithm is well modelled by the simulations. The \htt performance (efficiency, rejection, mass resolution) can be optimized for a given analysis by varying the algorithm parameters.

From the studies presented here, groomed jets and substructure variables are ready to be used in further ATLAS physics analyses.  These techniques will become extremely beneficial tools in upcoming searches for boosted physics objects in supersymmetric and exotic models, measurements of boosted Higgs topologies, and in detailed precision Standard Model measurements of QCD and electroweak processes with jets and boosted hadronic objects.

\clearpage
\section*{Acknowledgments}
\label{sec:Ack}
We thank CERN for the very successful operation of the LHC, as well as the
support staff from our institutions without whom ATLAS could not be
operated efficiently.

We acknowledge the support of ANPCyT, Argentina; YerPhI, Armenia; ARC,
Australia; BMWF and FWF, Austria; ANAS, Azerbaijan; SSTC, Belarus; CNPq and FAPESP,
Brazil; NSERC, NRC and CFI, Canada; CERN; CONICYT, Chile; CAS, MOST and NSFC,
China; COLCIENCIAS, Colombia; MSMT CR, MPO CR and VSC CR, Czech Republic;
DNRF, DNSRC and Lundbeck Foundation, Denmark; EPLANET, ERC and NSRF, European Union;
IN2P3-CNRS, CEA-DSM/IRFU, France; GNSF, Georgia; BMBF, DFG, HGF, MPG and AvH
Foundation, Germany; GSRT and NSRF, Greece; ISF, MINERVA, GIF, DIP and Benoziyo Center,
Israel; INFN, Italy; MEXT and JSPS, Japan; CNRST, Morocco; FOM and NWO,
Netherlands; BRF and RCN, Norway; MNiSW, Poland; GRICES and FCT, Portugal; MERYS
(MECTS), Romania; MES of Russia and ROSATOM, Russian Federation; JINR; MSTD,
Serbia; MSSR, Slovakia; ARRS and MIZ\v{S}, Slovenia; DST/NRF, South Africa;
MICINN, Spain; SRC and Wallenberg Foundation, Sweden; SER, SNSF and Cantons of
Bern and Geneva, Switzerland; NSC, Taiwan; TAEK, Turkey; STFC, the Royal
Society and Leverhulme Trust, United Kingdom; DOE and NSF, United States of
America.

The crucial computing support from all WLCG partners is acknowledged
gratefully, in particular from CERN and the ATLAS Tier-1 facilities at
TRIUMF (Canada), NDGF (Denmark, Norway, Sweden), CC-IN2P3 (France),
KIT/GridKA (Germany), INFN-CNAF (Italy), NL-T1 (Netherlands), PIC (Spain),
ASGC (Taiwan), RAL (UK) and BNL (USA) and in the Tier-2 facilities
worldwide.


\clearpage


\providecommand{\href}[2]{#2}\begingroup\raggedright\endgroup

\clearpage
\onecolumn
\begin{flushleft}
{\Large The ATLAS Collaboration}

\bigskip

G.~Aad$^{\rm 48}$,
T.~Abajyan$^{\rm 21}$,
B.~Abbott$^{\rm 112}$,
J.~Abdallah$^{\rm 12}$,
S.~Abdel~Khalek$^{\rm 116}$,
A.A.~Abdelalim$^{\rm 49}$,
O.~Abdinov$^{\rm 11}$,
R.~Aben$^{\rm 106}$,
B.~Abi$^{\rm 113}$,
M.~Abolins$^{\rm 89}$
% ...
\bigskip
\\
% ...
$^{11}$ Institute of Physics, Azerbaijan Academy of Sciences, Baku, Azerbaijan\\
$^{12}$ Institut de F{\'\i}sica d'Altes Energies and Departament de F{\'\i}sica de la Universitat Aut{\`o}noma de Barcelona and ICREA, Barcelona, Spain\\
% ...
$^{21}$ Physikalisches Institut, University of Bonn, Bonn, Germany\\
% ...
$^{48}$ Fakult{\"a}t f{\"u}r Mathematik und Physik, Albert-Ludwigs-Universit{\"a}t, Freiburg, Germany\\
$^{49}$ Section de Physique, Universit{\'e} de Gen{\`e}ve, Geneva, Switzerland\\
% ...
$^{89}$ Department of Physics and Astronomy, Michigan State University, East Lansing MI, United States of America\\
% ...
$^{106}$ Nikhef National Institute for Subatomic Physics and University of Amsterdam, Amsterdam, Netherlands\\
% ...
$^{112}$ Homer L. Dodge Department of Physics and Astronomy, University of Oklahoma, Norman OK, United States of America\\
$^{113}$ Department of Physics, Oklahoma State University, Stillwater OK, United States of America\\
% ...
$^{116}$ LAL, Universit{\'e} Paris-Sud and CNRS/IN2P3, Orsay, France\\
% ...
\end{flushleft}


\end{document}